\documentclass[final]{pd-tdr}
\pdfoutput=1            
\graphicspath{ {figures/}{/generated/} }

\def\expshort{DUNE\xspace}
\def\explong{The Deep Underground Neutrino Experiment\xspace}

\def\thedocsubtitle{The Single-Phase ProtoDUNE}%
\def\pdtdrtitle{Technical Design Report}

\newcommand{\pdsp}{ProtoDUNE-SP\xspace}

\def\Ar39{$^{39}$Ar}

\DeclareSIUnit \kton {\kilo\tonne}
\DeclareSIUnit \kt {\kilo\tonne}
\DeclareSIUnit \Mt {\mega\tonne}
\DeclareSIUnit \eV {\electronvolt}
\DeclareSIUnit \keV {\kilo\electronvolt}
\DeclareSIUnit \MeV {\mega\electronvolt}
\DeclareSIUnit \GeV {\giga\electronvolt}
\DeclareSIUnit \km {\kilo\meter}
\DeclareSIUnit \kW {\kilo\watt}
\DeclareSIUnit \MW {\mega\watt}
\DeclareSIUnit \MHz {\mega\hertz}
\DeclareSIUnit \mrad {\milli\radian}
\DeclareSIUnit \year {year}
\DeclareSIUnit \POT {POT}
\DeclareSIUnit \sig {$\sigma$}
\DeclareSIUnit\parsec{pc}
\DeclareSIUnit\lightyear{ly}
\DeclareSIUnit\foot{ft}
\DeclareSIUnit\ft{ft}

\newcommand{\SIadj}[2]{\SI[number-unit-product = -]{#1}{#2}}
\newcommand{\ktadj}[1]{\SIadj{#1}{\kt}}

\hypersetup{
    pdftitle={\expshort CDR \thedocsubtitle},
    pdfauthor={\expshort Collaboration},
    final=true,
    colorlinks=false,
    linktocpage=true,
    linkbordercolor=blue,
    citebordercolor=green,
    urlbordercolor=magenta,
    filecolor=black,
    pdfpagemode=UseOutlines,
    pdfborderstyle={/S/U},  
}

\def\titleextra{\includegraphics[width=0.7\textwidth]{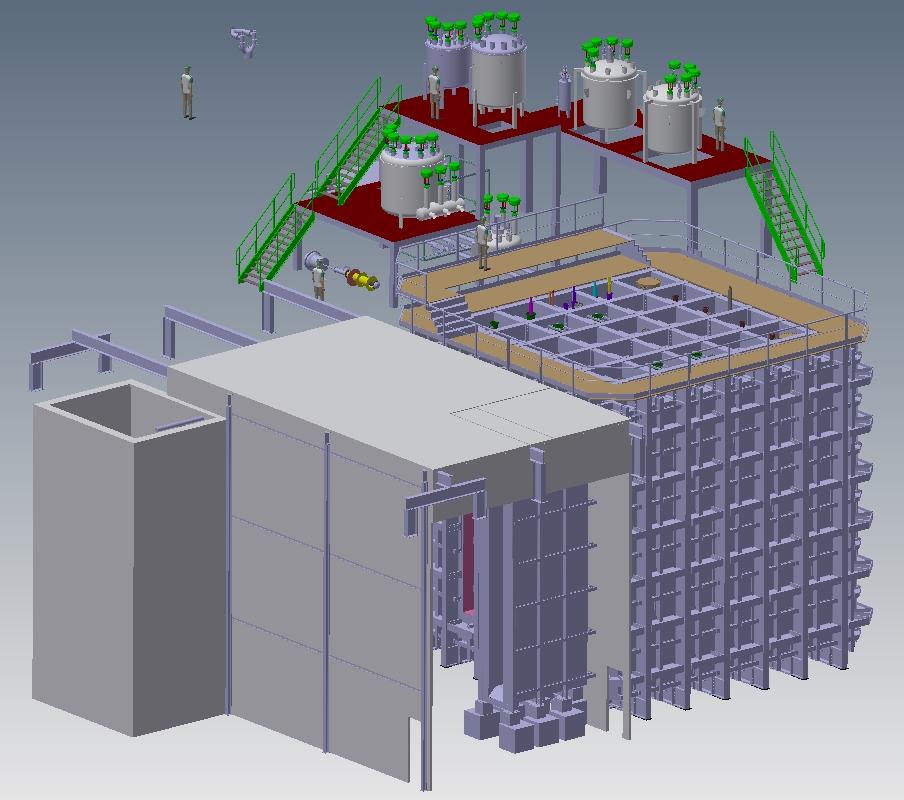}}

\begin{document}


\pagestyle{titlepage}

\begin{center}
   {\Huge  \thedocsubtitle}  

  \vspace{5mm}

  {\Huge  \pdtdrtitle}  

  \vspace{5mm}
  


  \vspace{15mm}

\titleextra

  \vspace{10mm}
  \today
    \vspace{15mm}
    
\end{center}

\cleardoublepage



# The Single-Phase ProtoDUNE Technical Design Report

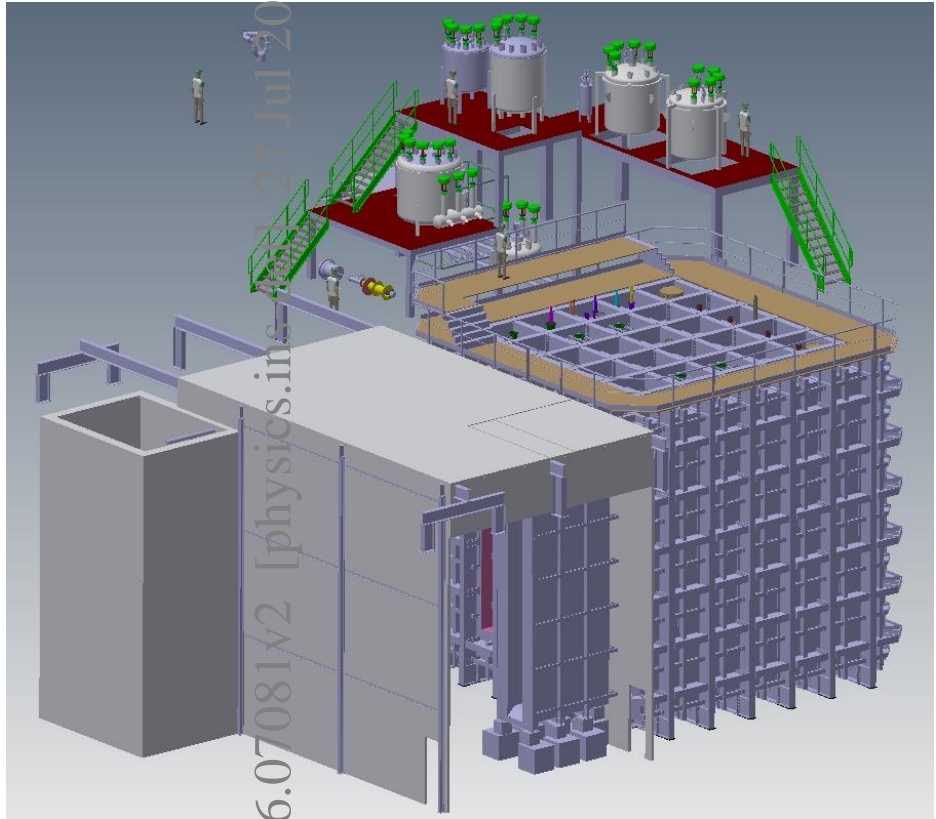

July 31, 2017

DUNE — DEEP UNDERGROUND NEUTRINO EXPERIMENT

# The DUNE Collaboration

B. Abi[105], R. Acciarri[45], M. A. Acero[7], M. Adamowski[45], C. Adams[149], D. L. Adams[12], P. Adamson[45], M. Adinolfi[11], Z. Ahmad[139], C. H. Albright[45], T. Alion[131], J. Anderson[5], K. Anderson[45], C. Andreopoulos[81,121], M. P. Andrews[45], R. A. Andrews[45], J. dos Anjos[14], A. Ankowski[141], J. Anthony[23], M. Antonello[47], A. Aranda Fernandez[30], A. Ariga[9], T. Ariga[9], E. Arrieta Diaz[129], J. Asaadi[135], M. Ascencio[115], D. M. Asner[106], M. S. Athar[2], M. Auger[9], A. Aurisano[28], V. Aushev[77], D. Autiero[66], F. Azfar[105], J. J. Back[144], H. O. Back[106], C. Backhouse[22], X. Bai[126], M. Baird[141], A. B. Balantekin[147], S. Balasubramanian[149], B. Baller[45], P. Ballett[39], B. Bambah[54], H. Band[149], M. Bansal[109], S. Bansal[109], G. J. Barker[144], G. Barr[105], J. Barranco Monarca[49], N. Barros[112], A. Bashyal[104], M. Bass[105], F. Bay[137], J. L. Bazo[115], J. F. Beacom[103], B. R. Behera[109], G. Bellettini[113], V. Bellini[24], O. Beltramello[16], N. Benekos[16], P. A. Benetti[110], A. Bercellie[119], E. Berman[45], H. Berns[19], R. Bernstein[45], S. Bertolucci[16], V. Bhatnagar[109], B. Bhuyan[59], J. Bian[20], K. Biery[45], M. Bishai[12], A. Bitadze[87], T. Blackburn[131], A. Blake[79], F. d. M. Blaszczyk[10], E. Blaufuss[88], G. C. Blazey[95], M. Blennow[73], E. Blucher[26], V. Bocean[45], F. Boffelli[110], J. Boissevain[83], S. Bolognesi[15], T. Bolton[74], M. Bonesini[61], T. Boone[32], C. Booth[122], S. Bordoni[16,16], P. Bour[35], B. Bourguille[55], S. B. Boyd[144], D. Boyden[95], D. Brailsford[79], A. Brandt[135], J. Bremer[16], S. J. Brice[45], C. Bromberg[91], G. Brooijmans[33], G. Brown[135], G. Brunetti[45,108], N. Buchanan[32], H. Budd[119], P. Calafiura[80], A. Calatayud[115], E. Calligarich[110], E. Calvo[17], L. Camilleri[33], M. Campanelli[82], C. Cantini[40], B. Carls[45], M. Cascella[82], E. Catano-Mur[68], M. Cavaili-Sforza[55], F. Cavanna[45,149], S. Centro[108], A. Cervera Villanueva[56], T. Cervi[110], M. Chalifour[16], A. Chappuis[78], A. Chatterjee[135], S. Chattopadhyay[45,95], S. Chattopadhyay[139], L. Chaussard[66], S. Chembra[54], H. Chen[12], M. Chen[20], D. Cherdack[32], C. Chi[33], S. Childress[45], S. Choubey[50], B. C. Choudhary[36], G. Christodoulou[81], C. Christofferson[126], E. Church[106], T. E. Coan[129], A. Cocco[96], P. Cole[63], G. Collin[89], J. M. Conrad[89], M. Convery[120], R. Corey[126], L. Corwin[126], J. I. Crespo-Anadón[33], J. Creus Prats[16], P. Crivelli[40], D. Cronin-Hennessy[94], C. Crowley[45], A. Curioni[61], A. Cussans[11], H. da Motta[14], D. Dale[63], T. Davenne[121], G. S. Davies[65], J. Davies[131], J. Dawson[1], K. De[135], I. De Bonis[78], A. De Gouvea[101], P. C. de Holanda[41], P. De Jong[100], P. De Lurgio[5], J. J. de Vries[23], M. P. Decowski[100], P. Dedin-Neto[41], A. Delbart[15], D. Delepine[49], M. Delgado[4], D. Demuth[140], S. Dennis[81], C. Densham[121], R. Dharmapalan[5], N. Dhingra[109], M. Diamantopoulou[6], J. S. Diaz[65], G. Diaz Bautista[119], P. Ding[45], M. Diwan[12], Z. Djurcic[5], M. J. Dolinski[37], G. Drake[5], D. Duchesneau[78], D. Dutta[50], M. Duvernois[147], H. Duyang[124], D. A. Dwyer[80], S. Dye[51], A. S. Dyshkant[95], S. Dytman[114], M. Eads[95], B. Eberly[120], D. Edmunds[91], S. Elliott[83], W. M. Ellsworth[52], M. Elnimr[20], S. Emery[15], S. Eno[88], A. Ereditato[9], C. O. Escobar[45], L. Escudero Sanchez[23], J. J. Evans[87], K. Fahey[45], A. Falcone[135], L. Falk[131], A. Farbin[135], C. Farnese[108], Y. Farzan[62], M. Fasoli[61], A. Fava[45], J. Felix[49], P. Fernandez Menendez[56], E. Fernandez-Martinez[86], L. Fields[45], F. Filthaut[100], A. Finch[79], M. Fitton[121], B.T. Fleming[149], T. Forest[63], J. Fowler[38], W. Fox[65], J. Freeman[45], J. Freestone[87], D. Friant[63], J. Fried[12], A. Friedland[120], S. Fuess[45], B. Fujikawa[80], A. Gago[115], H. Gallagher[138], S. Galvin[112], S. Galymov[66], T. Gamble[122], R. Gandhi[50], S. Gao[12], D. Garcia-Gamez[87], S. Gardiner[19], D. Gastler[10], A. Gendotti[40], D. Gibin[108], I. Gil-Botella[17], R. Gill[12], A. K. Giri[60], D. Goeldi[9], M. Gold[98], S. Gollapinni[133], R. A. Gomes[46], J. J. Gomez Cadenas[56], M. C. Goodman[5], D. Gorbunov[67], S. Goswami[107], N. Graf[114], N. Graf[120], M. Graham[120], E. Gramelini[149], R. Gran[93], C. Grant[10], N. Grant[144], V. Greco[24], H. Greenlee[45], L. Greenler[147], M. Groh[65,28], K. Grzelak[142], E. Guardincerri[83], V. Guarino[5], G. P. Guedes[44], R. Guenette[105], A. Guglielmi[108], K. K. Guthikonda[76], M. M. Guzzo[41], A. T. Habig[93], A. Hackenburg[149], R. W. Hackenburg[12], H. Hadavand[135], R. Haenni[9], A. Hahn[45], M. D. Haigh[144], T. Haines[83], T. Hamernik[45], P. Hamilton[132], T. Handler[133], S. Hans[12], D. Harris[45], J. Hartnell[131], T. Hasegawa[72], R. Hatcher[45], A. Hatzikoutelis[133], S. Hays[45], E. Hazen[10], M. Headley[127], A. Heavey[45], K. M. Heeger[149], J. Heise[127], K. Hennessy[81], S. Henry[119], J. Hernandez-Garcia[86], J. Hewes[87], A. Higuera[52], T. Hill[63], A. Himmel[45], A. Holin[82], E.W. Hoppe[106], S. Horikawa[40], G. Horton-Smith[74], M. Hostert[39], A. Hourlier[89], B. Howard[65], J. Howell[45], J. Hugon[84], P. Hurh[45], J. Huston[91], J. Hylen[45], J. Insler[37], G. Introzzi[110], A. Ioannisian[150], A. Izmaylov[56], D. E. Jaffe[12], C. James[45], E. James[45], C. H. Jang[27], F. Jediny[35], C. Jen[141], A. Jhingan[109], S. Jiménez[17], M. Johnson[45], R. Johnson[28], J. Johnstone[45], J. Joshi[12], H. Jostlein[45], C. K. Jung[130], T. Junk[45], A. Kaboth[121], I. Kadenko[77], Y. Kamyshkov[133], G. Karagiorgi[33], D. Karasavvas[6], Y. Karyotakis[78], P. Kaur[109], B. Kayser[45], N. Kazaryan[150], E. Kearns[10], P.T.Keener[112], E. Kemp[41], C. Kendziora[45], W. Ketchum[45], S. H. Kettell[12], M. Khabibullin[67], A. Khotjantsev[67], B. Kirby[12], M. Kirby[45], J. Klein[112], Y. J. Ko[27], T. Kobilarcik[45], B. Kocaman[137], L.W. Koerner[52], S. Kohn[18,80], G. Koizumi[45], A. Kopylov[67], M. Kordosky[146], L. Kormos[79], U. Kose[16,9], V. A. Kostelecký[65], M. Kramer[18,80], I. Kreslo[9], K. Kriesel[147], W. Kropp[20], Y. Kudenko[67], V. A. Kudryavtsev[122], S. Kulagin[67], A. Kumar[109], J. Kumar[51], L. Kumar[109], V. Kus[35], T. Kutter[84], K. Lande[112], C. Lane[37], K. Lang[136], T. Langford[149], F. Lanni[12], A. Laundrie[147], T. Le[138], J. Learned[51], P. Lebrun[45], D. Lee[83], W. M. Lee[45], M. A. Leigui de Oliveira[43], Q. Z. Li[45], S. Li[12], S. W. Li[103], X. Li[130], Y. Li[12], Z. Li[38], C. S. Lin[80], S. Lin[32], R. Linehan[120], J. Link[141], Z. Liptak[31], D. Lissauer[12],

1

L. Littenberg[12], B. R. Littlejohn[57], J. Liu[125], Q. Liu[83], T. Liu[129], S. Lockwitz[45], N. Lockyer[45], T. Loew[80], M. Lokajicek[13], L. LoMonaco[24], K. Long[64], K. Loo[71], J. P. Lopez[31], D. Lorca[9], J. M. LoSecco[102], W. Louis[83], M. Luethi[9], K. B. Luk[18,80], T. Lundin[45], X. Luo[149], T. Lux[55], J. Lykken[45], J. Maalampi[71], A. A. Machado[43], C. T. Macias[65], J. R. Macier[45], R. MacLellan[125], S. Magill[5], G. Mahler[12], K. Mahn[91], M. Malek[122], F. Mammoliti[24], S. K. Mandal[36], S. Mandodi[54], S. Manly[119], A. Mann[138], A. Marchionni[45], W. Marciano[12], D. Marfatia[51], C. Mariani[141], J. Maricic[51], A. D. Marino[31], M. Marshak[94], C. M. Marshall[80], J. Marshall[23], J. Marteau[66], J. Martin-Albo[105], D. A. Martinez Caicedo[57], M. Masud[56], S. Matsuno[51], J. Matthews[84], C. Mauger[112], K. Mavrokoridis[81], R. Mazza[61], A. Mazzacane[45], E. Mazzucato[15], N. McCauley[81], E. McCluskey[45], N. McConkey[122], K.T. McDonald[116], K. S. McFarland[119], C. McGivern[45], A. M. McGowan[119], C. McGrew[130], S. R. McGuinness[102], R. McKeown[146], D. McNulty[63], R. McTaggart[128], A. Mefodiev[67], P. Mehta[70], D. Mei[125], O. Mena[56], S. Menary[151], D. P. Mendez[131], H. Mendez[117], A. Menegolli[110], G. Meng[108], M. Messier[65], W. Metcalf[84], M. Mewes[65], H. Meyer[145], T. Miao[45], R. Milincic[51], W. Miller[94], G. Mills[83], O. Mineev[67], O. G. Miranda[29], C. S. Mishra[45], S. R. Mishra[124], B. Mitrica[53], D. Mladenov[16], I. Mocioiu[111], K. Moffat[39], R. Mohanta[54], N. Mokhov[45], L. Molina Bueno[40], C. Montanari[110,16], D. Montanari[45,16], J. Moon[89], M. Mooney[12], C. D. Moore[45], B. Morgan[144], C. Morris[52], W. Morse[12], C. Mossey[45], C. A. Moura[43], J. Mousseau[90], L. Mualem[22], M. Muether[145], S. Mufson[65], S. Murphy[40], J. Musser[65], Y. Nakajima[80], D. Naples[114], J. Navarro[7], D.Navas-Nicols[17], J. Nelson[146], M. Nessi[16], D. Newbold[11], M. Newcomer[112], K. T. T. Nguyen[135], R. Nichol[82], T. C. Nicholls[121], E. Niner[45], A. Norman[45], B. Norris[45], P. Novakova[12], P. Novella[56], E. Nowak[16], J. Nowak[79], M. S. Nunes[41], H. O'Keeffe[79], A. Olivares Del Campo[39], R. Oliveira[41], A. Olivier[119], Y. Onishchuk[77], T. Ovsjannikova[67], S. Pakvasa[51], O. Palamara[45,149], J. Paley[45], C. Palomares[17], E. Pantic[19], V. Paolone[114], V. Papadimitriou[45], S. Paramesvaran[11], J. Park[141], S. Parke[45], Z. Parsa[12], S. Pascoli[39], J. Pasternak[64], J. Pater[87], R. B. Patterson[22], S. J. Patton[80], T. Patzak[1], B. Paulos[147], L. Paulucci[43], Z. Pavlovic[45], G. Pawloski[94], D. Payne[81], S. J. M. Peeters[131], E. Pennacchio[66], G. N. Perdue[45], O. L. G. Peres[41], L. Periale[40], J. D. Perkin[122], K. Petridis[11], R. Petti[124], A. Petukhov[126], P. Picchi[110], F. Pietropaolo[108,16], P. Plonski[143], R. Plunkett[45], R. Poling[94], M. Popovic[45], R. Pordes[45], S. Pordes[45], M. Potekhin[12], R. Potenza[24], B. Potukuchi[69], S. Poudel[52], J. Pozimski[64], O. Prokofiev[45], N. Pruthi[109], P. Przewlocki[97], N. Pumulo[126], D. Pushka[45], X. Qian[12], J. L. Raaf[45], R. Raboanary[3], V. Radeka[12], J. Rademacker[11], B. Radics[40], A. Radovic[146], I. Rakhno[45], H. T. Rakotondramanana[3], L. Rakotondravohitra[3], Y. A. Ramachers[144], R. Rameika[45], J. Ramsey[83], A. Rappoldi[110], G. Raselli[110], P. Ratoff[79], B. Rebel[45], R. Regenfus[40], J. Reichenbacher[126], S. D. Reitzner[45], A. Remoto[78], A. Renshaw[52], S. Rescia[12], L. Rice[95], K. Rielage[83], K. Riesselmann[45], D. Rivera[112], M. Robinson[122], L. Rochester[120], O. B. Rodrigues[46], P. Rodrigues[119], B. Roe[90], R. M. Roser[45], M. Ross-Lonergan[39], M. Rossella[110], J. Rout[70], S. Roy[50], A. Rubbia[40], C. Rubbia[48], R. Rucinski[45], C. Rudolph von Rohr[9], B. Russell[149], D. Ruterbories[119], R. Saakyan[82], K. Sachdev[45], N. Sahu[60], P. Sala[92,16], N. Samios[12], F. Sanchez[55], M.C. Sanchez[68], W.R. Sands[116], S. Santana[117], R. Santorelli[17], L. M. Santos[41], G. Santucci[130], N. Saoulidou[6], G. Savage[45], A. Scaramelli[110], A. Scarpelli[1], T. Schaffer[93], H. Schellman[104], P. Schlabach[45], D. Schmitz[26], J. Schneps[138], K. Scholberg[38], A. Schukraft[45], E. Segreto[41], S. Sehrawat[50], J. A. Sepulveda-Quiroz[68], F. Sergiampietri[40], K. Sexton[12], L. Sexton-Kennedy[45], M. H. Shaevitz[33], J. Shahi[109], S. Shahsavarani[135], P. Shanahan[45], R. Sharma[12], R. K. Sharma[118], T. Shaw[45], D. Shooltz[91], R. Shrock[130], N. Simos[12], J. Sinclair[9], G. Sinev[38], I. Singh[109], J. Singh[109], J. Singh[85], V. Singh[8], F. W. Sippach[33], K. Siyeon[27], A. Smith[23], P. Smith[65], J. Smolik[35], M. Smy[20], E. Snider[45], P. Snopok[57], J. Sobczyk[148], H. Sobel[20], M. Soderberg[132], N. Solomey[145], W. Sondheim[83], M. Sorel[56], A. Sousa[28], K. Soustruznik[25], M. Spanu[110], J. Spitz[90], N. J. C. Spooner[122], M. Stancari[45], D. Stefan[16,97], A. Stefanik[45], H. M. Steiner[80,18], J. Stewart[12], J. Stock[126], S. Stoica[53], J. Stone[10], J. Strait[45], M. Strait[94], T. Strauss[45], S. Striganov[45], R. Sulej[97], G. Sullivan[88], Y. Sun[51], L. Suter[45], C. M. Sutera[24], R. Svoboda[19], B. Szczerbinska[134], A. M. Szelc[87], S. Söldner-Rembold[87], R. Talaga[5], S. Tariq[45], E. Tatar[63], R. Tayloe[65], K. Terao[33], M. Thiesse[122], L. F. Thompson[122], M. Thomson[23], C. Thorn[12], M. Thorpe[121], X. Tian[124], D. Tiedt[126], S. C. Timm[45], J. Todd[28], A. Tonazzo[1], T. Tope[45], F. R. Torres[41], M. Torti[110], M. Tortola[56], F. Tortorici[24], M. Toups[45], C. Touramanis[81], J. Trevor[22], M. Tripathi[19], I. Tropin[45], W. H. Trzaska[71], Y. Tsai[120], K. V. Tsang[120], A. Tsaris[45], R. Tsenov[123], S. Tufanli[149], C. Tull[80], J. Turner[39], M. Tzanov[84], E. Tziaferi[6], Y. Uchida[64], S. Uma Sankar[58], J. Urheim[65], T. Usher[120], M. R. Vagins[75,20], P. Vahle[146], G. A. Valdiviesso[42], L. Valerio[45], Z. Vallari[130], J. Valle[56], R. Van Berg[112], R. Van de Water[83], F. Varanini[108], G. Varner[51], G. Vasseur[15], K. Vaziri[45], G. Velev[45], S. Ventura[108], A. Verdugo[17], M. A. Vermeulen[100], E. Vernon[12], T. Viant[40], T. V. Vieira[41], C. Vignoli[47], S. Vihonen[71], C. Vilela[130], B. Viren[12], P. Vokac[35], T. Vrba[35], T. Wachala[99], D. Wahl[147], M. Wallbank[122], B. Wang[129], H. Wang[21], T. Wang[39], T.K. Warburton[68], D. Warner[32], M. Wascko[64], D. Waters[82], A. Weber[121,105], M. Weber[9], W. Wei[125], A. Weinstein[68], D. Wenman[147], M. Wetstein[68], A. White[135], L. H. Whitehead[16], D. Whittington[65], M. J. Wilking[130], J. Willhite[45], P. Wilson[45], R. J. Wilson[32], P. Wittich[34], J. Wolcott[138], H. H. Wong[18,80], T. Wongjirad[89], K. Wood[130], L.S. Wood[106], E. Worcester[12], M. Worcester[12], S. Wu[40], W. Xu[125], C. Yanagisawa[130], S. Yang[28], T. Yang[45], J. Ye[129], M. Yeh[12], N. Yershov[67], K. Yonehara[45], B. Yu[12], J. Yu[135], J. Zalesak[13], L. Zambelli[78], B. Zamorano[131], L. Zang[122], A. Zani[16], K. Zaremba[143], G.P. Zeller[45], C. Zhang[12], C. Zhang[125], Y. Zhou[39], E. D. Zimmerman[31], M. Zito[15], J. Zuklin[13], V. Zutshi[95], and R. Zwaska[45]

[1]APC - University Paris Diderot, Sorbonne Paris Cité, 75205 Paris, France
[2]Aligarh Muslim University, Department of Physics, Aligarh-202002, India
[3]University of Antananarivo, BP 566, Antananarivo 101, Madagascar
[4]Universidad Antonio Nariño, Cra 3 Este No 47A-15, Bogota, Colombia
[5]Argonne National Laboratory, Argonne, IL 60439, USA
[6]University of Athens, University Campus, Zografou GR 157 84, Greece
[7]Universidad del Atlantico, Km 7 antigua via Puerto Colombia, Barranquilla, Colombia
[8]Banaras Hindu University, Department of Physics, Varanasi - 221 005, India
[9]University of Bern, Sidlerstrasse 5, CH-3012 Bern, Switzerland
[10]Boston University, Boston, MA 02215, USA
[11]University of Bristol, H. H. Wills Physics Laboratory, Tyndall Avenue Bristol BS8 1TL, United Kingdom
[12]Brookhaven National Laboratory, Upton, NY 11973, USA
[13]Institute of Physics CAS, Na Slovance 2, 182 21 Praha 8, Czech Republic
[14]Centro Brasileiro de Pesquisas Físicas, Rio de Janeiro, RJ 22290-180, Brazil
[15]CEA/Saclay, IRFU (Institut de Recherche sur les Lois Fondamentales de l'Univers), F-91191 Gif-sur-Yvette CEDEX, France
[16]CERN, European Organization for Nuclear Research 1211 Geneve 23, Switzerland, CERN
[17]CIEMAT, Avenida Complutense 40, E-28040 Madrid, Spain
[18]University of California (Berkeley), Berkeley, CA 94720, USA
[19]University of California (Davis), Davis, CA 95616, USA
[20]University of California (Irvine), Irvine, CA 92697, USA
[21]University of California (Los Angeles), Los Angeles, CA 90095, USA
[22]California Institute of Technology, Pasadena, CA 91125, USA
[23]University of Cambridge, JJ Thomson Avenue, Cambridge CB3 0HE, United Kingdom
[24]University of Catania, INFN Sezione di Catania, Via Santa Sofia 64, I-95123 Catania, Italy
[25]Institute of Particle and Nuclear Physics of the Faculty of Mathematics and Physics of the Charles University in Prague, V Holešovičkách 747/2, 180 00 Praha 8-Libeň, Czech Republic
[26]University of Chicago, Chicago, IL 60637, USA
[27]Chung-Ang University, Dongjak-Gu, Seoul 06974, South Korea
[28]University of Cincinnati, Cincinnati, OH 45221, USA
[29]Cinvestav, Apdo. Postal 14-740, 07000 Ciudad de Mexico, Mexico
[30]Universidad de Colima, 340 Colonia Villa San Sebastian Colima, Colima, Mexico
[31]University of Colorado (Boulder), Boulder, CO 80309, USA
[32]Colorado State University, Fort Collins, CO 80523, USA
[33]Columbia University, New York, NY 10027, USA
[34]Cornell University, Ithaca, NY 14853, USA
[35]Czech Technical University in Prague, Břehová 78/7, 115 19 Prague 1, Czech Republic
[36]University of Delhi, Department of Physics and Astrophysics, Delhi 110007, India
[37]Drexel University, Philadelphia, PA 19104, USA
[38]Duke University, Durham, NC 27708, USA
[39]University of Durham, South Road, Durham DH1 3LE, United Kingdom
[40]ETH Zurich, Institute for Particle Physics, Zurich, Switzerland
[41]Universidade Estadual de Campinas, Campinas - SP, 13083-970, Brazil
[42]Universidade Federal de Alfenas, Poços de Caldas - MG, CEP 37715-400, Brazil
[43]Universidade Federal do ABC, 5001 Santo André - SP, 09210-580, Brazil
[44]UEFS/DFIS - State University of Feira de Santana, Feira de Santana - BA, 44036-900, Brazil
[45]Fermi National Accelerator Laboratory, Batavia, IL 60510, USA
[46]Universidade Federal de Goias, Goias, 74690-900, Brazil
[47]Laboratori Nazionali del Gran Sasso, I-67010 Assergi, AQ, Italy
[48]Gran Sasso Science Institute, Viale Francesco Crispi 7, L'Aquila, Italy
[49]Universidad de Guanajuato, Gto., C.P.37000, Mexico
[50]Harish-Chandra Research Institute, Jhunsi, Allahabad 211 019, India
[51]University of Hawaii, Honolulu, HI 96822, USA
[52]University of Houston, Houston, TX 77204, USA
[53]Horia Hulubei National Institute of Physiscs and Nuclear Engineering, Strada Reactorului 30, Măgurele, Romania
[54]University of Hyderabad, Gachibowli, Hyderabad - 500 046, India
[55]Institut de Fisica d'Altes Energies, Barcelona i.e. Catalan ersitat Autonoma de Barcelona, E-08193 Cerdanyola del Valles, Barcelona, Spain
[56]Instituto de Fisica Corpuscular, C/Catedratico Jose Beltran, 2 E-46980 Paterna (Valencia), Spain

[57] Illinois Institute of Technology, Chicago, IL 60616, USA
[58] Indian Institute of Technology Bombay, Department of Physics Mumbai 400 076, India
[59] Indian Institute of Technology Guwahati, Guwahati, 781 039, India
[60] Indian Institute of Technology Hyderabad, Hyderabad, 502285, India
[61] INFN Sezione di Milano Bicocca, Piazza della Scienza 3, 20126 Milano, Italy
[62] Institute for Research in Fundamental Sciences (IPM), Farmanieh St. Tehran, 19538-33511, Iran
[63] Idaho State University, Department of Physics 921 South 8th Ave. Pocatello, ID 83209, USA
[64] Imperial College of Science Technology & Medicine, Blackett Laboratory Prince Consort Road, London SW7 2BZ, United Kingdom
[65] Indiana University, Bloomington, IN 47405, USA
[66] Institut de Physique Nucleaire de Lyon (IPNL), Rue E. Fermi 4 69622 Villeurbanne, France
[67] Institute for Nuclear Research of the Russian Academy of Sciences (INR RAS), 60th October Anniversary Prosp. 7a Moscow, 117312, Russia
[68] Iowa State University, Ames, Iowa 50011, USA
[69] University of Jammu, Physics Department, JAMMU-180006, India
[70] Jawaharlal Nehru University, Jawaharlal Nehru University, New Delhi 110067, India
[71] University of Jyvaskyla, P.O. Box 35, FI-40014, Finland
[72] High Energy Accelerator Research Organization (KEK), Ibaraki, 305-0801, Japan
[73] KTH Royal Institute of Technology, Roslagstullsbacken 21, SE-106 91 Stockholm, Sweden
[74] Kansas State University, Manhattan, KS 66506, USA
[75] Kavli Institute for the Physics and Mathematics of the Universe (WPI), Kashiwa, Chiba 277-8583, Japan
[76] K L University, Guntur - 522 502, AP, India
[77] Kyiv National University, 64, 01601 Kyiv, Ukraine
[78] Laboratoire d'Annecy-le-Vieux de Physique des Particules, CNRS/IN2P3 and Université Savoie Mont Blanc, 74941 Annecy-le-Vieux, France
[79] Lancaster University, Bailrigg, Lancaster LA1 4YB, United Kingdom
[80] Lawrence Berkeley National Laboratory, Berkeley, CA 94720, USA
[81] University of Liverpool, L69 7ZE, Liverpool, United Kingdom
[82] University College London, London, WC1E 6BT, United Kingdom
[83] Los Alamos National Laboratory, Los Alamos, NM 87545, USA
[84] Louisiana State University, Baton Rouge, LA 70803, USA
[85] University of Lucknow, Lucknow 226007, Uttar Pradesh, India
[86] Madrid Autonoma University, Ciudad Universitaria de Cantoblanco 28049 Madrid, Spain
[87] University of Manchester, Oxford Road, Manchester M13 9PL, United Kingdom
[88] University of Maryland, College Park, MD 20742, USA
[89] Massachusetts Institute of Technology, Cambridge, MA 02139, USA
[90] University of Michigan, Ann Arbor, MI 48109, USA
[91] Michigan State University, East Lansing, MI 48824, USA
[92] INFN Milano, INFN Sezione di Milano, I-20133 Milano, Italy
[93] University of Minnesota (Duluth), Duluth, MN 55812, USA
[94] University of Minnesota (Twin Cities), Minneapolis, MN 55455, USA
[95] Northern Illinois University, Department of Physics DeKalb, Illinois 60115, USA
[96] Istituto Nazionale di Fisica Nucleare - Sezione di Napoli, Complesso Universitario di Monte S. Angelo, I-80126 Napoli, Italy
[97] National Centre for Nuclear Research, A. Soltana 7, 05 400 Otwock, Poland
[98] University of New Mexico, 1919 Lomas Blvd. N.E. Albuquerque, NM 87131, USA
[99] H. Niewodniczanski Institute of Nuclear Physics, Polish Academy of Sciences, Cracow, Poland
[100] Nikhef National Institute of Subatomic Physics, Science Park, Amsterdam, Netherlands
[101] Northwestern University, Evanston, Il 60208, USA
[102] University of Notre Dame, Notre Dame, IN 46556, USA
[103] Ohio State University, 191 W. Woodruff Ave. Columbus, OH 43210, USA
[104] Oregon State University, Corvallis, OR 97331, USA
[105] University of Oxford, Oxford, OX1 3RH, United Kingdom
[106] Pacific Northwest National Laboratory, Richland, WA 99352, USA
[107] Physical Research Laboratory, Ahmedabad 380 009, India
[108] University of Padova, Dip. Fisica e Astronomia G. Galilei and INFN Sezione di Padova, I-35131 Padova, Italy
[109] Panjab University, Chandigarh, 160014 U.T., India
[110] University of Pavia, INFN Sezione di Pavia, I-27100 Pavia, Italy
[111] Pennsylvania State University, University Park, PA 16802, USA

[112] University of Pennsylvania, Philadelphia, PA 19104, USA
[113] University di Pisa, Theor. Division; Largo B. Pontecorvo 3, Ed. B-C; I-56127 Pisa, Italy
[114] University of Pittsburgh, Pittsburgh, PA 15260, USA
[115] Pontificia Universidad Católica del Perú, Apartado 1761, Lima, Peru
[116] Princeton University, Princeton, New Jersey 08544, USA
[117] University of Puerto Rico, Mayaguez, 00681, USA
[118] Punjab Agricultural University, Department of Math. Stat. & Physics, Ludhiana 141004, India
[119] University of Rochester, Rochester, NY 14627, USA
[120] SLAC National Accelerator Laboratory, Menlo Park, CA 94025, USA
[121] STFC Rutherford Appleton Laboratory, OX11 0QX Harwell Campus, Didcot, United Kingdom
[122] University of Sheffield, Department of Physics and Astronomy, Sheffield S3 7RH, United Kingdom
[123] University of Sofia, 5 James Bourchier Blvd., Sofia, Bulgaria
[124] University of South Carolina, Columbia, SC 29208, USA
[125] University of South Dakota, Vermillion, SD 57069, USA
[126] South Dakota School of Mines and Technology, Rapid City, SD 57701, USA
[127] South Dakota Science And Technology Authority, Lead, SD 57754, USA
[128] South Dakota State University, Brookings, SD 57007, USA
[129] Southern Methodist University, Dallas, TX 75275, USA
[130] Stony Brook University, Stony Brook, New York 11794, USA
[131] University of Sussex, Brighton, BN1 9RH, United Kingdom
[132] Syracuse University, Syracuse, NY 13244, USA
[133] University of Tennessee at Knoxville, TN, 37996, USA
[134] Texas A&M University (Corpus Christi), Corpus Christi, TX 78412, USA
[135] University of Texas (Arlington), Arlington, TX 76019, USA
[136] University of Texas (Austin), Austin, TX 78712, USA
[137] TUBITAK Space Technologies Research Institute, TR-06800, Ankara, Turkey
[138] Tufts University, Medford, MA 02155, USA
[139] Variable Energy Cyclotron Centre, 1/AF, Bidhannagar Kolkata - 700 064 West Bengal, India
[140] Valley City State University, Valley City, ND 58072, USA
[141] Virginia Tech, Blacksburg, VA 24060, USA
[142] University of Warsaw, Faculty of Physics ul. Pasteura 5 02-093 Warsaw, Poland
[143] Warsaw University of Technology, Nowowiejska 15/19 00-665 Warszawa Poland, Poland
[144] University of Warwick, Coventry CV4 7AL, United Kingdom
[145] Wichita State University, Physics Division 1845 Fairmount St. Wichita, KS 67260, USA
[146] College of William and Mary, Williamsburg, VA 23187, USA
[147] University of Wisconsin (Madison), Madison, WI 53706, USA
[148] Wroclaw University, Plac Maxa Borna 9, 50-204 Wroclaw, Poland
[149] Yale University, New Haven, CT 06520, USA
[150] Yerevan Institute for Theoretical Physics and Modeling, Halabian Str. 34; Yerevan 0036, Armenia
[151] York University, Physics and Astronomy Department, 4700 Keele St. Toronto M3J 1P3, Canada

2016-06-01

  

\renewcommand{\familydefault}{\sfdefault}
\renewcommand{\thepage}{\roman{page}}
\setcounter{page}{0}

\pagestyle{plain} 


\setcounter{tocdepth}{2}
\textsf{\tableofcontents}

\textsf{\listoffigures}

\textsf{\listoftables}

\printnomenclature

\iffinal\else
\textsf{\listoftodos}
\clearpage
\fi

\renewcommand{\thepage}{\arabic{page}}
\setcounter{page}{1}

\pagestyle{fancy}

\renewcommand{\chaptermark}[1]{%
\markboth{Chapter \thechapter:\ #1}{}}
\fancyhead{}
\fancyhead[RO,LE]{\textsf{\footnotesize \thechapter--\thepage}}
\fancyhead[LO,RE]{\textsf{\footnotesize \leftmark}}

\fancyfoot{}
\fancyfoot[RO]{\textsf{\footnotesize ProtoDUNE Single-Phase Technical Design Report}}
\fancypagestyle{plain}{}

\renewcommand{\headrule}{\vspace{-4mm}\color[gray]{0.5}{\rule{\headwidth}{0.5pt}}}

\nomenclature{$\mathcal{O}(n)$}{of order $n$}
\nomenclature{3D}{3 dimensional (also 1D, 2D, etc.)} 
\nomenclature{DAQ}{data acquisition}
\nomenclature{DOE}{U.S. Department of Energy}
\nomenclature{DUNE}{Deep Underground Neutrino Experiment}
\nomenclature{ESH}{Environment, Safety and Health}
\nomenclature{eV}{electron-Volt, unit of energy (also keV, MeV, GeV, etc.)}
\nomenclature{FD}{far detector}
\nomenclature{FRP}{fiberglass-reinforced plastic}
\nomenclature{L}{level, indicates depth in feet underground at the DUNE far site, e.g., 4850L}
\nomenclature{LAr}{liquid argon}
\nomenclature{LArTPC}{liquid argon time-projection chamber}
\nomenclature{LBNF}{Long-Baseline Neutrino Facility}
\nomenclature{POT}{protons on target}
\nomenclature{t}{metric ton, written \textit{tonne} (also kt)}
\nomenclature{TDR}{Technical Design Report}
\nomenclature{TPC}{time-projection chamber 




\nomenclature{CPA}{cathode plane assembly}
\nomenclature{DP}{dual-phase}
\nomenclature{DUNE-DP}{dual-phase far detector module for DUNE}
\nomenclature{DUNE-SP}{single-phase far detector module for DUNE}
\nomenclature{FC}{field cage}
\nomenclature{GAr}{gaseous argon}
\nomenclature{GEANT4}{GEometry ANd Tracking, a platform for the simulation of the passage of particles through matter using Monte Carlo methods} 
\nomenclature{GENIE}{Generates Events for Neutrino Interaction Experiments (an object-oriented neutrino Monte Carlo generator)} 
\nomenclature{HV}{high voltage}
\nomenclature{MC}{Monte Carlo (detector simulation methods)}
\nomenclature{ProtoDUNE-DP}{dual-phase ProtoDUNE}
\nomenclature{ProtoDUNE-SP}{single-phase ProtoDUNE}
\nomenclature{QA/QC}{quality assurance/quality control}
\nomenclature{QE}{quantum efficiency}
\nomenclature{SP}{single-phase}


\nomenclature{ADC}{analog-to-digital converter}
\nomenclature{APA}{anode plane assembly} 
\nomenclature{ASIC}{application-specific integrated circuit}
\nomenclature{BGR}{band-gap reference}
\nomenclature{CASTOR}{Cern Advanced STORage system, a hierarchical storage management system developed at CERN}
\nomenclature{CE}{cold electronics}
\nomenclature{CMOS}{complementary metal-oxide semiconductor}
\nomenclature{COB}{cluster on-board (motherboards)} 
\nomenclature{CTE}{coefficient of thermal expansion} 
\nomenclature{CR}{capacitance-resistance (board)}
\nomenclature{DCS}{detector control system}
\nomenclature{DSS}{detector support system}
\nomenclature{DRAM}{dynamic random access memory}
\nomenclature{EHN1}{extension to hall in North Area at CERN in which both ProtoDUNEs reside}
\nomenclature{ENC}{electronic noise charge}
\nomenclature{FEMB}{front-end mother board}
\nomenclature{FE}{front-end (electronics)}
\nomenclature{FR4}{a composite material composed of woven fiberglass cloth with an epoxy resin binder that is flame resistant (Wikipedia)}
\nomenclature{FELIX}{Front-End-Link-EXchange}
\nomenclature{Fermilab (also FNAL)}{Fermi National Accelerator Laboratory (in Batavia, IL, the Near Site)}
\nomenclature{FNAL}{see Fermilab}
\nomenclature{FPGA}{field programmable gate array} 
\nomenclature{GP}{ground plane}
\nomenclature{GTT}{engineering company in containment systems for the shipping and storage in cryogenic conditions of LNG (liquefied natural gas)}
\nomenclature{H4}{H4 beamline, the beamline extension to ProtoDUNE-SP}
\nomenclature{MIP}{minimum ionizing particle}
\nomenclature{LN2}{liquid nitrogen}
\nomenclature{LV}{low voltage}
\nomenclature{MPOD}{a multi-channel LV/HV power supply system from Wiener\textsuperscript{TM}}
\nomenclature{MUX}{multiplex}
\nomenclature{NP04}{the CERN experiment designation for ProtoDUNE-SP}
\nomenclature{PDS}{photon detection (system)}
\nomenclature{pLAPPD}{a time-of-flight system from Argonne Nat'l Laboratory}
\nomenclature{RCE}{reconfigurable computing element}
\nomenclature{RT}{room temperature}
\nomenclature{SiPM}{silicon photomultiplier}
\nomenclature{S/N}{signal-to-noise (ratio)}
\nomenclature{SSP}{SiPM signal processor}
\nomenclature{TCO}{temporary cryostat opening}
\nomenclature{TPB}{tetraphenyl-butadiene}
\nomenclature{ToF}{time of flight}
\nomenclature{UHMW PE}{ultra-high-molecular-weight polyethylene}
\nomenclature{WEC}{warm electronics crate}
\nomenclature{WIB}{warm interface board}
\nomenclature{WLS}{wavelength shifting}
\nomenclature{CNN}{convolutional neural networks}

\chapter{Introduction}

\section{ProtoDUNE-SP in the context of DUNE/LBNF}

ProtoDUNE-SP is the single-phase DUNE Far Detector prototype that is under construction and will be operated at the CERN Neutrino Platform (NP) 
starting in 2018. It was proposed to the CERN SPSC in June 2015 (SPSC-P-351), and following positive recommendations by SPSC and the CERN Research Board in December 2015, was approved at CERN as experiment NP-04 (ProtoDUNE). The Fermilab Director and the CERN Director of Research and Scientific Computing signed a Memorandum of Understanding (MoU) for this experiment in December 2015 that is initially valid until December 2022, 
and may be extended by mutual agreement. 

ProtoDUNE-SP, a crucial part of the DUNE effort towards the construction of the first DUNE \ktadj{10} fiducial mass far detector module (17\,kt total LAr mass), is a significant experiment in its own right. With a total liquid argon (LAr) mass of 0.77\,kt, it represents the largest monolithic single-phase LArTPC detector to be built to date.  
It is housed in an extension to the EHN1 hall in the North Area, where the CERN NP is providing a new dedicated charged-particle test beamline. ProtoDUNE-SP aims to take its first beam data before the LHC long shutdown (LS2) at the end of 2018.

ProtoDUNE-SP prototypes the designs of most of the single-phase DUNE far detector module (DUNE-SP) components at a 1:1 scale, with an extrapolation of about 1:20 in total LAr mass. This is similar to the scaling factor adopted 
by ICARUS; its T600 detector, split into two half-modules of about 375\,t total LAr mass each, was preceded by the 14-t ``10-m$^3$'' prototype. 

The detector elements, consisting of the time projection chamber (TPC), the cold electronics (CE), and the photon detection system (PDS), are housed in a cryostat that contains the LAr target material. The cryostat, a free-standing steel-framed vessel with an insulated double membrane, is based on the technology used for liquefied natural gas (LNG) storage and transport. 
A cryogenics system maintains the LAr at a stable temperature of about 89\,K and at the required purity level  through a closed-loop process that recovers the evaporated argon, recondenses and filters it, and returns it to the cryostat. 

The construction and operation of ProtoDUNE-SP will serve to validate the membrane cryostat technology and associated cryogenics, and the networking and computing infrastructure that will handle the data and simulated data sets.
A charged-particle beam test will enable critical calibration measurements necessary for precise calorimetry. It will also enable the collection of invaluable data sets for optimizing the event reconstruction algorithms -- i.e., for finding interaction vertices and for particle identification -- and ultimately for quantifying and reducing systematic uncertainties for the DUNE far detector. These measurements are expected to significantly improve the physics reach of the DUNE experiment.

The timescale for the validation of the basic TPC design is driven by the schedule for the major LBNC and DOE reviews of the DUNE TDR that will take place in 2019. This sets the requirement for ProtoDUNE-SP data collection in 2018. 

Given its technical challenges, its importance to the DUNE experiment and the timeframe in which it must operate, ProtoDUNE-SP has put in place a strong organizational structure incorporating 
contributions from a large number of DUNE collaboration institutes. 

\section{The ProtoDUNE-SP detector}
\label{intro:detector}

The ProtoDUNE-SP TPC, illustrated in Figure~\ref{fig:protodune-sp-tpc} comprises two drift volumes, defined by  a central cathode plane that is flanked by two anode planes, each at a distance of 3.6~m, and a field cage (FC) that surrounds the entire active volume. The active volume is 6~m high, 7~m wide and 7.2~m deep (along the drift direction).
Each anode plane is constructed of three adjacent Anode Plane Assemblies (APAs) that are each 6~m high by 2.3~m wide in the installed position. Each APA consists of a frame that holds three parallel planes of sense and shielding wires on each of its two faces; the wires of each plane
  are oriented at different angles with respect to those making up the
  other planes to enable 3D reconstruction. The first two planes of
  wires wrap around to cover both faces of the APA and their wires have
  a 4.67~mm pitch.  The third planes (on both faces) are not
  electrically connected and have a 4.79~mm pitch.  Signals from the
  wires of each APA are read out via a total of 2,560 electronics
  channels.

The cathode plane, called the Cathode Plane Assembly (CPA) is an array of 18 (six wide by three high) CPA modules, which consist of  flame-retardant FR4 frames, each 1.18~m wide and 2~m high, that hold thin panels with a resistive coating on both sides. 
The CPA is held at $-$180\,kV providing the 500-V/cm drift field in the 3.6-m-deep drift regions. 
Uniformity of the electric field is guaranteed by the surrounding FC.
 
\begin{cdrfigure}[Major components of the ProtoDUNE-SP TPC]{protodune-sp-tpc}{The major components of the ProtoDUNE-SP TPC. }
\includegraphics[width=0.9\textwidth]{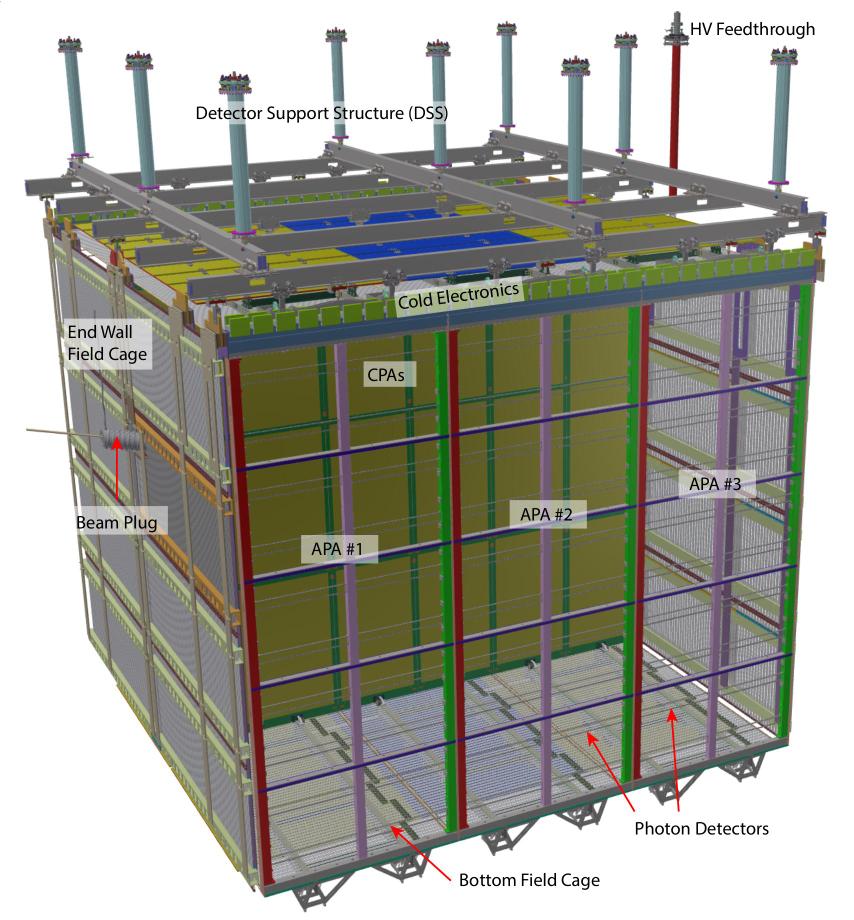}
\end{cdrfigure}

The CE, mounted onto the APA frame, and thus immersed in LAr, amplifies and continuously digitizes the induced signals on the sense wires at several MHz, and transmits these waveforms to the Data Acquisition system (DAQ). From the DAQ the data are  transmitted through the buffer to disk, then to the central CERN Tier-0 Computing Center, and finally to other partner sites for processing and analysis.  

The modular PDS is integrated into the APAs. Each PDS module (referred to as a PD) consists of a bar-shaped light guide and
a wavelength-shifting layer (surface-coating or mounted radiator plate). The wavelength-shifting layer converts incoming VUV (128 nm) 
scintillation photons to longer-wavelength photons, in the visible blue range. Some of the converted photons are emitted into the bar, a fraction of which are then internally reflected to the bar's end where they are detected by silicon photomultipliers (SiPMs).
Each APA frame is designed with ten bays into which PDs are inserted after the TPC wires have been strung. This  allows for final assembly at the integration area (in a clean room) at the CERN NP prior to installation inside the cryostat.

\section{Goals of ProtoDUNE-SP}
\label{intro:goals}
ProtoDUNE-SP has four principal goals, all of which are essential parts of the DUNE far detector development program:
\begin{itemize}
\item Prototype the production and installation procedures for the single-phase far detector design.
\item Validate the design from the perspective of basic detector performance; this can be achieved with cosmic-ray data. 
This is the critical step from the perspective of establishing the design for the DUNE TDRs, which will undergo LBNC and DOE reviews in 2019.
\item Accumulate large samples of test-beam data to understand/calibrate the response of the detector to different particle species. 
\item Demonstrate the long-term operational stability of the detector as part of the risk mitigation program ahead of the construction of the first \ktadj{10} far detector module.
\end{itemize}
Commissioning and successfully bringing the detector into operation and maintaining it in a stable operating condition over a period of 
one-to-three months would allow for validation of the design through a full characterization of the detector components, including the membrane cryostat and the cooling and purification circuit, the APA design and the layout of its cold read-out electronics, the HV system, and the PDS and its warm read-out electronics. This can be achieved by collecting cosmic-ray data. 

The large samples of test-beam data of different particle species would allow for a more in-depth understanding the response of the detector and could ultimately be used for calibration purposes. The analysis of the test-beam data will involve strong feedback 
between reconstruction, detector simulation, and hadronic modeling. Three types of systematic effects will be addressed: (1) a detailed understanding of energy scale and energy resolution for electromagnetic and hadronic showers; (2) the detector response to different particle species, including 
the impact of recombination effects as a function of angle; and (3) improved modeling of the interactions of hadrons in argon, which will provide constraints to the GEANT4 physics models.  A run period of approximately three months in the test beam is considered sufficient, assuming stable operation under good beam conditions. The aim is to accumulate these data before the start of LS2 (Jul-Oct, 2018). 

The liquid argon technology is, however, sufficiently new that it would be highly desirable to perform a long-term test of ProtoDUNE-SP, extending beyond the start of LS2, to reduce risks associated with the extended operation of the DUNE far detector modules. This argues for cosmic-ray operation in 2019 and most likely 2020. This extended running period would be an important component of our risk mitigation strategy for the overall DUNE detector construction effort. 




Ongoing detailed studies aim to quantify estimates regarding the level at which systematics in DUNE oscillation measurements can be constrained through analysis of ProtoDUNE test-beam data. The precise number of beam events required to meet the DUNE oscillation analysis physics goals is not yet determined - but this is not the primary driver of the beam time request.
It should be noted that the run duration will depend heavily upon actual detector performance and the beam quality (i.e., precision in particle momentum, impact position, beam PID), and that large uncertainties will remain until both ProtoDUNE-SP and the H4 beamline are commissioned. For all these reasons, beam operation after LS2 cannot be excluded.



\subsection{Detector engineering validation}  


ProtoDUNE-SP is designed such that it will provide information on the actual far detector performance in as close as possible a configuration to the actual far detector layout, given the practical considerations imposed by time, space, and cost. To achieve this, the cold components are identical to the components proposed for the far detector, to the extent possible. 

The APAs are full-scale pre-production modules for the far detector.  The ground connections between the electronics, the APA mechanical structure, the photon detectors, and the detector support structure are the same as those proposed for DUNE-SP. ProtoDUNE-SP is instrumented with three APAs along each wall; this will test for cross talk between the middle APA and its neighbors. Whereas DUNE-SP is designed with a 12-m-high TPC, based on a two-APA-high layout, ProtoDUNE-SP TPC is one APA high, due to practical limitations. The collaboration will test the mechanical process of installing the two-APA detector configuration, along with the related cabling, separately.


 The cold readout electronics  is based on the DUNE-SP cryogenic front-end pre-amp/shaper chip and ADC design. The front-end ASIC is essentially the final design {
while the cold ADC ASIC is not the final version and the dedicated ASIC (COLDATA) for serializing the data and providing a 1-GB/s link is not yet available. Given the schedule constraints for ProtoDUNE-SP, an FPGA that emulates the COLDATA functionality is used and mounted on a dedicated mezzanine board. 
All the analog components, the conversion to digital, and the grounding/power distribution for the final electronics can be tested in this configuration.  

This configuration of the electronics chain is the only option for meeting the requirement for protoDUNE-SP data collection in 2018 (requirement set by the schedule for the major LBNC and DOE reviews of the DUNE TDR that will take place in 2019 and the need of validation of the basic TPC design by then).
In the event further optimization of the ASICs (ADC and COLDATA) is required based on the ProtoDUNE-SP findings, it can be implemented before the start of production in 2020. 

Since there is no charge amplification in the liquid, the electronics is required to be extremely sensitive. The grounding and shielding are therefore critical, and the designs are as similar as is practical to those of DUNE-SP. The building ground in ENH1, with the interconnected rebar in the concrete floor 
connected to the building ground bus, will provide a fairly good ground. The cryostat itself is isolated from the building ground, and all the mechanical/electrical connections have dielectric breaks. DUNE-SP, in contrast, will be installed a mile underground in a very dry mine, and better isolation from the environment is expected. ProtoDUNE-SP will test the ground isolation and shielding under conservative conditions.

The FC and cathode planes are full-scale prototypes, 
with the same maximum drift distance and corresponding high voltage.
This allows ProtoDUNE-SP to use the same HV feedthrough and drift-field configuration as is planned for DUNE-SP. The cryostat dimensions are selected to be the same for both ProtoDUNEs 
to leverage the same cryostat and cryogenics system designs. 
In order to fit the ProtoDUNE-SP TPC in the cryostat, the wall-to-cathode plane distance is slightly smaller than in DUNE-SP, making this setup a conservative test of the HV design. 
Testing the TPC components under the most likely operating conditions for DUNE-SP is extremely important, as this is the first test of the updated LArTPC design, which incorporates a resistive cathode 
and FC construction with metal profiles and fiberglass I-beam support. 

ProtoDUNE-SP also offers a unique platform to validate and possibly optimize the cryogenics design for DUNE-SP. 
All cryostat penetrations are designed with gas purge to prevent contaminates from migrating from warm surfaces to the ullage volume, and the the liquid and gas flows inside the cryostat are being modeled. This is expected to provide an excellent test bed to validate the cryogenics design for DUNE-SP. 


ProtoDUNE-SP will prototype the tooling and procedures for transporting APAs and transferring them to a suspended rail system. Similarly, the assembly and transport processes of the CPAs and top/bottom FC assemblies are being developed.  Due to space constraints inside the cryostat, the rail structure on which the detector is hung will most likely be different for DUNE-SP. Despite these differences, the experience gained in installing the ProtoDUNE-SP detector will be invaluable in planning the DUNE-SP installation. 


Integration of the test beam with ProtoDUNE-SP is quite different from the neutrino beam integration planned for DUNE-SP. To minimize the material interactions of the particle beam in the ProtoDUNE-SP cryostat upstream of the TPC, a volume of LAr along the beam path (between the cryostat membrane and the FC) is displaced, and replaced by a less dense volume of dry nitrogen gas. This requires a penetration into one of the FC assemblies to install a  beam plug. To ensure that the displacement plug does not compromise the ProtoDUNE-SP operation, a dedicated HV test at Fermilab is planned that will test the final beam plug in the exact field configuration planned for ProtoDUNE-SP.

\subsection{Physics}

The use of the CERN charged-particle beams of known particle type and incident energy provide the means to achieve a better understanding of the interaction processes occuring within a argon target and help to optimize event reconstruction techniques, particle identification algortihms, and calorimetric energy measurements.   
The beam measurements will also provide a calibration data set for tuning the Monte Carlo simulations and a reference data set for the DUNE experiment. 

Pion and proton beams in an energy range from about one to a few GeV will be used primarily to study hadronic interaction mechanisms and secondary particle production.  At higher energies, these beams will be used to study shower reconstruction and energy calibration. Electrons will be used to benchmark and tune electron/photon separation algorithms, to study electromagnetic cascade processes and to calibrate electromagnetic showers at higher energies. Charged kaons produced in the tertiary beamline are rare but are copiously produced by the pion beam interactions inside the detector. These will be extremely useful for characterizing kaon identification efficiency for proton decay sensitivity studies.  Samples of stopping muons with Michel electrons from muon decay (or without them, in the case of negative muon capture) will be used for energy calibrations in the low-energy range of the SN neutrino events and for the development of charge-sign determination methods. 

A cumulative ProtoDUNE-SP test-beam run period of 16 weeks is assumed, but it depends on the extent of beamline sharing with other users at EHN1. The run will take place prior to the long shutdown of the LHC in late 2018 (LS2). 

ProtoDUNE-SP will acquire cosmic data during periods with no beam. A dedicated external trigger system consisting of arrays of scintillator paddles, suitably positioned and arranged in \textit{coincidence} trigger logic, 
will select specific classes of cosmic muon events. Dedicated runs, e.g., runs looking for long muon tracks crossing the entire detector at large zenith angles, allow for an overall test of the detector performance and the DAQ. Runs looking for muons stopping inside the LAr volume and the accumulation of accurate Michel electron spectra may be useful for energy calibration purposes in the low-energy range.

It is important to note that ProtoDUNE-SP offers much beyond calibration  and detector performance characterization.  The LArTPC simultaneously features precise 3D tracking and accurate measurement of energy deposited. Its large active volume allows for good containment of the hadronic and electromagnetic interaction products in the few GeV range. These capabilities have never before been combined in one detector.  The unprecedented event reconstruction capability combined with the exposure of the detector's large active volume to the CERN charged-particle beams open the way to a truly rich program of new physics investigations into particle interaction processes. 

  Hadronic ($\pi$, K and $p$) interactions on an Ar target around one GeV produce low-multiplicity final states rather than ``hadron showers,'' and similarily 1-GeV electrons  (with critical energy $\simeq 30$ MeV in Ar) produce low-populated cascades, with only a few tens of secondary energetic electrons (positrons).
``TPC/imaging-aided calorimetric measurements'' will allow investigations of energy deposition mechanisms in this energy range where the standard hadronic and electromagnetic shower concepts and features are either not well defined or cannot be applied.
The calorimetric measurement of the total energy deposited can, in fact, be accomplished, by detecting and summing over the energy deposited by each individual secondary particle/track thanks to the imaging capabilities of the LArTPC.
In particular, the determination of the electromagnetic  content in hadron-initiated cascades, $\pi^0$ multiplicity, and the energy fraction carried as a function of primary hadron incident energy will be of interest.

\section{The ProtoDUNE-SP Run plan}
\label{sec:runplan}


Beam simulations show that the hadron rates at 
energies below 1~GeV/c are low. Moreover, low-energy beams are more
subject to degradation by materials in the
beamline.  The optimization of the run plan factors in the beam composition and particle rates of the H4 beamline, and 
also particle interaction topologies in the ProtoDUNE-SP detector. Full 
simulations of particle transport in the ProtoDUNE-SP detector, including the
beam window, have been performed.

At a beam momentum of 1~GeV/c, 35\% of protons are stopped before reaching the active TPC region, while the percentage reduces to 0.5\% at 2~GeV/c.  The kinetic energy distributions of protons and pions at the entrance point of the TPC for different beam momenta are shown in Figure~\ref{fig:pandpiint}. 
The fraction of stopping $\pi$'s for one $\pi$
produced at the secondary target is 3\% at $p=0.4$~GeV/c and decreases to 1.3\% at $p=0.7$~GeV/c.
The long distance (37~m) between the secondary target and the front of the LAr cryostat has a significant impact on the pion and kaon rates in the TPC. Due to pion lifetime, many of the  low-energy pions produced at the secondary target decay in the beam pipe before reaching the cryostat. The situation is even more significant for kaons; most kaons below 2~Gev/c do not make it to the cryostat.
Consequently the H4 beamline will not be operated much below 1~GeV/c in the hadron mode.
For electrons, 
the beam momentum should go 
as low as possible to study the topology of very low-energy electron-initiated 
showers.
\begin{cdrfigure}[Kinetic energy at interaction point for different beam momenta]{pandpiint}{Kinetic energy of
    particles at the point of interaction in the ProtoDUNE-SP active
    volume, for different beam momenta. Histograms are normalized to one particle injected in the
    beamline acceptance. Simulations include the beam window
    materials, beams are considered as monochromatic and
    parallel. Left: protons, Right: pions.}
  \includegraphics[width=0.49\textwidth]{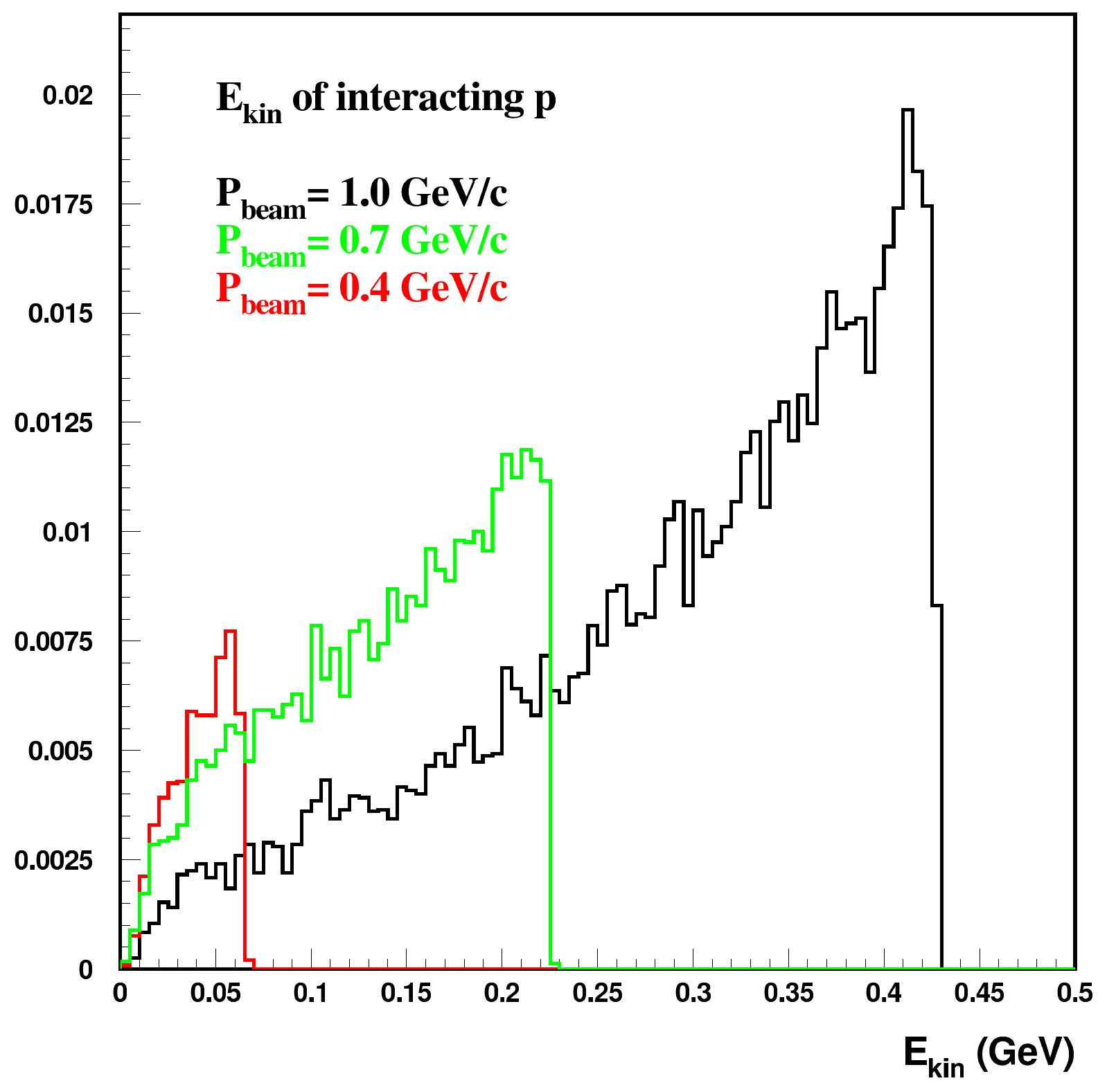}
  \includegraphics[width=0.49\textwidth]{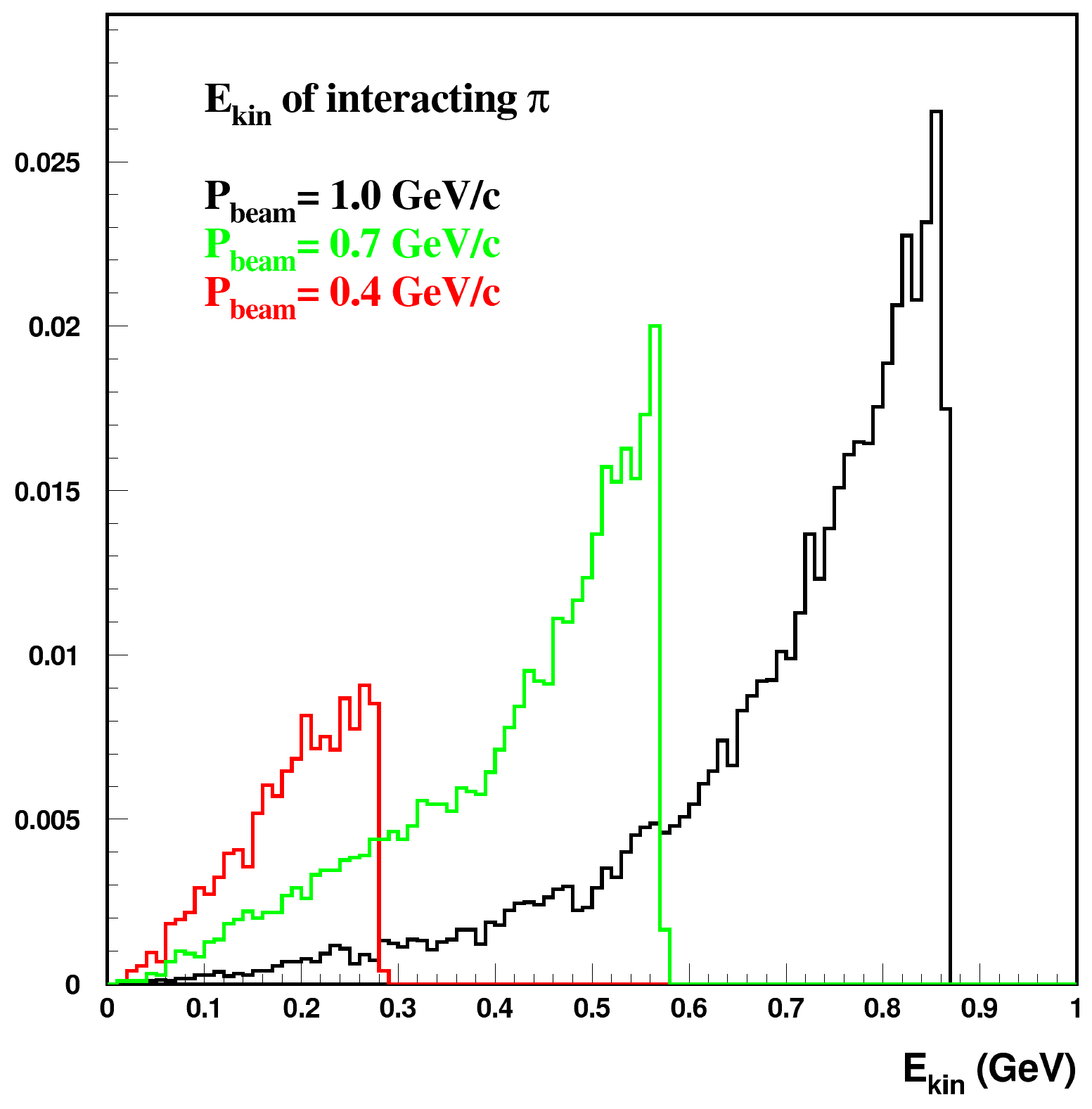}
\end{cdrfigure}

To formulate a preliminary run plan, 
the hadron beam spectrum and rates are assumed as given in Tables~\ref{tab:beampartcomp} and~\ref{tab:beampartrates}.   For the purpose of estimating the sample composition and beam time request, the following assumptions are used:
\begin{itemize}
\item { Trigger rate = 25~Hz}
\item { Two 4.8 sec spills per SPS Super Cycle }
\item { SPS Super Cycle = 48 sec}
\item { $10^6$ ($10^4$) secondary particles on target per spill for hadron (electron) beam}
\item { Particle ID trigger for electrons from 0.5 to 7 GeV/c}
\item { Trigger rate for electron in hadron beam is prescaled to 0.5~Hz}
\item { Data collection efficiency = 50\%}
\end{itemize}

It has not yet been determined whether the H2 and H4 tertiary beamlines at EHN1 can run simultaneously, or 
 whether the secondary beam (upstream the target of the H4 beamline) will be shared with other users. Therefore, a collection efficiency of 50\% has been assumed. 
 
It is planned to run the H4 beamline in two modes: the first configuration is optimized for the production of hadrons and the second configuration is optimized for the production of high-purity electrons. Even in the hadron mode, the beam is still dominated by electrons, especially for low beam momenta. However, the electrons in the hadron beam are not particularly ``clean'' due to the amount of materials in the beamline from the particle identification (PID) instrumentations.  The proposal is to heavily prescale the electron events using PID (e.g., Threshold Cherenkov counters) trigger while running in hadron mode. The PID systems that contribute significantly to the material budget will be removed when 
the beamline is reconfigured for electron beam.  Various run plan scenarios are under investigation. One of the scenarios is shown in Tables~\ref{tab:HadRunPlan} and ~\ref{tab:ElecRunPlan}. Similar values are expected for the negative beam sample. 
\begin{cdrtable}[Preliminary run plan for ProtoDUNE-SP hadron beam]{ccccccccc}{HadRunPlan}{A preliminary run plan for ProtoDUNE-S\
P hadron beam. The expected sample (positive beam) as a function of momentum is shown. }
P & \# of  &\# of $e^+$ & \# of $K^+$ & \# of $\mu^+$ & \# of $p$ & \# of $\pi^+$ & Total \# & Beam Time \\
(GeV/c) & Spills  & &  &  &  &  & of Events & (days) \\ \toprowrule
1 & 70K & 84K & $\approx$ 0 & 70K  & 689K & 625K & 1.5M & 19.4 days\\ \colhline
2 & 16K & 19K & 9K & 36K     & 336K & 572K & 1.0M &4.4 days\\ \colhline
3 & 13K & 16K & 26K  & 17K   & 181K & 540K  & 780K & 3.6 days\\ \colhline
4 & 11K & 13K & 19K & 16K    & 107K  & 510K & 660K & 3.1 days\\ \colhline
5 & 11K & 13K  & 29K  & 13K   & 96K  & 510K & 660K & 3.1 days\\ \colhline
6 & 11K & 13K & 36K  & 12K    & 94K  & 510K & 660K & 3.1 days\\ \colhline
7 & 11K & 13K & 42K & 8K     & 87K  & 510K & 660K & 3.1 days\\ \toprowrule
Total & 143K & 171K & 161K & 172K & 1.6M & 3.8M & 5.9M & 39.7 days\\
\end{cdrtable}
\begin{cdrtable}[Preliminary run plan for ProtoDUNE-SP electron beam]{cccc}{ElecRunPlan}{A preliminary run plan for ProtoDUNE-SP electron beam. The expected sample for positive beam configuration is shown. }
Momentum Bins & \# of Spills per Bin & \# $e^+$ per Bin & Beam Time per Bin \\ 
(GeV/c) & & & (days) \\ \toprowrule
0.5, 06, 0.7, 0.8, 0.9, 1, 2, 3, 4, 5, 6, 7 & 5000 & 300K & 1.4 \\
\end{cdrtable}

Based on the information available, the total estimated beam time needed to carry out the physics program in this proposal, with the assumptions stated earlier, is on the order of 16 weeks.

\chapter{Detector components} 

\section{Overview}

The elements composing the detector, listed in Section~\ref{intro:detector}, include the time projection chamber (TPC), the cold electronics (CE), and the photon detection system (PDS).  The TPC components, e.g., anode planes, a cathode plane and a field cage, are designed in a modular way.  
The six APAs are arranged into two APA planes, each consisting of three side-by-side APAs. Between them,  
a central cathode plane, composed of 18 CPA modules, splits the TPC volume into two electron-drift regions, one on each side of the cathode plane. 
A field cage (FC) completely surrounds the four
open sides of the two drift regions to ensure that the electric field within is uniform and unaffected by the presence of the cryostat walls and other nearby conductive structures. The sections in this chapter describe the components individually.

Figure~\ref{fig:fc-overview} illustrates how these components fit together.

Table~\ref{tab:tpc-components} lists the principal detection elements of ProtoDUNE-SP along with their approximate dimensions and their quantities. 

\begin{cdrtable}[TPC detection components, dimensions and quantities]{lll}{tpc-components}{TPC detection components, dimensions and quantities}
Detection Element & Approx Dimensions  & Quantity   \\  \toprowrule
APA          & 6~m H by 2.4~m W  & 3 per anode plane, 6 total  \\  \colhline
CPA module  & 2~m H by 1.2~m W  & 3 per CPA column,   \\  
  &  & 18 total in cathode plane    \\  \colhline
 Top FC module & 2.4~m W by 3.6~m along drift & 3 per top FC assembly, 6 total   \\  \colhline
 Bottom FC module & 2.4~m W by 3.6~m along drift & 3 per bottom FC assembly, 6 total   \\  \colhline
End-wall FC module & 1.5~m H by 3.6~m along drift & 4 per end-wall assembly (vertical   \\  
&  & drift volume edge), 16 total   \\  \colhline
PD module  & 2.2 m $\times$ 86 mm $\times$ 6 mm & 10 per APA, 60 total  \\ 
\end{cdrtable}

\begin{cdrfigure}[The field cages]{fc-overview}{A view of the TPC field cage and the central cathode plane (CPA). The APAs, which would be positioned at both open ends of the FC, are not shown.}
\includegraphics[width=0.6\linewidth]{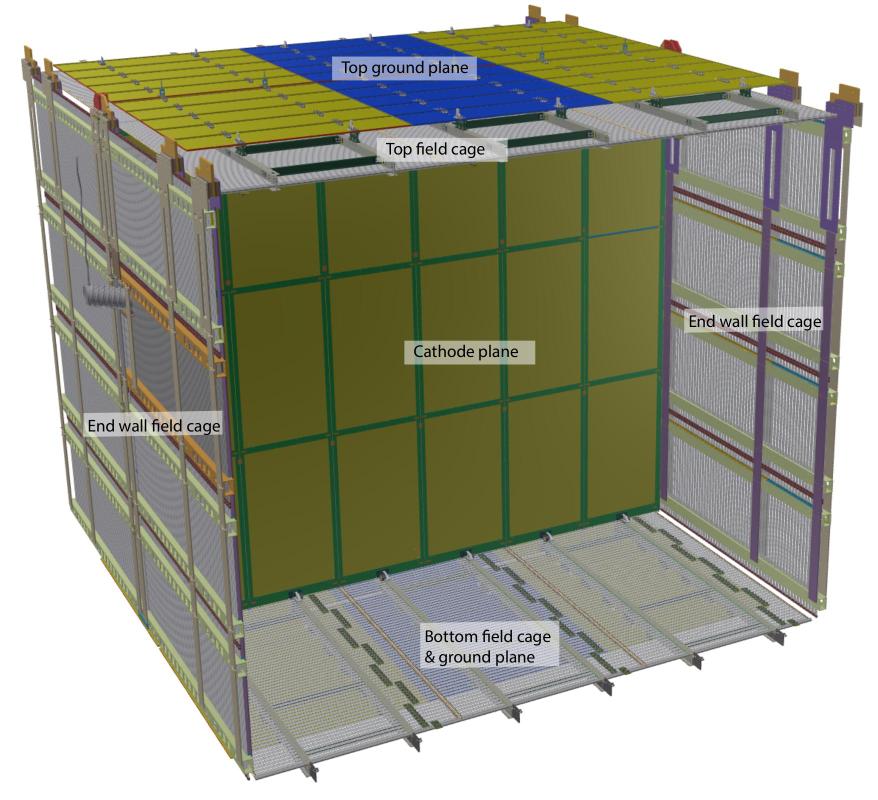}
\end{cdrfigure}


\section{Anode Plane Assemblies (APA)}

\subsection{Scope and requirements}

Anode Plane Assemblies (APAs) are the detector elements utilized to sense ionization created by charged particles traversing the liquid argon volume inside the single-phase TPC.  The scope of the APAs includes:
\begin{itemize}
\item a framework of lightweight, rectangular stainless steel tubing;
\item a mesh layer attached directly to both sides of the APA frame;
\item layers of sense and shielding wires wrapped at varying angles relative to each other;
\item stacked head electronics boards, which are wire boards for anchoring the wires at the top (head) of the APA;
\item capacitive-resistive (CR) boards that link the wire boards to the CE;
\item side and foot boards along the other three edges of the APA with notches and pins to hold the wires in place;
\item modular boxes to hold the CE;
\item comb wire supports, mounted on cross braces distributed along the length of the APA, to prevent wire deflection; and
\item pin/slot pairs on the side edges of adjacent APAs to maintain coplanarity.
\end{itemize}

The initial physics performance requirements that drive the design of the APA are listed in Table~\ref{tab:physicsrequirements}.  These are chosen to enable high-efficiency reconstruction throughout the entire active volume of the LArTPC.  

\begin{cdrtable}[Prelim physics requirements that motivate APA design parameters]{lr}{physicsrequirements}{Preliminary physics requirements that motivate APA design parameters.}   
Requirement & Value  \\ \toprowrule
MIP Identification & 100$\%$ efficiency \\ \colhline
High efficiency for charge reconstruction & $>$90$\%$ for $>$100 MeV \\ \colhline
Vertex Resolution (x,y,z) & (1.5 cm, 1.5 cm, 1.5 cm)\\ \colhline
\textbf{Particle Identification} & \\ 
Muon Momentum Resolution & $<$18$\%$ for non-contained \\
            & $<$5$\%$ for contained\\ 
Muon Angular Resolution & $<$1$^{\circ}$\\            
Stopping Hadrons Energy Resolution & 1-5$\%$\\
Hadron Angular Resolution & $<$10$^{\circ}$ \\ \colhline
\textbf{Shower identification} & \\
Electron efficiency & $>$90$\%$\\
Photon mis-identification & $<$1$\%$\\
Electron Angular Resolution & $<$1$^{\circ}$ \\
Electron Energy Scale Uncertainty & $<$5$\%$\\
\end{cdrtable}


The ability to identify minimum-ionizing particles (MIPs) is a function of several detector parameters, including argon purity, drift distance, diffusion, wire pitch, and Equivalent Noise Charge (ENC).  It is required that MIPs originating anywhere inside the active volume of the detector be reconstructed with 100$\%$ efficiency.   The choice of wire pitch combined with the design values of the other high-level parameters, listed in Table~\ref{tab:apaparameters},  is expected to enable  this  efficiency.

The fine wire spacing of the APA enables excellent precision in identifying the location of any vertices in an event (e.g., the primary vertex in a neutrino interaction, or gamma conversion points in a $\pi^{0}$ decay), which has a direct impact on reconstruction efficiency. It is required to reach a vertex resolution of $\sim$1.5 cm along each coordinate direction.  In practice, the resolution on the drift-coordinate ($x$) of a vertex or hit will be better than that on its location in the $y$-$z$ plane, due to the combination of drift-velocity and electronics sampling-rate uncertainties.

\subsection{APA design overview}
\label{sec:apa-design-overview}

An APA is constructed from a framework of lightweight, rectangular stainless steel tubing, with a fine mesh covering the rectangular area within the frame, on both sides, that defines a uniform ground across the frame. Along the length of the frame and around it, over the mesh layer, layers of sense and shielding wires are strung or wrapped at varying angles relative to each other, as illustrated in  Figure~\ref{fig:tpc_apa1}. The wires are terminated on  boards that anchor them and also provide the connections to the cold electronics. The APAs are 2.3\,m wide, 6.3\,m high, and 12\,cm thick.  
The size of the APAs is chosen for fabrication purposes, compatibility with over-the-road shipping, and for eventual transport to the 4850 level at SURF and installation into the membrane cryostat of a detector module. Sufficient shock absorption and clearances are taken into account at each stage.  The dimensions are also chosen such that an integral number of electronic readout channels and boards fill in the full area of the APA. The modularity of the APAs allows them to be built and tested at off-site production facilities, decoupling their manufacturing time from the construction of the membrane cryostat. 
As mentioned above, the principal design parameters are listed in Table~\ref{tab:apaparameters}.

\begin{cdrtable}[APA design parameters]{lr}{apaparameters}{APA design parameters}   
Parameter & Value  \\ \toprowrule
Active Height & 5.984 m\\ \colhline
Active Width & 2.300 m\\ \colhline
Wire Pitch (U,V) & 4.669 mm\\ \colhline
Wire Pitch (X,G) & 4.790 mm\\ \colhline
Wire Position Tolerance & 0.5 mm \\ \colhline
Wire Plane Spacing & 4.75 mm\\ \colhline
Wire Angle (w.r.t. vertical) (U,V) & 35.7$^{\circ}$\\ \colhline
Wire Angle (w.r.t. vertical) (X,G) & 0$^{\circ}$\\ \colhline
Number Wires / APA & 960 (X), 960 (G), 800 (U), 800 (V) \\ \colhline
Number Electronic Channels / APA & 2560 \\ \colhline
Wire Tension & 5.0 N \\ \colhline
Wire Material & Beryllium Copper \\ \colhline
Wire Diameter & 150 $\mu$m \\ \colhline
Wire Resistivity & 7.68 $\mu\Omega$-cm $@$ 20$^{\circ}$ C \\ \colhline
Wire Resistance/m & 4.4 $\Omega$/m $@$ 20$^{\circ}$ C \\ \colhline
Frame Planarity & 5 mm \\ \colhline
Photon Detector Slots & 10 \\
\end{cdrtable}

Starting from the outermost wire layer, 
there is first a shielding (grid) plane, followed by two induction planes, and finally the collection plane. All wire layers span the entire height of the APA frame. The layout of the wire layers is illustrated in  Figure~\ref{fig:tpc_apa1}.

\begin{cdrfigure}[APA diagram]{tpc_apa1}{Sketch of a ProtoDUNE-SP APA. This shows only portions of each of the three wire layers, U (green), V (magenta), the induction layers; and X (blue), the collection layer, to accentuate their angular relationships to the frame and to each other.  The induction layers are connected electrically across both sides of the APA.  The grid layer (G) wires (not shown), run vertically, parallel to the X layer wires;  separate sets of G and X wires are strung on the two sides of the APA.  The mesh is not shown.}
\includegraphics[width=0.8\textwidth, angle=90]{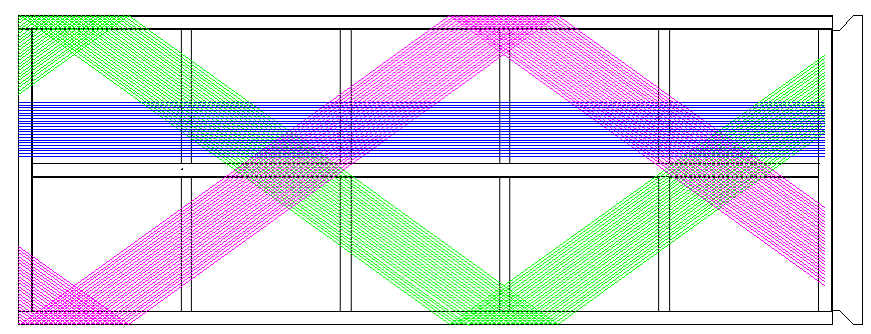} 
\end{cdrfigure}

The angle of the induction planes in the APA ($\pm$35.7$^{\circ}$) is chosen such that each induction wire only crosses a given collection wire one time, reducing the ambiguities that the reconstruction must address.  The design angle of the induction wires, coupled with their pitch, was also chosen such that an integer multiple of electronics boards reads out one APA.

The wires of the grid (shielding) layer, G,  are not connected to the electronic readout; the wires run parallel to the long edge of the APA frame; there are separate sets of G wires on the two sides of the APA. 
 The two planes of induction wires (U and V) wrap in a helical fashion around the long edge of the APA, continuously around both sides of the APA.  The collection plane wires (X) run vertically, parallel to G.   The ordering of the layers, from the outside in, is G-U-V-X, followed by the mesh.   

The operating voltages of the APA layers are listed in Table~\ref{tab:bias}.  When operated at these voltages, the drifting ionization follows trajectories around the grid and induction wires, ultimately terminating on a collection plane wire; i.e., the grid and induction layers are completely transparent to drifting ionization, and the collection plane is completely opaque.  The grid layer is present for pulse-shaping purposes, effectively shielding the first induction plane from the drifting charge and removing the long leading edge from the signals on that layer; again, it is not connected to the electronics readout. The mesh layer serves to shield the sense planes from pickup from the Photon Detection System and from ``ghost'' tracks that would otherwise be visible when ionizing particles have a trajectory that passes through the collection plane. 

\begin{cdrtable}[Baseline bias voltages for APA wire layers]{lr}{bias}{Baseline bias voltages for APA wire layers}   
Anode Plane & Bias Voltage  \\ \toprowrule
Grid (G) & -665 V\\ \colhline
Induction (U) & -370 V\\ \colhline
Induction (V) & 0 V\\ \colhline
Collection (X) & 820 V\\ \colhline
Mesh (M) & 0 V\\
\end{cdrtable}

The wrapped style allows the APA plane to fully cover the active area of the LArTPC, minimizing the amount of dead space between the APAs that would otherwise be occupied by electronics and associated cabling.   

In the current design of the DUNE-SP far detector module, a central row of APAs is flanked by  drift-fields, requiring sensitivity on both sides. The wrapped APAs allow the induction plane wires to sense drifting ionization originating from either side of the APA.  This double-sided feature is not strictly necessary for the ProtoDUNE-SP arrangement, which has APAs located against the cryostat walls and a drift field on one side only, but it is compatible with this setup as the grid layer facing the wall effectively blocks any ionization generated outside the TPC from drifting in to the wires on that side of the APA.

The choices of wire tension and wire placement accuracy are made to ensure proper operation of the LArTPC at voltage, and to provide the precision necessary for reconstruction.  The tension of 5~N, when combined with the intermediate support combs (described in Section~\ref{subsec:apa_combs}) ensure that the wires are held taught in place with no sag.  Wire sag can impact the precision of reconstruction, as well as the transparency of the TPC.  The tension of 5~N is low enough that when the wires are cooled, which increases their tension due to thermal contraction, they will stay safely below the break load of the beryllium copper wire, as described in Section~\ref{subsec:apa_wires}.  To further mitigate wire breakage and its impact on detector performance, each wire in the APA is anchored twice on both ends, with both solder and epoxy.  

\subsection{Wire properties}
\label{subsec:apa_wires}

Beryllium copper (CuBe) wire is known for its high durability and yield strength. It is composed of $\sim$98$\%$ copper, 1.9$\%$ beryllium, and a negligible amount of other elements. The APA wire has a diameter of 150$\mu$m (.006~in), and is strung in varying lengths across the APA frame. Three key properties for its usage in the APA are: low resistivity, high tensile or yield strength, and coefficient of thermal expansion suitable for use with the APA's stainless steel frame.

Tensile strength of the wire describes the wire-breaking stress (see Table~\ref{tab:wire}).  The yield strength is the stress at which the wire starts to take a permanent (inelastic) deformation, and is the important limit stress for this case, though most specifications give tensile strength.  Fortunately, for the CuBe alloys of interest, the two are fairly close to each other.  Based on the tensile strength of wire purchased from Little Falls Alloy (over 1,380~MPa or 200,000~psi), the yield strength is greater than 1,100~MPa.  Given that the stress while in use is around 280~MPa, this leaves a comfortable margin.

The coefficient of thermal expansion (CTE) describes how material expands and contracts with changes in temperature.  The CTEs of CuBe alloy and 304 stainless steel are very similar.  Integrated down to 87~K, they are 2.7e-3 for stainless and 2.9e-3 for CuBe~\cite{cryo-mat-db}.
Since the wire contracts slightly more than the frame during cool-down the wire tension increases.  If it starts at 5~N, the tension rises to about 5.5~N when everything is cool.  

The change in wire tension during cool-down could also be a concern.  In the worst case, the wire
 cools quickly to 87\,K before any significant cooling of the frame  -- a realistic case because of the differing thicknesses.  In the limiting case, with complete contraction of the wire and none in the frame, the tension would be expected to reach $\sim$11.7 N.  This is still well under the $\sim$20 N yield tension.
In practice, the cooling will be done gradually to avoid this tension spike as well as other thermal shock to the APA.

\begin{cdrtable}[CuBe wire tensile strength and CTE]{lr}{wire}{Tensile strength and coefficient of thermal expansion (CTE) of beryllium copper (CuBe) wire.}
Parameter & Value \\ \toprowrule
Tensile Strength (from property sheets) (psi) & 208,274 \\ \colhline
Tensile Strength (from actual wire) (psi) & 212,530 \\ \colhline
CTE of CuBe, integrated to 87 K (m/m) & 2.9e-3 \\ \colhline
CTE of 304 stainless steel, integrated to 87 K (m/m) & 2.7e-3 \\
\end{cdrtable}

\subsection{APA frame and mesh}
\label{subsec:apa_frame}


The stainless steel frame of the APA (Figure~\ref{fig:tpc_apa_frame}) is 6.06~m long, not counting electronics and mounting hardware, and 2.30~m wide.  It is 76.2~mm thick, made from imperial size 3-in $\times$ 4-in $\times$ 0.120-in wall rectangular tubing.  The cross pieces have a cross-sectional area of 2\,in $\times$ 3\,in, and are connected to edge pieces using joints, as in Figure~\ref{fig:tpc_apa_boltedjointdrawing}.  It is mounted in the cryostat with its long axis vertical; multiple APAs are mounted edge-to-edge to form a continuous plane. An electron deflection technique, described in Section~\ref{sec:apa:electrondiverter}, is used to ensure that electrons drawn towards a joint between two APAs will be deflected to one or the other, and not lost.

\begin{cdrfigure}[APA dimensions]{tpc_apa_frame}{An APA showing overall dimensions and main components. }
\includegraphics[width=0.9\textwidth]{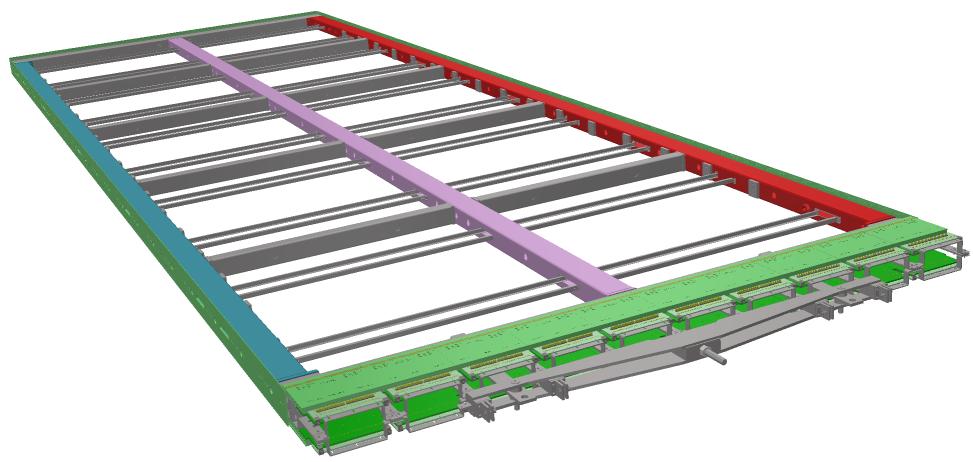} 
\end{cdrfigure}

\begin{cdrfigure}[APA bolted joint drawing]{tpc_apa_boltedjointdrawing}{A model of the bolted joint.  The holes on the top of the tube are for access to tighten the screws.  The heads actually tighten against the lower hole, inside the tube.}
\includegraphics[width=0.4\textwidth]{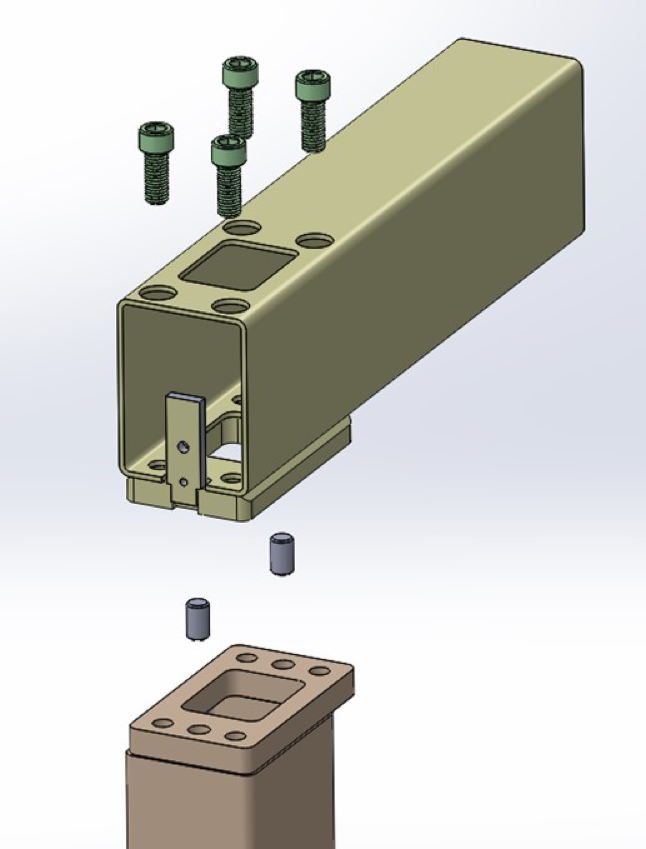} 
\end{cdrfigure}

A fine mesh screen is glued directly to the steel frame surface, over the frame on both sides.  It creates a uniform ground layer beneath the wire planes.  

The mesh is clamped around the perimeter of the opening and then pulled tight (by opening and closing clamps as needed during the process).  When the mesh is taut, a 25-mm-wide strip is masked off around the opening and glue is applied through the mesh to attach it to the steel.  Although measurements have shown that this gives good electrical contact between the mesh and the frame, a deliberate electrical connection is also made.  Figure~\ref{fig:tpc_apa_fullsizemeshdrawing} depicts the mesh application setup for a full-size ProtoDUNE-SP APA.

\begin{cdrfigure}[APA full-size mesh drawing]{tpc_apa_fullsizemeshdrawing}{The mesh clamping jig for the full size APA. }
\includegraphics[width=0.8\textwidth]{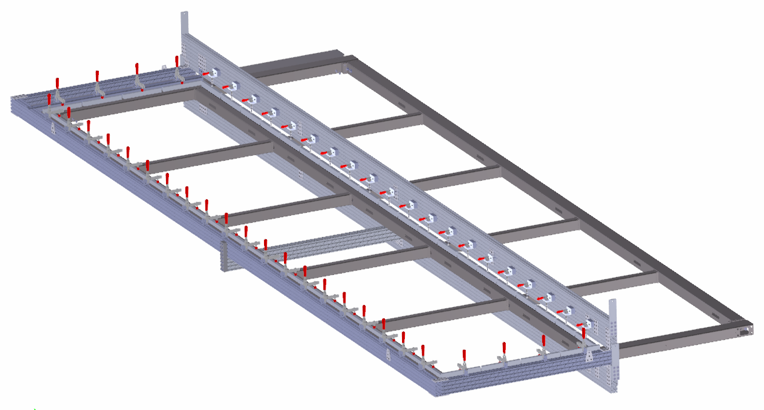} 
\end{cdrfigure}

\subsection{Anchoring elements and wire boards}
\label{subsubsec:apa_wire_anchor}

\subsubsection{Head electronics boards}

At the head end of the APA, stacks of electronics boards (referred to as ``wire boards'') are arrayed to anchor the wires.  They also provide the connection between the wires and the 
cold electronics.

All APA wires are terminated on the wire boards, which are stacked along the electronics end of the APA frame; see Figure~\ref{fig:tpc_apa_boardstack}. 
Attachment of the wire boards begins with the X plane (innermost). After the X-plane wires are strung top to bottom along each side of the APA frame, they are soldered and epoxied to their wire boards and trimmed. The remaining wire board layers are attached as each layer is wound.  The main CR boards (capacitive-resistive), which provide DC bias and AC coupling to the wires, are attached to the bottom of the wire board stack. 

\begin{cdrfigure}[APA board stack]{tpc_apa_boardstack}{Left: View of the APA wire board stack, as seen from the top/side. The wire board layers can be seen at the bottom-left of the illustration, X on the bottom (it doesn't go all the way back, but extends farther forward and has the main CR board attached), followed by U, V, then G (which doesn't go all the way forward, and has its own CR board attached). Right: the same stack viewed from below. }
\includegraphics[width=0.45\textwidth]{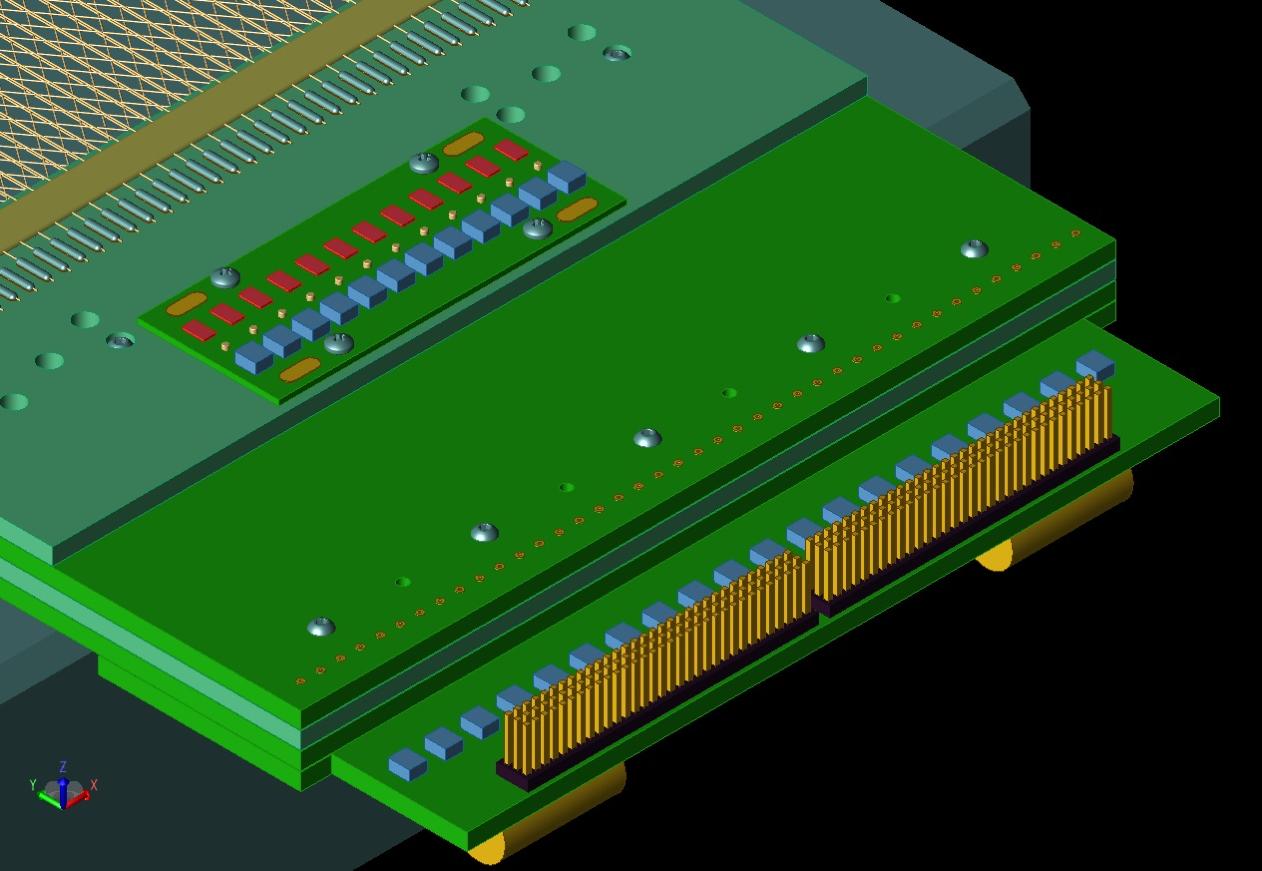}
\includegraphics[width=0.45\textwidth]{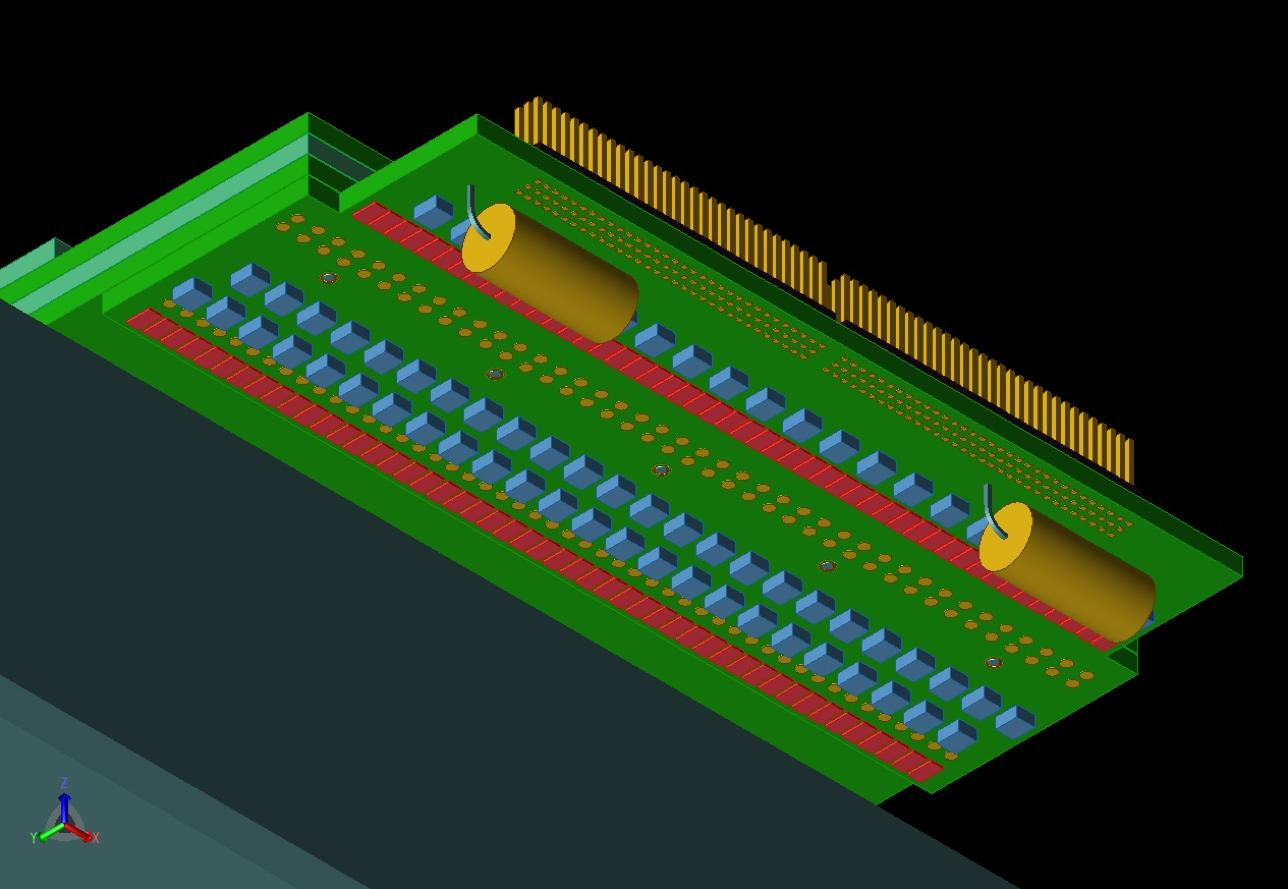}
\end{cdrfigure}


The outermost G-plane wire boards connect adjacent groups of four wires together, and bias each group through an R-C filter whose components are placed on special CR boards  
that are attached after the wire plane is strung. The X, U and V layers of wires are connected to the CE (housed in boxes mounted on the APA) either directly or through DC-blocking capacitors. The X and U planes have wires individually biased through 50-M$\Omega$ resistors. Electronic components for the X- and U-plane wires are located on a common CR board. 

Mill-Max pins and sockets provide electrical connections between circuit boards within a stack. They are pressed into the circuit boards and are not repairable if damaged. To minimize the possibility of damaged pins, the boards are designed so that the first wire board attached to the frame has only sockets. All boards attached subsequently contain pins that plug into previously mounted boards. This process eliminates exposure of any pins to possible damage during winding, soldering, or trimming processes.

Ten stacks of wire boards are installed across the width of each side along the head of the APA.  The X-layer board in each stack has room for 48 wires, the V layer has 40 wires, the U layer 40 wires and the G layer 48 wires.  Each board stack, therefore, has 176 wires but only 128 signal channels since the G wires are not read out.  
With a total of 20 stacks per APA, this results in 2,560 signal channels per APA and a total of \SI{3520} wires starting at the top of the APA and ending at the bottom.  There is a total of $\sim$23.4 km of wire on the two surfaces of each APA.  Many of the capacitors and resistors that in principle could be on these wire boards are instead
placed on the attached CR boards to improve their accessibility in case of component failure.   Figure~\ref{fig:tpc_apa_electronics_connectiondiagram} depicts the connections between the different elements of the APA electrical circuit. 

At the head end of the APA, the wire-plane spacing is set by the thickness of these wire boards.  The first layer's wires solder to the surface of the first board, the second layer's wires to the surface of the second board, and so on.  For installation, temporary toothed-edge boards beyond these wire boards align and hold the wires until they are soldered to pads on the wire boards.  After soldering, the extra wire is snipped off. 

\begin{cdrfigure}[APA wire board connection to electronics]{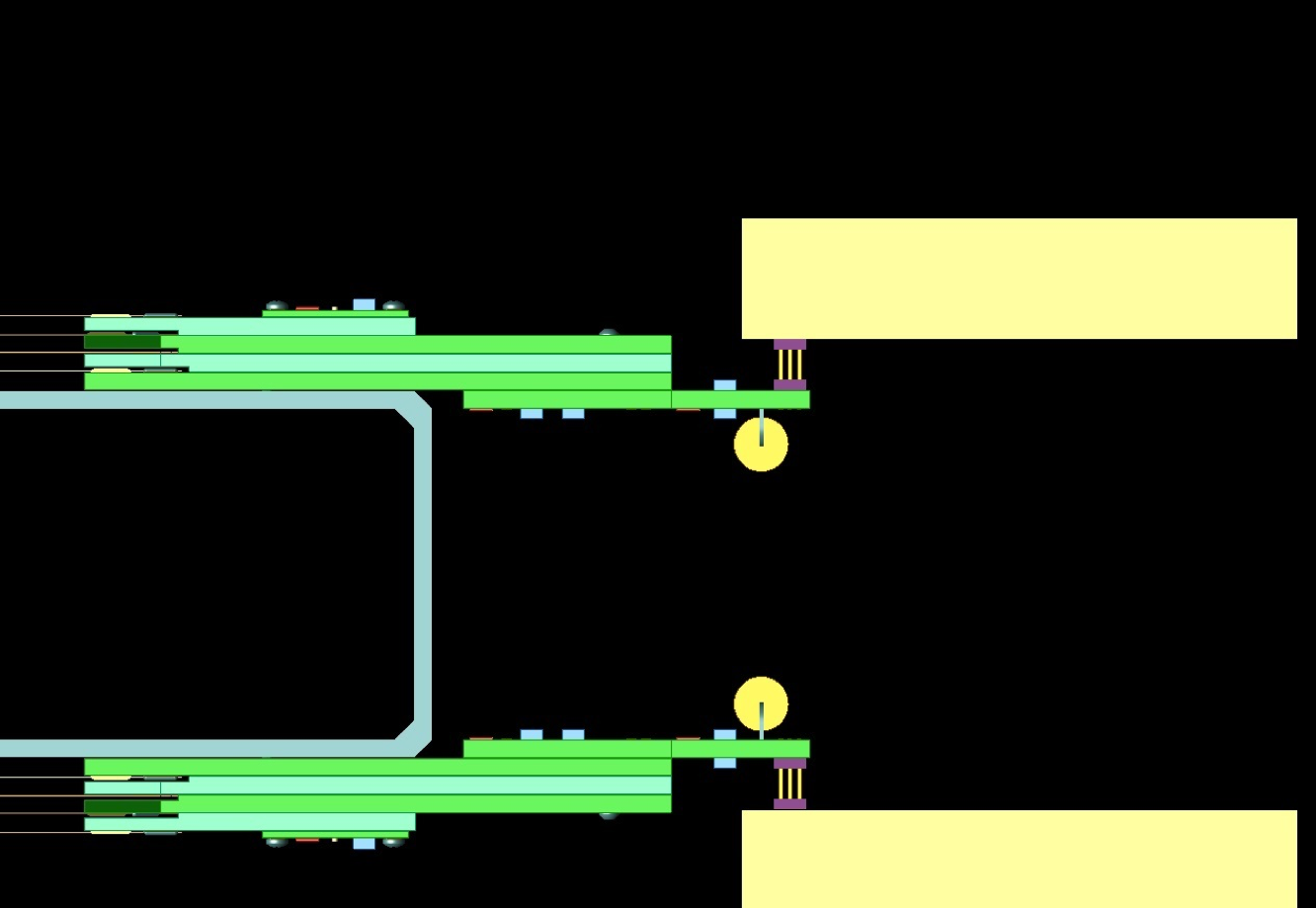}{Diagram of the connection between the APA wires, viewed from the APA edge. The set of wire boards within a stack can be seen on both sides of the APA, with the CR board extending further to the right, providing a connection to the cold electronics, which are housed in the boxes at the far right of the figure. }
\includegraphics[width=0.7\textwidth]{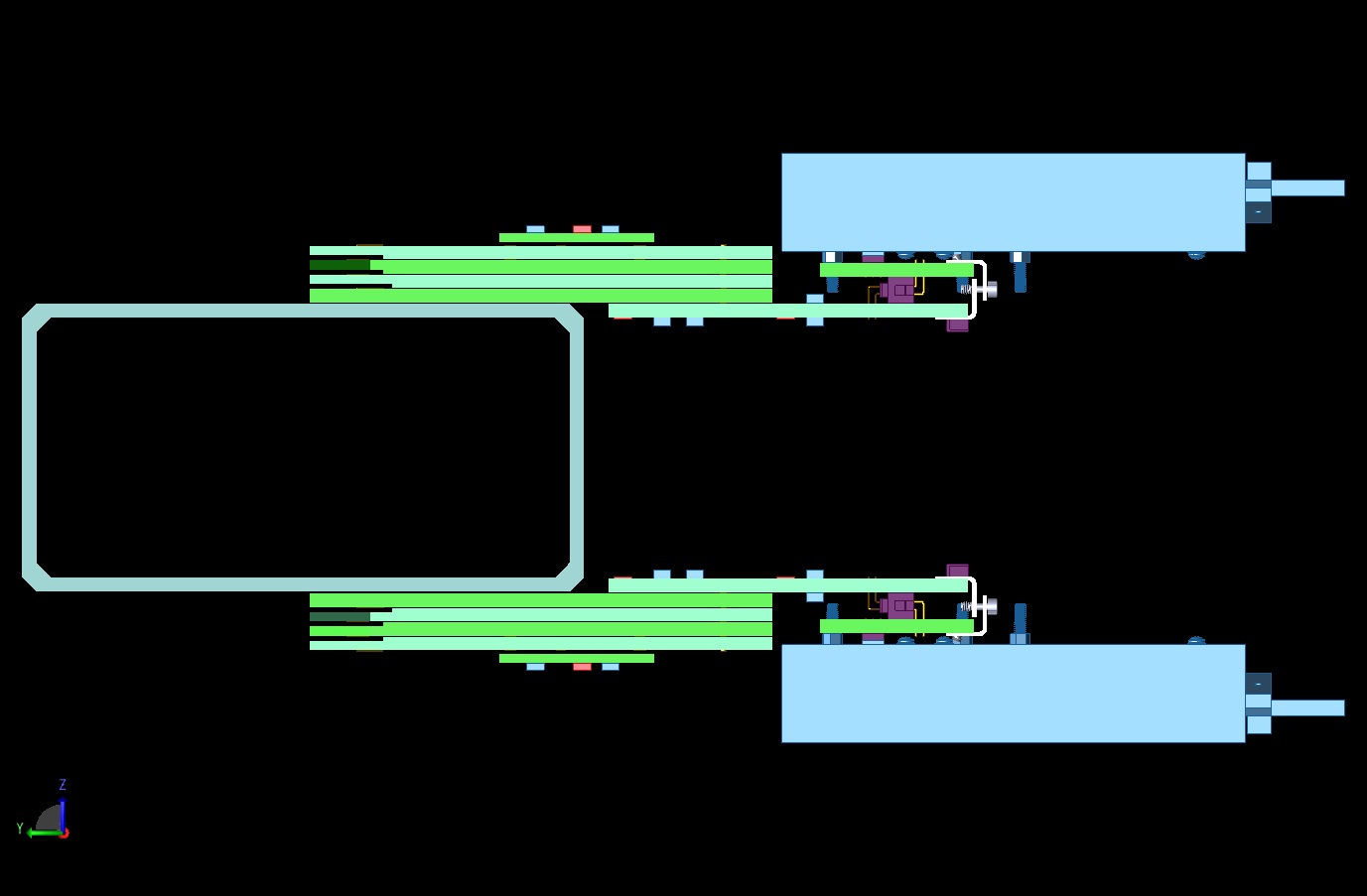}
\end{cdrfigure}

\subsubsection{CR boards}
\label{sec:crboards}

The CR boards carry a bias resistor and a DC-blocking capacitor for each wire in the X and U planes. These boards are attached to the board stacks after fabrication of all wire planes.  Electrical connections to the board stack are made though Mill-Max pins that plug into the wire boards. Connections from the CR boards to the CE are made through a pair of 96-pin Samtec connectors.

Surface-mount bias resistors on the CR boards have resistance of 50\,M$\Omega$ are constructed with a thick film on a ceramic substrate. Rated for 2.0-kV operation, the resistors measure 0.12 $\times$ 0.24 inches. Other ratings include operation from $-$55 to +155 C, 5\% tolerance, and a 100-ppm/C temperature coefficient.
Performance of these resistors at LAr temperature is verified through additional bench testing.

The selected DC-blocking capacitors have capacitance of 3.9\,nF and are rated for 2.0-kV operation. Measuring 0.22 $\times$ 0.25\,inches across and 0.10\,inches high, the capacitors feature flexible terminals to comply with PC board expansion and contraction. They are designed to withstand 1,000 thermal cycles  
between the extremes of the operating temperature range. Tolerance is also 5\%.


In addition to the bias and DC-blocking capacitors for all X- and U-plane wires, the CR board includes two R-C filters for the bias voltages. The resistors are of the same type used for wire biasing except with a resistance of 2\,M$\Omega$. Capacitors are 47\,nF at 2\,kV. Very few choices exist for surface-mount capacitors of this type, and they are exceptionally large. 
Polyester or Polypropylene film capacitors that are known to perform well at cryogenic temperatures are used.

\subsubsection{Side and foot boards}

The boards along the sides and foot of the APA have notches, pins and other location features to hold the wires in the correct position as they wrap around the edge from one side of the APA to the other.

G10 circuit board material is ideal for these side and foot boards due to its physical properties alone, but it has an additional advantage: a number of hole or slot features in the edge boards provide access to the underlying frame.  In order that these openings are not covered by wires, the sections of wire that would go over the openings are replaced by traces on the boards.  After the wires are wrapped, the wires over the opening are soldered to pads at the ends of the traces, and the section of wire between the pads is snipped out (Figure~\ref{fig:tpc_apa_sideboardmodel}).  These traces are easily and economically added to the boards by the many commercial fabricators who make circuit boards. 

\begin{cdrfigure}[APA side board model]{tpc_apa_sideboardmodel}{Model of board with wires showing how traces connect wires around openings in the side boards.  The wires are wound straight over the openings, then soldered to pads at the ends of the traces.  After soldering the sections between the pads are trimmed away.}
\includegraphics[width=0.9\textwidth]{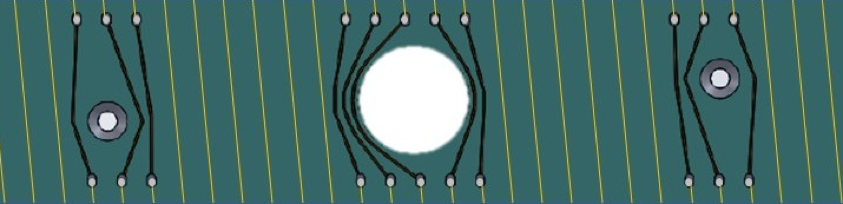} 
\end{cdrfigure}

\begin{cdrfigure}[APA side board photo]{tpc_apa_sideboardphoto}{Boards with injection molded tooth strips glued on.  The left shows an end board with teeth for fixing the position of the longitudinal wires.  The teeth there form small notches. The right is a side board for fixing the position of the angled wires where the wires are angled around a pin. (These boards are prototype test pieces and are not used in the production APAs.)}
\includegraphics[width=0.7\textwidth]{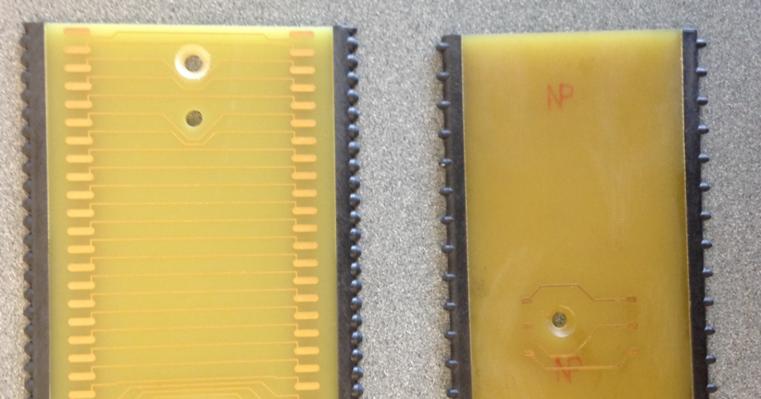} 
\end{cdrfigure}

The placement of the angled wires are fixed by pins 
as shown in the right-hand picture of Figure~\ref{fig:tpc_apa_sideboardphoto}.  The wires make a partial wrap around the pin as they change direction from the face of the APA to the edge.  The X- and G-plane wires are not pulled to the side so they cannot be pulled against a pin.  Their positions are fixed 
by teeth with slots, as shown in the left-hand picture in Figure~\ref{fig:tpc_apa_sideboardphoto}. 
	
The polymer used for the strips is Vectra e130i (a trade name for 30$\%$ glass filled liquid crystal polymer or LCP). It retains its strength at cryogenic temperature and has a CTE similar enough to G10 that differential expansion/contraction is not a problem.

\subsubsection{Glue and solder}
The ends of the wires are soldered to pads on the edges of the wire boards.  Solder provides both an electrical connection and a physical anchor to the wires.  As an additional physical anchor, roughly 10~mm of the wires are glued near the solder pads.  For example, in Figure~\ref{fig:tpc_apa_sideboardphoto}, in addition to soldering the wires on the pads shown in the left-hand photograph, an epoxy bead is applied on the wires in the area between the solder pads and the injection-molded tooth strips.

Gray epoxy 2216 by 3M is used for the glue.  It is strong, widely used (therefore much data is available), and it retains good properties at cryogenic temperatures.  A 62$\%$ tin, 36$\%$ lead and 2$\%$ silver solder was chosen.  A eutectic mix (63/37) is the best of the straight tin/lead solders but the 2$\%$ added silver gives better creep resistance.

\subsubsection{Comb wire supports on inner frame members}
\label{subsec:apa_combs}


Some wire segments are quite long; for instance, the X- and G-plane wires 
extend from one end of the APA to the other without going around a side -- a length of 6\,m.  Even the diagonal wires across the middle of the APA are 3.9\,m long.  To prevent deflection from gravity, electrostatic forces, or liquid drag from moving argon, the wires are supported at regular locations along the length of the APA.  This is done with \textit{combs} mounted on each of the four cross braces that are  
regularly spaced along the length of the APA.  This keeps the longest unsupported wire length under 1.6\,m.

The nominal wire tension is 5\,N but even the 1.6-m-long wires could fall to 3\,N of tension before the wire, held horizontally, would deviate 150\,microns -- one wire diameter.  During operation the wires are either vertical or 35.7$^{\circ}$ from vertical, so the actual deviation would be less.


The combs are made from 0.5-mm-thick G10 with slots cut into it.  The comb for the lowest layer is glued to a base strip that is glued to the frame.  After each layer is wound, another comb strip is glued to the tips of the teeth of the previous one to position the wires in the next layer.  Each successive comb holds the previous layer of wires in the bottom of its slots (Figure~\ref{fig:tpc_apa_supportcombmodel}).

\begin{cdrfigure}[APA support comb model]{tpc_apa_supportcombmodel}{A model of the combs showing how they stack.  After winding a layer, the comb for the next layer is put in place.  Each comb holds the wires from the previous layer in its slots.}
\includegraphics[width=0.7\textwidth]{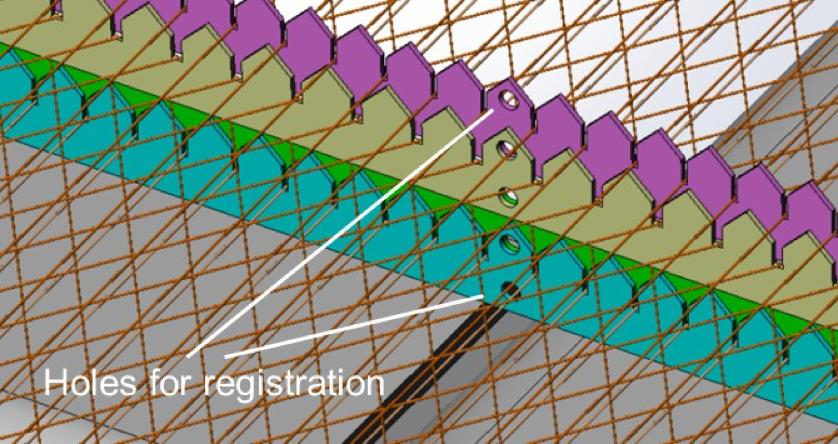} 
\end{cdrfigure}

Periodic holes along the length of the strip allow the use of pins to accurately position each successive strip with respect to the previous one.  A series of jigs is used to create and install these combs.  One jig aligns the first strip to the base strip during gluing.  Another jig locates this assembly on the frame as it is glued in place. A third jig locates each successive comb using these holes (labeled registration holes in Figure~\ref{fig:tpc_apa_supportcombmodel}).

The wire openings in the comb stack are small enough that the wires are accurately positioned at the combs, and therefore the gluing of wires into the combs is not required.

\subsubsection{Electron diverter}
\label{sec:apa:electrondiverter}
 
The active aperture of the collecting plane wires is 2300\,mm wide, while the APA modules are installed at a 2320\,mm pitch.  This leaves a 20\,mm gap between two adjacent APAs that is occupied by the stacks of wire wrapping boards and some clearance space. Electrons from the ionizing tracks drifting over the gap have an ill-defined trajectory; some get lost and some land on the wrong wire and create unexpected signal waveforms.  A set of \textit{electron diverters} installed between APAs would nudge the incoming electrons away from the gap, towards 
an active collection wire. It is planned to install electron diverters between some of the ProtoDUNE-SP APAs to determine whether they should be included in the DUNE-SP detector design.
 
The electron diverter is a set of thin printed circuit boards with two strip electrodes sticking up from the first wire plane.  The two electrodes are biased at 
negative voltages with respect to to the natural potential at their locations.  A repulsive force from these electrodes pushes the incoming electrons away 
from the inactive gap.  Figure~\ref{fig:tpc_apa_e_diverter} illustrates the principle. The left plot shows the E-field lines near the symmetry plane between two APAs, where a roughly 10-mm band of field lines land on the edge of the boards holding the wrapped wires.  The right plot shows the field lines with the electron diverter installed (at the left edge of the plot); here, all field lines are pushed to the active region.

\begin{cdrfigure}[Field maps of the region near the inactive gap of an APA]{tpc_apa_e_diverter}{Left: field map of the region near the inactive gap of an APA without the electron diverter; Right: field map with the electron diverter in place.
Electric field lines are shown in black, equipotential contours are in white, and electric field strength is represented in color gradient.}
\includegraphics[width=0.9\textwidth]{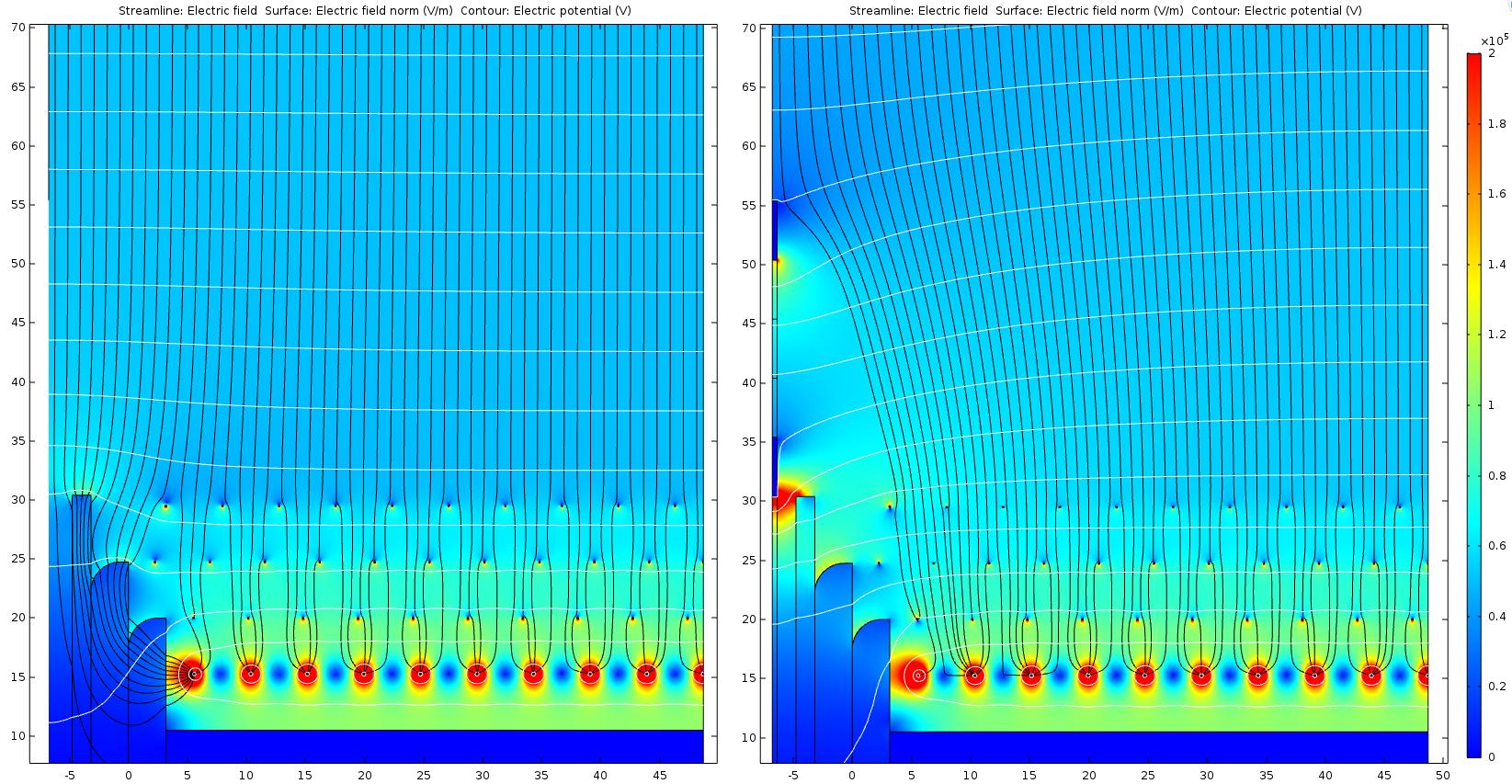} 
\end{cdrfigure}

The leading edge of the diverter board is 25\,mm above the grid plane.  The first electrode is flush with the leading edge of the board, and the second electrode just clears the grid plane.  Both of them are 5\,mm wide.  The leading electrode is biased at a nominal voltage of $-$2300\,V, and the inner is biased at -1300\,V.     Figure\ref{fig:tpc_apa_e_diverter_on_apa} shows the electron diverter boards mounted on a side of an APA.  A total of 10 boards are screwed onto the APA frame using existing tapped holes.  The topmost board has low pass RC network to filter the voltages fed to the electrodes.  These boards will be installed on two APAs on one side of the drift, leaving the other drift volume without such a feature, for comparison.
 
 \begin{cdrfigure}[Corner of APA with electron diverter boards installed]{tpc_apa_e_diverter_on_apa}{A view of a corner of an APA with a full set of electron diverter boards installed (along right-hand edge in diagram).  The first board, with its end at the top of the APA (the top of the APA is in the foreground) has the bias voltage connectors and RC filtering network (brown with blue connectors protruding).  The reddish-brown strips on the boards are the strip electrodes. The rest of the boards (delineated by blue lines) are interconnected 
 through copper wires soldered to the ends of the strip electrodes.}
\includegraphics[width=0.9\textwidth]{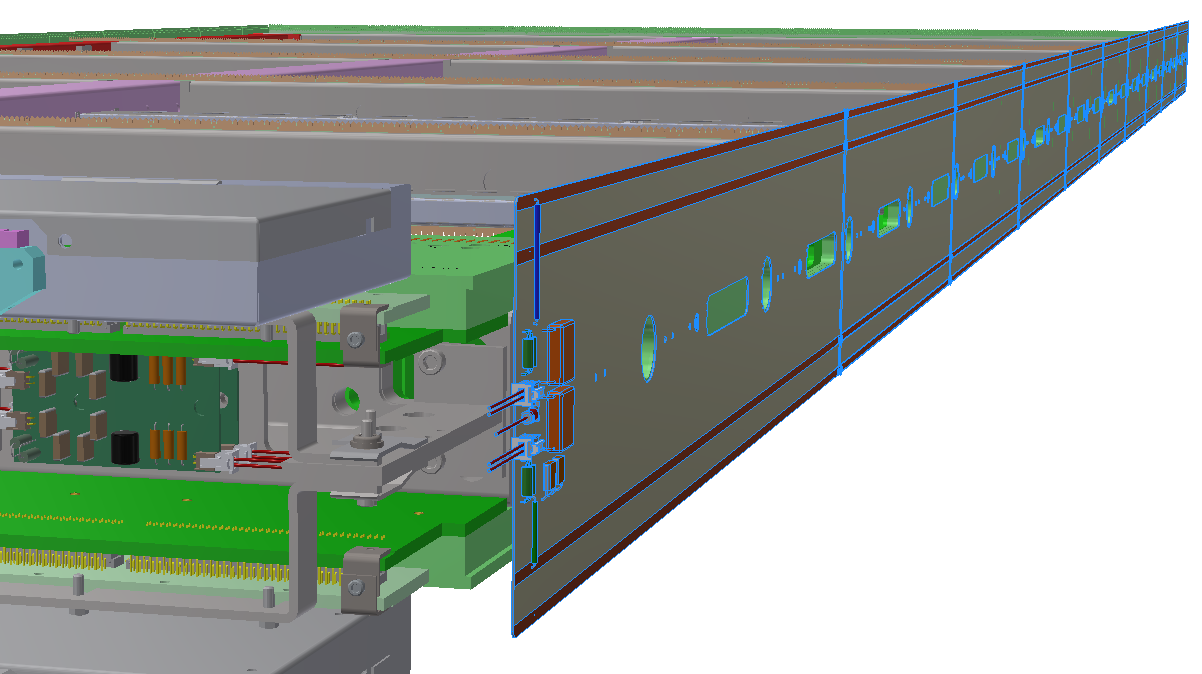} 
\end{cdrfigure}

\subsection{Interconnection features}

\subsubsection{CE boxes}

Pins extending outward from the CR boards provide connections from the APA to the modularly designed CE, and the CE modules are housed in small boxes that provide shielding.
The CE boxes are illustrated in Figure~\ref{fig:tpc_apa_electronicsmountingdiagram}. 
Each board stack has one CE box installed near it that holds the CE module for the stack.  

Putting the electronics in small boxes simplifies installation and replacement, and also helps with the dissipation 
of argon gas generated by the warm electronic components.  The CE modules are mounted in such a way that any of them can be removed from a single side of the APA after APA installation.

\begin{cdrfigure}[Solid model of modular CE boxes]{tpc_apa_electronicsmountingdiagram}{Solid model of 
modular CE boxes.}
\includegraphics[width=0.9\textwidth]{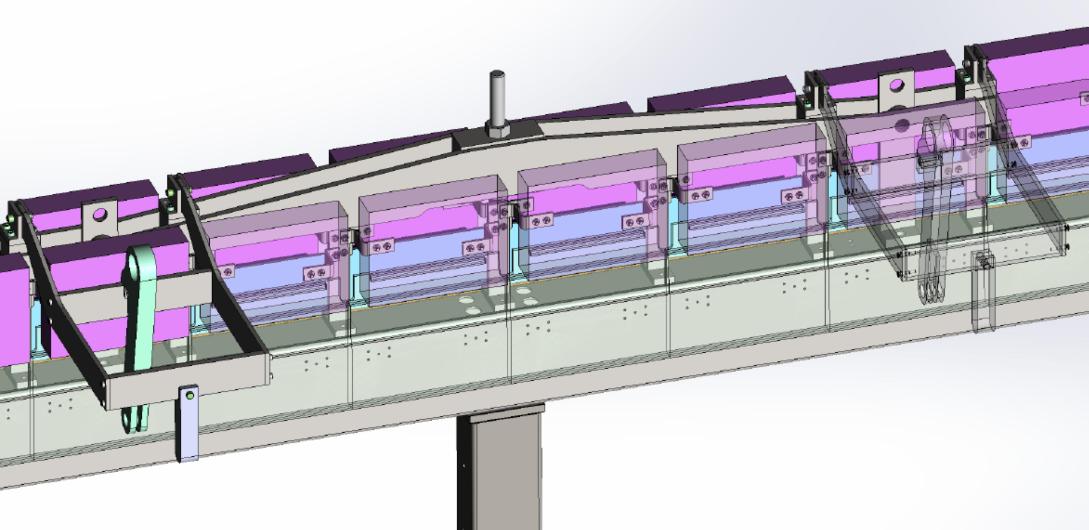} 
\end{cdrfigure}

\subsubsection{Adjacent APAs}

A constraint is needed between adjacent APAs to keep them co-planar.  It is also important that this constraint not apply a vertical load to adjacent APAs.  The constraint takes the form of 
 a pair of protruding pins on one edge of the APA (one high and one low) and a pair of matching slots 
on the edge of the adjacent APA
to engage the pins (Figure~\ref{fig:tpc_apa_pinslotdrawing}). The pins have steel cores for strength.

\begin{cdrfigure}[APA interconnect drawing]{tpc_apa_pinslotdrawing}{The pin/slot constraint.  The pin screws into an insert in the outside frame member of one APA and engages a slot in the outside frame member of the adjacent APA. An insulating sleeve surrounds the guiding pin to ensure electrical isolation between the APAs.}
\includegraphics[width=0.9\textwidth]{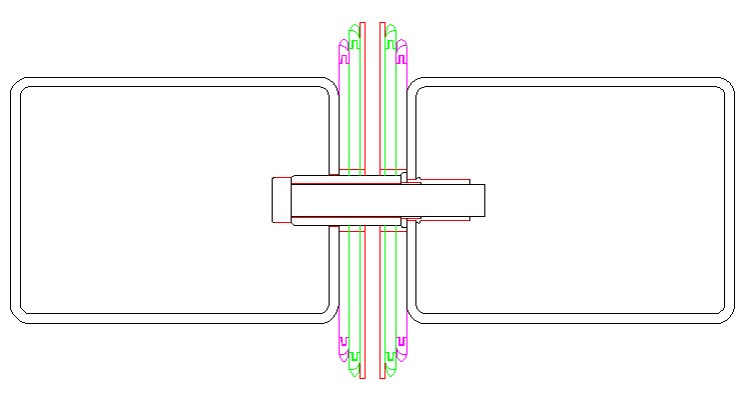} 
\end{cdrfigure}

Electronics noise concerns have made it desirable to isolate APAs from each other.  Therefore, the alignment pins have G10 sleeves covering their steel cores at places where they come into contact with the frame of the adjacent APA.



\section{Cathode Plane Assembly (CPA)}
\label{sec:cpa}

\subsection{Scope and requirements}

The cathode plane, also called the cathode plane assembly, or CPA, is located in the middle of the TPC, dividing the detector into two equal-distance drift volumes. The cathode plane's 7\,m $\times$ 6\,m area is made up of six \textit{CPA columns}, each of which is constructed of three vertically stacked \textit{CPA modules}. The CPA therefore consists of 18 CPA modules. 

The scope of the CPA includes:

\begin{itemize}
\item 18 CPA modules, each with a frame and resistive cathode panel, 
\item HV bus connecting the CPA columns and modules, and
\item HV cup for receiving input from the power supply.
\end{itemize}

The CPA plane is required to:
\begin{itemize}
\item provide equipotential surfaces at $-$180kV nominal bias voltage,
\item maintain a flatness better than 1~cm when submerged in the liquid argon,
\item be constructed of materials with comparable CTEs to that of stainless steel, 
\item limit the electric field exposed to LAr to under 30~kV/cm 
\item prevent damage to the TPC, including its readout electronics, in case of a HV discharge anywhere on the cathode,
\item provide constant bias voltage and current to all attached field cage (FC) resistor divider chains,
\item support the full weight of the four connected top/bottom FC modules plus a person on the bottom FC during installation,
\item accommodate cryostat roof movement between warm and LAr-filled states,
\item be constructed in a modular form that can be easily installed in the cryostat,
\item accommodate Photon Detection System (PDS) calibration features, and
\item avoid any trapped volume of LAr.
\end{itemize}

\subsection{Design considerations}

For the future DUNE-SP far detector, the cathode planes are planned to be 12\,m tall by nearly 60\,m long.  When biased to the nominal voltage of $-$180\,kV, each of these cathode planes stores more than 100\,J of energy. If this energy were to be released  suddenly and completely in a high-voltage discharge event, it could greatly affect the integrity of the detector elements, including the sensitive front-end electronics.  
Study has shown~\cite{cathode-hv-1320} that a cathode plane made of interconnected metallic electrodes would present significant risk to the front-end ASICs based on the potential charge injection through the capacitive coupling between the cathode and the anode wires originating from such an event.  

To address this issue in ProtoDUNE, the entire cathode plane is made out of highly resistive material such that it has a very long discharge time constant.  In the event of HV breakdown at a given location on the cathode plane, the sudden change in voltage is restricted to  a relatively localized area on the CPA module in question.  The rest of the CPA maintains its original bias voltage, and gradually discharges to ground through the large resistivity of the cathode material.  This greatly reduces the instantaneous charge injection to the front-end electronics.

\subsection{CPA design}

\subsubsection{CPA modules}


The cathode plane design chosen for the ProtoDUNE-SP TPC is an array of 18 moderately sized modules constructed from strong 6-cm-thick FR4 (the fire-retardant version of G10) frames. The frames hold 3-mm-thick FR4 panels laminated on both sides with a commercial resistive Kapton film.   Each CPA module is 1.16\,m wide and 2\,m high, and they stack to form six CPA columns of height 6\,m.  The six-column-wide CPA has the same dimensions as each of the two APA planes, with a width of about 7\,m.

 The surface of the frame facing the APAs is covered by a set of resistive FR4 strips with a different bias voltage, set such that 
 the frame itself causes no distortion in the drift field beyond the resistive surfaces.

Each CPA column is suspended under the cathode support rail by a single insulating FR4 bar.  On the top and bottom edges of a CPA, two hinges support the partial weight of the top and bottom field cage modules.   Adjacent CPAs are aligned through pin-and-slot connections to maintain co-planarity while allowing minor relative vertical shifts due to cryostat roof movement.

The electrical connectivity of the resistive panels within a CPA column is maintained by several tabs through the edge frames.  There is no direct electrical connection across the CPA columns. 
Instead, the voltage is passed from one column to another through embedded cables in the CPA panels referred to collectively as the HV bus.  Redundant connections in the HV bus between CPA columns are used to ensure reliability.  
The HV bus also provides a low-resistance path for the voltage needed to operate the FC resistive divider chains.
The required connections to the FC modules are made at the edges of the cathode plane. 
Along the perimeter of the CPA, the HV bus cables are hidden between the field-shaping strip overhang and the main cathode resistive sheet.  The cables must be capable of withstanding the full cathode bias voltage to prevent direct arcing to (and as a result, the recharging of) a CPA panel that discharges to ground. 

The outer edges of the cathode plane facing the cryostat wall are populated with the same metal profiles, with insulating polyethylene caps, as used in the field cage.  This eliminates the need for a special design of the most crucial regions of the cathode plane: the edges of the CPA now look just like a continuation of the FC.  Since these profiles are the only objects facing grounded surfaces, they are the most likely candidates to have HV discharges to ground.   To limit peak current flow, these edge profiles are resistively connected to the field shaping strips.


The resistive surface concept is illustrated in Figure~\ref{fig:cpa-concept}.

\begin{cdrfigure}[Resistive surface CPA concept]{cpa-concept}{The resistive surface CPA concept showing  
 a 3D model of a corner of the cathode with major components.} 
\includegraphics[width=\linewidth]{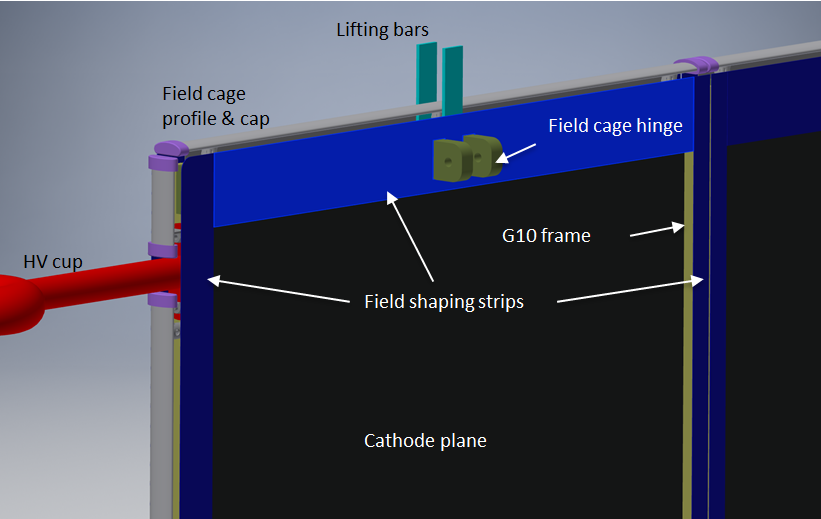}
\end{cdrfigure}

The CPA is connected to the HV feedthrough through a receptacle, called the \textit{HV cup}, at the back end of the cryostat (with respect to the beam entrance) and biased at $-$180\,kV.   It provides this voltage and the required current to all the FC modules (top, bottom and end walls) through electrical interconnects (Section~\ref{detcompsec-fc}). 



The main criteria for the selection of Kapton as the resistive material to be used to coat the CPA module panels are: 
\begin{itemize}
\item surface resistivity range,
\item compatibility with cryogenic temperatures,
\item robustness to HV discharges, 
\item material ageing,
\item radio-purity,
\item ability to coat a large surface area, and 
\item flatness, per the cathode flatness requirement. 
\end{itemize}


Figures~\ref{fig:cpa-geometry} and~\ref{fig:cpa-view2} show the basic geometry of the cathode plane. Figure~\ref{fig:cpa-view2}a shows the block at the top of a CPA that is secured to the top cross bar and extends to the top supporting I-beam.  This block must support the weight of four half FCs (4 $\times$ 150~lbs) and the weight of the CPA itself (160~lbs) for a total weight of approximately 750~lbs.  Design analysis was done with earlier, heavier FCs (220~lbs). Figure~\ref{fig:cpa-hinge1} shows how the FC is attached to the assembled cathode plane. 

\begin{cdrfigure}[CPA geometry]{cpa-geometry}{Basic geometry of the CPA array, close ups and a CPA column (on its side)} 
\includegraphics[width=\linewidth]{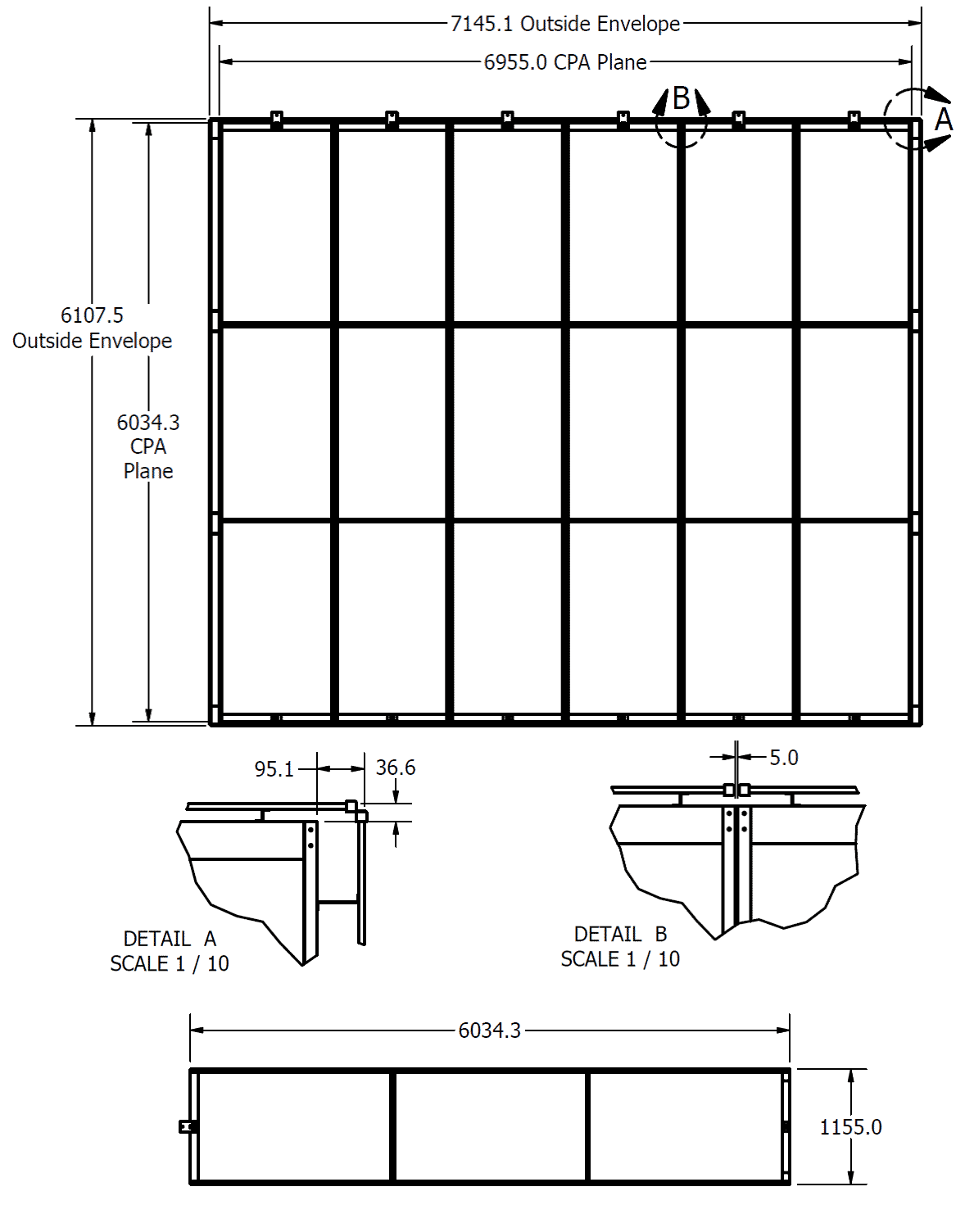}
\end{cdrfigure}

\begin{cdrfigure}[Views of various parts of a CPA]{cpa-view2}{Views of various parts of a CPA. Top: the block at the top of a CPA. Middle: hardware connecting two vertically stacked CPA modules. Bottom: connection between two adjacent CPA columns.} 
\includegraphics[width=0.8\linewidth]{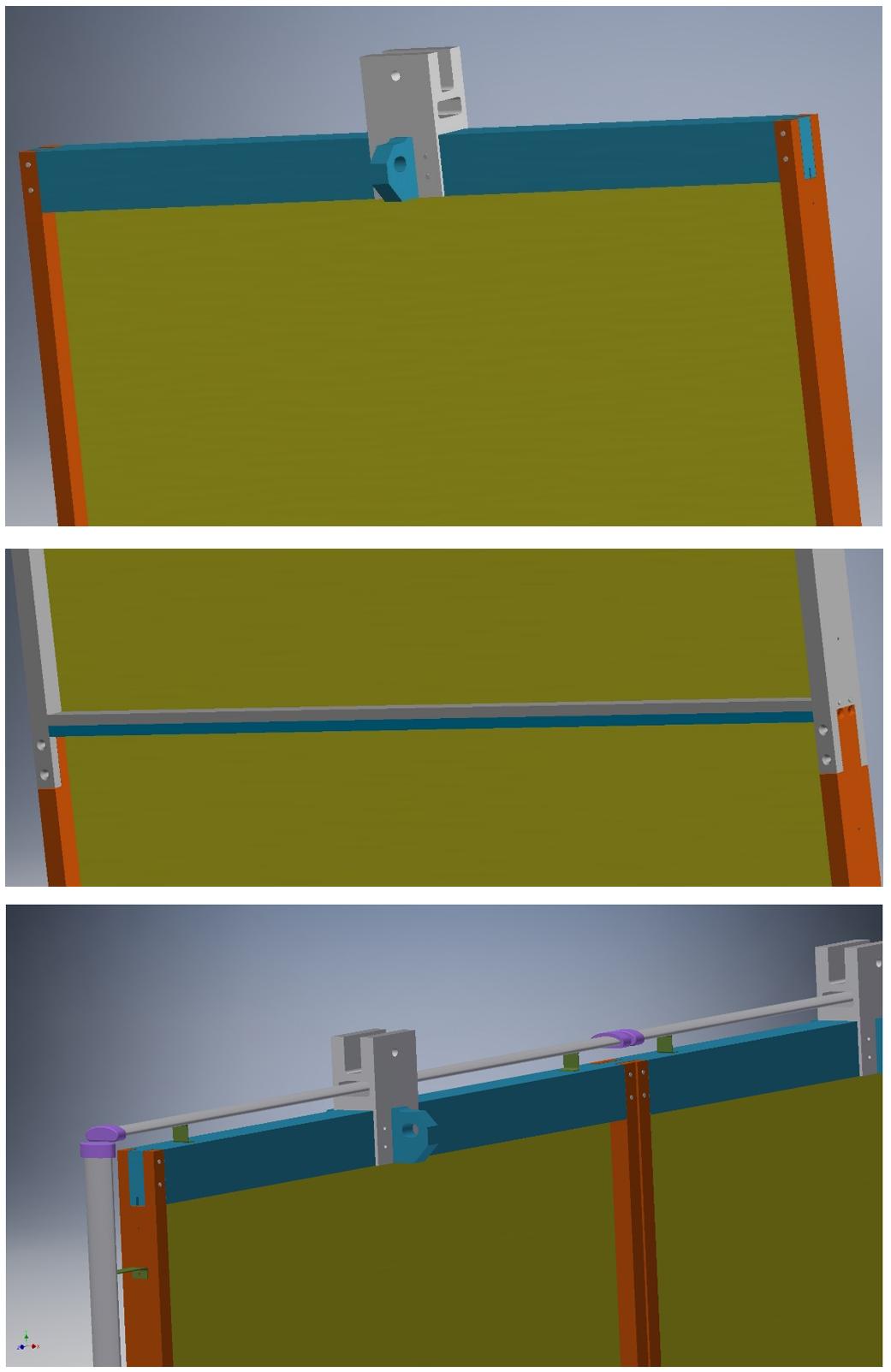}
\end{cdrfigure}

\begin{cdrfigure}[Hinged connection between CPA and FC module]{cpa-hinge1}{Hinged connection between CPA and FC module; the top field cage modules are hung vertically with the CPAs when moved into the cryostat, then rotated to horizontal to attach to the APA.} 
\includegraphics[width=\linewidth]{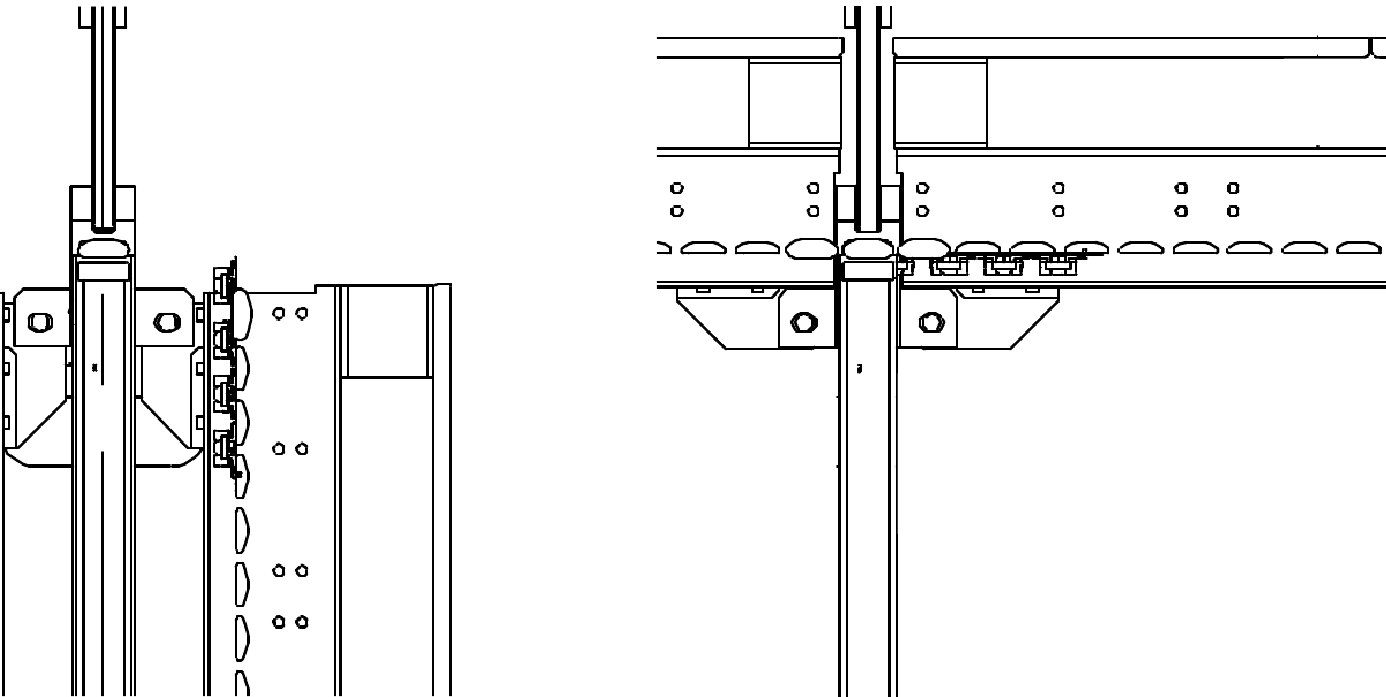}
\end{cdrfigure}

\textit{Deformation and stress due to pressure from circulating LAr}

Calculations indicate that a uniform 2-Pa pressure 
  applied to the resistive panels  during cooldown will result in 0.090~inch deflections of the panel at its center.  The entire TPC (i.e., the CPA/FC/APA assembly) will displace 8.8~mm laterally as a result of the net force from this pressure.  

\textit{Thermal considerations}

When the CPA modules are cooled, their width will shrink by 0.9~mm.  The supporting stainless steel beam will shrink by 1.6~mm over the width of the CPA.  If the CPA supports are rigidly attached to the supporting stainless steel beam, then an interference of 0.7~mm (the difference) will occur.  To prevent this interference and ensure contact between CPAs after cooldown, an initial gap of 0.7~mm between CPAs is required.  

The steel beam between the CPA and APA will shrink by 5.2~mm relative to the field cage length when cooled to LAr temperature.  The joint between the FC and the CPA is designed to accommodate this shrinkage.

\subsection{Mechanical and electrical interconnections between modules}

Three modules are stacked vertically to form the 6-m height of a CPA. 
The frames of these modules are bolted together using tongue-and-groove connections at the ends. The resistive cathode sheets and the field-shaping strips are connected using metallic tabs to ensure redundant electrical contact between the CPAs. 

Each CPA is suspended from the cathode rail using a central lifting bar.  Due to the roof movement between the warm and cold phases of the cryostat as it is cooled, each CPA is expected to move $\sim$2~mm relative to its neighbors.  Several pin-and-slot connections are implemented at the long edges of the CPA columns to ensure the co-planarity of the modules while allowing for a small vertical displacement.  

The HV bus provides the high voltage to the FC
circuits and CPA modules with a voltage drop much less than 0.1\% of the
default voltage. The location of the bus with respect to the CPA frame is shown in Figure~\ref{fig:HVbusmodel}. The field-shaping electrodes on the faces of the CPA module
frames are part of the FC circuit, described in Section~\ref{detcompsec-fc}. 
FC electrodes on the outer edges of the
CPA are held at the cathode potential to provide field
uniformity and to protect the HV bus from discharge.  

\begin{cdrfigure}[Model of the HV bus]{HVbusmodel}{A perspective view of CPA frame showing the location of the HV bus cable and attachments to the HV cup and resistive cathode, with CPA frame electrodes omitted to make HV bus visible.}
\includegraphics[height=0.35\textheight]{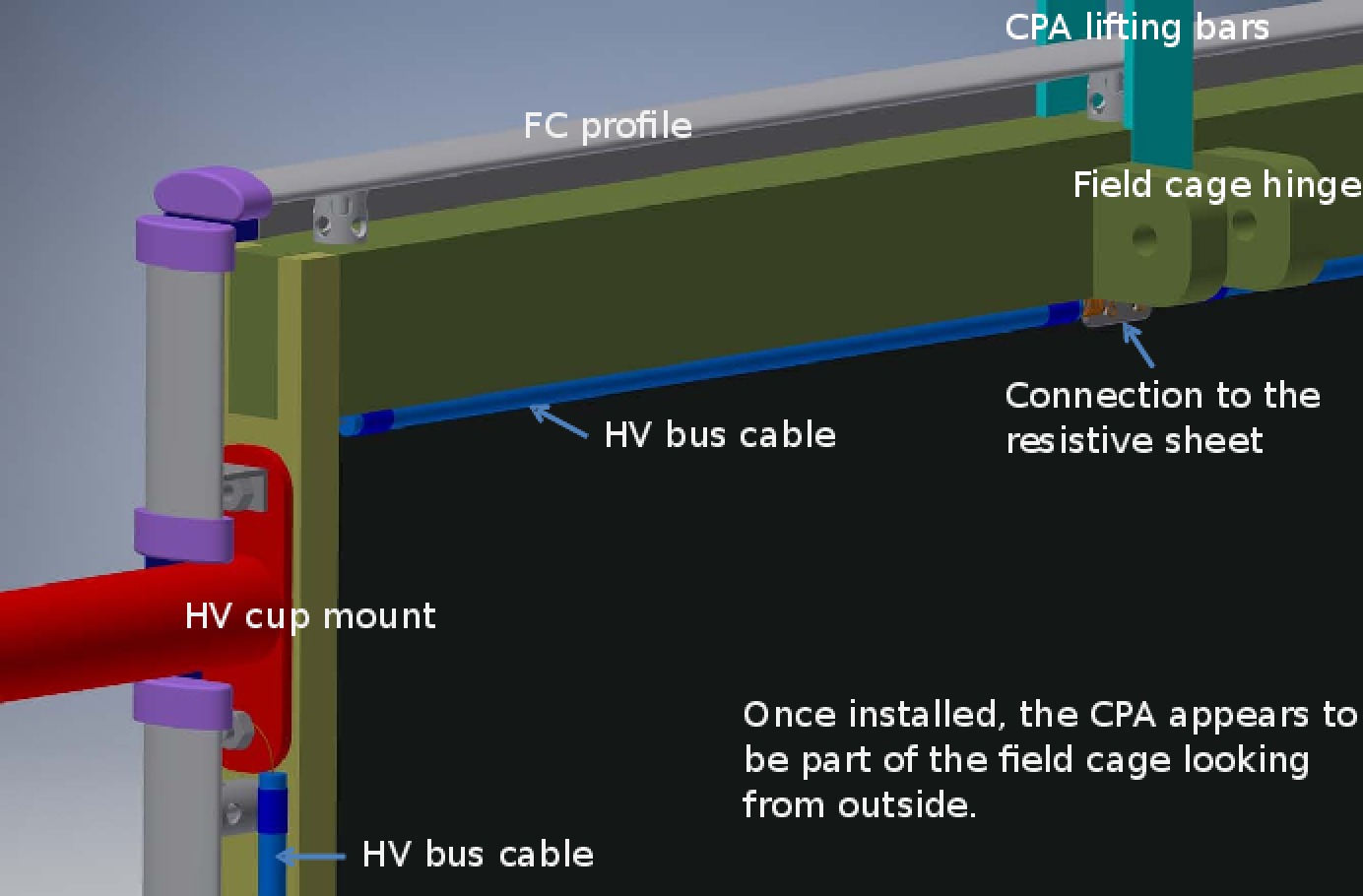}
\end{cdrfigure}




\section{Field Cage (FC)}
\label{detcompsec-fc}
\subsection{Scope and requirements} 

In the TPC, each pair of facing cathode and anode planes form an electron-drift region. A field
cage (FC) must completely surround the four open sides of this region to provide the necessary boundary
conditions to ensure a uniform electric field within, unaffected by the presence of the cryostat walls.  The scope of the FC includes:

\begin{itemize}
\item six top FC assemblies;
\item six bottom FC assemblies; and
\item four end-wall panels, each of which consist of four (side) assemblies.
\end{itemize}

Each assembly is made up of
\begin{itemize}
\item parallel metallic profiles;
\item a resistive divider chain that interconnects the parallel metal profiles  to provide a linear voltage gradient;
\item FRP I-beams or box beams that form an insulating mechanical support structure for field cage assemblies; and
\item ground planes (GP) for the top and bottom assemblies, five GP panels per FC module.
\end{itemize}

In addition, the FC assemblies include attachment fixtures to make the mechanical connections to the CPA and APAs.
One end-wall assembly is specially configured to accommodate the
cylindrical beam plug, which displaces the LAr in the region where the beam enters the cryostat.

The FC is required to:
\begin{itemize}
\item provide the nominal drift field of 500\,V/cm;
\item withstand $-$180\,kV near the cathode;
\item define the drift distance between the APAs and CPAs to <1\,cm;
\item limit the electric field in the LAr volume to under 30\,kV/cm;
\item miminize the peak energy transfer in case of a HV discharge anywhere on the field cage or cathode;
\item provide redundancy in the resistor divider chain;
\item maintain the divider current much greater than the ionization current in the TPC drift cell, yet less than the power supply current limit when all dividers are connected in parallel;
\item be modular in form such that it can be easily installed in the cryostat;
\item provide support for the beam plug; 
\item support a 200-lb. person standing on the support beam of the bottom FC module; and
\item prevent any trapped volume of liquid.
\end{itemize}

\subsection{Mechanical design}

The FC has six top and six bottom FC assemblies, arranged in sets of three along each horizontal boundary of the two drift regions. It has 
four end-wall panels (each made up of four end-wall assemblies), one covering each vertical boundary of the two drift regions, shown in Figures~\ref{fig:fc-overview} and~\ref{fig:fc-endwall_module}.
Each endwall panel consists of four assemblies in ``landscape'' orientation, stacked vertically.
FC assemblies are constructed from pultruded FRP I-beams and box beams that support extruded field-shaping aluminum profiles. The support structure for each of the top and bottom FC assemblies consists of two main I-beams that are 3.6~m long, and three cross I-beams that brace the main I-beams for structural stability. The main I-beams have cutouts to hold the field-shaping profiles. 

\begin{cdrfigure}[An end-wall field cage assembly]{fc-endwall_module}{A view of a the top most end-wall field cage assembly. An end-wall panel consists of a stack of four of these assemblies (each in landscape orientation).  This top assembly has attachment features connecting to the DSS support beams.}
\includegraphics[width=0.6\linewidth]{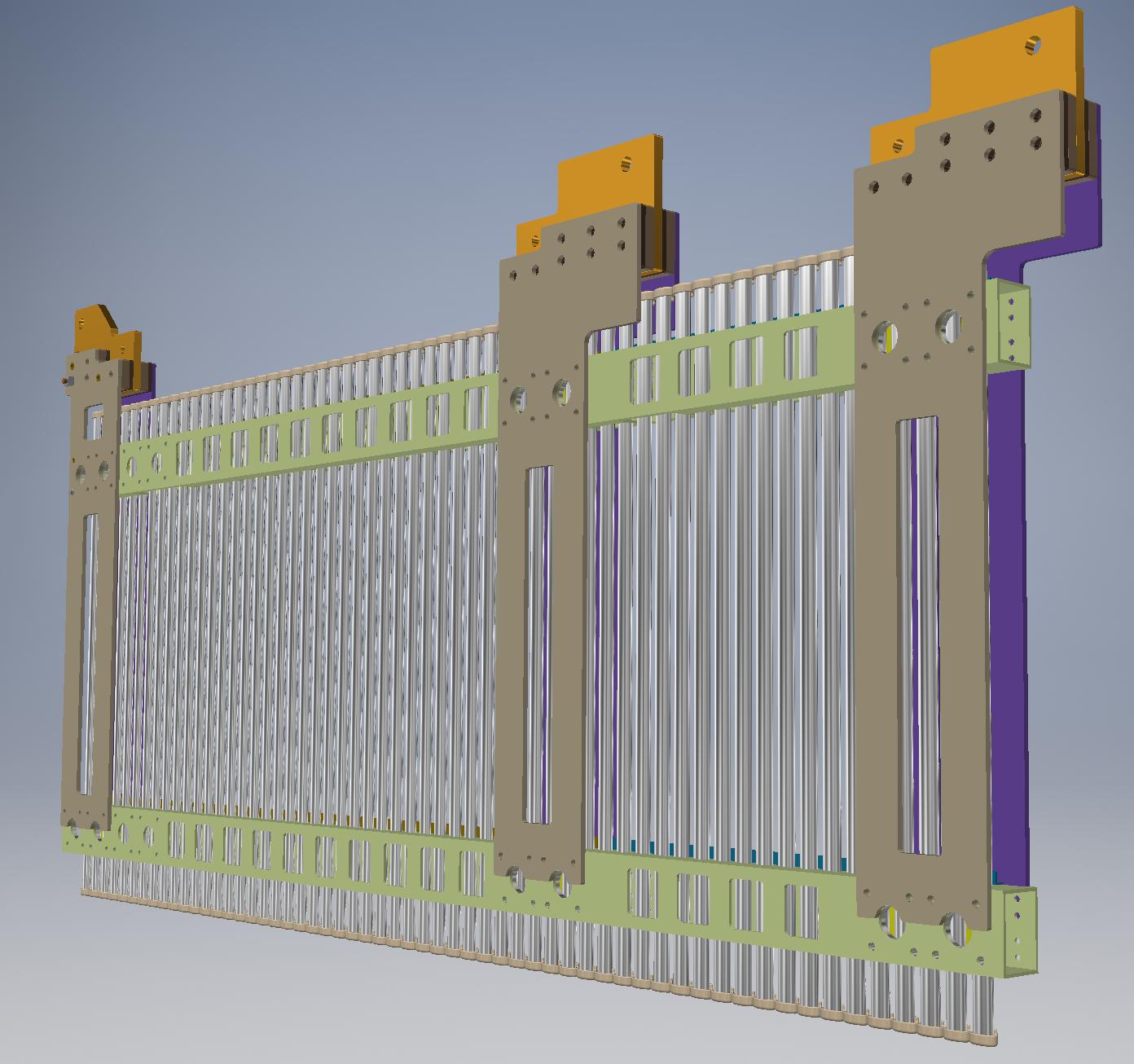}
\end{cdrfigure}

Aside from the profiles themselves, the nuts and bolts holding them, and the ground plane panels, all FC components are made of insulating material. The material selected for these structural components is fiberglass-reinforced plastic (FRP), which has good mechancal strength at cryogenic temperatures and low CTE. The ground plane panels are made of stainless steel, perforated to allow liquid argon circulation. 

The inward-facing side of the ground planes are approximately 20~cm away from the top of the field-shaping profiles, mounted at this fixed distance 
by standoffs. Figure~\ref{fig:fc-with-ground-planes} shows a set of ground planes situated over the I-beams and cross beams attached to a FC assembly.

\begin{cdrfigure}[The field cage with ground planes]{fc-with-ground-planes}{The field cage with ground planes}
\includegraphics[width=0.8\linewidth]{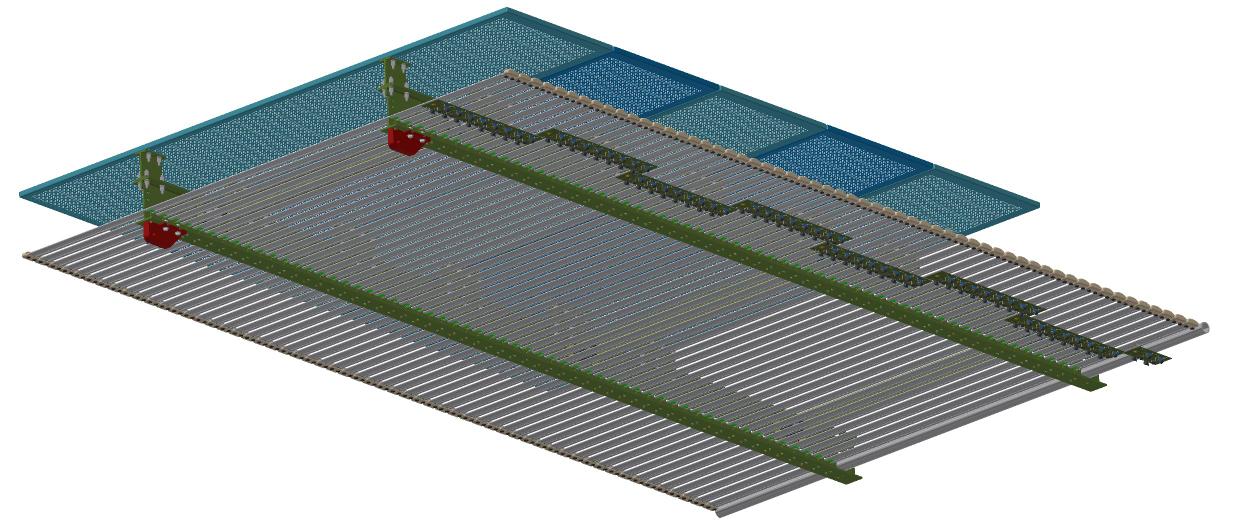}
\end{cdrfigure}

The parallel metal profiles in each FC assembly 
 are interconnected by a resistive divider chain, and supported by the FRP beams that span the drift distance.  Between adjacent field cage assemblies, however,  
the metal profiles are neither mechanically nor electrically connected. Gaps between assemblies that range from a few millimeters to a few centimeters are designed into the TPC assembly to ensure sufficient clearance for the installation.  The electrical isolation between the field cage modules minimizes the peak energy dump in case of a HV discharge.

\subsection{Electrical design}

Given a large standoff distance between the FC and the grounded cryostat wall, it is relatively easy to design a FC that meets the 30\,kV/cm electric field limit with 180\,kV bias.  However, as this distance is reduced to increase the active detector volume, the design becomes more challenging.  Optimizing the active detector volume requires the use of electrodes with low profiles, rounded edges, no trapped volumes, and low cost.  Several commercially available roll-formed metal profiles were studied and appear to meet these requirements. One of the profiles were tested in the Icarus 50 liter cryostat at a higher electric field to validate the design concept.  This shape is adopted to produce the FC profiles using aluminum extrusion.

Figure~\ref{fig:fc-schematic} is a simplified schematic of the electrical design of  the CPA and a top/bottom field cage module pair.

\begin{cdrfigure}[Field cage schematic diagram]{fc-schematic}{A schematic diagram of the CPA and a top/bottom field cage module pair}
\includegraphics[width=0.8\linewidth]{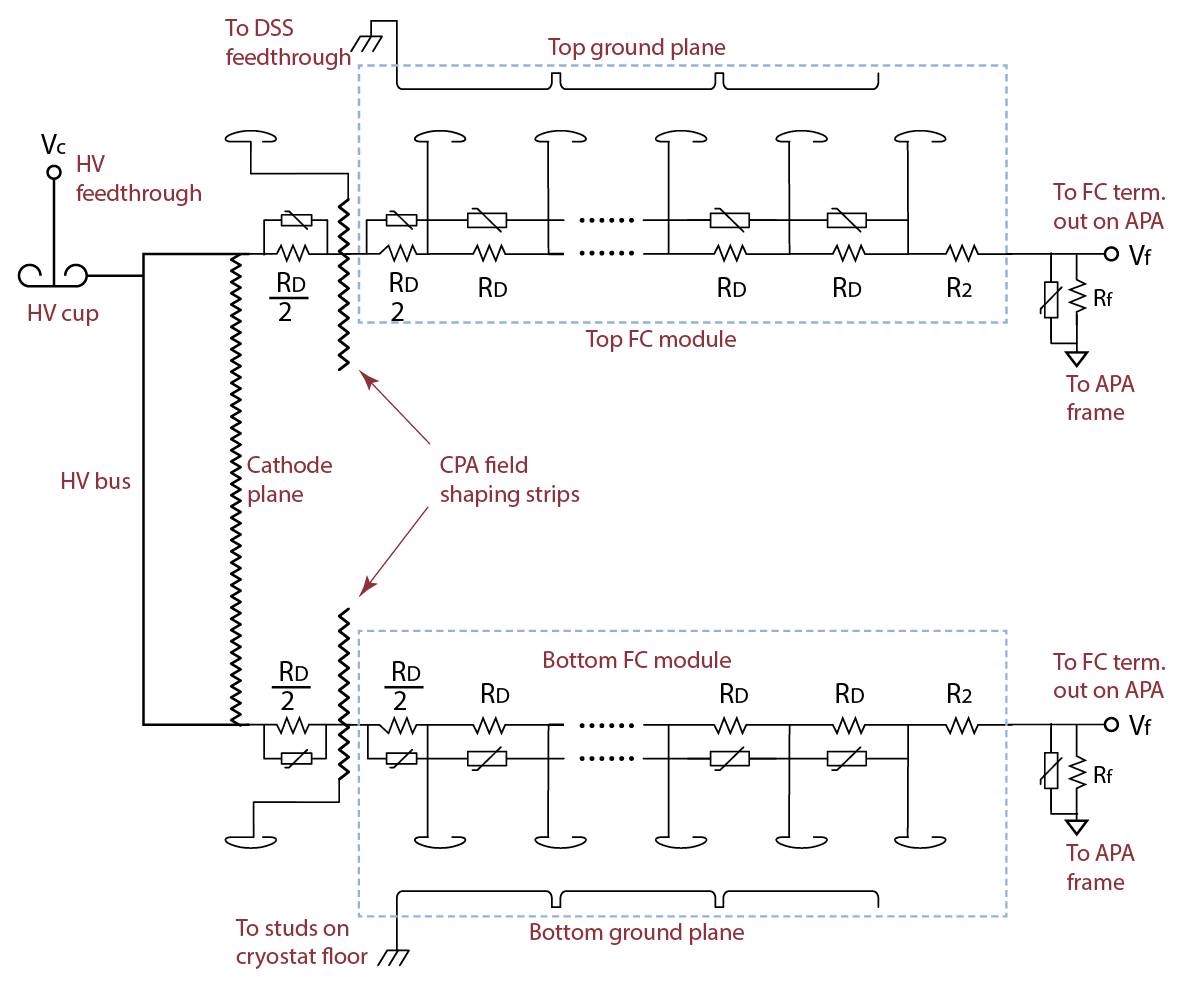}
\end{cdrfigure}

\subsection{Top/bottom FC assemblies and ground-plane modules}

In order to prevent high electric field entering the gas argon volume above the TPC, a grounded metallic plane is installed between the upper FC profiles and the liquid-gas interface. The same arrangement, flipped upside down, is used for the bottom FC assemblies to minimize the electric field on the cryostat floor and cryogenic services. 
Each of the six top FC assemblies has five ground plane (GP) panels attached to it, aligned 
along their long 
(2.3\,m) dimension. The GPs are connected to the FC I-beam 
with additional FRP standoffs that are also used to connect adjoining GP panels. 
Figure~\ref{fig:fc-panel-endwall-frame} shows the frame of a top/bottom FC assembly and 
Figure~\ref{fig:fc-with-ground-planes} shows a 3D model of a fully assembled FC module with attached GP panels.

\begin{cdrfigure}[Top/bottom FC panel with endwall frame]{fc-panel-endwall-frame}{The frame structure of a top/bottom FC assembly}
\includegraphics[width=0.8\linewidth]{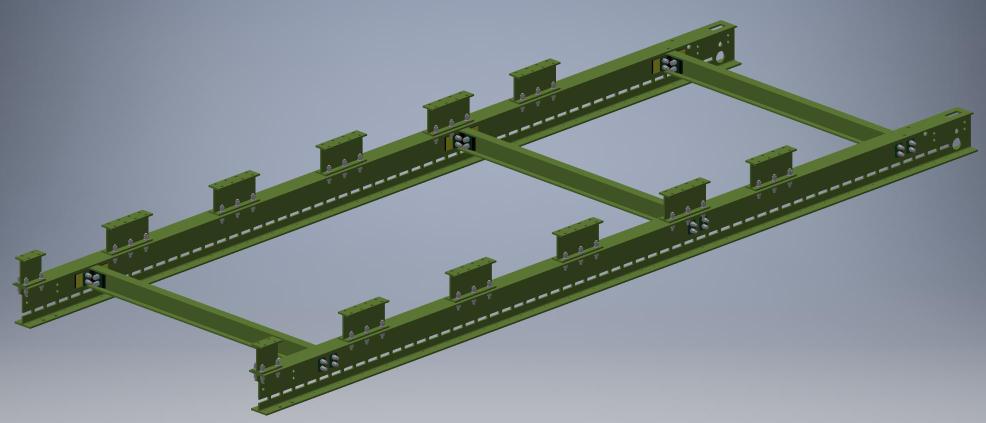}
\end{cdrfigure}

The electrical connections between adjacent GP panels are to be made 
with flexible copper braids to accommodate the relative movement of the structure during cool down.  The top GP panels are connected to the mechanical feedthroughs used by the DSS, while the bottom GP panels are connected to metal studs welded on the cryostat floor.

In addition to the six primary GP panels attached to the top FC modules,
Smaller panels are added to fill the gap between the FC modules directly above the CPA, and surround the CPA lifting bars. These smaller panels are connected to the modules on one side of the CPA so that, once in position, they 
cover the CPA frame. 


\subsection{Interfaces to other TPC components}

\subsubsection{FC to CPA and APA}

On each of the top and bottom FC assemblies, a pair of hinges connects each FC module to two CPA columns. These hinges allow the FC modules to be pre-attached to the CPAs during installation, preventing accidental damage to the APA wire planes when FC modules are raised or lowered and connected to the APAs.  The mechanical connection between a top/bottom FC module and the APA is maintained by a pair of custom-designed latches that transfer the load of the FC module to the APA without imparting a bending moment.

The end-wall FC assemblies are hung from the CPA and APA support rails.  There is no rigid mechanical connection 
between the end-wall FC assemblies and the CPAs or APAs. 

Each top and bottom FC module has one electrical connection each to the APA and the CPA. Even though each end-wall FC assembly has a resistive divider chain, once these four assemblies form the end-wall panel, the four divider chains are also consolidated into one connection each to the APA and the CPA to minimize \fixme{simplify?} access requirements at different heights. 

\begin{cdrfigure}[CPA to field cage connection]{cpa-fc-connection}{A top field cage module (grey) connected to two CPA modules underneath (brown)}
\includegraphics[width=0.8\linewidth]{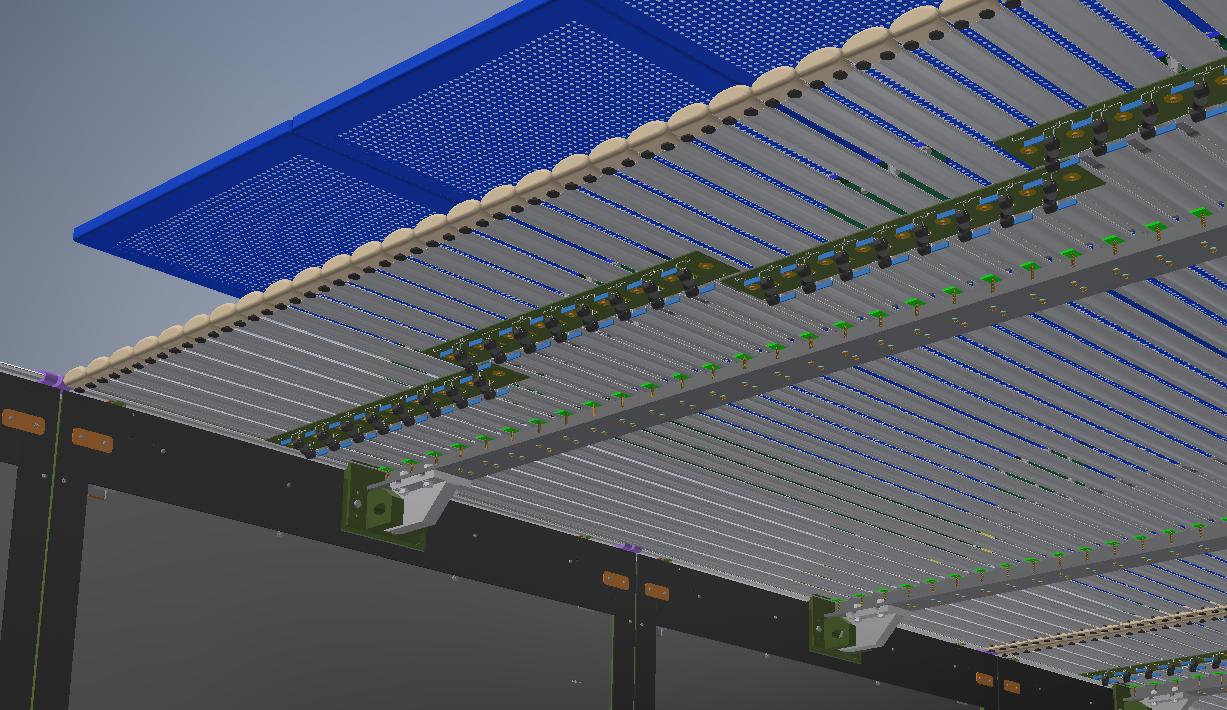}
\end{cdrfigure}

The voltage on the last field cage electrode near the APA is about $-$5.5\,kV at the nominal 500-V/cm drift field. A nominal step-down resistor is used to bring this voltage to about $-$1\,kV and form a dual-
purpose bias/monitoring point.  A HV cable connects this point to an SHV feedthrough on the TPC signal feedthrough flange, which enables either monitoring of the field cage divider current, or fine tuning of the drift field near the APAs.  

\subsubsection{FC to beam plug}
\label{subsec:fc-beamplug}



The LAr-displacement beam plug, a cylindrical glass-fiber composite pressure vessel, about 50\,cm in length and 22\,cm in diameter, is filled with nitrogen gas via a stainless steel line that extends to the top of the cryostat. It is illustrated in Figure~\ref{fig:beamplug}.
A pressure relief valve (or burst disk) is installed on the nitrogen fill line on the top of the cryostat (externally) to ensure the pressure inside the beam plug does not exceed the safety level of about 22\,psi. The nitrogen system schematic is shown in Figure~\ref{fig:beamplugN2}. 
The beam plug is secured to the FC support structure as illustrated in Figure~\ref{fig:beamplug-fc}. The front portion of the beam plug extends  5\,cm into the active region of the TPC  through an opening in the FC. The FC support is designed with sufficient strength and stiffness to support the weight of the beam plug before filling, while it is suspended in air.  When the cryostat is filled with LAr, the beam plug is roughly neutrally buoyant.  The total internal volume of the beam plug is about 16 liters. 


The requirements on the acceptable leak rate is between $7.8\times 10^{-5}$ scc/s and $15.6\times 10^{-5}$ scc/s. This is very conservative and is roughly equivalent to leaking  15\% of the nitrogen in the beam plug over the course of a year.
  In the worst-case scenario in which all the nitrogen in the beam plug leaks into the LAr cryostat, the increase in concentration is about 0.1\,ppm, which is still a factor of 10 below the maximum acceptable level, as specified by light detection requirements.
  At nominal operation, the voltage difference across the beam plug (between the first and the last grading ring) is 165\,kV. 

    To minimize risk of electrical discharges, the beam plug is divided into sections, each of which is bonded to stainless steel conductive grading rings. The seven grading rings are connected in series with two parallel paths of resistor chains. The ring closest to the FC is electrically connected to one of the FC profiles. 
  The ring nearest the cryostat wall is grounded to the cryostat inner membrane via a short grounding cable. 
The maximum total power dissipated by the resistor chain is about 0.6\,W.

\begin{cdrfigure}[Beam plug]{beamplug}{The beam plug is a  composite pressure vessel filled with dry nitrogen gas. The vessel is about 50\,cm in length and about 22\,cm in diameter. The pressure vessel is divided into sections with each section bonded to a stainless steel grading ring. The grading rings are connected by three parallel paths of resistor chain.}
  \includegraphics[width=0.75\textwidth]{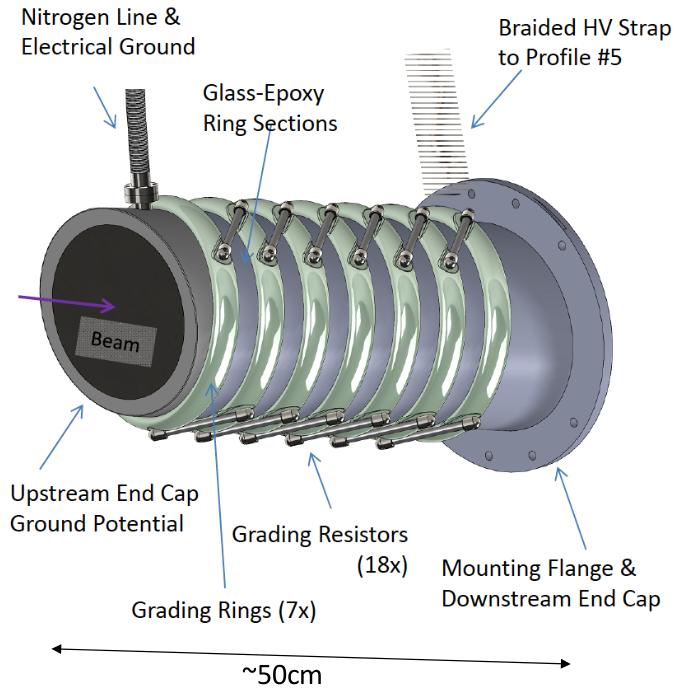}
\end{cdrfigure}


\begin{cdrfigure}[Beam plug nitrogen system]{beamplugN2}{Beam plug nitrogen gas system schematics. The Local Control Panel is mounted on top of the cryostat near the DN160 flange feedthrough. The nitrogen line enters the cryostat via the 6-way flange which also has a burst disk for emergency pressure relief and temperature/pressure sensors.  }
  \includegraphics[width=0.75\textwidth]{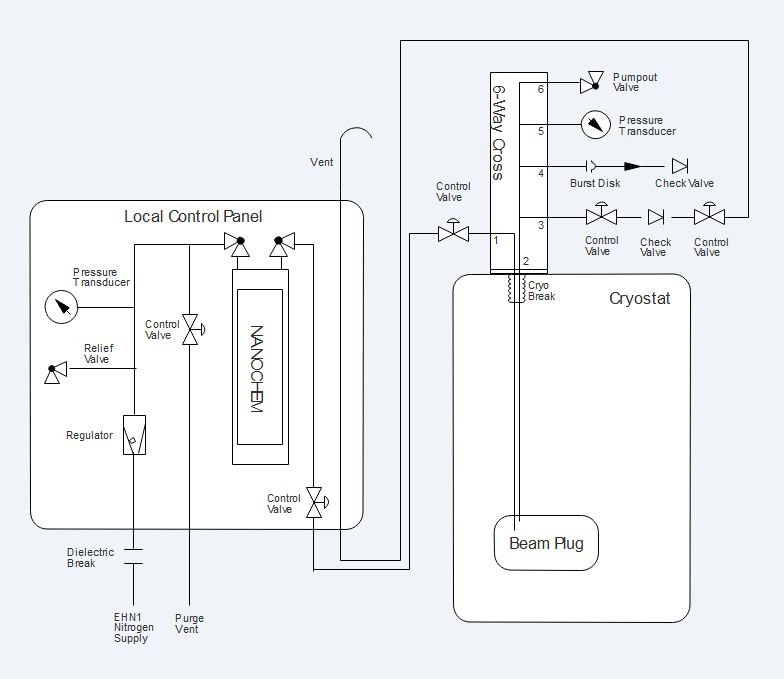}
\end{cdrfigure}

The metal electrode rings are spaced at regular intervals and interspersed with composite tube sections. The shape of the rings has been designed to minimize high electric field corners. The results of the field calculations are shown in Figures~\ref{fig:beamplug_ring1} and~\ref{fig:beamplug_ring2}. The average field in the vicinity of the beam plug is about 4.4\,kV/cm. The maximum field of 15.7\,kV/cm is on the electrode ring surface. In all regions the field is well below the 30-kV/cm limit.

\begin{cdrfigure}[Electric field calculation of beam plug electrode ring design]{beamplug_ring1}{Electric field calculation of the electrode ring design. The average field in the beam plug region is about 4.4\,kV/cm. The maximum field of 15.7\,kV/cm is on the electrode ring surface. }
  \includegraphics[width=0.85\textwidth]{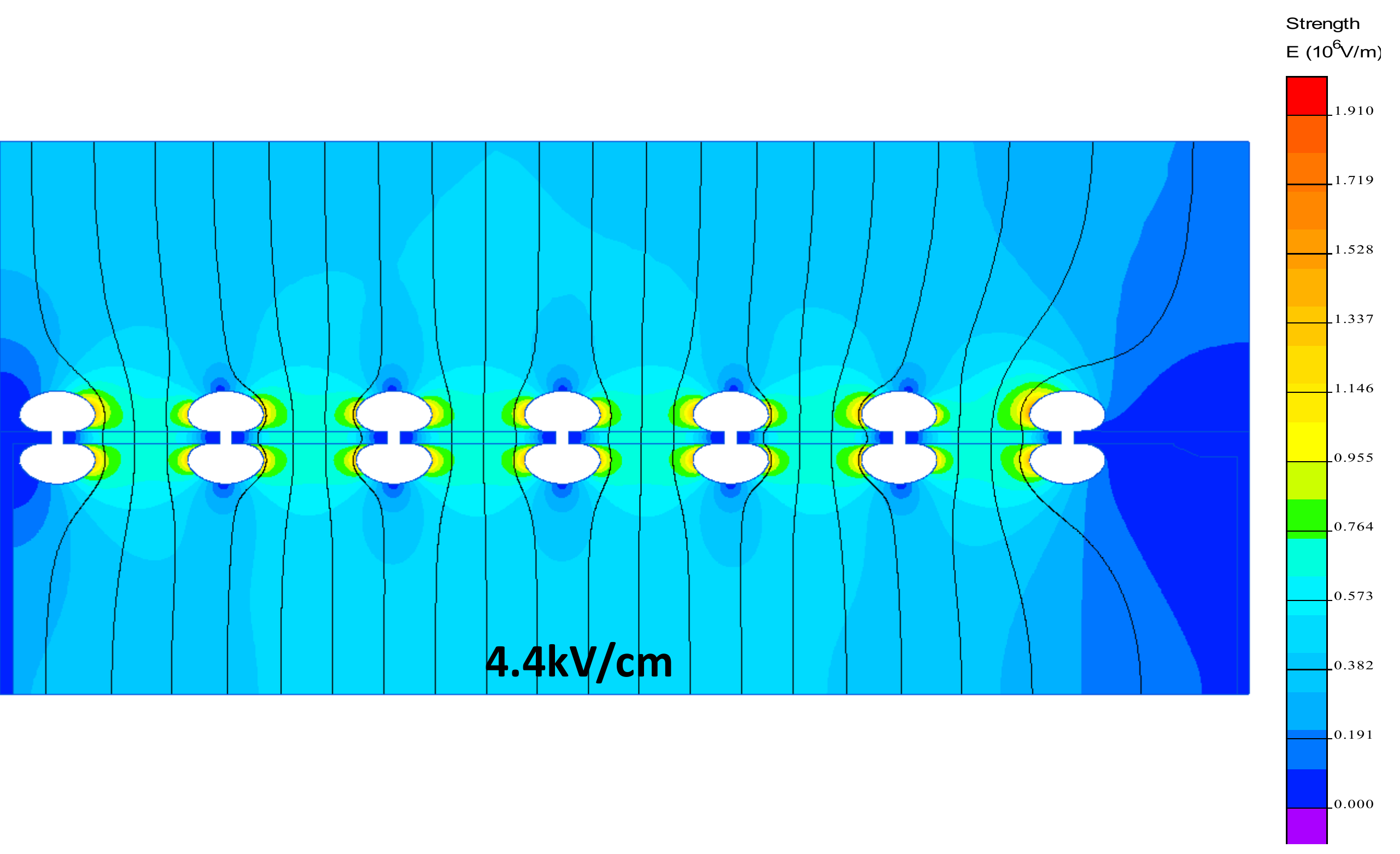}
\end{cdrfigure}

\begin{cdrfigure}[Electric field calculation near the vicinity of the beam plug electrode]{beamplug_ring2}{Electric field calculation near the vicinity of the electrode. The shape of the ring minimizes the high field region near the joints between the electrode, LAr, and composite shell. The field is well below the 30-kV/cm limit in all regions.}
  \includegraphics[width=0.5\textwidth]{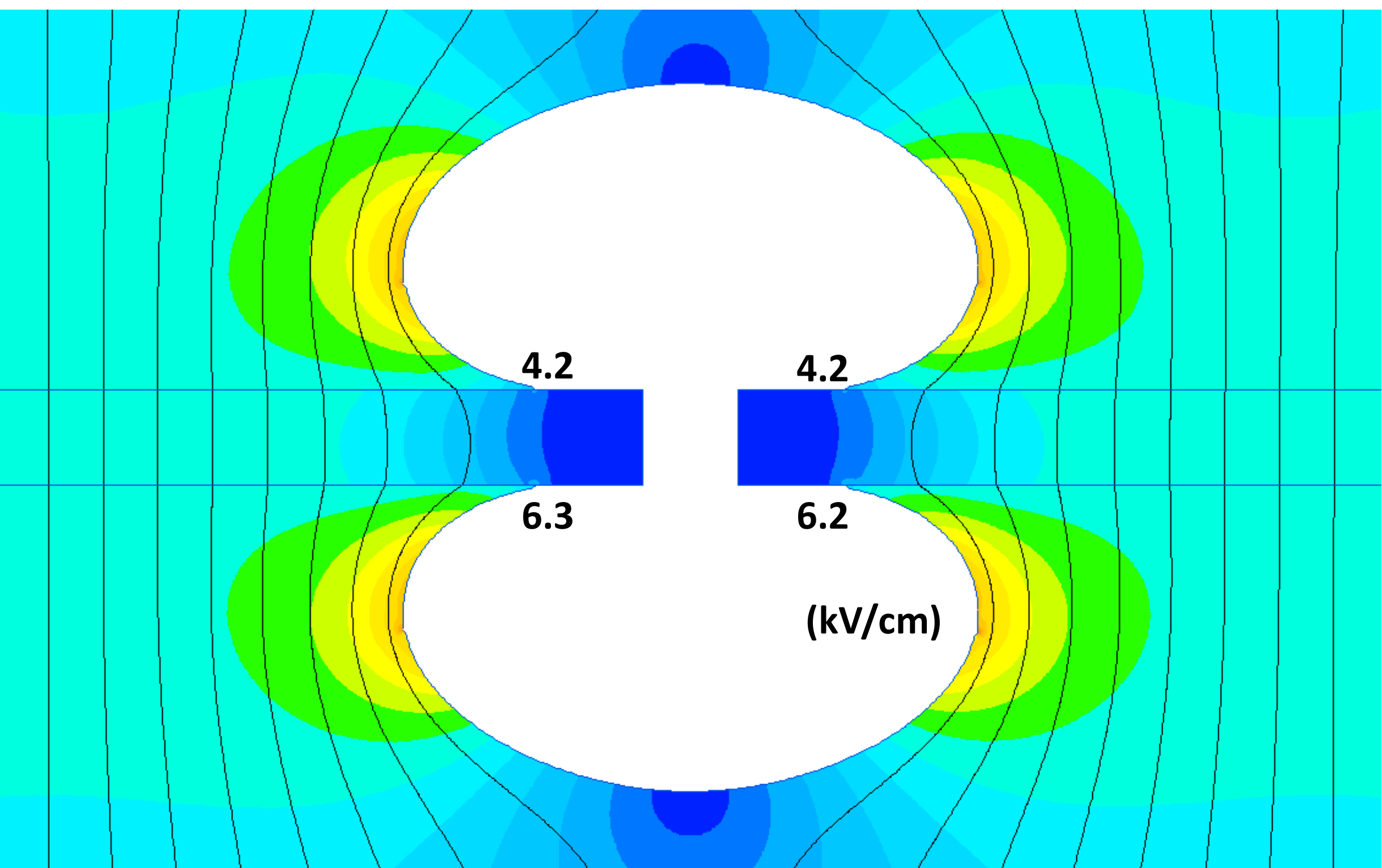}
\end{cdrfigure}

The beam plug is mounted onto one of the field cage support structures as shown in Figure~\ref{fig:beamplug-fc}. 
\begin{cdrfigure}[Beam plug to field cage interface]{beamplug-fc}{Beam plug to field cage interface.}
\includegraphics[width=0.75\linewidth]{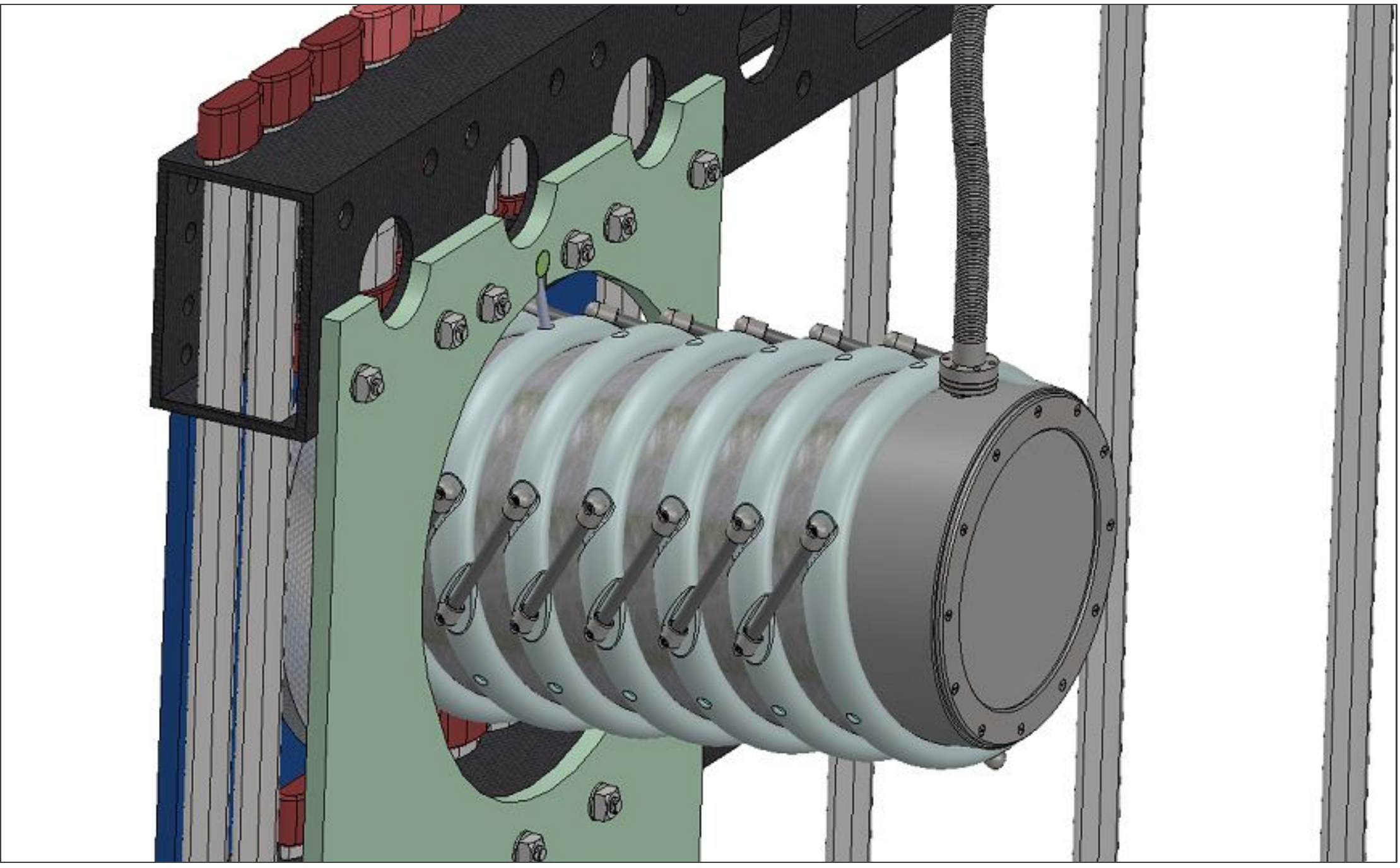}
\end{cdrfigure}





\section{TPC high-voltage (HV) components}

\subsection{Scope and requirements}

The TPC high voltage (HV) components include the HV power supply, cables,
filter circuit, HV feedthrough,  and monitoring for currents and voltages (both steady
state and transient).

A schematic of the complete TPC HV circuit is shown in Figure~\ref{fig:TPCHVcircuit}.

\begin{cdrfigure}[TPC HV circuit]{TPCHVcircuit}{A schematic of the TPC high voltage circuit.}
  \includegraphics[width=0.95\textwidth]{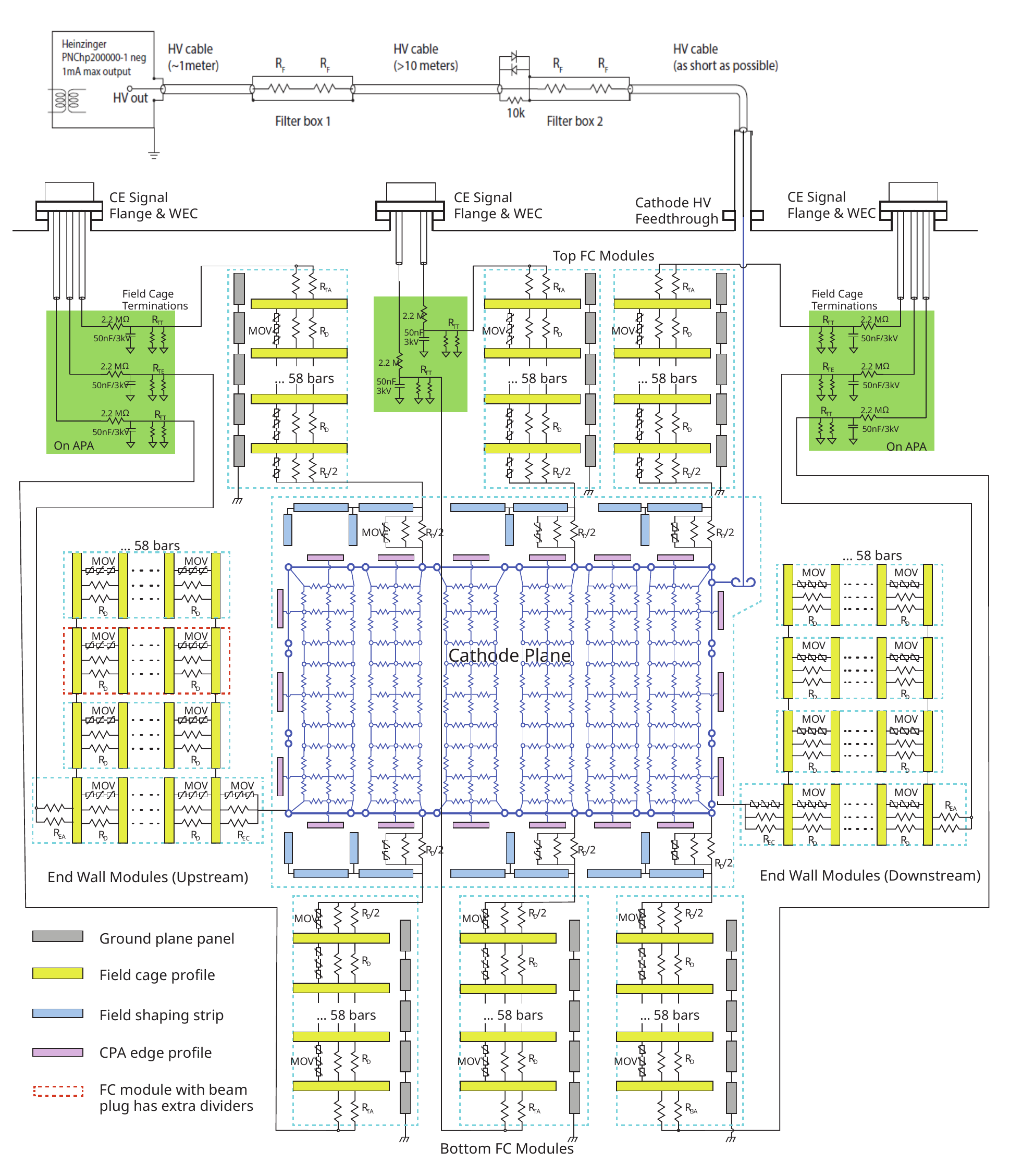}
\end{cdrfigure}

The cathode plane is biased at \SI{-180}{kV} to provide the
required \SI{500}{V/cm} drift field.  It is 
powered by a dedicated HV power supply through an RC filter and HV 
feedthrough.  The power supply for the cathode plane must be able
to provide \SI{-200}{kV}.  The output voltage
ripple must not introduce more than 10\% 
of the equivalent thermal
noise from the front-end electronics. The power supply must be
programmable to shut down its output at a certain current
limit. During power cycling, both controlled and uncontrolled, the voltage 
ramp rate at the feedthrough must be limited
to prevent damage to the in-vessel electronics from excess charge
injection. The HV feedthrough must be able to withstand \SI{-250}{kV}
in its center conductors in a \SI{1}{atm} argon gas environment when
terminated in liquid argon.

\subsection{HV feedthrough design, power supply and cabling}
The design of the HV feedthrough as well as the procurement of the HV power supply, cables, and possibly filter
circuits, are activities being jointly pursued by the \pdsp and ProtoDUNE-DP efforts. In particular:

\begin{itemize}	
\item The Heinzinger 300-kV power supply (residual ripple less than $10^{-5}$) and cable specified for ProtoDUNE-DP  are  well suited for \pdsp, which operates at a lower HV setting.
\item The present ProtoDUNE-DP HV feedthrough design is adaptable to \pdsp without any major modification in its dimensions or mechanical features.
\item The filtering scheme and the monitoring system required for \pdsp is more demanding than that for ProtoDUNE-DP, due to its more sensitive front-end electronics, and can be used for both detectors.
\item Common spare components are also being utilized.
\end{itemize}

The 
design of the 300-kV feedthrough is based on the very successful construction technique adopted for the ICARUS HV feedthrough, which was operated at 75\,kV without interruption for more than three years without any failure. The feedthrough was also successfully operated for several days as a test after the run at 150\,kV.  
The design is based on a coaxial geometry, with an inner conductor (HV) and an outer conductor (ground) insulated by ultra-high-molecular-weight polyethylene (UHMW PE)  as shown in Figure~\ref{fig:hv-feed-through}. 

The outer conductor, made of a stainless-steel tube, surrounds the insulator, extending down through the cryostat into the LAr. 
In this geometry, the electric field is 
confined within regions occupied by high-dielectric-strength media (UHMW PE and LAr).  The inner conductor is made of a thin-walled stainless steel tube to minimize the heat input and to avoid the creation of argon gas bubbles around the lower end of the feedthrough. A contact, welded at the upper end for the
connection to the HV cable, and a round-shaped elastic contact for the connection to the cathode, screwed at the lower end, completes the inner electrode. Special care has been taken in the assembly to ensure complete filling  of the space between the inner and outer conductors with the PE dielectric, and to guarantee leak-tightness at ultra-high vacuum levels.

\begin{cdrfigure}[HV feedthrough]{hv-feed-through}{Preliminary design of the HV feedthrough.}
\includegraphics[width=0.95\textwidth]{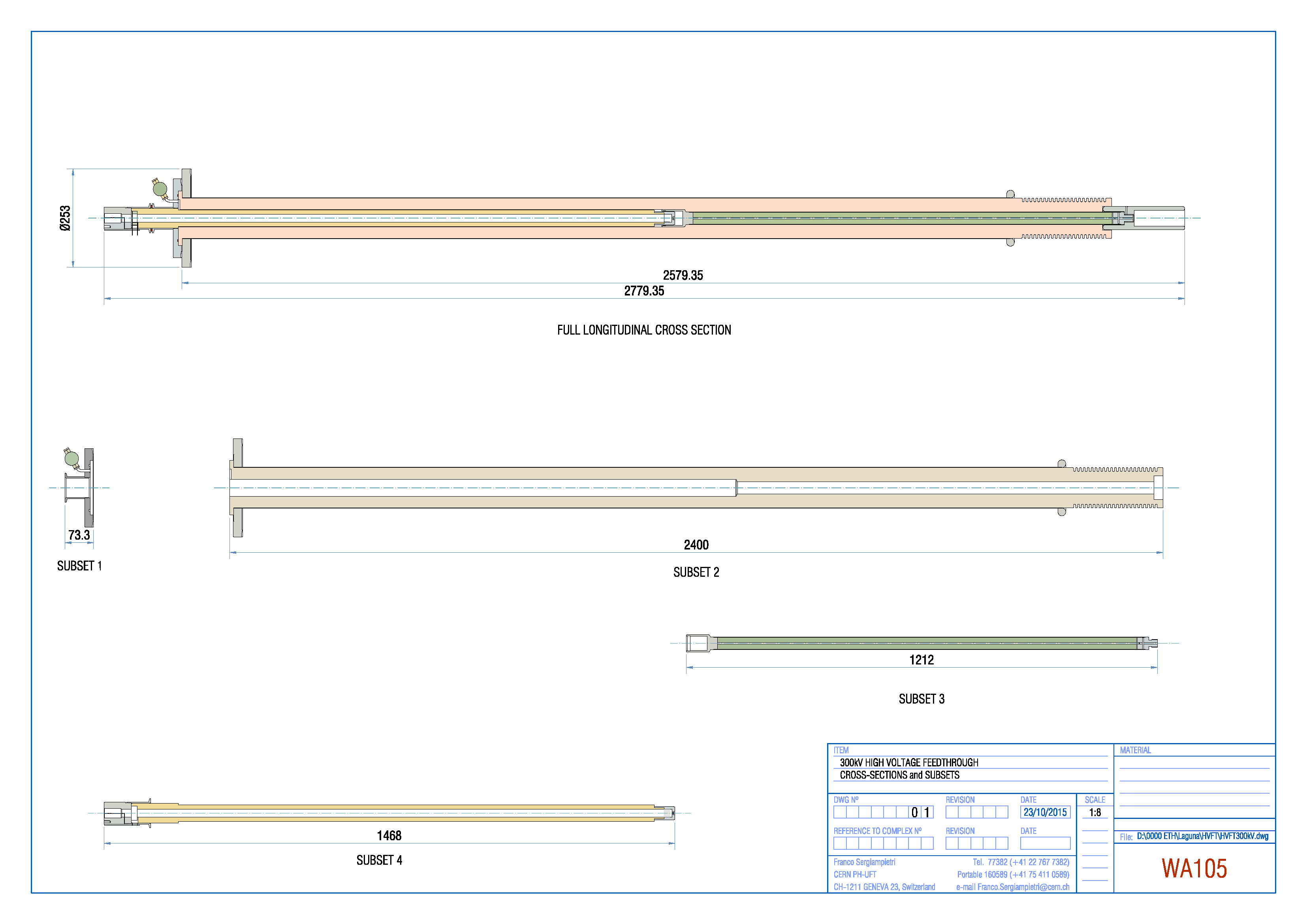}
\end{cdrfigure}

\subsection{HV monitoring}

HV-circuit monitoring devices include a toroid transformer to detect
spikes and noise in the current draw, and a monitoring point at the
end of the field cage resistor chain, which also provides a means to
control field shaping around the exterior APA edges by fine-tuning the
voltage at the last field cage element.






\section{TPC front-end electronics}
\label{ch:ce}

%
\subsection{Scope and requirements}
\label{subsec:ce_intro}

The DUNE single-phase TPC read-out electronics are referred to as the ``Cold Electronics'' (CE) because they reside in LAr,
mounted directly on the APA, 
thus reducing channel capacitance and noise by minimizing the length of the connection between an anode wire
and its corresponding electronics input.


The CE signal processing is implemented in ASIC chips using CMOS technology,
which has been demonstrated to perform well at cryogenic temperatures,
and includes amplification, shaping, digitization, buffering, and multiplexing (MUX) of the signals.
The CE is continuously read out,
resulting in a digitized ADC sample from each APA channel (wire) up to every 500\,ns (2-MHz maximum sampling rate). 

The 2,560 channels from each APA are read out by 20 Front-End Motherboards (FEMBs), each providing 
digitized wire read-out from 128 channels. One cable bundle 
connects each FEMB to the outside of the cryostat via a feedthrough (a \textit{CE feedthrough}) in the signal cable flange at the top of the cryostat, where a single flange services each APA, as shown in Figure~\ref{fig:tpcce_apa_flange}. 
Each cable bundle contains wires for low-voltage (LV) power, high-speed data readout,
and clock/digital-control signal distribution.
Eight separate cables carry the TPC wire-bias voltages from the signal flange to the APA wire-bias boards, as
shown schematically in Figure~\ref{fig:tpcce_cr_board}.

\begin{cdrfigure}[Connections between signal flange and APA]{tpcce_apa_flange}{Connections between
the signal flange and APA.}
\includegraphics[width=0.4\linewidth]{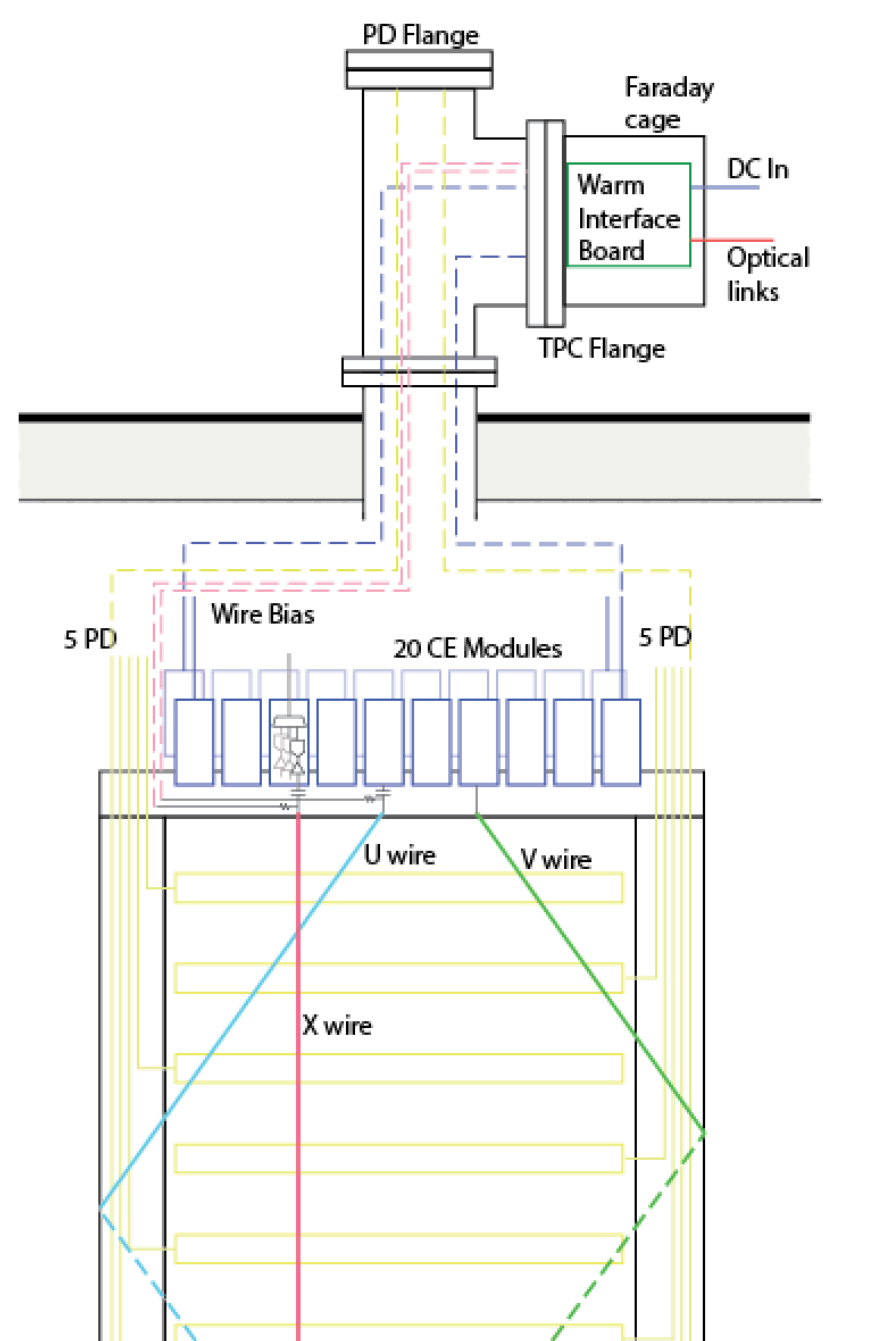}
\end{cdrfigure}

The components of the CE system are the:
\begin{itemize}
\item Front-end mother boards (FEMBs), which house the cold ASICs and are installed on the APAs;
\item Cables for the data, clock/control signals, LV power, and wire-bias voltages 
between the APA and the signal flanges (cold cables);
\item Signal flanges with a CE feedthrough to pass the data, clock/control signals, 
LV power, and APA wire-bias voltages between the inside and outside of the cryostat;
\item Warm interface electronics crates (WIECs) that are mounted on the signal flanges and contain the warm interface boards (WIBs)
and Power and timing cards (PTCs) for further processing and distribution of the signals entering/exiting the cryostat;
\item Fiber cables for transmitting data and clock/control signals between the WIECs and the data acquisition 
(DAQ) and slow control systems;
\item Cables for LV power and wire-bias voltages between the signal flange and external power supplies (warm cables); and
\item LV power supplies for the CE and bias-voltage power supplies for the APAs.
\end{itemize}

The electrical cables for each APA enter the cryostat through a single 
signal flange, creating an integrated unit that provides local diagnostics for noise and validation testing,
and follows the grounding guidelines in Section~\ref{subsec:groundshield}. The components, the quantity of each required for ProtoDUNE-SP, and the number of channels that each component has, are listed in Table~\ref{tab:ce-components}.

\begin{cdrtable}[Electronics components and quantities]{llr}{ce-components}{Electronics components and quantities}
Element                                                             &  Quantity                                  &  Channels per element   \\  \toprowrule
TPC                                                                   & 1                                               & 15,360    \\  \colhline
APA                                                                   & 6                                               & 2,560     \\  \colhline
Front-End Mother Board (FEMB)                         & 120, 20 per APA                       & 128     \\  \colhline
FE ASIC chip                                & 120 $\times$ 8, 8 per FEMB      & 16          \\   \colhline
ADC ASIC chip                             & 120 $\times$ 8, 8 per FEMB      & 16          \\   \colhline
FEMB FPGA                                  & 120, 1 per FEMB                         & 128          \\   \colhline
Cold cable bundles                     & 120, 1 per FEMB                        & 128      \\   \colhline
Signal flange                                & 6, 1 per APA                              & 128 $\times$ 20  (i.e., 2,560)  \\   \colhline
CE feedthrough                            & 6, 1 per APA                             &128 $\times$ 20         \\   \colhline
Warm interface boards (WIB)         & 30, 5 per APA                             & (128 $\times$ 20) /5 (i.e., 512)        \\   \colhline
 Warm interface electronics crates (WIEC)     & 6, 1 per APA                             & 128 $\times$ 20         \\   \colhline
 Power and timing cards (PTC)       & 6, 1 per APA                             & 128 $\times$ 20         \\   \colhline
Passive backplane (PTB)                & 6, 1 per APA                             & 128 $\times$ 20 \\   \colhline
LV power mainframe               & 2                        &   7,680   \\    \colhline
LV supply modules          &  6, 1 per APA       &  128 $\times$ 20            \\    \colhline
Wire-bias mini-crate               & 2                        &   7,680   \\    \colhline
Wire-bias supply modules    & 6, 1 per APA                        &   128 $\times$ 20   \\
\end{cdrtable}

The most significant requirements for the CE are listed here. The CE shall:

\begin{itemize}	
\item Provide the means to read out the TPC wires and transmit their data in a useful format to the DAQ;
\item Operate for the life of the facility without significant loss of function;
\item Dead or unusable channels at $<$~1$\%$, causing $>$~97$\%$ of the fiducial volume to be observed by all 3 active wire planes;
\item Record the channel waveforms continuously without dead time;
\item Be constructed only from materials that are compatible with high-purity LAr;
\item Provide sufficient precision and range in the digitization to:
\begin{itemize}
\item Discriminate electrons from photon conversions;
\item Optimize the reconstruction of high- and low-energy tracks from accelerator-neutrino interactions;
\item Distinguish a Minimum Ionizing Particle (MIP) from noise with a signal-to-noise ratio $>$ 9:1; and
\item Measure ionization up to 15 times that of a MIP particle, so that stopping kaons from proton decay can be identified;
\end{itemize}
\item Ensure that all power supplies have: 
\begin{itemize}
\item Local monitoring and control;
\item Remote monitoring and control through DAQ; and
\item Over-current and over-voltage protection circuits;
\end{itemize}
\item And ensure that the CE feedthroughs are able to withstand twice their nominal operating voltages 
with a maximum specified leakage current in 1-atm argon gas.
\end{itemize}

\subsection{Grounding and shielding}
\label{subsec:groundshield}

To avoid structural ground loops, the APA frames described in Section~\ref{subsec:apa_frame} 
are insulated from each other. Each frame is electrically connected to the cryostat at a single 
point on the CE feedthrough board in the signal flange where the cables exit the cryostat. Mechanical suspension of the APAs 
is accomplished using insulated supports. 

The analog portion of the FEMB contains eight front-end (FE) ASICs configured as 16-channel 
digitizing charge amplifiers. Input amplifiers on the ASICs have their ground terminals connected 
to the APA frame.  All power-return leads and cable shields 
are connected to both the ground plane of the FEMB and to the signal flange.

Filtering circuits for the APA wire-bias voltages are locally referenced to the ground plane of the FEMBs through low-impedance 
electrical connections. This approach ensures a ground-return path in close proximity to the 
bias-voltage and signal paths. The close proximity of the current paths minimizes the size of potential loops to further 
suppress noise pickup.

Photon detector signals, described in Section~\ref{sec:pd_system}, are carried directly on shielded, 
twisted-pair cables to the signal flange. The cable shields are connected to the 
cryostat at a second feedthrough, the PDS feedthrough, and to the PCB shield layer on the photon detectors. There is no 
electrical connection between the cable shields and the APA frame except at the signal flange.

The frequency domain of the TPC wire and photon detector signals are separate. The wire readout digitizes at 2~MHz 
with $<$~500~kHz bandwidth at 1 $\mathrm{\mu}$sec peaking time, while the photon readout operates at 150~MHz 
with $>$~10~MHz bandwidth. They are separated from the clock frequency (50~MHz) and common noise frequencies 
through the FE ASIC and cabling designs. All clock signals are transmitted differentially with individual shield 
to avoid the interference to power lines. 

%
\subsection{Distribution of APA wire-bias voltages}
\label{subsec:ce_wire_bias}

Each side of an APA includes four wire layers as described in Section~\ref{sec:apa-design-overview}. 
The innermost X-plane layer of wires is nominally biased at +820 Volts, with each wire AC coupled 
to one of the 128 charge amplifier circuits on the FEMB. The V-plane wire layer is effectively biased at zero volts, 
with each wire directly connected to one of the charge amplifier circuits. The U-plane wire layer is nominally 
biased at $-$370 Volts with each wire AC-coupled 
to one of the 128 charge amplifier circuits. The outermost G-plane wire layer,
which has no connection to the charge amplifier circuits, is biased at $-$665 Volts.

Electrons passing through the wire grid must drift unimpeded until they reach the X-plane 
collection layer. The nominal bias voltages are predicted to result in this electrically 
transparent configuration.

As described in Section~\ref{subsubsec:apa_wire_anchor}
 the filtering of wire-bias voltages and AC coupling of wire signals passing
onto the charge amplifier circuits is done on Capacitance-Resistance (CR) boards that plug in between the APA wire-board stacks and FEMBs.
Each CR board includes single R-C filters for the X- and U-plane wire-bias voltages. In addition, each board has 48 
pairs of bias resistors and AC coupling capacitors for X-plane wires, and 40 pairs for the U-plane wires. The coupling capacitors block DC while passing AC 
signals to the CE motherboards.

\begin{cdrfigure}[APA wire bias schematic diagram]{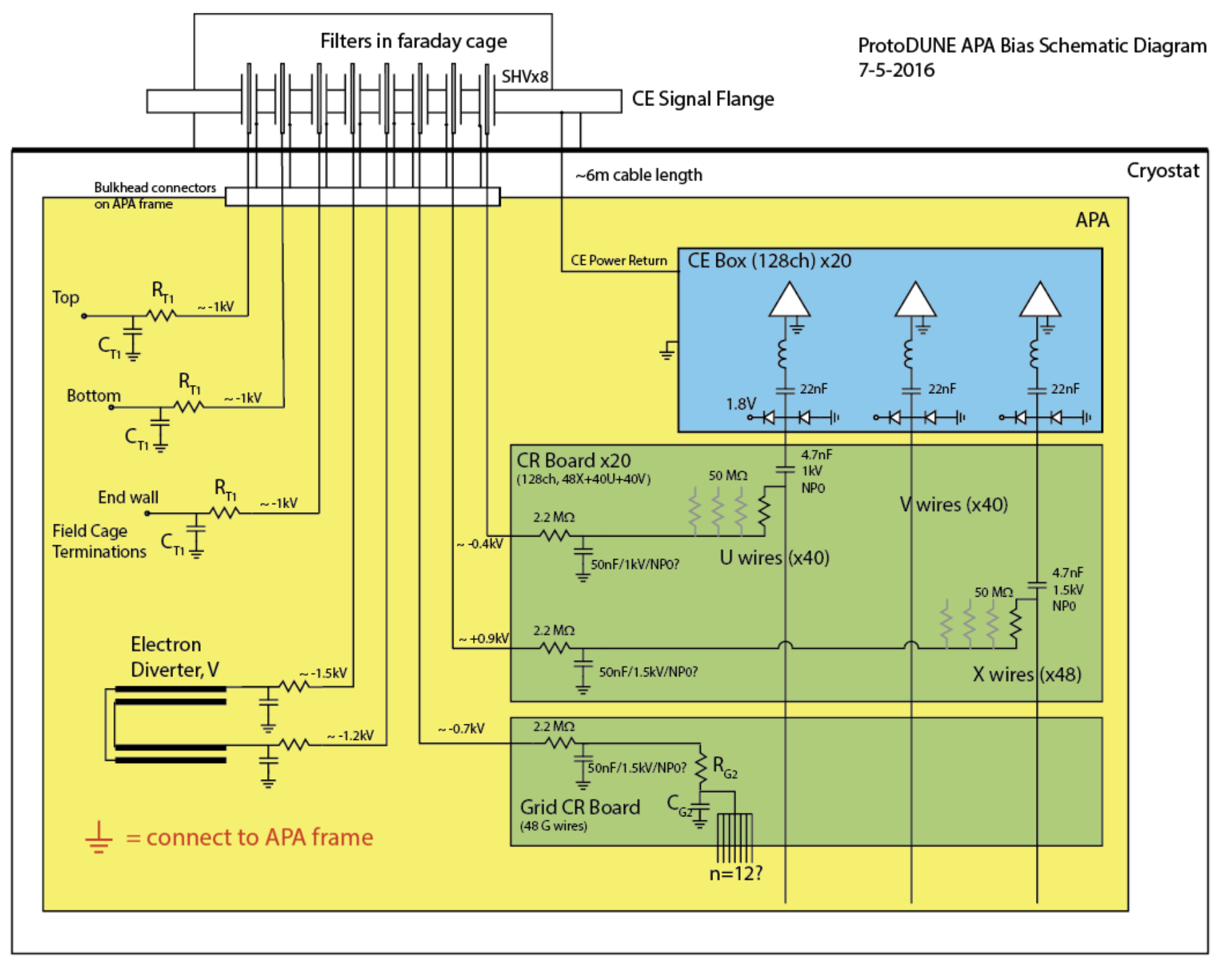}{APA wire bias 
schematic diagram, including the Capacitance-Resistance (CR) board.}
\includegraphics[width=0.9\linewidth]{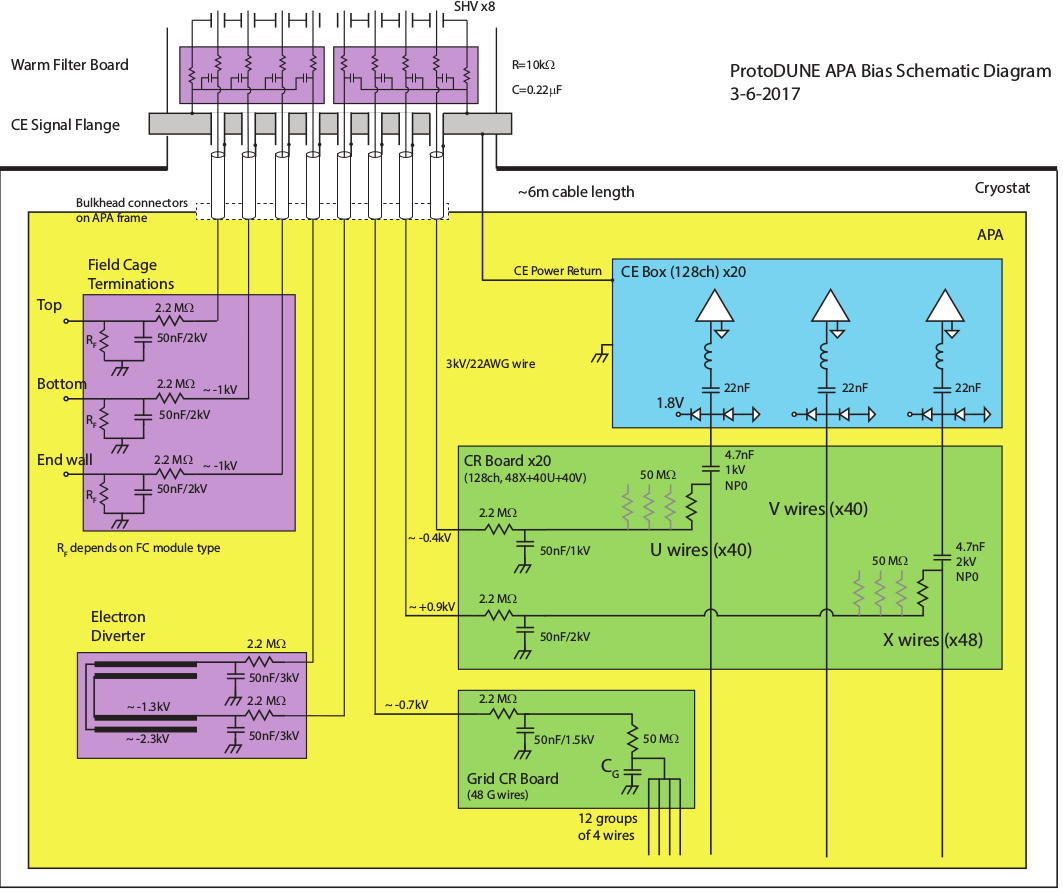}
\end{cdrfigure}

Separate CR boards include a single R-C filter for the G-plane wires and 12 pairs of bias resistors and
coupling capacitors.
Groups of four wires are tied together to share single
bias resistors and filter capacitors. These CR boards do not connect to the charge amplifier circuits on the FEMB.

Clamping diodes limit the input voltage received at the amplifier circuits to between 1.8V~$\pm$~U$\_$D, where U$\_$D
is the breakdown voltage of the diode $\sim$0.7V.
The amplifier circuit has a 22-nF coupling capacitor at input to avoid leakage current from the protection clamping diodes. 

Coupling capacitors for the X-plane and U-plane wires are required to block DC bias voltages.
However they also impact the efficiency of the detector circuits.
The sense wires are expected to have $\sim200$~pF of capacitance to the APA frame.
Induced or collected charges are effectively divided between the wire capacitance and the coupling capacitor.
To achieve a charge-calibration accuracy of 0.5 percent or better,
the coupling capacitors must be 4.7\,nF at ten percent tolerance, or 2.2\,nF at five percent tolerance.
Voltage ratings should be at least 1.5 times the expected operating voltages.

Bias resistance values should be at least 20\,M$\Omega$ to maintain negligible noise contributions.
A target value of 50\,M$\Omega$ is desired.
The higher value helps to achieve a longer time constant for the high-pass coupling networks.
Time constants should be at least 25 times the electron drift time so that the undershoot in the digitized waveform
is small and easily correctable.
However, leakage currents can develop on PC boards that are exposed to high voltages over extended periods.
If the bias resistors are much greater than 50\,M$\Omega$, leakage currents may affect the bias voltages applied to the wires.

The bias-voltage filters are R-C low-pass networks.
Resistance values should be much smaller than the bias resistances to control crosstalk between wires
and limit the voltage drop if any of the wires becomes shorted to the APA frame.
A value around 2.2\,M$\Omega$ is desired.
Smaller values may be considered although a larger filter capacitor would be required to maintain a given level of noise reduction.
A target value of 47\,nF has been established for the filter capacitors.

For the grid-plane bias filters, component values are less critical.
If possible they will be identical to those used for the bias resistors and coupling capacitors
(50~M$\Omega$ and 2.2 to 4.7\,nF).

\subsection{Front-end mother board (FEMB)}
\label{subsec:fe_arch}

The main component of the CE architecture, illustrated in Figure~\ref{fig:tpcce_schem}, is the 
128-channel FEMB, which itself consists of an analog motherboard and an attached FPGA 
mezzanine card for processing the digital outputs.
Each APA is instrumented with 20 FEMBs, for a total of 2,560 channels per APA.
The FEMBs plug directly into the APA CR boards, making the connections from the U- and V-plane induction wires and 
X-plane collection wires to the charge amplifier circuits as short as possible.

\begin{cdrfigure}[The CE architecture]{tpcce_schem}{The CE architecture. The basic unit is the 128-channel FEMB.}
\includegraphics[width=0.9\linewidth]{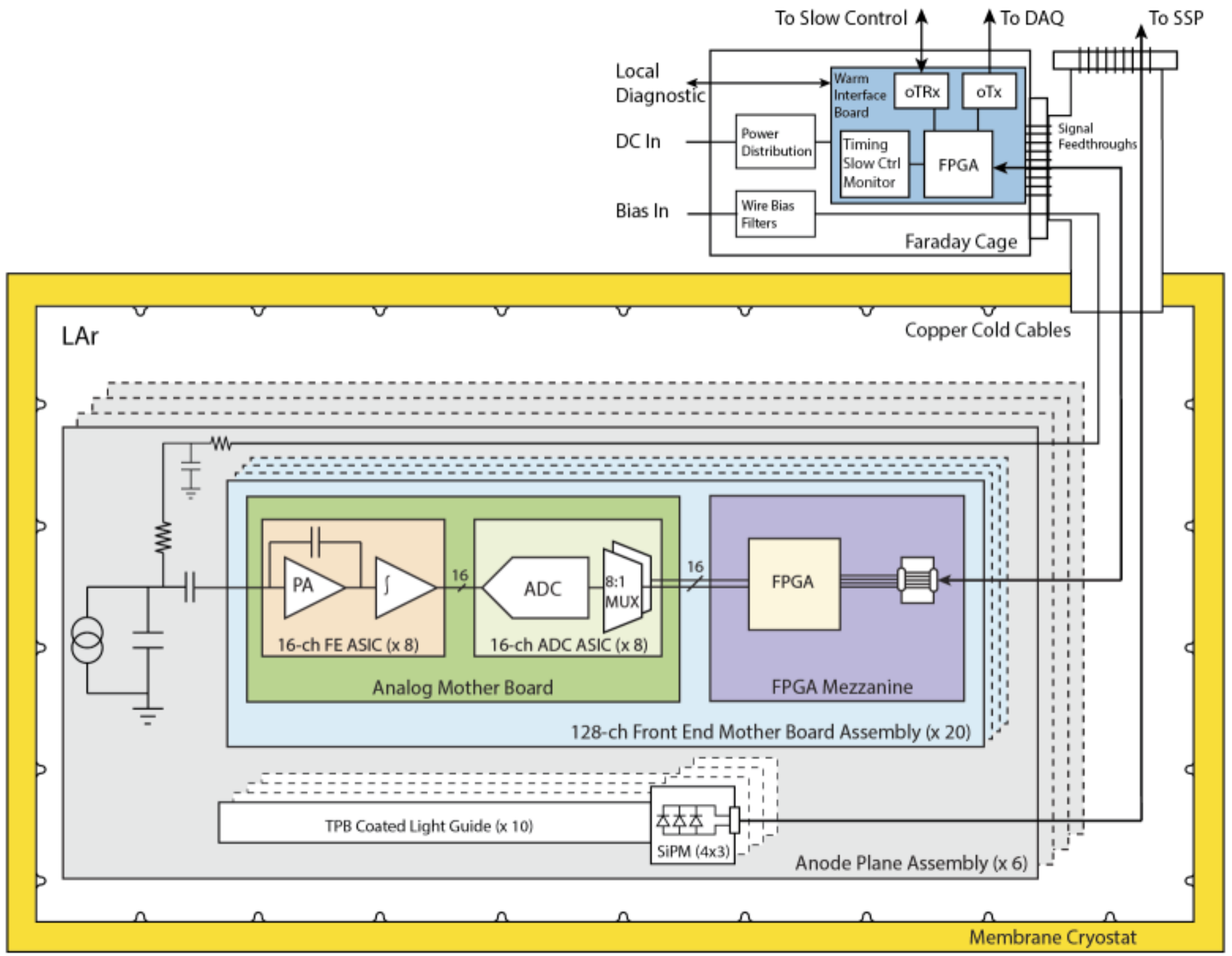}
\end{cdrfigure}

The analog mother board is instrumented with eight 16-channel FE ASICs,
eight 16-channel ADC ASICs, LV power regulators, and input-signal protection circuits.
The 16-channel FE ASIC provides amplification and pulse shaping.
The 16-channel ADC ASIC comprises 12-bit digitizers performaning at speeds up to 2 MS/s, local buffering,
and an 8:1 MUX stage with two pairs of serial readout lines in parallel. The 2016 prototype version of
the FEMB is shown in Figure~\ref{fig:tpcce_CMBpix}.

\begin{cdrfigure}[The front-end mother board (FEMB), as used in an early set of tests]{tpcce_CMBpix}
{The Front End Mother Board (FEMB), as used in the early set of tests.
  {\bf Top:} The analog mother board, showing four ADC ASICs and four FE ASICs surface mounted.
  The other side of the board has another four ADC and FE ASICs.
  Except for anticipated small modifications, this board is essentially the final version.
  {\bf Middle:} The FPGA mezzanine, used in place of the digital ASIC mezzanine for the early set of tests.
  {\bf Bottom:} The complete FEMB assembly as used in the early set of tests.
  The cable shown in the high-speed data, clock, and control cable. The middle and bottom photos are from an SBND version, which uses a different output data connector and a different input connection orientation.}
\includegraphics[width=0.65\linewidth]{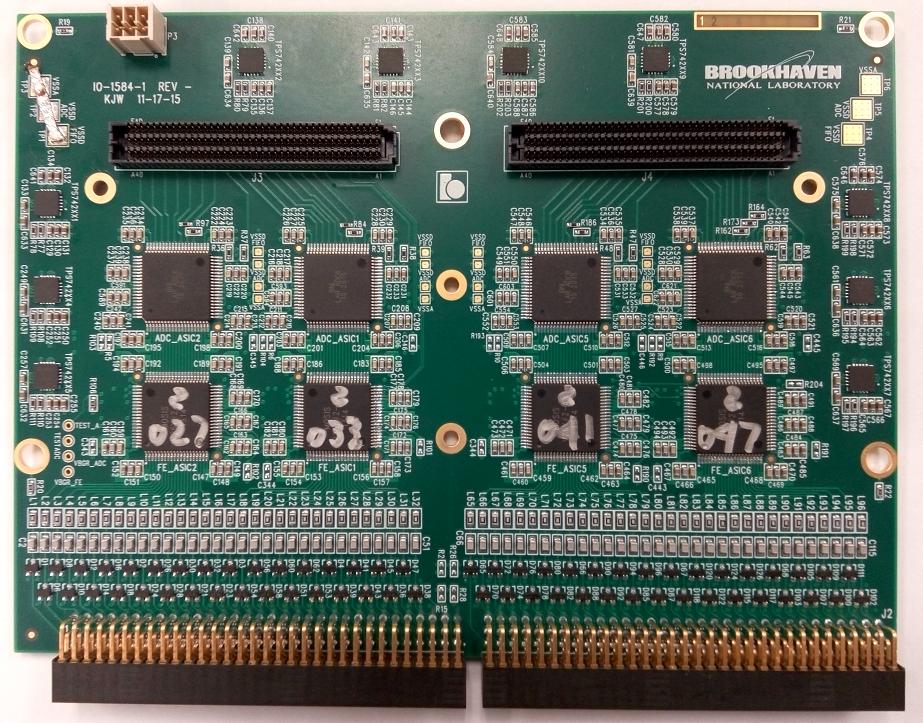}
\includegraphics[width=0.65\linewidth]{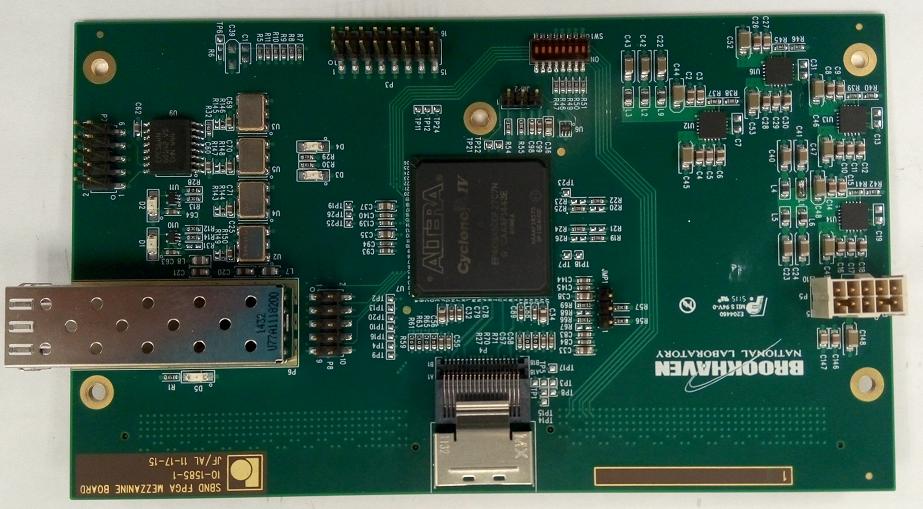}
\includegraphics[width=0.45\linewidth]{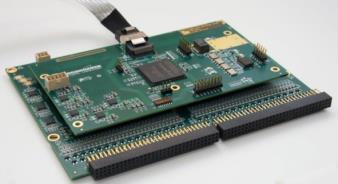}
\end{cdrfigure}

\subsubsection{Front-End ASIC}
 
Each FE ASIC channel has a charge amplifier circuit with a programmable gain selectable from one of 4.7, 7.8, 14 and 25~mV/fC
(full scale charge of 55, 100, 180 and 300~fC),
a high-order anti-aliasing filter with programmable time
constant (peaking time 0.5, 1, 2, and 3 $\mathrm{\mu}$s),
an option to enable AC coupling,
and a baseline adjustment for operation with either the collecting (200~mV) or the non-collecting (900~mV) wires.
Shared among the 16 channels in the FE ASIC are the bias circuits, programming registers,
a temperature monitor, an analog buffer for signal monitoring, and the digital interface.
The estimated power dissipation of FE ASIC is about 6~mW per channel at 1.8~V supply.

The FE ASIC layout is shown in Figure~\ref{fig:tpcce_FE_ASIC}.
The ASIC was implemented using the commercial CMOS process (0.18~$\mu$m and 1.8~V), which 
is expected to be available for at least another 10~years. 
The charge amplifier input MOSFET is a p-channel biased at 2~mA with a L/W (channel length/width) ratio
of 0.27~$\mu$m / 10~$\mu$m, followed by dual cascade stages.

\begin{cdrfigure}[Layout of the 16-channel FE ASIC]{tpcce_FE_ASIC}{The layout of the 16-channel FE ASIC}
\includegraphics[width=\linewidth]{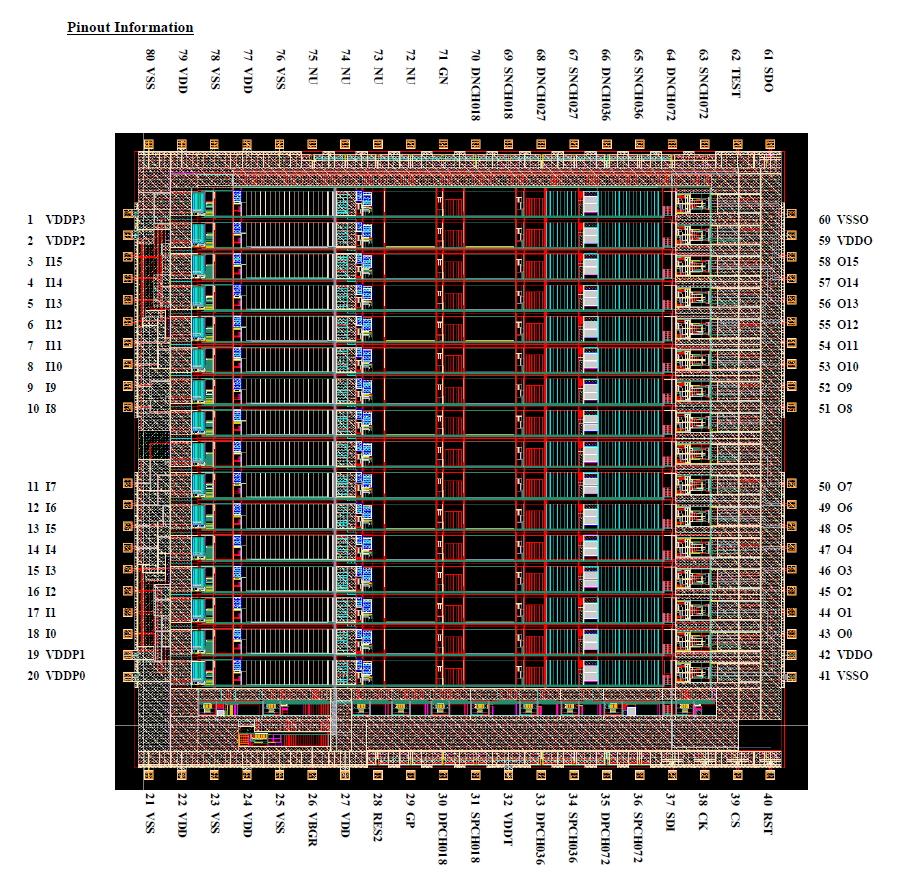}
\end{cdrfigure}

Each channel also implements a high-performance output driver,
which can be used to drive a long cable, but is disabled when interfaced to an ADC ASIC to reduce the power consumption.
The ASIC integrates a band-gap reference (BGR) to generate all the internal bias voltages and currents.
This guarantees a high stability of the operating point over a wide range of
temperatures, including cryogenic.
The ASIC is packaged in a commercial, fully encapsulated plastic QFP 80 package.

Each FE ASIC channel is equipped with an injection capacitor which can be used
for test and calibration and can be enabled or disabled through a
dedicated register. The injection capacitance has been measured using 
a calibrated external capacitor. The measurements show
that the calibration capacitance is extremely stable, changing from
184~fF at RT to 183~fF at 77~K. This result and the measured
stability of the peaking time demonstrate the high stability of the
passive components as a function of temperature. Channel-to-channel and chip-to-chip
variation in the calibration capacitor are typically less than 1\%. 

Prototype ASICs have been evaluated and characterized at RT (300~K) and LN2 (77~K) temperature.
During testing the circuits have been cycled multiple times
between the two temperatures and operated without any change in performance.
Figure~\ref{fig:tpcce_shaper_out} shows the measured pulse response, both as a function
of temperature and the programmable settings of the chip.
These results are in close agreement with simulations and indicate
that both the analog and the digital circuits and interface operate as
expected in a cryogenic environment.

\begin{cdrfigure}[Measured pulse response with details]{tpcce_shaper_out}{Measured pulse response with
 details on gain, peaking time and baseline adjustments}
\includegraphics[width=\linewidth]{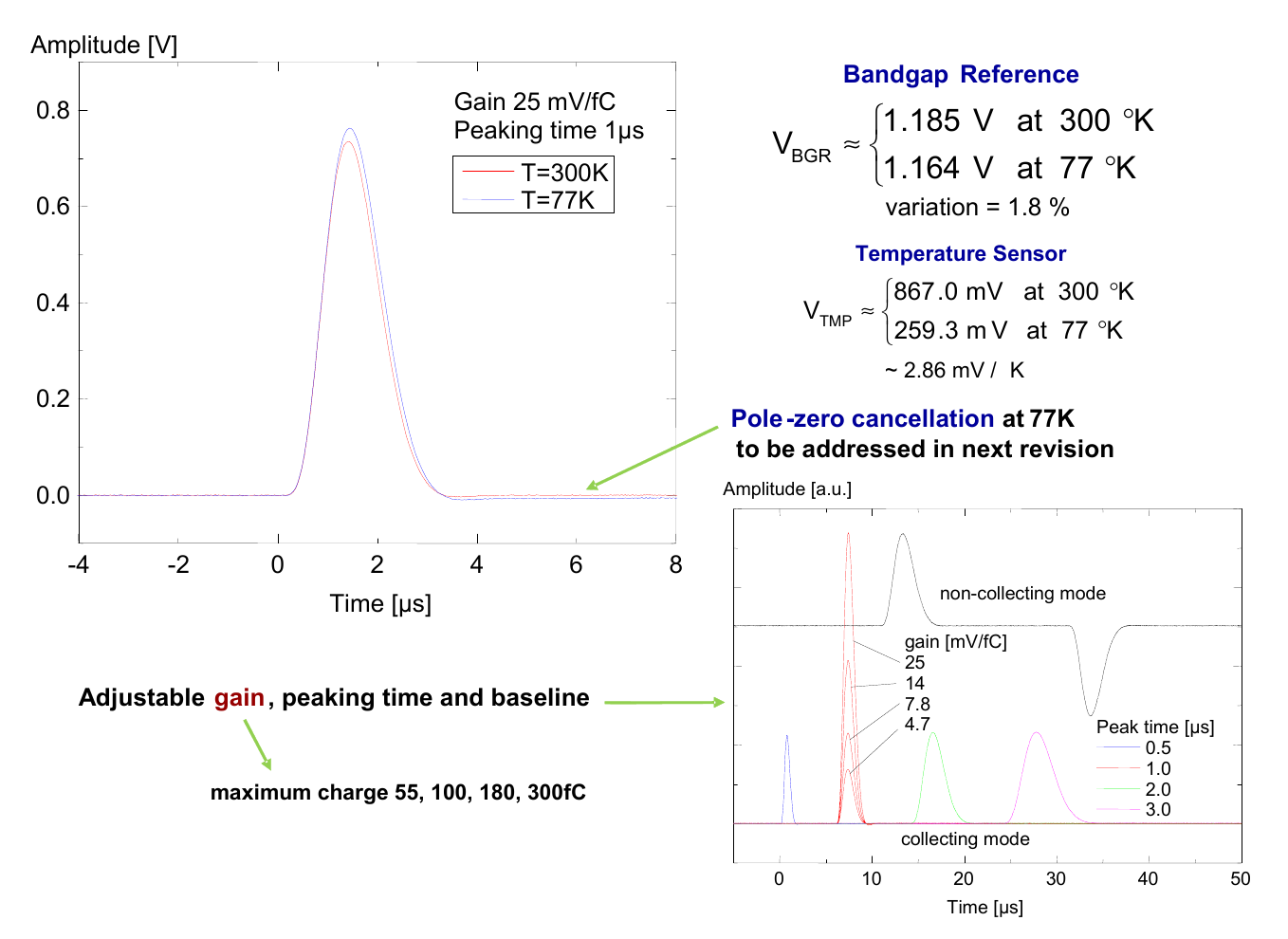}
\end{cdrfigure}

\subsubsection{ADC ASIC}

The ADC ASIC design is also implemented using the CMOS process (0.18~$\mu$m and 1.8V).
The layout of the ADC ASIC is shown in Figure~\ref{fig:tpcce_ADC_ASIC}. 
The ADC ASIC is a complex design with 320,000 transistors, while the FE ASIC has 16,000.
The transistor design work has been done following the rules for long cryo-lifetime.
Shared among the 16 channels in the ADC ASIC are the bias circuits, programming registers,
an 8:1 MUX, and the digital interface.
The estimated power dissipation of ADC ASIC is below 5~mW per channel at 1.8~V supply.
  
\begin{cdrfigure}[The layout of the 16-channel ADC ASIC]{tpcce_ADC_ASIC}{The layout of the 16-channel ADC ASIC}
\includegraphics[width=\linewidth]{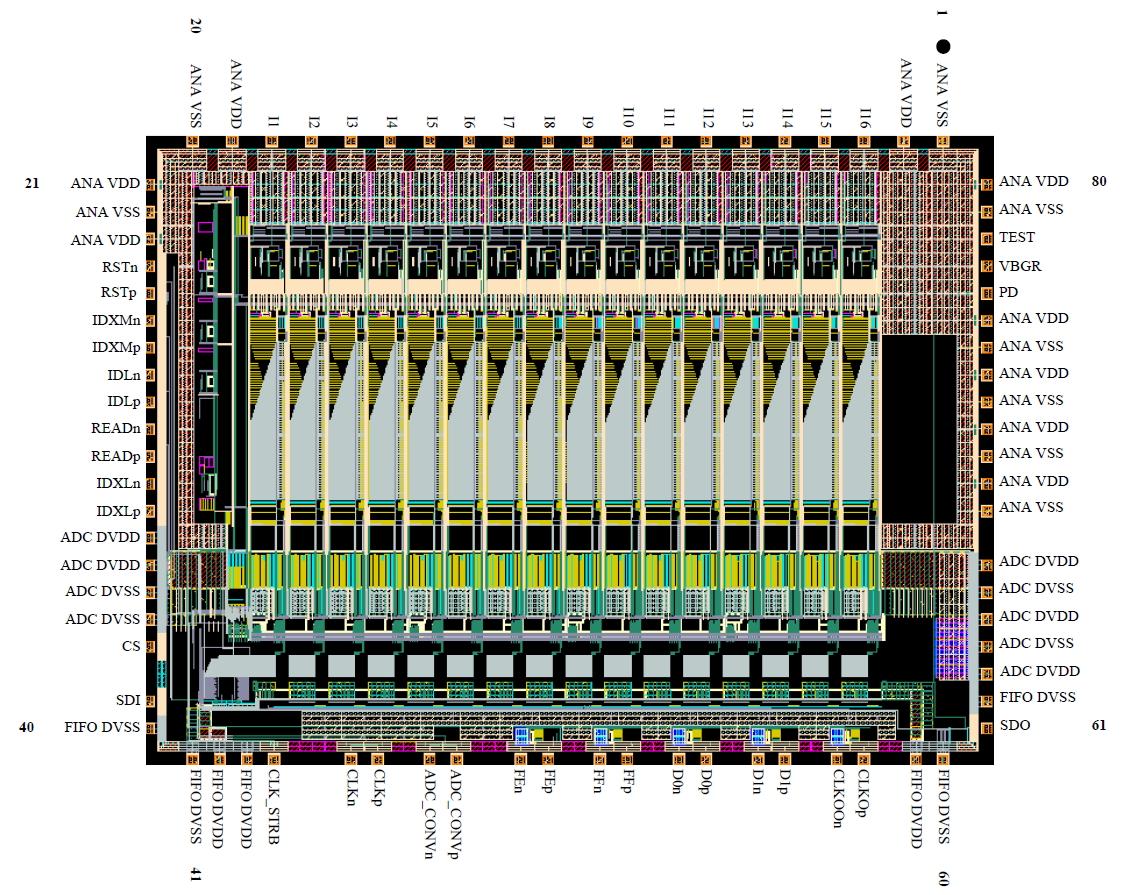} 
\end{cdrfigure}

The ADC ASIC has an input buffer with offset compensation to match the output of the FE ASIC.
The input buffer first samples the input signal (with a range of 0.2~V to 1.6~V),
then provides a current output after compensating for offset voltage error.
This current output is then supplied to the ADC which converts the input to digital in two phases.
The MSB (Most Significant Bit) 6~bits are first determined followed by the LSB (Least Significant Bit) 6~bits.

After the conversion the code is converted to binary and latched.
The output of ADC channel 16 can be monitored externally.
The data from the 16 ADCs are transferred in parallel to the FIFO block.
The built-in FIFO is 32~bits wide and 192~bits long,
and has full and empty indicator flags, needed for interfacing to the FPGA.
The ADC along with the input buffers are biased internally using a bias generator and a bandgap voltage reference.
The bandgap voltage (VBGR) can be monitored and/or controlled externally.
It can be put in the low-power sleep mode, and woken up in less than 1~$\mu$s.

Prototypes have been evaluated and characterized at RT (300~K) and LN2 (77~K) temperature.
During these tests the circuits have been temperature-cycled multiple times.
The effective resolution with reference to the input referred noise is $\sim$11.6~bits at both 300~K and 77~K.
The differential non-linearity (DNL) is less than 4 LSBs for 99\% of ADC bins at both 300~K and 77~K.

\subsubsection{Cold FPGA}

The ADC ASIC data are passed to the FPGA mezzanine board for transmission to the warm electronics
located on the outside of the signal flange.
The FPGA has four 4:1 MUX circuits that combine the 16 serial lines from the eight ADC
channels into four serial lines of 32 channels each, and 
four $\sim$1.2 Gigabit-per-second (Gbps) serial drivers that drive the data in each
line over cold cables to the WIBs.
The data are transmitted to the signal flange and WIBs on copper cables utilizing LV differential signaling (LVDS).
On the WIBs, the data is further MUXed by 4:1 and transmitted over optical
fibers to the DAQ system described in Section~\ref{sec:DAQ_online_interface}.

The FPGA on the mezzanine card is also responsible for communicating with the
control and timing systems from the WIB and providing the clock and control signals required by the FE and ADC ASICs. 
By default the FPGA will load firmware from a cold flash onboard the FEMB. The remote update of 
FPGA firmware via JTAG chain is available through the WIB, with 4 differential cables to each FEMB 
from the WIB. In addition, the cold flash can be programmed remotely. In the case that faulty firmware 
is identified in-situ by slow control and monitoring, an update of the firmware will be initiated.

The FPGA and all other electrical components on the FEMB
assembly have been evaluated and characterized at RT (300~K) and LN2 (77~K) temperature on 
prototype FEMB. During these tests the FEMB have been both temperature-cycled and power-cycled at
cryogenic temperature multiple times.

Figure~\ref{fig:tpcce_enc} shows the measured Equivalent Noise Charge (ENC) as a function of 
filter-time constant (peaking time) for two different gains as measured on a prototype FEMB. ENC is the value of charge 
(in electrons) injected across the detector capacitance that would produce at the output of the 
shaping amplifier a signal whose amplitude equals the output R.M.S. noise. These measurments
were made with the prototype FEMB at both RT and submerged in LN2 with a wire-simulating input capacitance of $C_f~=~150$~pF
(equivalent to approximately 7 meter sense wire load).
In LN2, for peaking times $>$1~$\mu$s, less than 600~e$^{-}$ was measured. For comparison,
a MIP travelling perpendicularly to the wire plane in the direction of wire spacing is
expected to deposit $\sim$~10,000~e$^{-}$ on the collection wires, for a worst-case
S:N$\sim$16:1.

\begin{cdrfigure}[Measured ENC vs filter time constant]{tpcce_enc}{
Measured ENC vs filter time constant from the latest prototype version of the FEMB
for two different gains, 14~mV/fC and 25~mV/fC. RT = room temperature and 
LN2 = liquid nitrogen}
\includegraphics[width=0.45\linewidth]{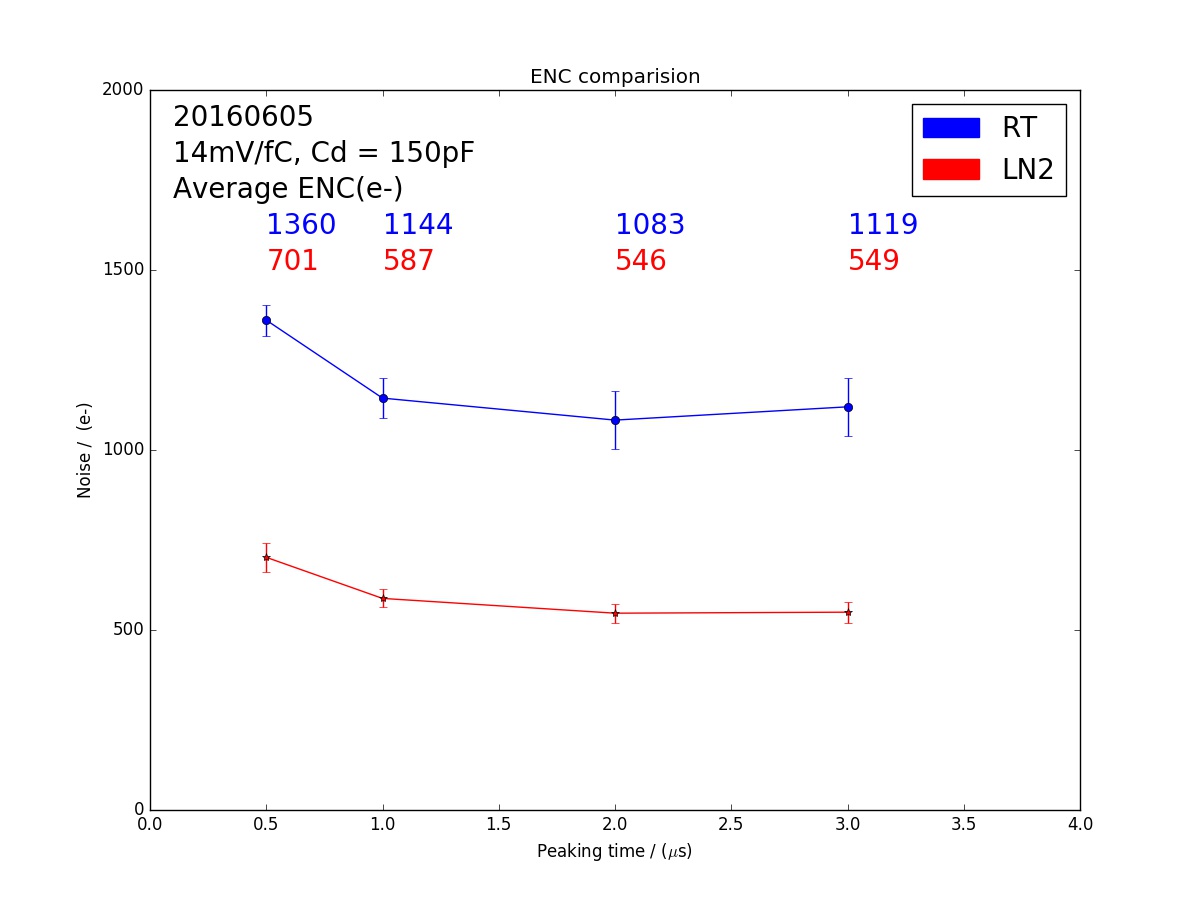}
\includegraphics[width=0.45\linewidth]{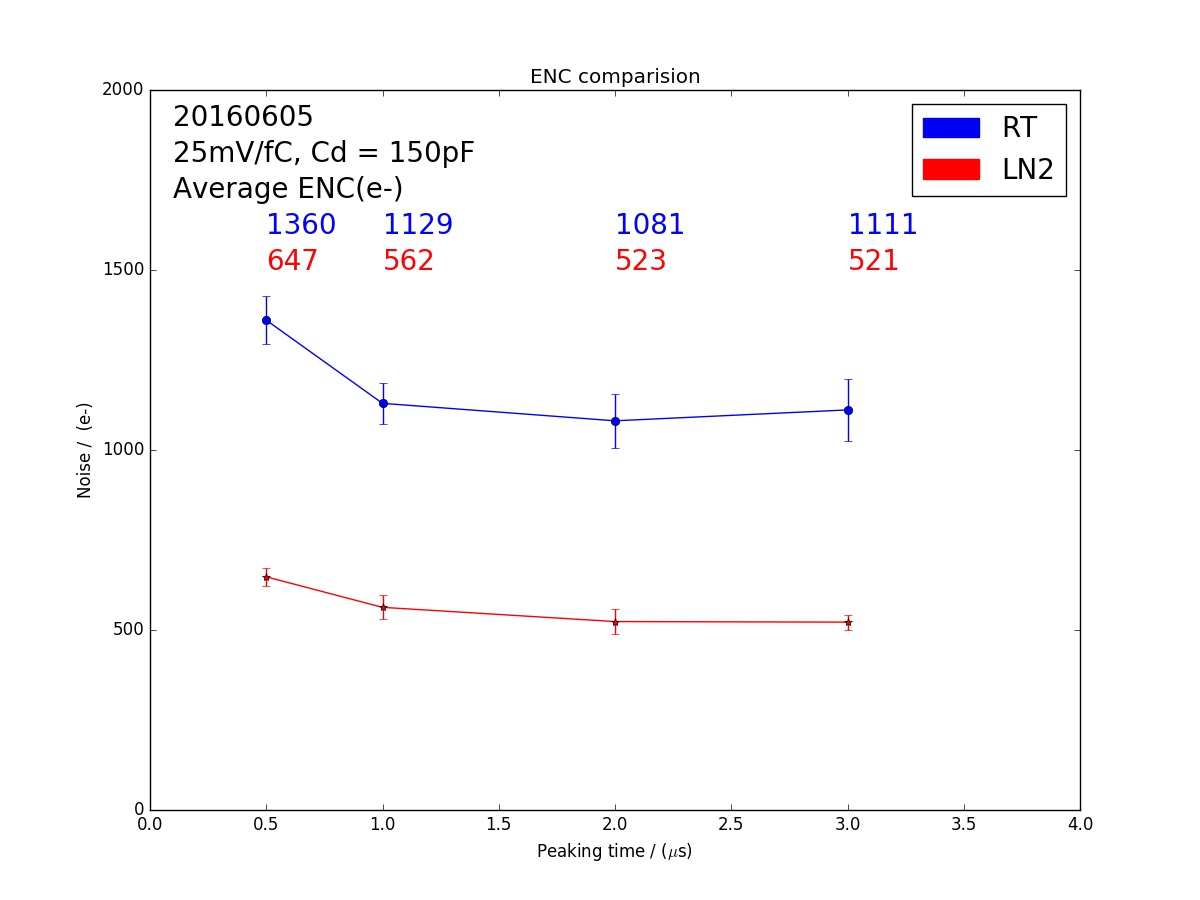}
\end{cdrfigure}

\subsubsection{CE Box}

Each FEMB is enclosed in a Faraday box to provide shielding from noise. 
As shown in Figure~\ref{fig:tpcce_box}, the Faraday box is designed to make the electrical connection 
between the FEMB and the APA frame, as defined in Section~\ref{subsec:groundshield}. Mounting 
hardware inside the Faraday box connects the ground plane of the FEMB to the box casing. The
box casing is electrically connected to the APA frame via twisted conducting wire (not 
shown in Figure~\ref{fig:tpcce_box}). This is the only point of contact between the FEMB and
APA, except for the input amplifier circuits connected to the CR board, which also terminate to
ground at the APA frame, as shown in Figure~\ref{fig:tpcce_cr_board}.

\begin{cdrfigure}[Faraday box for the FEMB]{tpcce_box}{Faraday box for the FEMB.}
\includegraphics[width=3in]{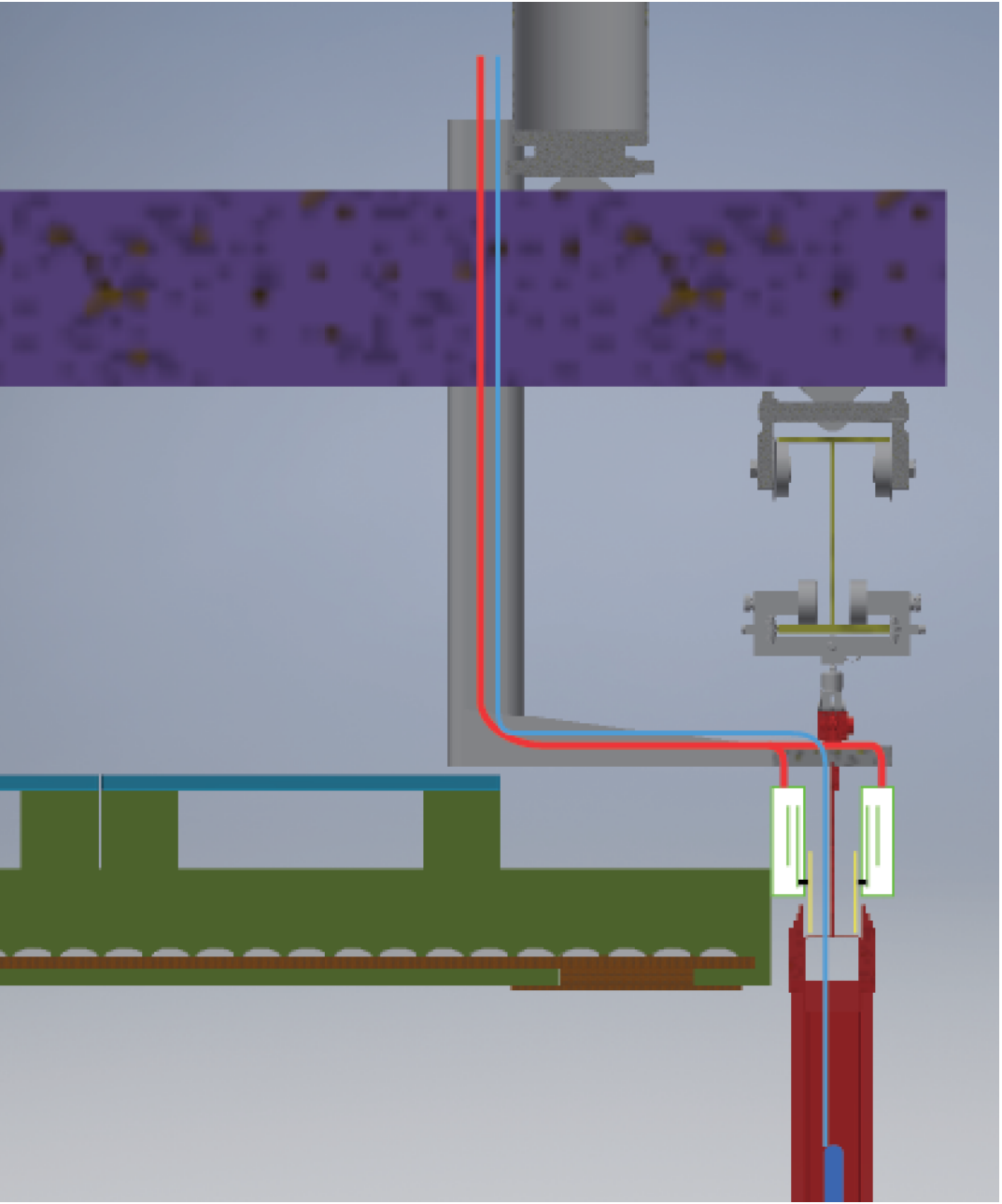}
\includegraphics[width=3in]{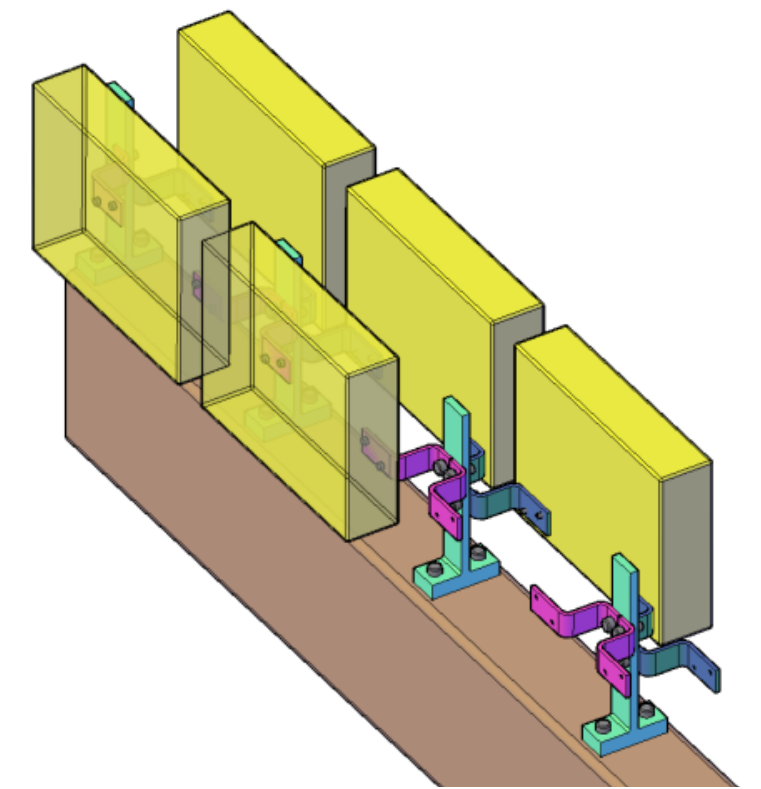}
\end{cdrfigure}

%
\subsection{CE feedthroughs and cold cables}
\label{subsec:ce_feedthrough}

All cold cables originating from inside the cryostat connect to the outside warm electronics through PCB board feedthroughs
installed in the signal flanges that are distributed along the cryostat roof (Figure~\ref{fig:tpcce_FT_InternalCableRoute}).
The TPC data rate per APA, with an overall 32:1 MUX and 80 $\sim$1~Gbps data channels per APA,
is sufficiently low that the LVDS signals can be driven over copper twin-axial transmission lines.
Additional transmission lines are available for the distribution of LVDS clock signals and I$^2$C control information,
which are transmitted at a lower bit rate.
Optical fiber is employed externally from the WIBs on the signal flange to the DAQ and slow control systems.

\begin{cdrfigure}[CE feedthrough configuration and internal cable routing]{tpcce_FT_InternalCableRoute}{
The CE feedthrough configuration and internal cable routing. The left panel shows a cutaway view of the cryostat.
The right panel shows more detail at the Faraday boxes.}
\includegraphics[width=7in]{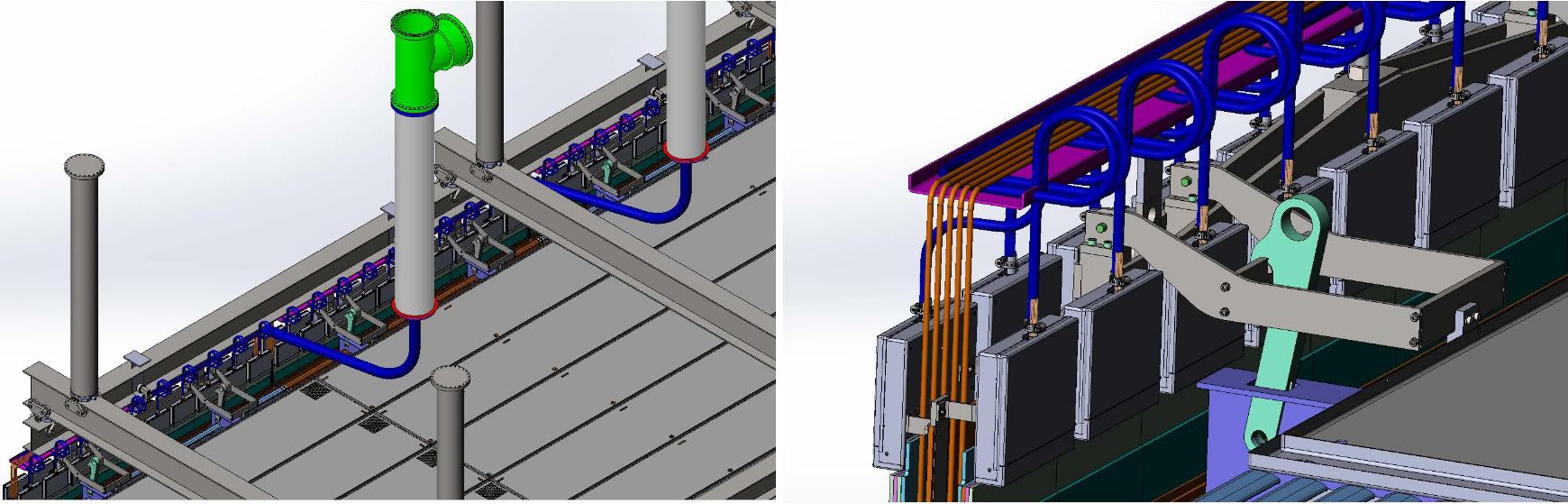}
\end{cdrfigure}

\begin{cdrfigure}[Conceptual design of TPC CE feedthrough]{tpcce_signal_FT}{
TPC CE and PD signal feedthroughs mounted on the top of the ProtoDUNE-SP cryostat. Also shown are the warm interface electronics crate and boards.}
\includegraphics[width=0.6\linewidth]{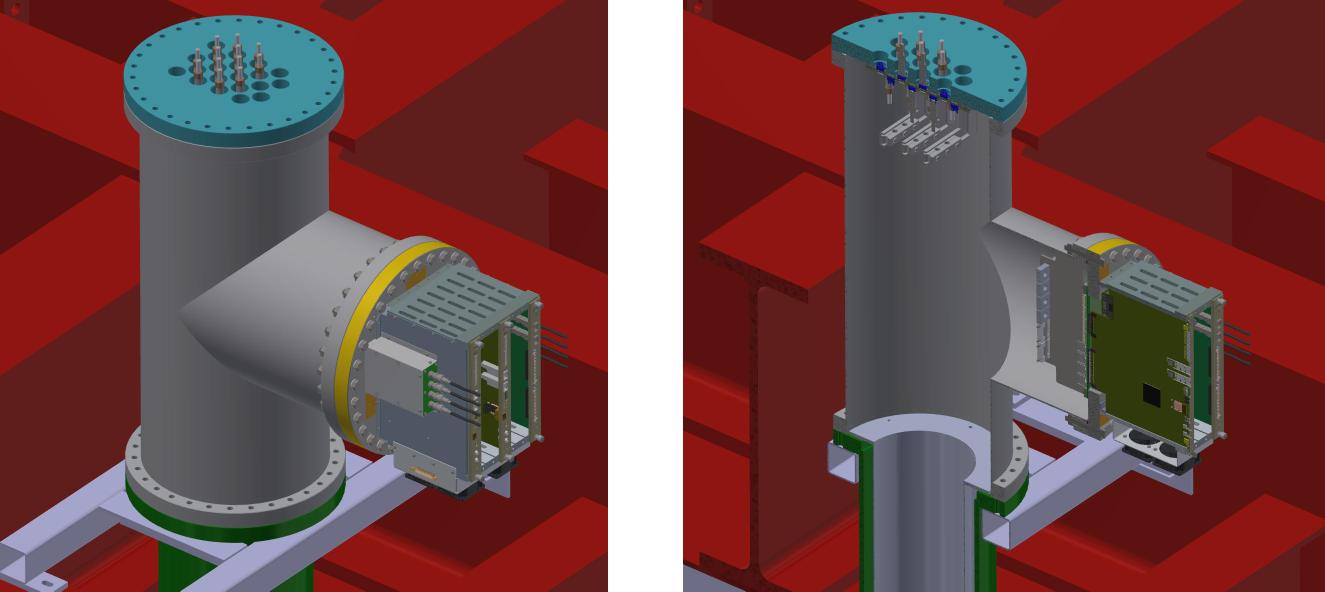}
\end{cdrfigure}

The design of the signal flange includes a T-shaped pipe, separate PCB feedthroughs for the CE and PDS cables, and
an attached crate for the TPC warm electronics, as shown in Figure~\ref{fig:tpcce_signal_FT}.
The wire-bias voltage cables connect to standard SHV (safe high voltage) connectors machined directly into the CE feedthrough,
ensuring no electrical connection between the wire-bias voltages and other signals passing through the signal flange.
Each CE feedthrough serves the bias/power/digital IO needs of one APA, as shown 
in Figure~\ref{fig:tpcce_cable_routing}.  

\begin{cdrfigure}[TPC cable routing scheme]{tpcce_cable_routing}{TPC cable routing scheme for three APA section.}
\includegraphics[width=0.9\linewidth]{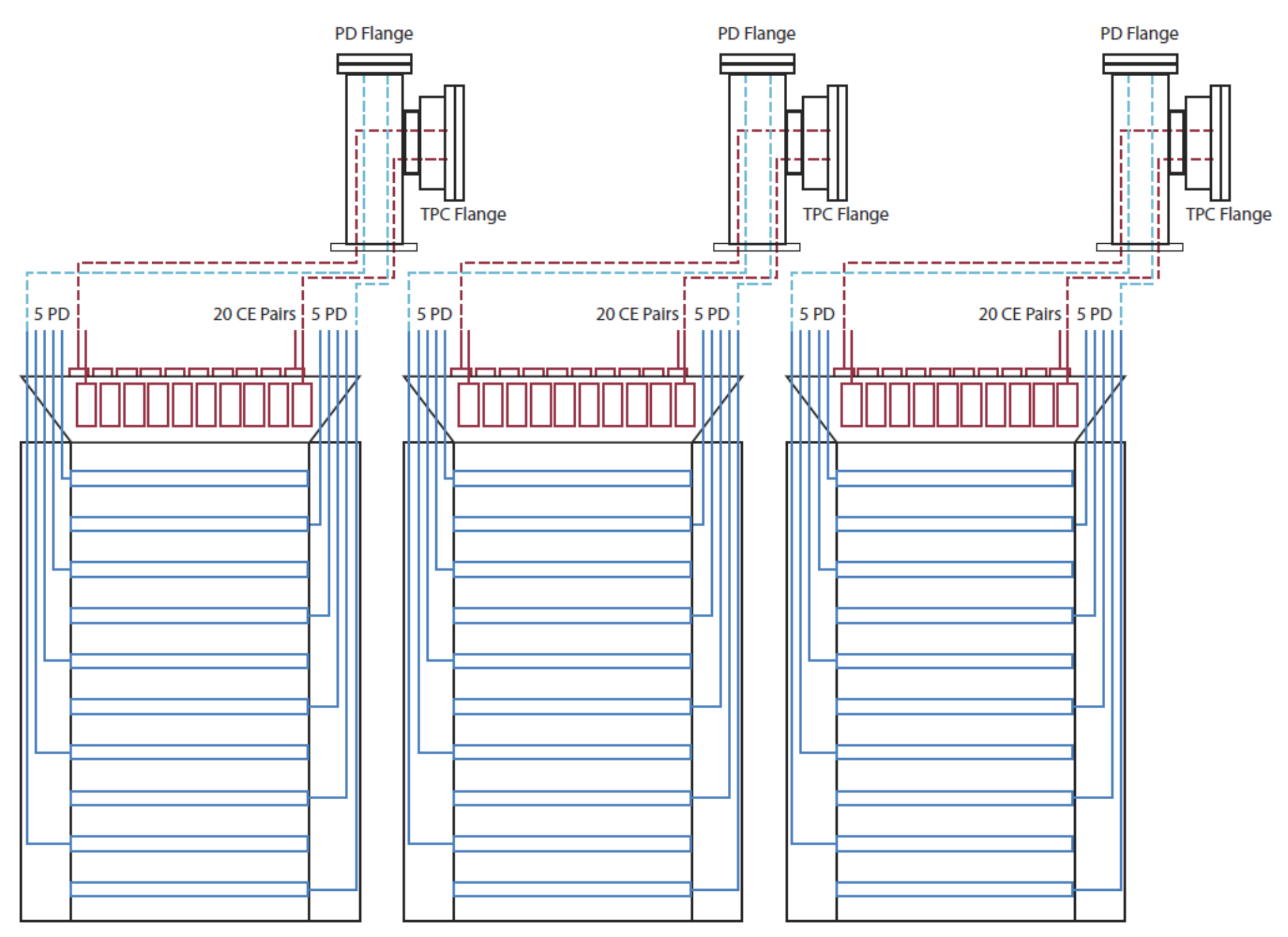}
\end{cdrfigure}

A program for minimizing potential contamination of the LAr from the cable plant contained within the ullage
(the warmer gas phase at the top of the cryostat) 
is being carefully followed.

Data/control cable bundles are used to send system clock and control signals from the 
signal flange to the FEMB, stream the $\sim$1~Gbps high-speed data from the FEMB to the signal flange, and 
provide backup JTAG programming to the cold FPGA, in case the power-up programming from the onboard 
flash EEPROM fails. As described in Section~\ref{subsec:ce_intro}, each FEMB 
connects to a signal flange via one data cable bundle, leading to 20 bundles between one APA and one flange.
Each data bundle contains 12 low-skew copper twin-axial cables with a drain wire, 
to transmit the following differential signals:

\begin{itemize}
    \item 4$\times$1.2~Gbps high-speed data
    \item One 100~MHz clock
    \item One 2~MHz CONVERT clock
    \item 2 I$^2$C control and configure
    \item 4 single-ended JTAG programming for the FPGA
\end{itemize}

The selected cables are Samtec 26~AWG twin-axial bundles with Samtec HSEC08 connectors to both
the FEMB mezzanine board and the signal flange. Each twin-axial pair is separately shielded.
The HSEC08
connectors lock into place with tabs on each side of the connector. A sample of the Samtec cable with
THV outer jacket has passed outgassing tests in the LAr Materials Test Stand at Fermilab.

The Samtec 26~AWG cable has been
tested and demonstrated to have low enough dispersion such that both the LVDS 50~MHz system clock and
$\sim$1~Gbps high-speed data can be recovered over 25~meters of RT cable, 
significantly longer than the required seven meters needed to run cables between the FEMBs and signal flanges.

\begin{cdrfigure}[Results from cable validation testing]{tpcce_samtec_results}{Eye diagrams 
from cable validation testing. {\bf Top Left:} 50~MHz system clock over 25~m RT  
(RT) Samtec 26AWG cable. For comparison, {\bf Bottom Left} shows the same clock over 
the heavier, prohibitively expensive Gore 24AWG cable. {\bf Top Right:} 1~Gbps data over 
7~m (ProtoDUNE length) RT Samtec 26AWG cable without active recovery by equalizers. {\bf Bottom Right} 1~Gbps
data over 25~m (DUNE length) RT Samtec 26AWG cable with active recovery.}
\includegraphics[width=0.9\linewidth]{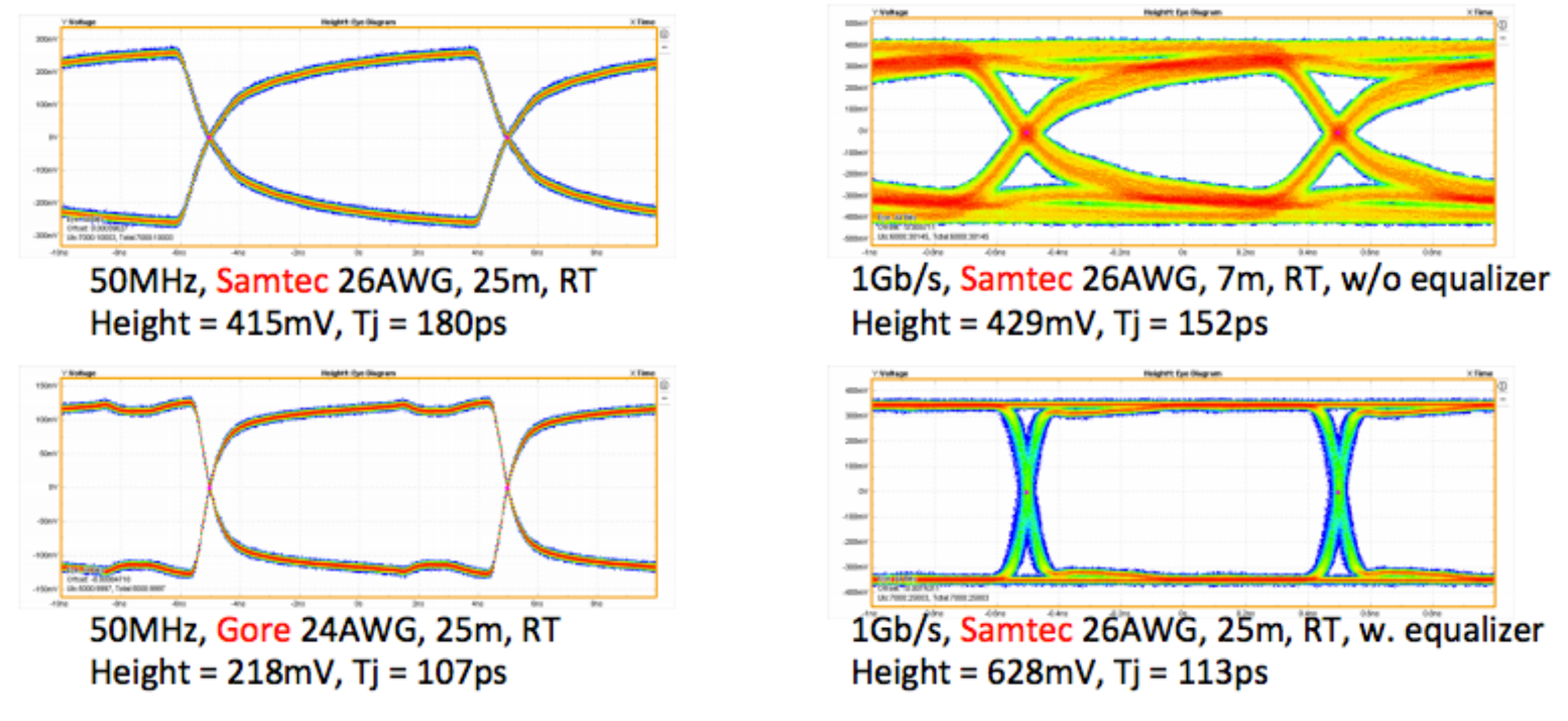}
\end{cdrfigure}

Figure~\ref{fig:tpcce_samtec_results} shows results from the cable 
validation testing. The cable connectors through the signal feed-through are emulated in the test stand with 
proper connectors and a test PCB. The eye diagrams show the edges of the differential signals after 
LVDS transmission over the specified cable types and lengths. The height of eye diagram shows the size 
of the recovered signal in mV and the slope of the rising and falling edges are jitter in picoseconds (ps). 
An eye diagram is sufficient to show that the edges of the differential signals can
be recovered, but not enough to demonstrate the bit error rate (BER). However, the Samtec 26~AWG cable has 
also passed a BER test, transmitting $10^{13}$ bits without error.

LV power is passed from the signal flange to the FEMB by bundles of 18 Samtec 
20~AWG twisted-pair wires, as shown in Figure~\ref{fig:tpcce_cable_routing}. One IPD1 connector
attaches all 18 wires at the signal flange, and two IPD1 connectors are attached to the FEMB (one to the
analog motherboard and one to the FPGA mezzanine). In total, 20 wire bundles 
 bring LV power to the FEMBs associated with one APA.

Nine of the 18 wires are power feeds; the other nine wires
are attached to the grounds of the input amplifier circuits, as described in Section~\ref{subsec:ce_wire_bias}.
For a single FEMB, the resistance is measured to be $<30$~m$\Omega$ at RT or $<10$~m$\Omega$ at 
LAr temperature. Each APA has a copper cross-section of approximately $80~\mathrm{mm}^2$, with a 
resistance $<1.5$~m$\Omega$ at RT or $<0.5$~m$\Omega$ at LAr temperature. The power loss 
on the 7 meter LV cables to each FEMB is $\sim$0.1W at room temperature, or $\sim$2W per APA, and will be 
further reduced when operating in LAr.

The wire-bias voltage cables are required to deliver voltages up to a few thousand Volts and currents up to a few
milliAmps.

The bias voltages are applied to the X-, U-, and G-plane wire layers, three field cage terminations, 
and an electron diverter, as shown in Figure~\ref{fig:tpcce_cr_board}. The voltages are supplied 
through eight SHV connectors mounted on the signal flange. RG-316 coaxial cables carry the voltages 
from the signal flange to a patch panel PCB which includes noise filtering mounted on the top 
end of the APA. 

From there, wire-bias voltages are carried by single wires to 
various points on the APA frame, including the CR boards, a small PCB mounted on or near 
the patch panel that houses a noise filter and termination circuits for the field cage voltages, and 
a small mounted board near the electron diverter that also houses wire-bias voltage filters.

\subsection{Warm interface electronics}
\label{subsec:warm_interface_elec}

The warm interface electronics are housed in warm interface electronics crates (WIECs)
attached directly to the signal flange.  The WIEC shown in Figure~\ref{fig:tpcce_ceflange_dune} 
contains one
Power and Timing Card (PTC), up to five Warm Interface Boards (WIBs) and a passive
Power and Timing Backplane (PTB), which fans out signals and LV power from the PTC to the WIBs.

\begin{cdrfigure}[Exploded view of the signal flange]{tpcce_ceflange_dune}{Exploded view of the ProtoDUNE-SP signal flange.}
\includegraphics[width=0.9\linewidth]{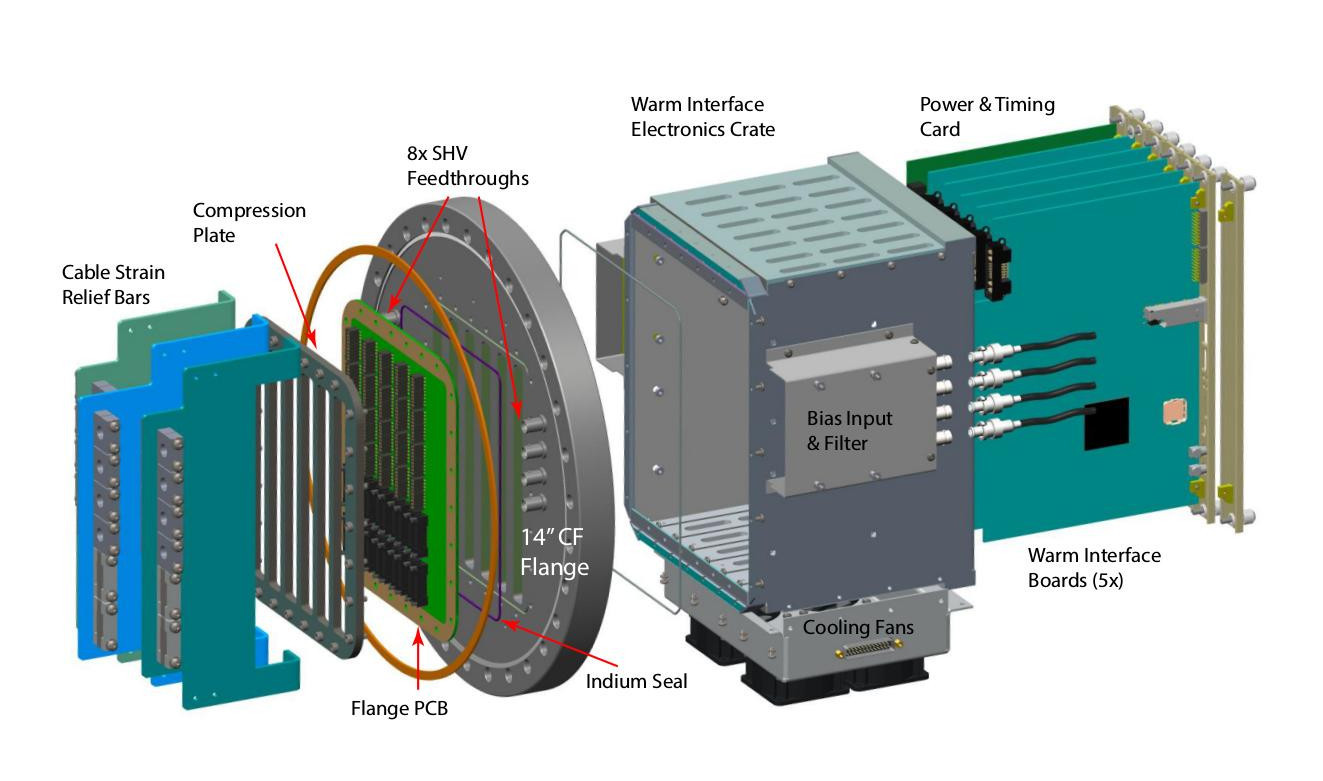}
\end{cdrfigure}

The WIB is the interface between the
DAQ system and up to four
FEMBs. It receives the system clock and control signals from the
timing system and provides for processing and fan-out of those signals to the four
FEMBs. The WIB also receives the high-speed data signals from the four 
FEMBs and transmits them to the DAQ system over optical
fibers.  The WIBs are attached directly to the TPC
CE feedthrough on the signal flange. The feedthrough
board is a PCB with connectors to the cold signal and LV power cables fitted
between the compression plate on the cold side, and sockets for
the WIB on the warm side. Cable strain relief for the cold cables is 
supported from the back end of the feedthrough.


\begin{cdrfigure}[Power and Timing Card (PTC) and timing distribution]{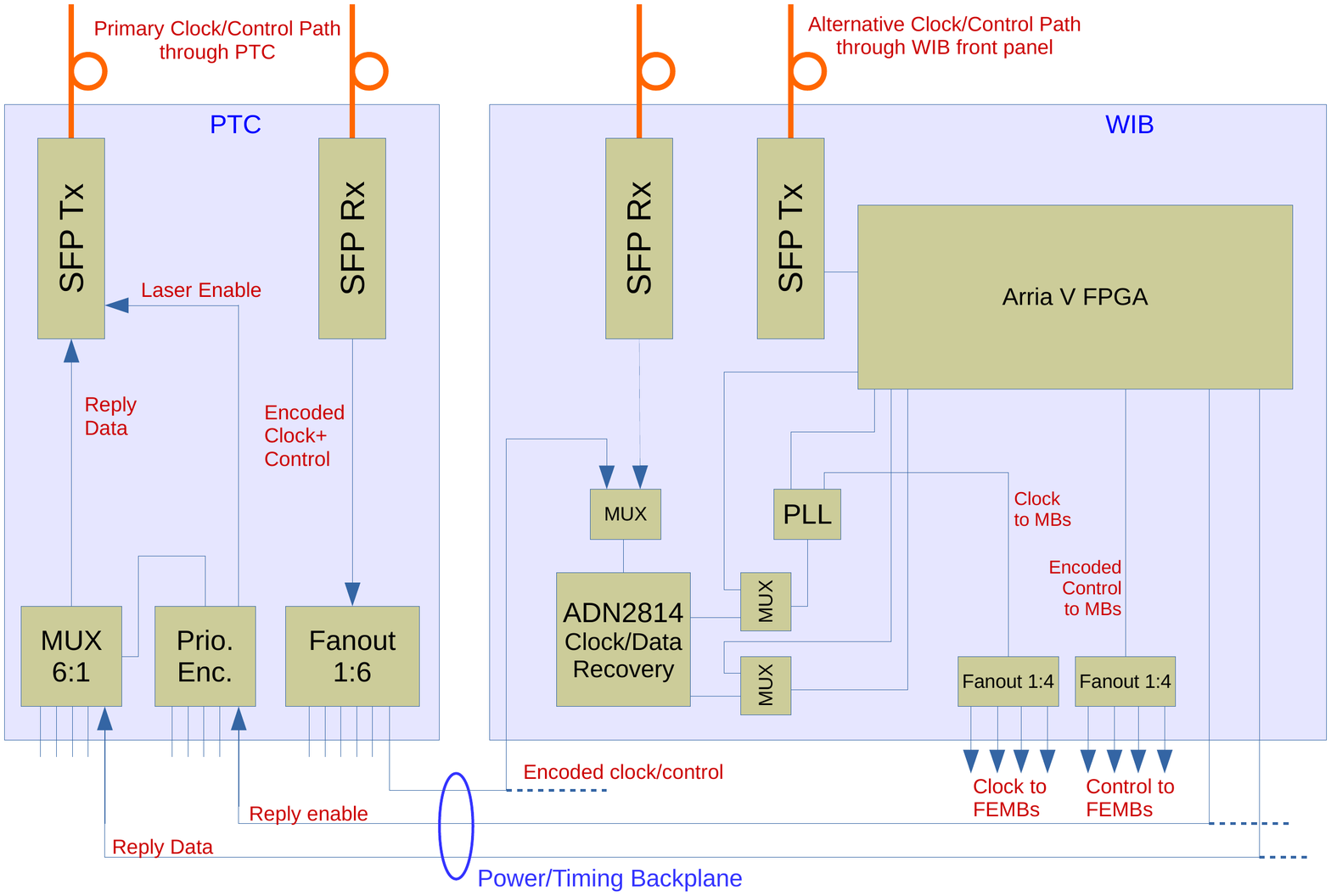}{Power and Timing Card (PTC) 
and timing distribution to the WIB and FEMBs.}
\includegraphics[width=0.5\linewidth]{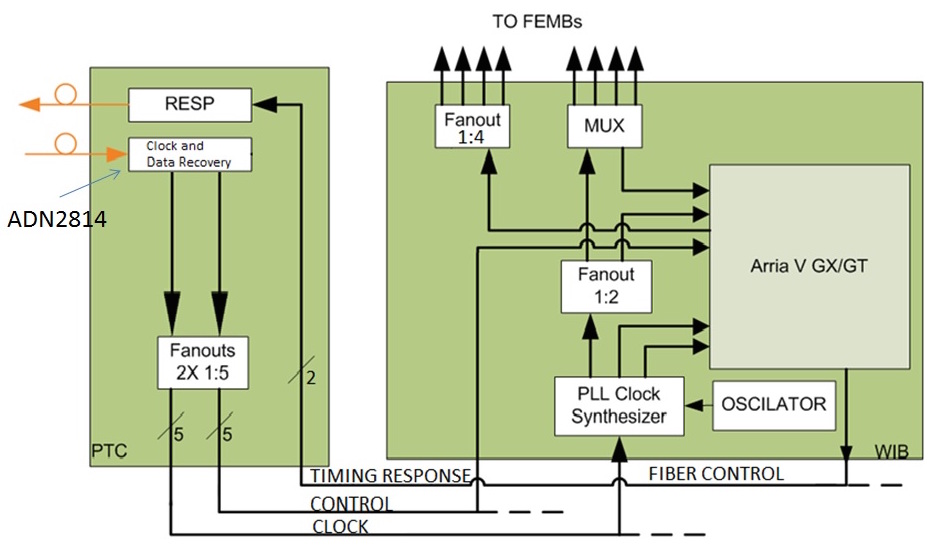}
\end{cdrfigure}

The PTC provides a bidirectional fiber interface to the
timing system.  The clock and data
streams are separately fanned-out to the five WIBs as shown in
Figure~\ref{fig:tpcce_wib_timing}. The PTC fans the clocks out to the WIB over the
PTB, which is a passive backplane attached directly to the PTC and
WIBs.  The received clock on the WIB is separated into clock and
data using a clock/data separator.

\begin{cdrfigure}[LV power distribution to the Warm Interface Boards and FEMBs]{tpcce_wib_power}{LV power distribution 
to the WIB and FEMBs. 250\,W is for a fully-loaded crate 
with the majority of the power dissipated by the 20 cold FEMBs in the LAr.}
\includegraphics[width=0.6\linewidth]{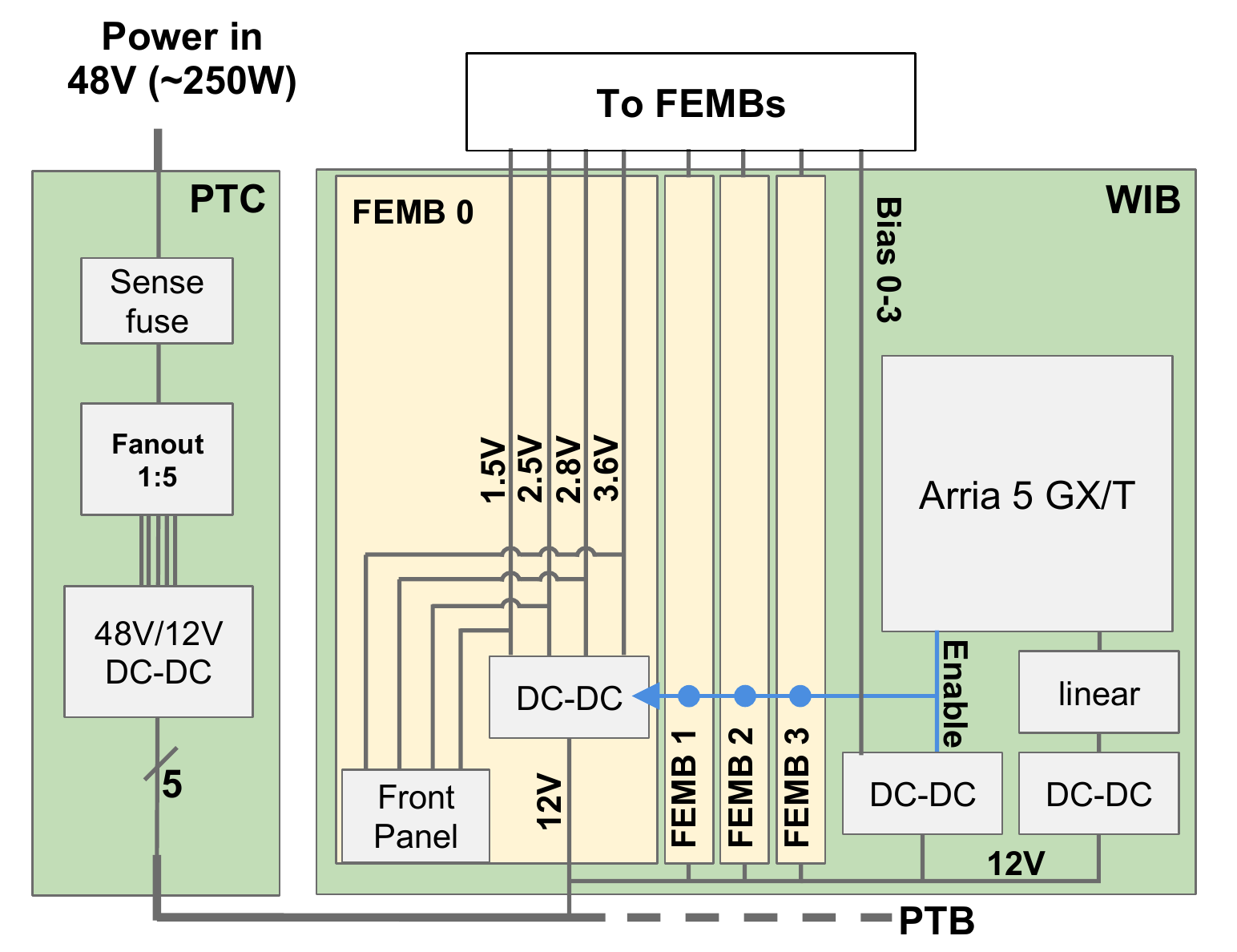} 
\end{cdrfigure}

The PTC also receives LV power for all cold
electronics connected through the signal flange, approximately 250W at 48V for a
fully-loaded flange (one~PTC, five~WIB, and 20~FEMB). The LV power is then stepped down
to 12V via a DC/DC converter onboard the PTC and fanned out
on the PTB to each WIB, which provides the necessary 12V DC/DC conversions and fans
the LV power out to each of the cold FEMBs supplied by that WIB, 
as shown in Figure~\ref{fig:tpcce_wib_power}. The 
majority of the 250W drawn by a full flange is dissipated in the LAr
by the cold FEMB.


Each WIB contains a 
unique IP address for its UDP slow control interface. The IP address for the WIB is 
derived from a crate and slot address: the crate address is generated on the PTC 
board via dipswitches and the slot address is generated by the PTB slot, numbered 
from one to five. Note that the WIBs also have front-panel
connectors for receiving LV power; these can be used in place of 
the LV power inputs on the PTB generated by the PTC.

The WIB is also capable of
receiving the encoded system timing signals over bi-directional optical
fibers on the front panel, and processing these using either
the on-board FPGA or clock synthesizer chip to provide the 50~MHz
clock required by the cold electronics.  

\begin{cdrfigure}[Warm Interface Board]{tpcce_dune_wib}{Warm interface board (WIB). Note 
that front panel inputs include a LEMO connector and alternate inputs for LV power.}
\includegraphics[width=0.9\linewidth]{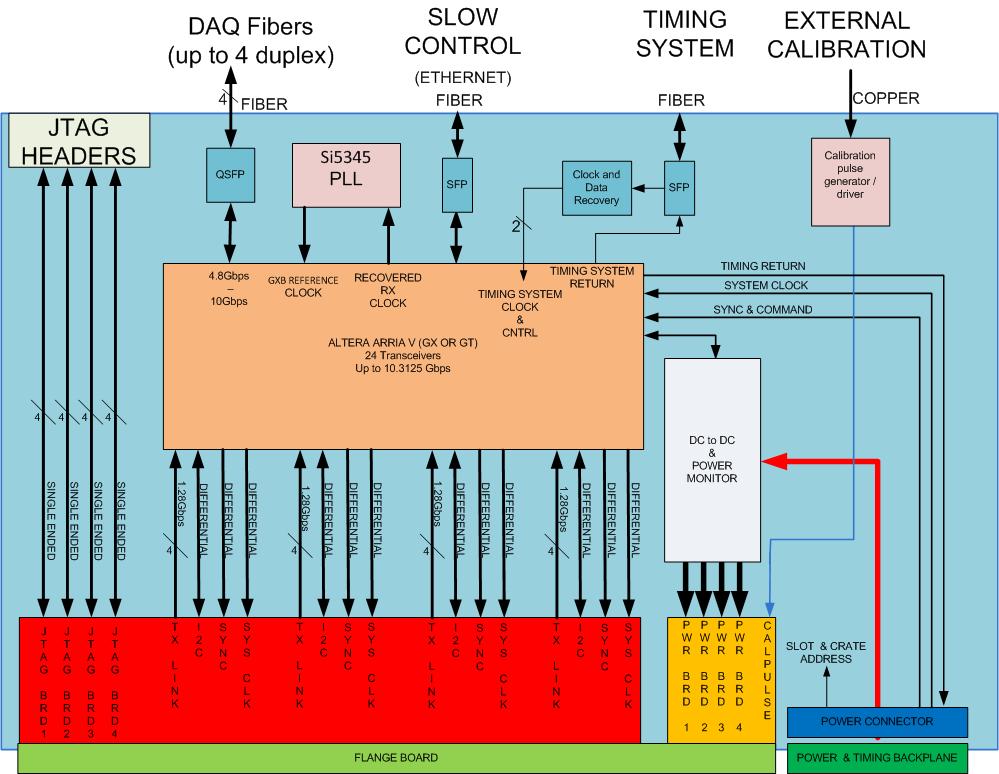}
\end{cdrfigure}

The FPGA on the WIB is an Altera Arria V GT variant, which requires a
125~MHz clock for its state machine that is provided by an on-board crystal
oscillator. The GT variant of the Arria V
transceivers can drive the high-speed data to the DAQ system up to
10.3125~Gbps per link,  implying that all data from
two FEMB (2$\times$5~Gbps) could be transmitted on a single link. However, it is planned to
use a QSPF socket on the WIB to deliver $\sim$5~Gbps on four optical fiber pairs 
(one fiber pair per FEMB: 20~Gbps total) to two Reconfiguable Computing Element (RCE) DAQ modules 
described in Section~\ref{sec:daq}. The WIB will also be capable of sending $\sim$10~Gbps on 
two optical fiber pairs to the Front-End-Link-EXchange (FELIX) DAQ system also described in Section~\ref{sec:daq}.
The FPGA has an additional Gbps Ethernet transceiver I/O based on the 125~MHz clock, which 
provides real-time digital data readout to the slow control system.

%
\subsection{External power and cables}
\label{subsec:ce_feedthrough_power}

The LV power to the FEMB and WIB is supplied by two Weiner PL506 mainframe power supplies.
The CE power-per-channel is about 25~mW in the LAr.
Including power for the WIB, a fully loaded WIB (one WIB plus four FEMBs) requires
12V and draws up to approximately 4~Amps. Therefore, the full electronics for one APA (one PTC, five WIBs, and 20 FEMBs) 
requires 12V and draws approximately 20A, for a total power of almost 250W, as 
described in Section~\ref{subsec:warm_interface_elec}.

Each PL506 LV power mainframe is configured with 3 MEH-30/60 modules which
operate at 30-60V/13.5A/650W maximum capacity. Using 10~AWG cable, an 0.8V drop is expected along the cable with a
required power of 306.12W out of 650W available. 
Therefore, one MEH-30/60 module will supply the 250W to one APA at 48V, with the 48/12V conversion done by the PTC. 
Four wires will be used for each module, two 10~AWG,
shielded, twisted pair for the power and return, two 20~AWG, shielded, twisted pair for the sense. Sense line fusing will be
provided on the PTC card. This fusing would serve as
a final protection. The primary protection would come from the Over Current protection
on the LV supply modules, which is set above the $\sim$20~Amps. The LV power cable uses FCi micro TCA connectors, 
shown in Figure~\ref{fig:tpcce_power_conn}.




\begin{cdrfigure}[Micro TCA power connectors]{tpcce_power_conn}{FCi micro TCA power connector at the PTC end of the cable.}
\includegraphics[width=0.7\linewidth]{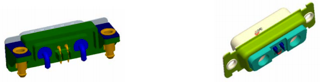} 
\end{cdrfigure}

Two wire-bias mini-crates supply the wire-bias voltages to all 6 signal feedthroughs. Each mini-crate 
contains three wire-bias HV modules, with each module supplying all the wire-bias power to one APA via
8 SHV connector feedthroughs at the CE flange.

Each APA requires three wire-bias voltage connections 
at $+$820V, $-$370V, and $-$665V, as described in Section~\ref{subsec:ce_wire_bias}.
The remaining five wire-bias voltage lines supply between 1 and 1.5~kV to the field cage terminations (3)
and electron diverters (2).
The current on each of these supplies is expected to be zero at normal operation.
However the ripple voltage on the supply must be carefully controlled 
to avoid noise injection into the front-end electronics.

RG-58 coaxial cables connect the wire bias voltages from the mini-crate to the standard SHV
connectors machined directly into the CE feedthrough, so there is no electrical connection between 
the LV power and data connectors and wire-bias voltages. The length of the cables from the Weiner mainframe
to the signal flanges is estimated to be 18~meters.

Optical fibers provide the connections between the WIECs, which act as
Faraday-shielded boxes, to the DAQ and slow control systems.
Each WIB uses QSFP sockets for 
four pairs of fiber, 
implying a total of 120 optical data lines for the 30 WIB boards in the system. The optical fibers from
the signal flanges to the DAQ room are estimated to be 30$-$40~m in length.

Duplex LC optical fiber is under consideration for transmitting the one GIG-E connection from each
WIB to the slow control system. The WIB reports the current draw from each FEMB to the slow control system, while the 
current draw for each APA is monitored at the mainframe itself.

\section{Photon detection system (PDS)}
\label{sec:pd_system}

LAr is an excellent scintillating medium and the photon detection
system (PDS) is used to obtain additional event information from
the photons produced by particles traversing the detector.
 With an
average energy of 19.5~eV needed to produce a photon (at zero field),
a typical particle depositing 1~MeV in LAr generates
40,000~photons with a wavelength of 128~nm. In higher electric fields this number is 
reduced, but at 500~V/cm the yield is still $\sim$20,000~photons
per MeV. Roughly 1/4 of the photons are promptly emitted with a
lifetime of about 6\,ns, while the rest are produced with a lifetime of
1100--1600~ns. Prompt and delayed photons are detected in
  precisely the same way by the photon detection system. LAr is
highly transparent to the 128-nm VUV photons with a Rayleigh
scattering length of (66~$\pm$~3)\,cm~\cite{Rayleigh} 
and absorption
length of $>$200~cm; assuming a LN$_2$
  content of less than 20~ppm. The relatively large light output makes
the scintillation process an excellent candidate for determining the
$t_0$ for non-beam related events. Detection of the scintillation
light can also be helpful in performing background rejection and triggering on
non-beam events.

\subsection{Scope and requirements}

The photon detector system (PDS) 
includes the following components:

\begin{itemize}
\item Light collection system including wavelength shifter and light guides
\item Light sensors: Silicon photo-multipliers (SiPMs)
\item Readout electronics
\item Monitoring system
\item Related infrastructure (frames, mounting boards, etc.).
\end{itemize}

The primary requirement is the detection of light from proton decay
candidates (as well as beam neutrino events) with high efficiency to
enable 3D spatial localization of candidate events. The assumed light yield necessary for
meeting this requirement is 0.1\,pe/MeV for a particle track near the cathode plane
(the farthest possible location from the PDS) after the application of all relevant detection efficiencies. In DUNE, 
the TPC  provides supernova neutrino detection, while the detection of light
from supernova neutrino interactions localize the events and disentangles
them from background noise in the TPC.
The photon system provides the $t_0$  of
events relative to TPC timing with a resolution better than 1~$\mu$s
(providing position resolution along the drift direction on the order of a couple of mm). 
Measurements 
will determine the absolute
light yield by measuring light from beam particles and cosmic ray muons
tracked in the TPC or identified by external muon trigger counters.

Figure~\ref{fig:PD_overview} shows the layout for the photon detector
system described in this section. 

\begin{cdrfigure}[Photon detection system overview]{PD_overview}{Overview of the PDS
    system showing a cartoon schematic (a) of a single PDS module
    in the LAr and the channel ganging scheme used to reduce the
    number of readout channels. Panel (b) shows how each PDS module
   is inserted into an APA frame. Ten photon detectors (PDs) are inserted
    into an APA frame.}
\includegraphics[width=1.0\linewidth]{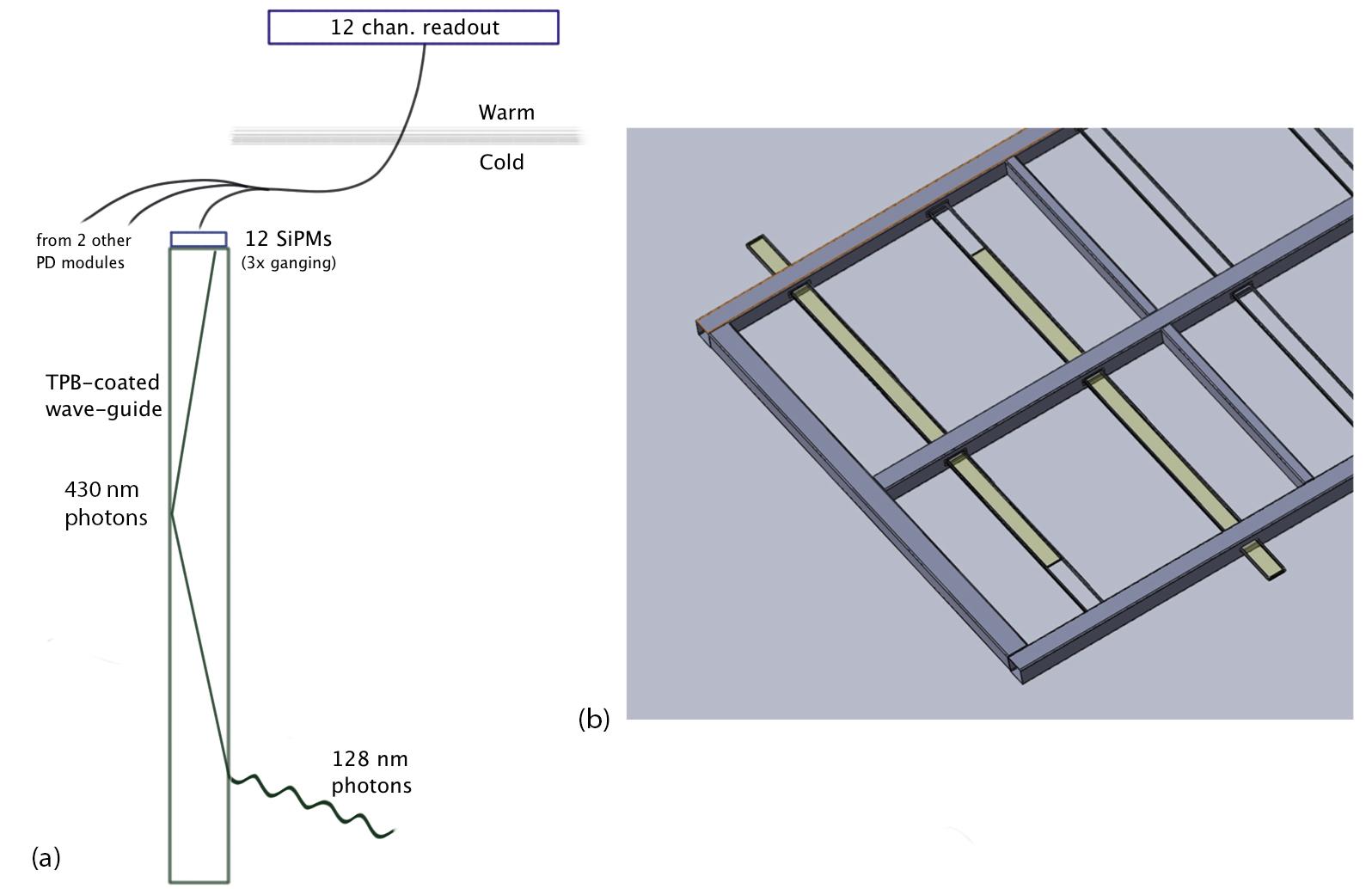}
\end{cdrfigure}

\subsection{Photon detector modules}

Two different styles of PDS 
modules are planned to be installed in the detector.  The two designs test similar light-collection strategies, which differ primarily in  the number of times the LAr scintillation 
light is shifted.  

The first design, shown schematically in Figure~\ref{fig:PD_overview}. is based on
wavelength-shifting radiator plates mounted to wavelength-shifting light guides.
The plates are coated 
with tetraphenyl-butadiene (TPB) to produce blue ($430nm$) light from the 128-nm VUV 
scintillation light.  
This blue light is absorbed by the commercially produced wavelength shifting (WLS)
polystyrene bar with Y-11 fluor, producing green light, which is transmitted
through the light guide to the photosensor 
mounted at its end.
The radiator plates are held captive in mounting blocks that are glued to the WLS bar
at regular intervals as shown in Figure~\ref{fig:PD_radiator_mount}.

\begin{cdrfigure}[Radiator plate mounting Blocks]{PD_radiator_mount}
  {Mounting of the radiator plates to the WLS bar for the reference design scheme}
\includegraphics[width=0.5\linewidth]{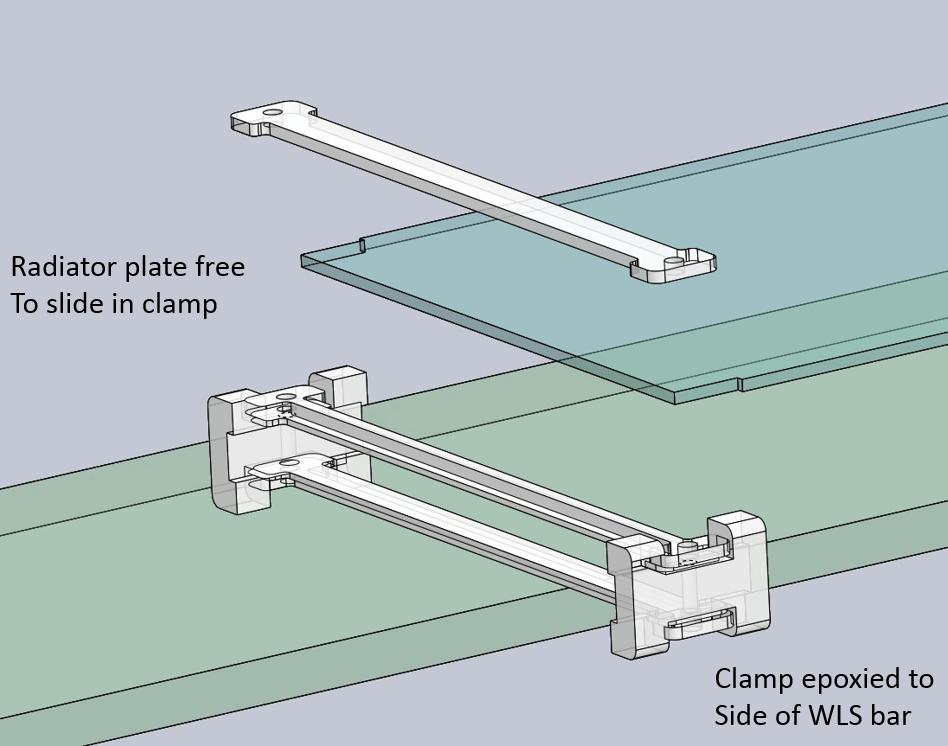}
\end{cdrfigure}

The second design uses the same mounting, but with no radiator plates installed.  
Instead, an acrylic light-guide bar is dip-coated  with a solution
of TPB, solvents, and a surfactant to produce a wavelength-shifting layer on the
outside surface of the bar.
The 128-nm VUV scintillation light is shifted to blue light within the surface coating and transmitted directly
through the lightguide to the photosensor. Because the design uses only one wavelength shifting step, potential efficiency increases are possible.

\subsection{Sensors}

The planned photodetector is a SiPM, model 
SensL C-Series 6~mm$^2$
(MicroFB-60035-SMT). 
This model of SiPM has a detection efficiency of
41\%; the quoted detection efficiency incorporates Quantum Efficiency (QE) and 
the effective area
  coverage accounting for dead space between pixels.   At LAr temperature (89~K) the dark rate is of order 10~Hz
(0.5 p.e. threshold), and  after-pulsing has not been observed. An on-going testing program is in place to ensure 
that the SiPMs can reliably survive the stresses associated with 
any thermal cycling in LAr and long-term operation at LAr temperature.

All photodetectors 
are subjected to testing to determine
 forward and reverse bias I-V curves,
 breakdown voltage, dark current and dark count rate, photodetector gain, crosstalk estimation, response, and bias dependence of parameters.
 
Each SiPM is tested before mounting on the readout boards to determine
if the part meets the specifications in a warm test.  After mounting to
the readout board all items are tested both warm and cold (cyrogenic 
temperature) to determine the operating characteristics.

In addition to these tests, the photodetectors are tested for their
response to light signals from an LED of appropriate wavelength.
These tests will be sensitive enough to determine if one of the three SiPM
elements operating in parallel is not functioning.

\subsection{Mechanical design and installation}

The PDS is configured as a set of \textit{modules} that are mounted on the APA frames.  A PDS module is
the combination of one light guide (also called a ``bar'' due to its
shape) and 12 SiPMs, as shown in Figure~\ref{fig:PD_overview}~(a). 
The APA frames hold ten PDS modules, approximately 2.2-m long,
86-mm wide and 6-mm thick, equally spaced along the full length of the
APA frame, as shown in Figure~\ref{fig:PD_overview}~(b). 
The light guides are inserted into the APA frame on rails gliding on their radiator
plate mounting blocks, as shown in Figure~\ref{fig:PD_mounting_inslide}. 

\begin{cdrfigure}[Diagram of PDS installation into APA frame and installed SiPM mounting board]
  {PD_mounting_inslide}{(left) Rendering of the installation of a PDS module
    into an APA frame, shown just before it comes to rest on the inside face
    of the APA tube. (right) Rendering of the the SiPM mounting board
    installed on the end of the PDS module before insertion.}
\includegraphics[width=0.456\linewidth]{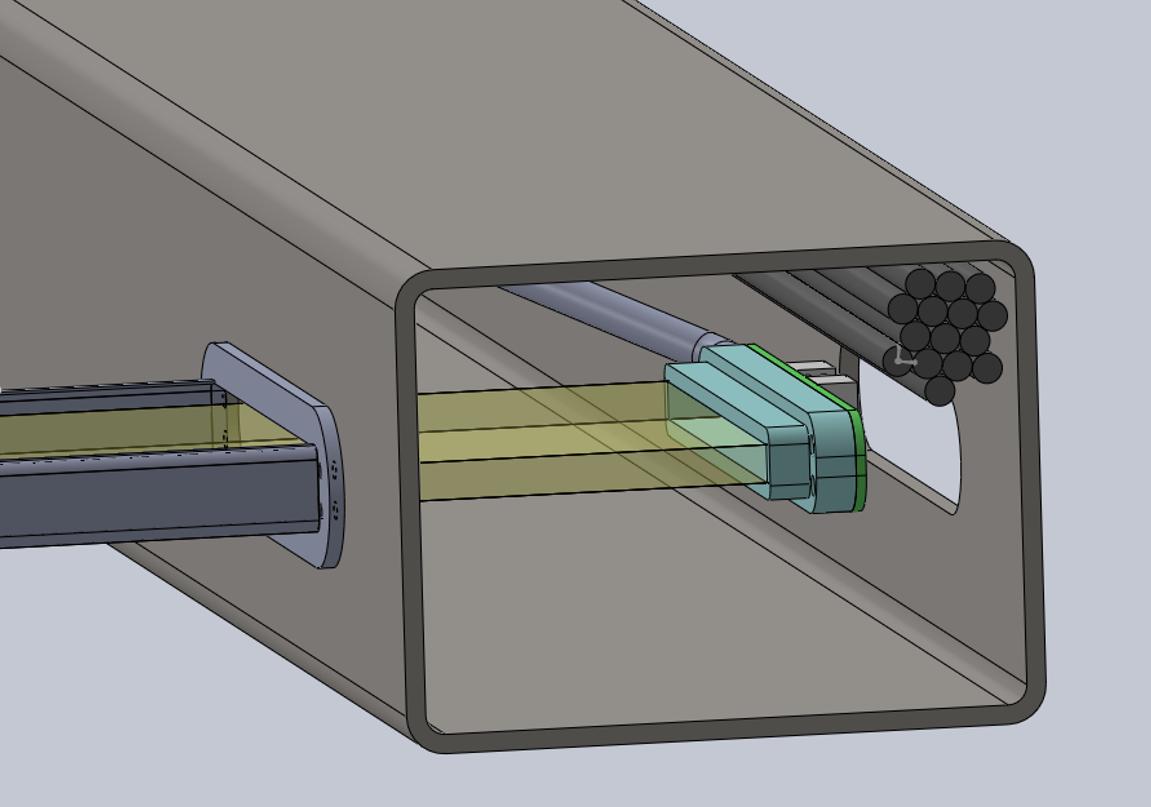}
\includegraphics[width=0.444\linewidth]{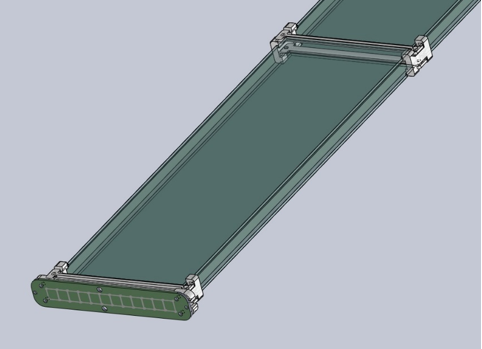}
\end{cdrfigure}

The mounting system has been tested using a prototype PDS module  
as shown in Figure~\ref{fig:PD_flat_installtest}.
\begin{cdrfigure}[Photo of PD mock installation]
  {PD_flat_installtest}{Photograph of the installation
    test of a mock PDS module in a 1/5 section of an APA frame.}
\includegraphics[width=0.50\linewidth]{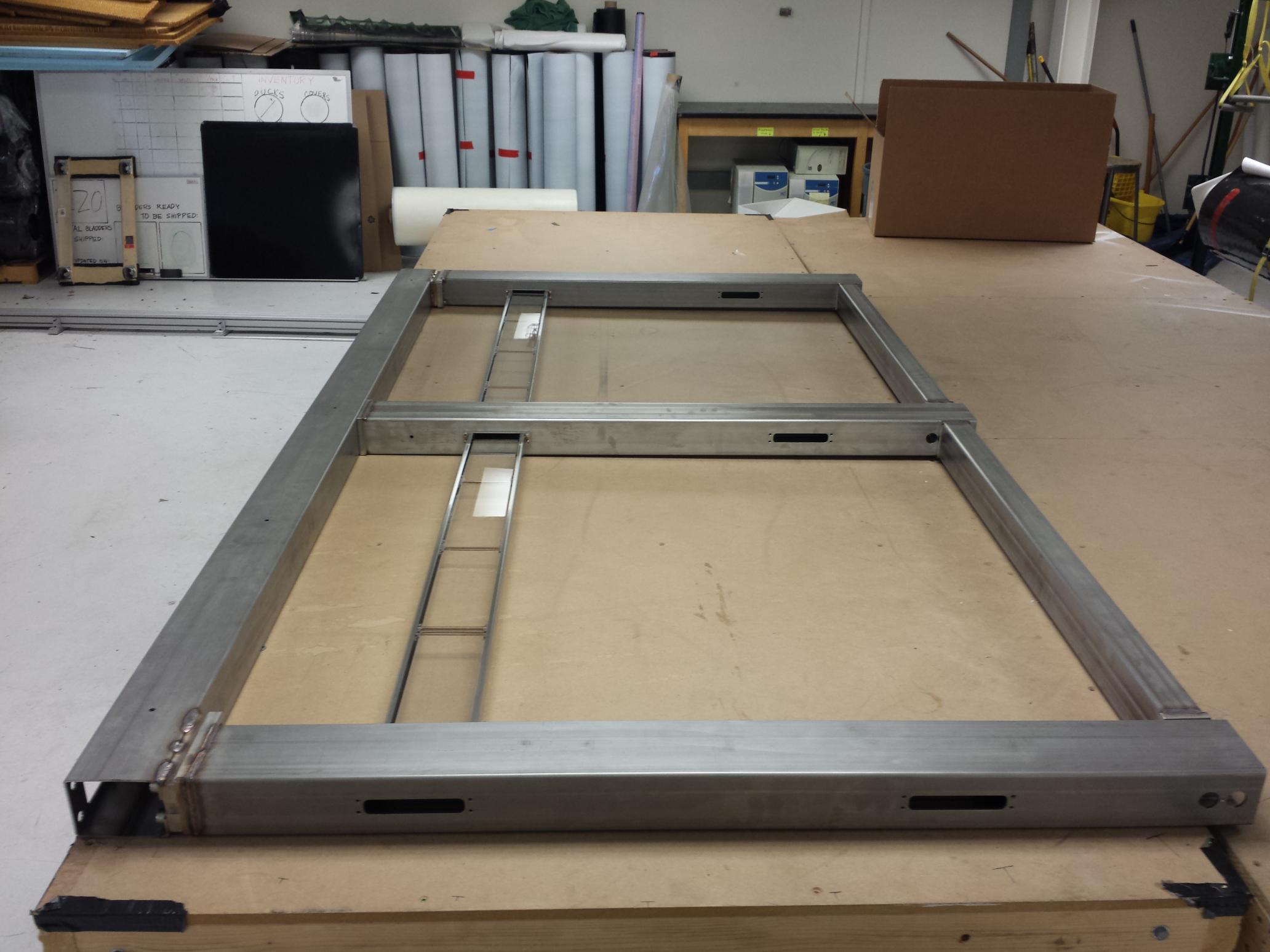}
\end{cdrfigure}

Each photon detector has a single SiPM mounting board with 12 surface-mount SiPMs 
mounted on its face as shown in Figure~\ref{fig:PD_SiPM_PCB_front} (left).
Four groups of $3$ SiPM elements go to single 
channels of the readout electronics in order to reduce the overall system cost.
The board is held close to the bar, without touching, using four screws that go into 
tapped holes on the end mounting block that is glued to the bar.  
The mounting block assembly is shown in Figure~\ref{fig:PD_mounting_inslide} (right) 
The circuit board also has holes at each end for mounting to the APA frame.  

\begin{cdrfigure}[SiPM mounting board with 12 SiPMs and with Rj-45 connector]
  {PD_SiPM_PCB_front}{Photograph of a SiPM mounting board
    with the full complement of 12 SiPMs installed on the board (left), and with the 
    RJ-45 connector for the cable (right).}
\includegraphics[width=0.55\linewidth]{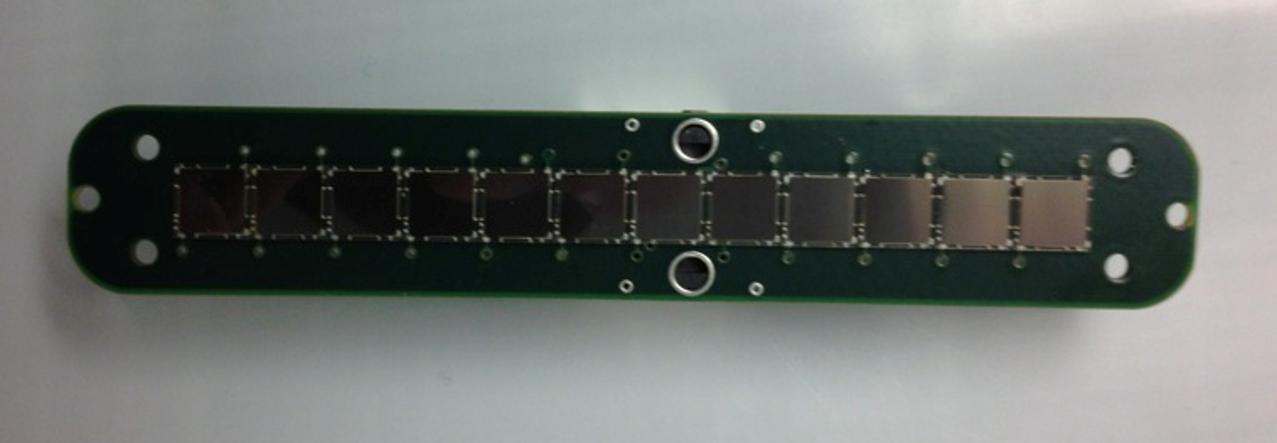}
\includegraphics[width=0.4\linewidth]{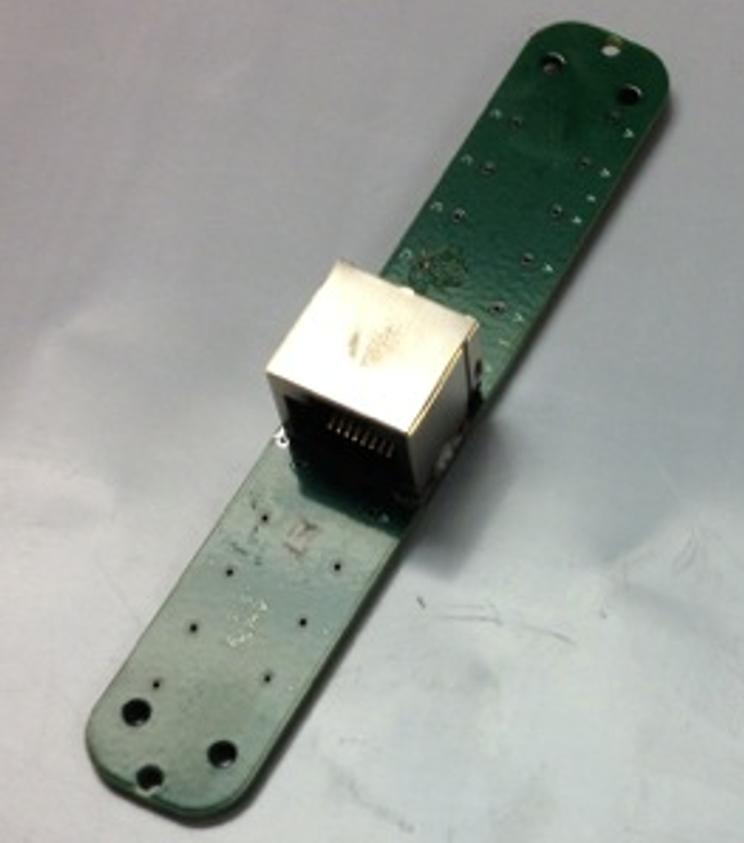}
\end{cdrfigure}


The cabling plan for the system has one cable with four shielded twisted pairs 
connected to each SiPM mounting board via the surface mount RJ-45 connector
shown mounted on the back of the readout PCB in 
Figure~\ref{fig:PD_SiPM_PCB_front} (right).  
The cables run through the APA tubing to the top of the APA frame as seen
in Figure~\ref{fig:PD_cable_intube}.
The cable bundles are installed and connected 
after the PD has been installed into the slot.
\begin{cdrfigure}[Cables in APA frame]
  {PD_cable_intube}{Diagram showing the routing of the PDS cables
    through the APA frame.}
\includegraphics[width=0.50\linewidth]{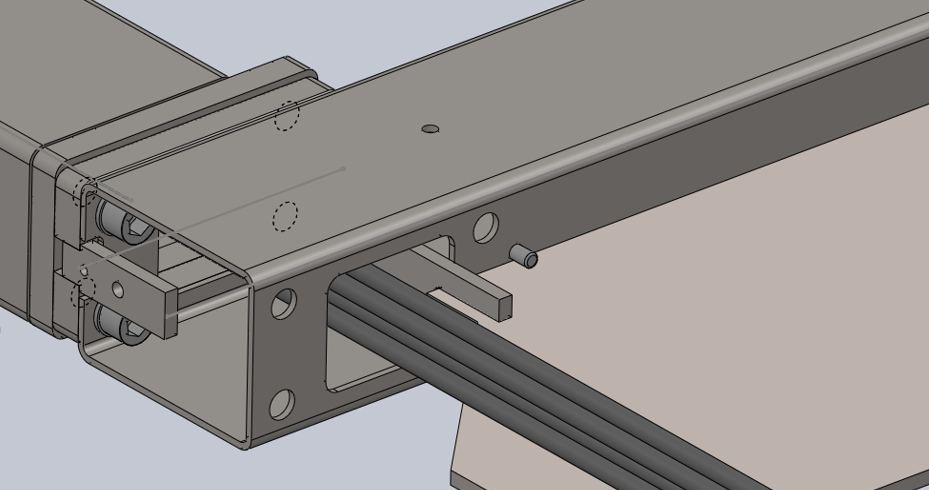}
\end{cdrfigure}

\subsection{Alternative photodetector under development}

While a sufficient number of the two types of PDS modules are being produced to fully outfit the detector, it is possible that a few of the 60 APA slots will be used to house experimental photon detectors. One type of detector that is being developed in an attempt to increase the light detection eficiency is referred to as the ARAPUCA design.

The ARAPUCA design is based on a new technology that allows for the collection of photons in a 5$\times$5-cm$^2$ window  with detection efficiencies at the level of several percent. The trapped light is detected by two SIPMs (SensL 60035 - 6$\times$6 mm$^2$ active area each). The basic concept behind the ARAPUCA design is to trap photons inside a teflon box, of  dimensions of 5$\times$5$\times$1 cm$^3$, with highly reflective internal surfaces, such that the detection efficiency of the trapped photons remains high even with limited sensor coverage on these internal surfaces~\cite{Machado:2016jqe}.

Photon trapping is achieved by using a wavelength-shifting technique coupled with the technology of the dichroic shortpass optical filters. The latter are multilayer acrylic films 
with the  property of being highly transparent to photons with a wavelength below a tunable cut-off while being almost perfectly reflective to photons with wavelength above the cut-off. 
A dichroic shortpass filter deposited with two different wavelength shifters (one on each side) is the core of the device. In particular, it serves as the acceptance window for the 
ARAPUCA device. The rest of the device is a flattened box with highly reflective internal surfaces (PTFE, 3M-VIKUITI ESR, \dots), closed on the top by the 
dichroic  filter deposited with the two shifters. A fraction of the internal surface of the box is occupied by the active photo-sensors (Silicon Photomultipliers, or SiPMs) that detect the trapped photons.

It is envisioned that several of these devices could be installed on the detector to directly compare their performances with those of the
other PDS modules. A mechanical solution has been developed such that 16 of the ARAPUCA devices can be installed on the same mounting structure used for the other PDS modules as shown in Figure~\ref{fig:arapuca_array}.
\begin{cdrfigure}[ARAPUCA array installed in a APA frame]
  {arapuca_array}{One ARAPUCA array of eight devices installed in a APA frame in place of a scintillating bar.}
\includegraphics[width=0.50\linewidth]{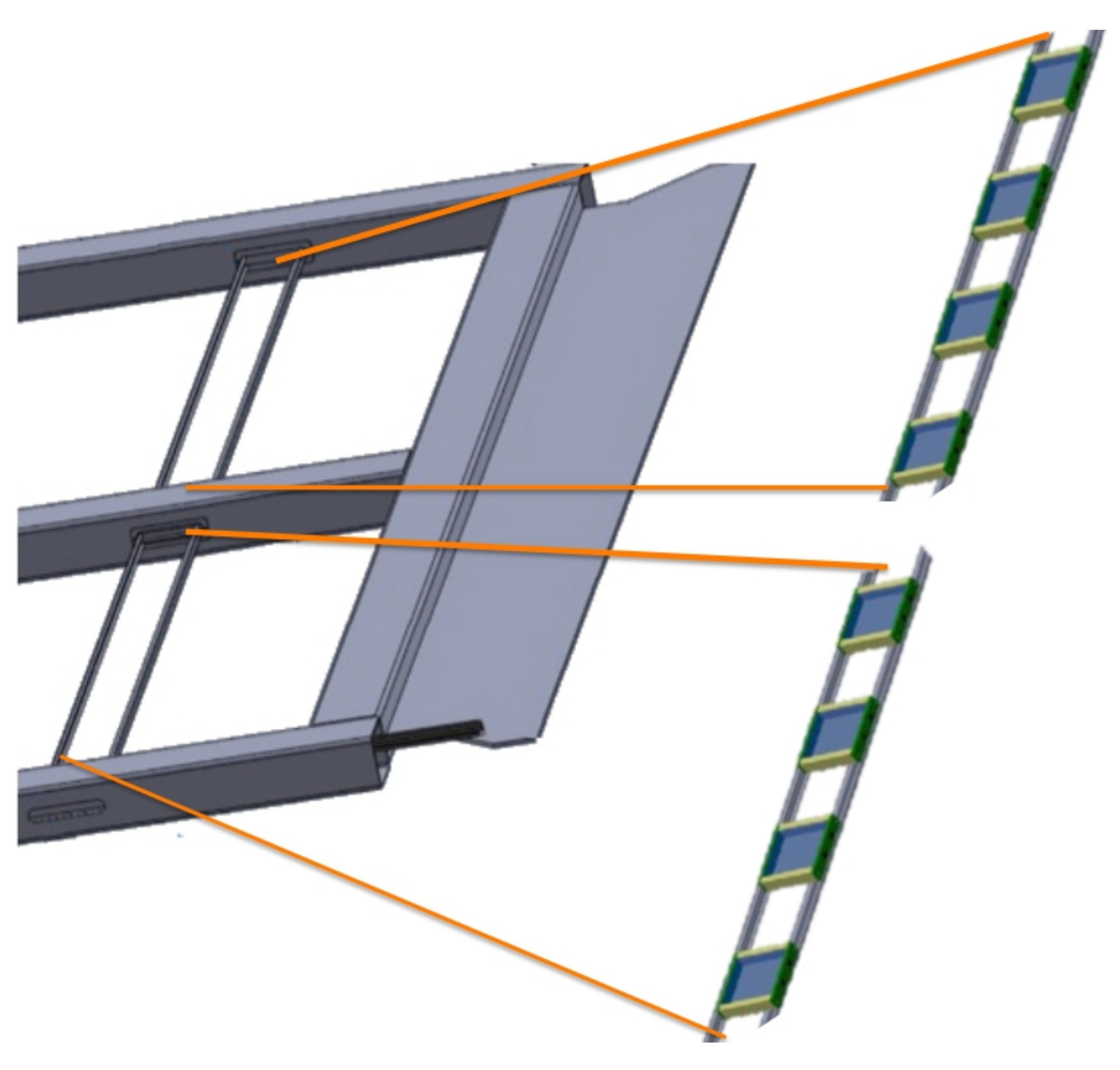}
\end{cdrfigure}

The readout scheme foresees the ganging of the SiPMs from two ARAPUCA devices (four sensors total). The 16 devices on one mounting structure then have a total of four readout channels, which can take advantage of the same cabling and readout scheme used for the other PDS modules.

\subsection{Photon detector UV-light monitoring system}
\label{sec_pd_calib}

A UV-light-based monitoring system is used to monitor the relative performance and time resolution of the system.
The system uses external UV LEDs (245-280\,nm) as light sources in the VUV wavelength range, which are coupled to quartz fibers to transmit light from outside the detector volume to desired locations on the CPA plane.
Light diffusers located on the CPA surface uniformly illuminate the APA area 
containing the  PDS.
The UV light system is used in association with cosmic ray muon 
tracks and Michel electrons as means of calibration.
The UV light essentially mimics physics, although at a different wavelength starting from the wavelength-shifter conversion, 
light guide propagation, photo-sensor detection and the front-end electronics readout.
	
The external UV-light monitoring system is designed with the following goals:
				
\begin{itemize}
\item No active components within PD/APA;
\item Provide uniform illumination over the APA surfaces;
\end{itemize}

In terms of technical requirements the system needs to:
\begin{itemize}
\item provide light levels down to a single p.e. for individual photon-detector channels,
\item provide higher light levels to test linearity of the PDS, and
\item provide variable pulse width to test the time resolution of the photon detector response.
\end{itemize}

Figure~\ref{fig:fig-c-1} illustrates the system design schematically. The system consists of a 1U rack mount Light Calibration Module (LCM) sitting outside the cryostat. The LCM generates light pulses that propagate through a quartz fiber-optic cable to diffusers at the CPA to distribute the light uniformly across the photon detectors mounted within the APA.  Five light 
diffusers on the CPA plane are used: one in the center and four diffusers close to the CPA corners. 
 \begin{cdrfigure}[UV-light monitoring system]{fig-c-1}{Concept of the UV-light monitoring system for the photon detector in liquid argon.}
\includegraphics[angle=0,width=0.7\textwidth]{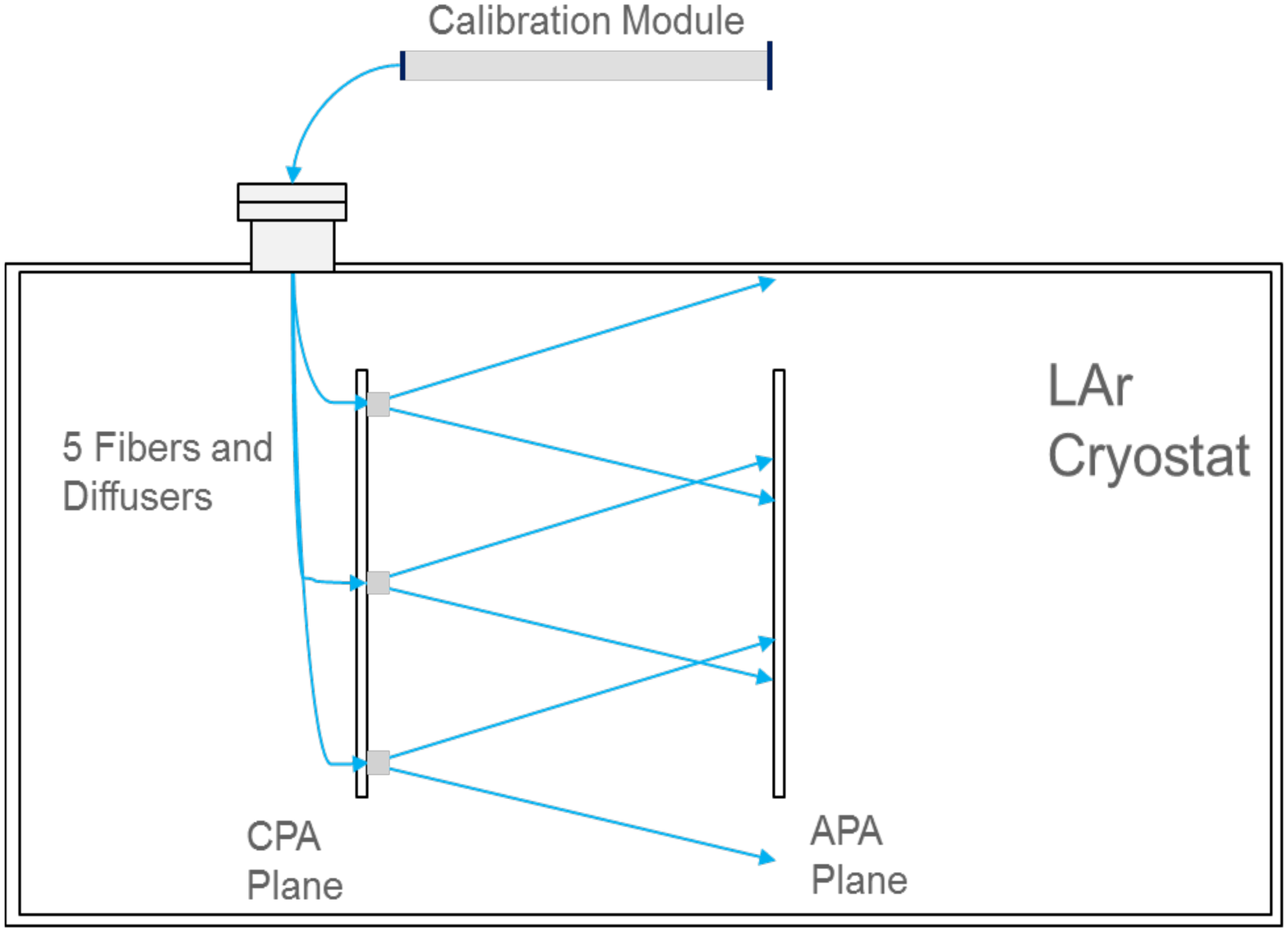}
\end{cdrfigure}

The LCM utilizes the logic and timing control of the photon-detector readout electronics (\textit{SSP}) unit, described in Section~\ref{sec:pds-elec-daq}.  
A single SSP board was repackaged into a deeper rack mount chassis that accommodates a new internal 
LED Pulser Module (LPM) and an additional bulk power supply. The LPM utilizes five digital outputs from the SSP board to control the LPM pulse and its duration.  
These outputs are derived from the charge injection control logic within the SSP's FPGA.  
The even-channel SiPM bias Digital to Analog Converters (DACs)
are used to control the LPM pulse amplitude.  
The adjacent odd channels are used to read out a reference photodiode used for pulse-by-pulse monitoring of the LED light output.  
The output of the monitoring diode is available for normalizing 
the response of the SiPMs in the detector to the monitoring pulse.

The controlled source of light  
in this monitoring system is used to perform time offset and time resolution measurements.  
Many effects contribute to a finite time resolution, including the relative time offset of photon-detector channels, scintillation time constants, 
photon conversion with wavelength shifter, photon propagation through photon-detector paddle, SiPM jitter, and the resolution of the readout electronics.  
Most of these effects are constant and can be individually 
measured on the bench.  The UV light monitoring system monitors overall stability of the photon detector in both time
and amplitude.

\section{PDS electronics}\label{sec:pds-elec-daq}

Scintillation light from LAr comes from two different excited 
states with lifetimes of about 6 ns and 1.6 $\mu$s. 
Only a limited amount of light is collected, so the electronics are designed to collect the light 
from both excited states. A summary of the general requirements 
for the system, including initial requirements from a 
physics performance perspective, are given in Table~\ref{tab:fee_req}.

\begin{cdrtable}[Physics requirements for the PDS electronics]{ll}{fee_req}{Physics requirements for the PDS electronics}
 Performance Parameter       & Target   \\ \toprowrule
Time Resolution                   & Better than 30 ns wrt event time zero (``t0'')      \\ \colhline
 Charge Resolution               & 0.25 photo-electron equivalent                    \\ \colhline
 Dynamic Range                   & $\sim \times$10 better than detector (1000:1)         \\ \colhline
 Linearity                               & Sufficient to resolve 1 photo-electron signals   \\ \colhline
 Multi-Hit Capability              & Sufficient to measure Triplet (late) Photons          \\ \colhline
 Dead Time                           & Live up to 2 drift times either side of beam spill         \\ \colhline
 Bias Control                        & 0.1 V resolution up to 30 V per channel  \\ \colhline
 Calibration                          & On-board Charge Injection  \\ \colhline
 Timing                                 & Events time-stamped via ProtoDUNE Timing System  \\    \end{cdrtable}

There is no PDS front-end electronics in the LAr cold volume.  
The un-amplified analog signals from the SiPMs are transmitted directly to outside the cryostat
for processing and digitization, with the advantage that the infrastructure required for inside the cryostat is  
reduced (power, data cables, precision clocks, data protocols).  
A custom module, called the SiPM Signal Processor (SSP), receives the SiPM signals outside the cryostat.

As noted previously, three PDS SiPM signals are summed together into a single readout channel. 
A 20-m long multi-conductor cable with four twisted pairs is used to read out PDS modules,
each of which incorporates  12 SiPMs i.e., four readout channels per PDS module.
A total of ten PDS modules are inserted within a single APA, resulting in ten readout cables
using 40 SSP readout channels distributed over four SSP modules. A total of 24 SSPs serve
to read out the ProtoDUNE-SP photon-detector modules in all six APAs.

%
An SSP consists of 12 readout channels packaged in a self-contained 1U module.  
Each channel contains a fully-differential voltage amplifier and a 14-bit, 150-MSPS analog-to-digital converter (ADC) that 
digitizes the waveforms received from the SiPMs.  
The front-end amplifier is configured as fully-differential, and receives the SiPM signals into a termination resistor that 
matches the characteristic impedance of the signal cable. 
Currently there is no shaping of the signal, since the SiPM response 
is slow enough relative to the speed of the digitization to obtain 
several digitized samples of the leading edge of the pulse for the determination of signal timing.  

The processing is pipelined, and performed by a Xilinx Artix Field-Programmable Gate Array (FPGA).  
The FPGA implements an independent Data Processor (DP) for each channel.  
The processing incorporates a leading edge discriminator for detecting events
and a constant fraction discriminator (CFD) for sub 
clock timing resolution.  

In the standard mode of operation, the module performs waveform capture, 
using either an external or internal trigger.  
In the latter case the module self-triggers to capture only waveforms with an amplitude greater than
a specified threshold.
Up to 2048 waveform samples may be read out for each event, with the current firmware 
configuration.

Because  the Xiling Artix FPGA  is programmable  and accessible, it is possible to  explore  
different data processing algorithms  and  techniques,  and then  summarize
information  on  processed  waveforms  in 
a header output. 
It is also possible to customize the readout for a given type of event (e.g., a supernova).  
When waveform readouts overlap, the device can be configured to offset, 
truncate or completely suppress the overlapping waveform.  
Pile-up events can also be suppressed.  A picture of the prototype module is shown in Figure~\ref{fig:PD_fig-e-2}.  
%
\begin{cdrfigure}[SSP module prototypes]{PD_fig-e-2}{SSP module prototypes} 
\includegraphics[angle=90,width=0.41\textwidth]{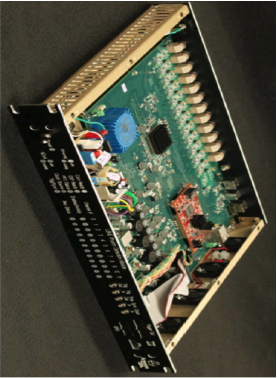} \includegraphics[angle=0,width=0.4\textwidth]{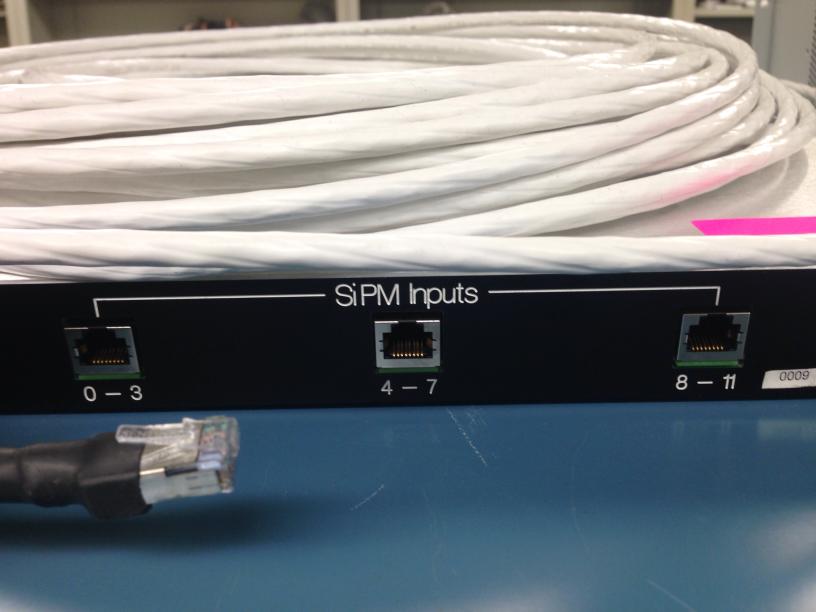}
\end{cdrfigure}
%

In order for the events measured in the photon detector to be matched up 
with the corresponding events in the TPC, the front-end electronics 
attaches a timestamp to the data as it is acquired.  
The timestamp is unique, and has a correspondence with the timestamps in 
the TPC electronics processing.  
The timestamp in the SSP is applied to the event data as it is digitized. 
To achieve this, the TPC and PD electronics must be synchronized, 
including timestamp counter resets, based on a known and stable calibration 
for the timing resolution of the ADC conversion between the two systems.  
In the ProtoDUNE-SP the photon readout is configured to read waveforms when triggered by a beam event,
and/or to provide header information when self-triggered by cosmic muons.
The header portion summarizes pulse amplitude, integral, and time-stamp information of events.

A Xilinx Zynq FPGA handles the slow control and event data transfer.  
The SSP for ProtoDUNE-SP uses Gb Ethernet communication implemented over an optical interface.
The 1 Gb/s Ethernet supports full TCP/IP protocol.  
The module includes a separate 12-bit high-voltage DAC for each channel to 
provide up to 30\,V of bias to each SiPM.  
The module also features charge injection for performing diagnostics and linearity 
monitoring, and also voltage monitoring.

In tests to date, the SSP has been able to measure single photo-electron signals 
coming from the SiPMs over a cable length of 25 meters, when three SiPMs 
are summed together and operated at LAr temperatures.  
The timing resolution of the signals has been measured to be better than 3\,ns.  
The full-differential signal processing in the front-end circuitry 
is important in achieving this result.

The SSP provides a trigger output signal from internal discriminators in firmware based on programmable
coincidence logic, with a standard ST fiber interface to the central trigger board (CTB).
Input signals are provided to CTB from the beam instrumentation, the SSPs, and the beam TOF system.
The CTB receives timing information from the ProtoDUNE-SP timing system and the CTB trigger inputs are distributed to 
the experiment via the timing system.
To that end the SSP implements the timing receiver/transmitter endpoint hardware to receive trigger inputs and clock signals from the timing system.
A block diagram of the system is shown in Figure~\ref{fig:PD_fig-e-3}.
\begin{cdrfigure}[Block diagram of the ProtoDUNE SSP module]{PD_fig-e-3}{Block diagram of the ProtoDUNE SSP module} 
\includegraphics[width=\textwidth]{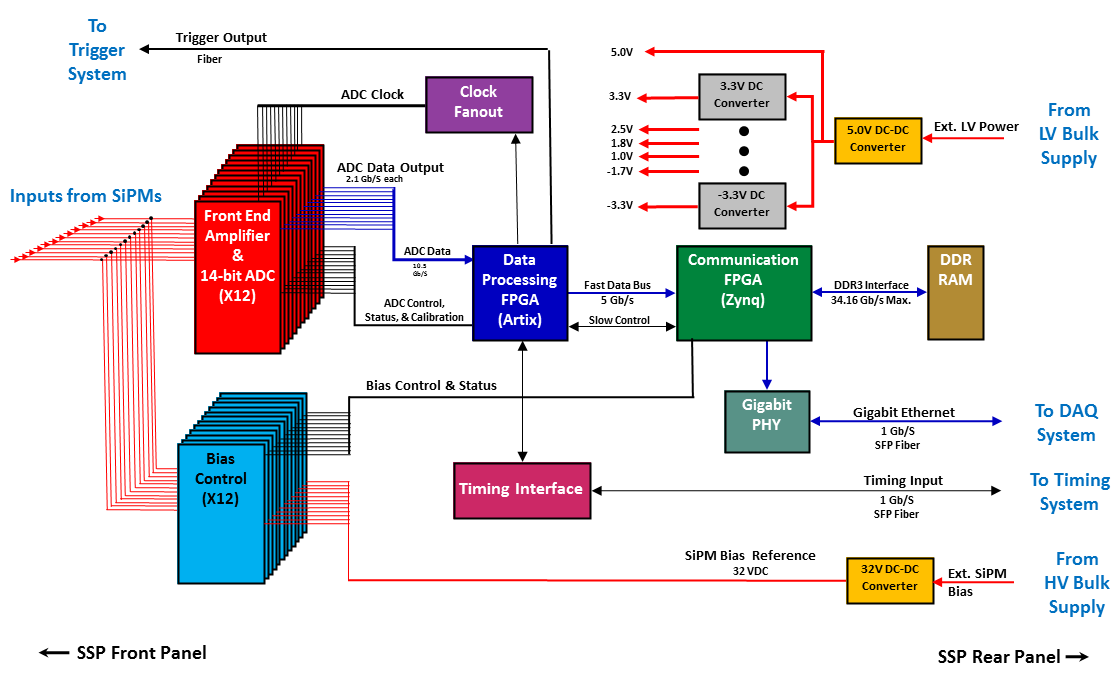}
\end{cdrfigure}

\section{Data acquisition (DAQ)} 
\label{sec:daq}

\subsection{Scope and requirements}

The data acquisition (DAQ) system is shown in Figure~\ref{fig:daq-overview} along with its
interfaces to the cold electronics, beam instrumentation, and offline
computing systems.  

\begin{cdrfigure}[DAQ Overview]{daq-overview}{Overview of the
DAQ system, its interconnections, data flow, timing and trigger signals,
and the interfaces to the electronics and offline computing systems.  DAQ elements are shown in bold. The TPC Readout (RCE and FELIX) accept
data from the TPC Warm Electronics at total rate of 480\,Gb/s, where it is compressed and 
selected based on trigger information.  Triggered data is then sent to the artDAQ event builder farm, and subsequently 
stored to disk with a parallel sample sent for online monitoring.  Triggers are formed from inputs 
from the beam instrumentation, photon detection and cosmic ray tagger systems and forwarded
to the timing system which broadcasts the synchronous trigger and clock signals to the electronics and DAQ systems.
}
        \includegraphics[width=1.0\textwidth]{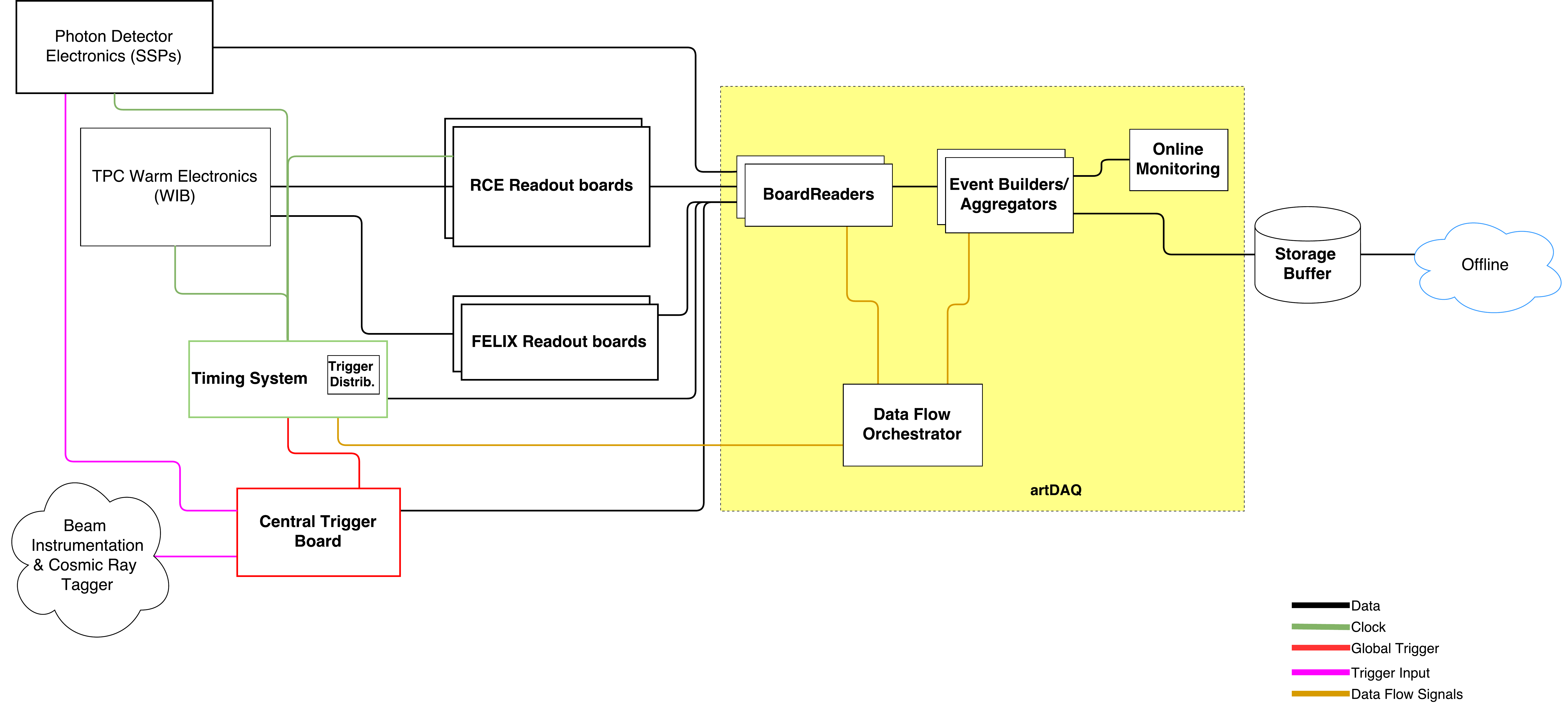}
\end{cdrfigure}

 
The physics requirements of  ProtoDUNE-SP are the primary drivers of the
DAQ system requirements.
The front-end electronics and assumed bandwidth and storage requirements
from the online and offline computing systems impose additional constraints.  

The run plan (see Section~\ref{sec:runplan}) calls for about 25\,M analyzable beam events
to be collected in the first run of \pdsp. Data sets may be enhanced  in desirable particle types and energies
with dedicated triggers (such as PID)  from the beam instrumentation. The latter is described in Section~\ref{sec:beaminstruments}.

Parameters of the data collection plan are listed in Table~\ref{tab:daqgoldi}. The lossless 
compression factor cited
in the table is based on the assumption that the signal-to-noise level 
is achievable when compared that achieved by MicroBooNE\cite{Acciarri:2017sde}.

\begin{cdrtable}[Parameters defining data rate and volume in the ``most likely'' scenario]{ll}{daqgoldi}{Parameters defining data rate and volume in the ``most likely'' scenario v5\,\cite{data_spreadsheet}. The buffer depth includes both
  the in-spill and out-of-spill data.}
Parameter & Value \\ \toprowrule
    Trigger rate & \SI{25}{\Hz} \\  \colhline
    Spill duration & \SI{4.8}{\second} \\ \colhline
    SPS Cycle & \SI{22.5}{\second} \\ \colhline
    Readout time window & \SI{5}{\milli\second} \\ \colhline
    \# of APAs to be read out & 6 \\ \colhline
    Single readout size (per trigger) & \SI{230.4}{\mega\byte} \\ \colhline
    Lossless compression factor & 4 \\ \colhline
    Instantaneous data rate (in-spill) & \SI{1440}{\mega\byte\per\second} \\ \colhline
    Average data rate & \SI{576}{\mega\byte\per\second} \\ \colhline
    3-Day buffer depth & \SI{300}{\tera\byte} \\

\end{cdrtable}

 The baseline trigger
rate during the SPS spill is taken to be 25 Hz.  Cosmic data are also
acquired at an appropriate rate such that bandwidth and processing priority are given 
to beam data.
The data rate from the electronics is dominated by the
TPC data.  However, the photon detection system (PDS) can produce a signifcant
amount of calibration data (where full waveforms are extracted), during
commissioning and special runs (up to 24\,Gb/s maximum).

The TPC data are sent via the Warm Interface Boards  (WIB) on the cryostat flanges
for the six APAs, un-triggered at a total rate of 480\,Gb/s.  
The PDS is estimated to send data
at a total rate of 1.2\,Gb/s from the 24 SSPs (detailed further in Section~\ref{subsec:pds-readout}).
The maximum bandwidth from the ProtoDUNE-SP online system to CERN IT (and
hence, the offline world), is 20\,Gb/s at a maximum.
Therefore, the DAQ system must reduce the data by a significant fraction
before they are sent offline.  This is achieved by a combination of
data compression and triggering.

\subsection{Timing, trigger and beam interface}
\label{sec:daq_time}

The timing and trigger are two distinct subsystems.  The timing
system provides the distribution for the trigger signals over the same
fabric as the clock and calibration signals.  

\subsubsection{Timing}

The timing system is required to provide a stable and phase-aligned
master clock to all DAQ components; synchronize external signals into the
ProtoDUNE-SP clock domain and time-stamp them; distribute synchronization,
trigger and calibration commands to the DAQ system; and conduct continuous
checks of its own function. In addition, the timing system acts as a
data source, providing a record of triggers received, distributed, or
throttled. The system is designed to meet the full eventual requirements
of the DUNE experiment, but needs only a subset of that functionality
for ProtoDUNE-SP. 

An FPGA-based master unit 
receives a high-quality clock signal 
(from a quartz crystal oscillator or external source) 
and external signals from the
trigger system and SPS accelerator. It interfaces to the ProtoDUNE-SP control
and DAQ via a gigabit Ethernet interface. The master unit multiplexes
synchronization and trigger commands, along with arbitrary command
sequences requested by software, into a single encoded data stream,
which is broadcast to all timing endpoints, and decoded into separate clock
and data signals. A uniform phase-aligned cycle counter, updating at the
ProtoDUNE-SP system frequency of 50\,MHz, is maintained at all endpoints,
allowing commands to take effect simultaneously at all endpoints
regardless of cable lengths or other phase delays.

The timing signal is broadcast via multi-mode optical fiber (for
medium-distance connection to the WIB crates on the detector) and LVDS
signals are sent over twisted-pair cable (for short-distance connection to RCEs,
SSPs and FELIX modules). Optical signals are fanned out and recombined
using commercial 32:1 passive splitters, and active optical--LVDS
converter boards further split the signals for local distribution to
endpoints. 

Endpoints decode the timing signal into separate clock and data
signals using a commercial clock-data recovery ASIC~\cite{siliconlabs:Si5344}, which in turn
feeds a low-bandwidth phase-locked loop (PLL) in order to remove any remaining jitter in
the clock and provide phase adjustment. The data stream employs 8b/10b
encoding, ensuring sufficient transitions in the timing signal for clock
recovery and correct operation of optical links, and uses scrambling
of idle patterns to minimize electromagnetic interference (EMI).
 A common firmware block is used to
decode the timing protocol, which is incorporated into the overall
firmware design of the receiving FPGA in each DAQ component. 

\subsubsection{Trigger}

The baseline trigger solution for ProtoDUNE-SP is the Central Trigger Board
(CTB). The triggers expected for the CTB are a beam minimum-bias trigger,
an electron beam trigger, and, if possible, a $\pi$/K/p trigger will also be added.
The CTB provides a means to select the most interesting physics events
during the beam spill and suppress inter-spill background events (or
events so busy to prohibit accurate reconstruction).  The final trigger
selection will be driven by ProtoDUNE's physics goals.  The CTB is
designed to receive triggers from various subsystems (Photon Detection
System, Beam Instrumentation, SPS spill signal, etc.).
Up to 100 input channels are provided.  It combines these into a global
trigger based on a configurable input mask (or more sophisticated
algorithm, if desired).  It provides functionality to globally
time-stamp triggers, keep event counts, and provide artDAQ-compatible
(see Section~\ref{sec:daq-ev-build}) header information with trigger type
and error conditions.  Internally generated triggers and calibration
pulses allow for testing the board itself and the end-points receiving
the trigger signals.

The CTB is based on the MicroZed development board~\cite{avnet:microzed},
which comprises a Xilinx Zynq-7000 System-on-Chip (SoC), 1-GB DDR3 RAM,
Gigabit Ethernet, and 115 I/O ports.  The Programmable Logic of the
Zynq-7000 is used to perform fast triggering operations, whilst the
Processing System is used to interface to the readout and controls
systems.

Configuration and operation is performed using an XML file which is sent
to the CTB.  This allows for fast reconfiguration of the CTB without
the need for new firmwares.  The file is used to both configure the board
and send start/stop/reset, etc., commands from the run control.
The trigger output format consists of a trigger type (physics/calibration/random),
a time-stamp, the trigger word, counter information and the values 
of the inputs causing the triggers.  
The trigger system uses the timing system clock and global triggers are
distributed via the timing system.

\subsubsection{Beam Interface}

The beam instrumentation is described in Section~\ref{sec:beaminstruments}.
The beam instrumentation DAQ and the DAQ for ProtoDUNE-SP
have separate timing systems.  A common GPS clock is used to keep both
systems synchronized relative to each other.  Global GPS timestamps are used
in both systems to match data sets.  
Data from the beam
instrumentation are acquired continuously via a separate DAQ path.  Triggered
data for ProtoDUNE-SP includes TPC, PDS, and beam instrumentation data, as well as
the timestamps and trigger information associated with each. 

\subsection{TPC data readout}

The readout of the TPC wires, prior to being received by PCs in
the back-end DAQ, consists of  CE on the APAs
inside the cryostat and the warm electronics outside
the cryostat on the flange.  CE data are received on the WIBs
which are housed in the WECs, situated on the top of the flanges.

Each WIB receives the
data from four FEBs over sixteen 1.25-Gbps data lines, and multiplexes
these data to four 5-Gbps (or two 10 Gbps) lines that are sent over
optical fibers to the DAQ.

Two systems are used to receive data from the WIBs.  The default system
 is based on Reconfigurable Computing Elements (RCE) and a proposed alternative system is based on
the Front-End-Link-EXchange (FELIX) technology. 

\subsection{RCE-based readout}
The data from the WIB are received by processing units called RCEs (Reconfigurable Cluster Element), 
\cite{slac:rce}
which are housed in industry-standard
ATCA shelves on COB (cluster-on-board) motherboards that are designed
at SLAC for a wide range of applications.   The RCE is a SoC 
from the
Xilinx Zynq family and contains a full Linux processor system on the chip
accompanied by 1\,GByte of DRAM.   The primary processing functions of the
RCEs are compression (and/or zero-suppression) and buffering
of the raw data and then sending data to the back-end upon the receipt of
an external trigger.  Each COB carries eight RCEs for data processing, all connected to each
other via an on-board 10-Gbps Ethernet switch, which also sends data out
of the COB to the back-end DAQ PCs.

The interface with the WIB is provided via the ATCA compliant rear-board, the RTM (Rear Transition Module).  
This application-specific board uses a set of QSFP transceivers to receive
the data from the WIB and an SFP+ (small form-factor pluggable)
 optical interface for communication
with the timing and trigger distribution system.

As the multiplexed data from the WIB come into the RCE FPGA fabric,
it is de-multiplexed and buffered into per-channel, fixed-time-length 
chunks (for instance 512- or 1024-ticks).  These chunks are
compressed and written to the DRAM where the RCE processor waits
for a trigger (also handled by the FPGA) to arrive.  Upon a trigger, the
processor sends data for a fixed window in time, including pre- and post-trigger time chunks
for all channels, to the back-end PCs.  

For ProtoDUNE-SP, 256 wires worth of data (2 FEBs) are sent to each RCE.
Given that there are 120 FEBs in ProtoDUNE-SP, 60 RCEs are needed to
readout the full detector.  These fit into eight COBs which in turn
reside in a single 14-slot ATCA shelf.

\subsection{FELIX-based readout}
The FELIX is a PCIe card receiving data on point-to-point links from
the detector electronics and routing those through a switched network
to computers.  The aim is to reduce to a minimum any specific hardware
developments and to fully rely on commercial networks and servers to
perform the DAQ tasks.  For ProtoDUNE-SP, data from five WIBs (20 FEBs) are read out over ten 9.6-Gbps links into two FELIX cards.  Grouping
time slices around a trigger signal, as well as data compression, is
dealt with in software. Similar to the RCE-based readout, the FELIX 
generates artDAQ fragments to be sent to the event builder.

\subsection{PDS and beam instrumentation data readout}
\label{subsec:pds-readout}

A combination of externally triggered events
and self-triggered events make up the PDS data.  The external triggers
come from the beam instrumentation via the trigger system at 25\,Hz.
This amounts to 118\,Mb/s.  The self-triggered data
are induced by cosmic rays.  A cosmic rate of 10\,kHz is assumed,
totalling 1106 Mb/s.
The combined rate comes to $\approx$1.2\,Gb/s.  An alternative scheme with
just self-triggered header-only data with a resultant rate of $\approx$1.1\,Gb/s is considered for implementation if
the former proves difficult.

\subsection{Event-building software }
\label{sec:daq-ev-build}

Developed within the Fermilab Scientific Computing Division and
already used for the 35-t prototype, \textit{artDAQ} provides data
transfer, event building, and event analysis functionality. This
latter feature includes built-in support for the \textit{art} event analysis
framework,
also developed at Fermilab~\cite{fnal:art}, allowing experiments to run art modules for real-time
filtering, compression, disk-writing and online monitoring. As art is also used for offline analysis, a major
advantage of artDAQ is that it allows developers to easily switch
between developing online and offline software.

ArtDAQ provides three types of processes, each of which
fulfills a specific role. In the order of upstream-to-downstream, these
are boardreader processes, eventbuilder processes, and aggregator
processes. A given boardreader process is intended to be associated 
with a particular geographical region of the detector, and provides
hooks (in the form of C++ base classes) for an experiment's developers
to embed experiment-specific code (called ``fragment generators''),
designed both to upload configuration values to hardware and to read
out the hardware. For ProtoDUNE-SP, the full DAQ consists of 87+ 
boardreaders, in charge of the 60 RCEs, 24 SSPs, the timing system, 
the Central Trigger Board, and at least one for the beam instrumentation.
For testing purposes, fragment generators can perform useful
functions such as providing a ``playback" mechanism,'' and modeling sudden or unexpected data flow events.

Downstream of the boardreader processes are the eventbuilder
processes. An eventbuilder receives data from every boardreader (a
chunk of data from one boardreader corresponding to an event is
referred to as a ``fragment''), and assembles the fragments for a given
event into a raw, complete data event. Optionally, filtering via art
modules can be performed at this stage.

The most downstream process type is the aggregator. Traditionally in
artDAQ-based DAQ systems, there are two aggregators, one in charge of
writing data to disk and reporting aggregate statistics (e.g., MB/sec), and one in which 
experiments can run art analysis modules for
real-time online monitoring. 
%
%
For ProtoDUNE-SP this model will change as
artDAQ becomes more flexible and throughput
capability increases. The functionality
of aggregators may be replicated in eventbuilders. While this solution  reduces
the number of interprocess connections in the DAQ software, the number of
processes assembling raw events is the same as the number of processes
writing to disk.

 
For the 35-t prototype, artDAQ processes were controlled by a program called
\textit{DAQInterface}. DAQInterface takes 
charge of launching the artDAQ processes, checking for error states, and
shutting down
 processes in an orderly fashion as needed, to avoid improperly closed output files,
zombie processes, etc. For ProtoDUNE-SP, some of the functionality of DAQInterface (e.g., querying  status) is shifting to JCOP (Joint Controls Project);
DAQInterface code is reused as appropriate/possible, to minimize duplication of effort.

\subsection{Control, configuration and operational monitoring}

The artDAQ software used for all applications dealing with the movement,
processing and storage of data is interfaced with software of the
Joint Controls Project (JCOP) for the purpose of control, configuration
and operational monitoring.  JCOP provides a toolkit to implement run
control (finite state machine (FSM), distribution of commands, error
propagation and handling) as well as graphics tools that allow for the
implementation of user interfaces and monitoring dashboards.  

In order to
minimize the software development needs, the same FSM as defined by artDAQ
is implemented and commands are sent to the applications using
the already supported XML-RPC protocol.  Monitoring data are pushed
into the JCOP framework by implementing the appropriate artDAQ monitoring
plugin.  Log and error messages will be most probably collected and
processed using an implementation of the ELK (elastic search, logstash,
kibana \cite{elastic:kibana}) stack. 
 The internal configuration of DAQ applications are
carried out using the mechanisms provided by artDAQ. The overall system
is modeled and configured using the JCOP paradigm of data-points.

\subsection{Interface of the DAQ to the online storage}
\label{sec:DAQ_online_interface}

The software framework for
interfacing with the electronics, building events, writing data files,
and providing an interface to online monitoring of data as they are
acquired is \textit{artdaq}~\cite{artdaq}. 

Computers running BoardReader processes read out
the RCEs and SSPs and transmit data to a set of computers running EventBuilder processes.
These computers and a
pair of 10\,Gbit/sec Network Interface Cards (NICs) provide the CPU and networking needed to
build events, collect basic metadata, and send the data to storage and
online monitoring.  The Event Builders assemble data fragments into self-consistent events and perform basic data integrity checks before writing records out.

Table~\ref{tab:daqgoldi} indicates a nominal trigger rate of 25\,Hz for
the mid-range scenario. 
 Data are assumed to be
collected based on prompt trigger signals generated by the beamline
instrumentation in order to purify samples of desired particles.

Current estimates put the PDS data at approximately 10\% 
of the TPC data rate.  Beam instrumentation data are expected to be lower still.
Although adding to the
total data rate only slightly, adequate resources must
be provisioned in order to acquire and store the data from these
systems.  
The network speed of all computers in the DAQ
chain is anticipated to be 20\,Gbit/sec. 

Given that each RCE reads out 256 channels of the TPC, 60 RCEs
need to be active.  For the PDS, 24 SSPs are used.  At least four computers 
running BoardReader processes are therefore used to read out
the RCEs and transmit the data to the EventBuilder processes.

The online buffer layer consists of $\sim$300\,TB of storage,
which is connected directly to the Event Builders.  The baseline
storage option consists of two SAS arrays DAS with $>40$Gbit/s bandwidth,
redundant paths, controllers, and power supplies.
A backup option for storage is an XRootD cluster~\cite{xrootd} taking
data directly from the Event Builders over the network.

After the data are written to disk by artDAQ, the data handling
system creates metadata files, and
transfers the data from EHN1 to the CERN Computing Center~\cite{data_managm_sys}.
The \textit{Fermi File Transfer Service} (F-FTS) software 
developed and maintained at Fermilab is the central element
of the data flow management at this level.

\subsection{Online monitoring}
\label{sec:daq_online_monitoring}

In addition to the monitoring of the operations of the DAQ system, the
quality of the data taken by the detectors has to be constantly monitored.
This assurance is provided by the online monitoring system.
This subsection describes the baseline monitoring framework for ProtoDUNE-SP.  
The final implementation is subject to change, but will likely be similarly
linked to artDAQ and LArSoft (introduced in Section~\ref{sec:comp:larsoft}) as described here.

The online monitoring framework runs as a DAQ process and therefore is
able to provide data quality assurance in real-time. ArtDAQ splits the  
data into distinct physics and monitoring streams via its aggregator processes.  
The data rate to the monitoring is tunable such that the monitoring 
can digest the data in a timely fashion.
The software framework used for online monitoring 
consists of an
\texttt{art::Analyzer} module which interfaces with the artDAQ framework and
owns instances of further classes, each designed to handle different aspects
of the monitoring.  

The DataReformatters restructure the data to allow for efficient subsequent
analysis and provide a standard interface to the methods, which look through
the events.  These reformatted objects are passed to MonitoringData, which
owns all of the data products (\texttt{TTree}s, \texttt{TH1}s, \texttt{TGraph}s,
etc.) and output from the monitoring software, and provides methods for filling
them when required.  Finally, the online event displays are
contained within the monitoring framework using the EventDisplay class.

The output is then saved in a common area for offline access and for syncing
with a web server. This is hosted at CERN and allows for remote
monitoring of the experiment.





\section{Cryostat and feedthroughs}

\subsection{Scope and requirements}

The cryostat consists of a steel warm outer structure, layers of insulation and an inner cold membrane.  The outer 
structure (shown in Figure~\ref{fig:warm-vessel-layout}), which provides 
the mechanical support for the  
membrane and its insulation,  consists of vertical beams that alternate with a web of metal frames. It is constructed to 
withstand the hydrostatic pressure of the liquid argon and the pressure of the gas volumes, and 
to satisfy the external constraints. 
In particular, this structure is to be constructed in EHN1 without any mechanical attachment to the floor or the building side walls.  
\begin{cdrfigure}[Warm vessel layout]{warm-vessel-layout}{Warm vessel layout showing the various major components}
  \includegraphics[width=1.\textwidth]{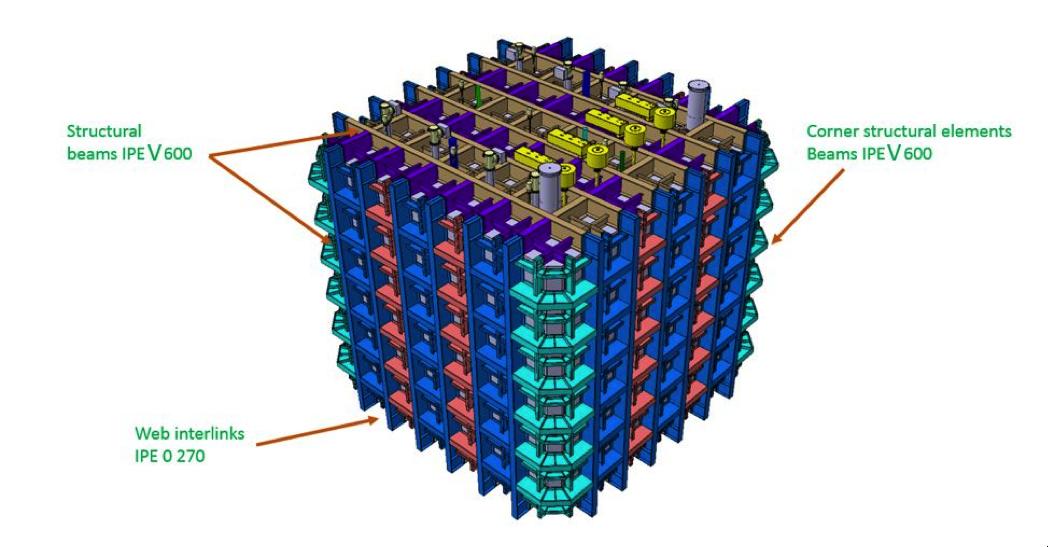}
\end{cdrfigure}
Inside the steel structure, a 10-mm thick skin of stainless steel plates are welded to provide a gas barrier to the outside.
The top of the cryostat is accessible for installation of the detector elements, the electrical/signal feedthrough, the detector supports and other cryogenics services.  The dimensions 
are dictated by the required 
active volume of LAr, constraints on the distances from the active volume to the cryostat inner walls and cryostat material thicknesses.  
The inner cryostat dimensions are: width = 8.548~m, length = 8.548~mm and height = 7.900~m. The dimensions 
ensure that all crossing penetrations are arranged as requested and that there is enough space for maintenance. 

A secondary membrane is located within the insulation layer. The cold vessel is based on the GTT membrane technology~\cite{gtt}.   Thermal requirements call for 
a thickness of 800\,mm, including the insulation, and the primary and secondary membranes. These two membranes provide a first and second level of containment. There is no requirement at this point for additional containment at the level of the warm steel structure. The SS skin of 10-mm thickness between the warm structure and the insulation provides an effective gas enclosure, which allows control of the argon atmosphere inside the insulation volume.
All necessary information can be found in~\cite{edms1}. 
The 3D detailed CAD model is visible in~\cite{edms2}. \fixme{Check if these two references are still valid (per David M)}

Prior to installation of the GTT insulation and 
inner membranes, the gas tightness of the SS 10-mm membrane was measured and verified by CERN using dye penetrant analysis, local vacuum-bag techniques, and He leak 
detection at the level of the natural He present in the atmosphere ($\sim2-3 \times 10^{-6} mbar/l/sec$). A report was presented to GTT. \fixme{David asks: is this correct?}

\subsection{Storage characteristics}

The 
cryostat is required to store LAr at a temperature between 86.7\,K and 87.7\,K with a pressure inside the 
tank of 950 < P < 1100\,mbar. 
The thermal fluxes must be tightly controlled, i.e., 
they will be kept under 5\,W/m2 on 
the inner membrane that is in contact with liquid, in order to prevent boiling of the LAr.

The storage parameters of the cryostat are as follows:
\begin{itemize} 
\item The inner dimensions are 7900\,mm high $\times$ 8548\,mm length $\times$ 8548\,mm wide.  This corresponds to a total volume of $\sim$580 m$^3$. 
\item Tank liquid capacity (assuming a $\sim$4\% ullage): $\sim$ 557 m$^3$
\item Residual Heat Input (RHI): 5-6\,W/m$^2$
\item Insulation weight: 90 kg/m$^3$  
\item Insulation thickness (all included): 0.8\,m 
\item Design pressure: Max \SI{1350} mBar / Min 950 mBar.  The \SI{1350} mBar is for an accident condition during the cryogenics operation.
\item Operating temperature: 86K-89\,K
\end{itemize}

Figure~\ref{fig:cryo-overall-dim} shows a side cross section of the cryostat with the inner dimensions of the cryostat, 
the thickness of the insulation 
and the overall outer dimensions of the warm structure.

\begin{cdrfigure}[Cryostat overall dimensions]{cryo-overall-dim}{Cryostat overall dimensions}
  \includegraphics[width=0.8\textwidth]{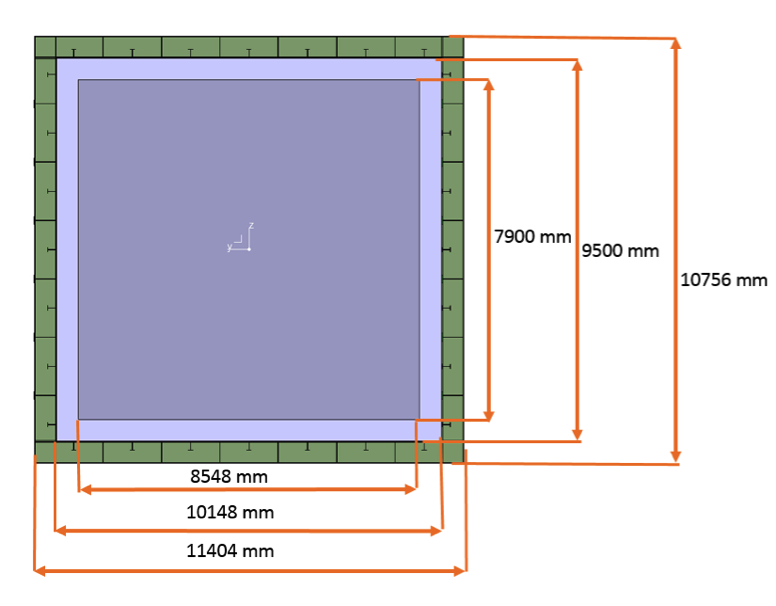}
\end{cdrfigure}

\subsection{Cold \textit{GTT} vessel}

The cold vessel from GTT is installed inside the warm support structure, which includes the stainless steel gas enclosure membrane. The cold vessel consists of a thermal insulation, a primary corrugated stainless steel membrane, as well as a secondary thin membrane, to provide primary and secondary liquid containment. A cross sectional view of the insulation and membrane layers is shown in Figure~\ref{fig:xsec-insulation-layers}.

\begin{cdrfigure}[Cross section of insulation layers on membranes]{xsec-insulation-layers}{Cross section of insulation layers on membranes}
  \includegraphics[width=0.8\textwidth]{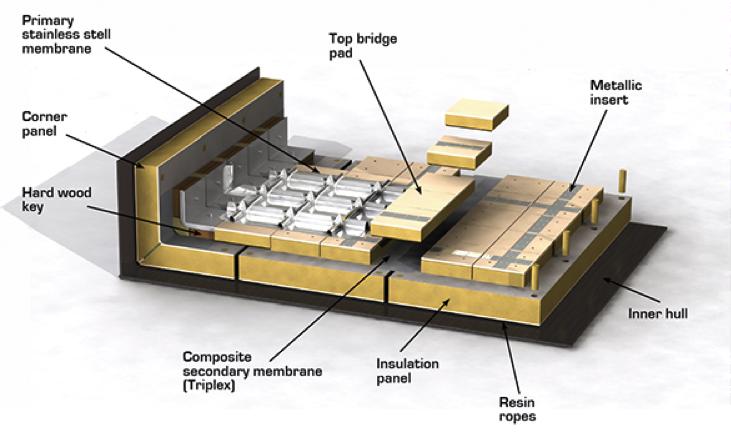}
\end{cdrfigure}

The primary membrane is made of corrugated stainless steel 304 L and is 1.2 mm in thickness.  The standard size of the sheets is 3\,m $\times$ 1\,m.  The secondary membrane is made of Triplex.  This is a composite laminated material of a thin sheet of aluminium between two layers of glass cloth and resin.  It is positioned inside the prefabricated insulation panels between two of the insulation layers.    The insulation is made from reinforced polyurethane foam.  The insulation panels are bonded to the inner 10 mm skin using resin ropes.  The insulation layers are instrumented with gas inlets, outlets, temperature and pressure sensors.

\subsection{Temporary Construction Opening (TCO)}

A dedicated access window is necessary to install the ProtoDUNE-SP detector.  This is referred to as the temporary construction opening (TCO) and shown in Figure~\ref{fig:front-tco-np04}.  This means that no insulation of membrane can be installed at the beginning in this location. Once the detector installation has progressed as far as possible and all of the large TPC components are inside the cryostat, the TCO will be closed.  The 10-mm SS skin, insulation and cold membranes are installed and welded in place. 

\begin{cdrfigure}[Front view of the cryostat with the TCO for the NP04 cryostat]{front-tco-np04}{Front view of the cryostat with the TCO for the NP04 cryostat shown in green.}
  \includegraphics[width=0.8\textwidth]{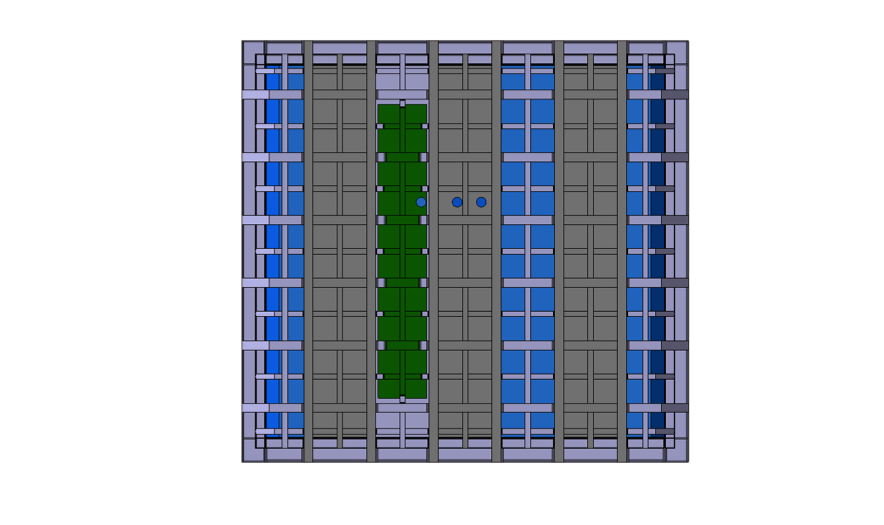}
\end{cdrfigure}

\subsection{LAr pump penetration}

To keep the high level of purity required, LAr is extracted from a point as low as possible in the cryostat and pushed by cryo-pumps 
to the external filtering system through the liquid recirculation circuit. A special penetration is thus 
required on one side wall of the cryostat to connect through a dedicated system of safety valves to the liquid argon pumps.
This penetration requires a local modification of the insulation panels and the SS primary membrane, and consists of a crossing tube with a diameter of 168\,mm for the insulation and the membrane, and a larger-diameter hole at the stainless steel plate.

\subsection{Beam window penetration}
\label{subsec:beamwindow}
Once constructed, the ProtoDUNE-SP detector will be exposed to the charged
particle beam from the SPS accelerator. To minimize energy loss and
multiple scattering of the beam particles in the dead material of the
cryostat and its insulation, a beam window is inserted at the primary beam position as defined  in Section~\ref{sec:beamrequirements}. 

  The vacuum pipe of the beamline has an external diameter of 219~mm. The beam window is being designed with a dimension of 250~mm in diameter to allow for alignment tolerances.  The direction of the beam window follows the one of the beam.
The outer portion of the
beam window penetration is a vacuum pipe that extends from the H4 beamline (see Section~\ref{sec:h4beamline})  through the outer insulation layer and ends at the secondary
membrane. A safety valve at the cryostat entrance ensures fast segmentation of the vacuum in case of accident.   The
portion of the foam insulation between the secondary and the primary
membrane is replaced with a lower density foam 
(9~kg/m$^3$).
To maintain structural integrity, the plywood supporting
the primary membrane in the vicinity of the beam window penetration is
replaced with a Nomex honeycomb plate sandwiched between thin G10 or Carbon layers. Nomex is a polymer material with high thermal resistance and Nomex sandwiches are well known for their structural resistance, and have already been used at cryogenic temperatures in the ATLAS detector.
 In this design, both the
primary and secondary stainless steel membranes remain intact. Care has been taken to position the beam window exit on the interior of the cryostat to match a flat section of the corrugated primary SS membrane.
Thermal and stress analyses are being conducted in collaboration with GTT. These will influence the detailed design of the first segment of the beam window. 
The total amount of material in this design, including the primary membrane, and assuming a 0.3~mm G10 thickness on both sides of the Nomex sandwich and a 0.3-mm-thick steel beam window, is equivalent to 10\% of a radiation length. 

\subsection{Roof signal, services and support penetrations}

The penetrations through the cryostat have been arranged by position and diameter. 
Most of the penetrations are placed on the ceiling of the cryostat. They have been differentiated into two main groups according to their function and the thermal stresses they will be submitted to. The classification determines whether penetrations can be used to support the weight of the detector or not.
The penetrations on the roof of the NP04 cryostat are detailed in Table~\ref{tab:roofpenetrations}.  
A 3D CAD model to identify all positions can be found at~\cite{edms4} and in an associated drawing~\cite{edms5}. \fixme{Check these two references per David}

\begin{cdrtable}[Cryostat penetrations, roof]{llll}{roofpenetrations}{Cryostat penetrations in roof and on side.}
Component &  Quantity & Value \\ \toprowrule
West TPC translation suspension:  & crossing tube diameter & \SI{200}{mm}\\ \colhline
Center TPC translation suspension: & crossing tube diameter & \SI{200}{mm}\\ \colhline
East TPC translation suspension: & crossing tube diameter & \SI{200}{mm}\\ \colhline
Signal cable chimney FTs:& crossing tube diameter & \SI{250}{mm}\\ \colhline
Spare on Signal cable row FTs: & crossing tube diameter & \SI{250}{mm}\\ \colhline
Laser FTs: &  crossing tube diameter & \SI{160}{mm}\\ \colhline
Calibration Fiber CPA FT:&  crossing tube diameter & \SI{250}{mm}\\ \colhline
Spare on CPA line FTs:&  crossing tube diameter & \SI{150}{mm}\\ \colhline
HV FT: &  crossing tube diameter & \SI{250}{mm}\\ \colhline
Manhole:&   crossing tube diameter & \SI{710}{mm}\\ \colhline
Angled beam windows -- west side:&   crossing tube diameter & \SI{250}{mm}\\ \colhline
 &   Vertical: \SI{11.342}{\degree} & \\ \colhline
 &    Horizontal: \SI{11.844}{\degree} & \\ \colhline
TCO - side:&   \num{1200}\si{mm} $\times$ \num{7300}\si{mm} & \\ \colhline
 Cryogenic pipes - roof: &    & \\ \colhline
 &   crossing tube diameter & \SI{250}{mm}\\ \colhline
 &   crossing tube diameter & \SI{304}{mm}\\ \colhline
 &  crossing tube diameter & \SI{152}{mm}\\ \colhline
 &   crossing tube diameter & \SI{125}{mm}\\ \colhline
 &   crossing tube diameter & \SI{250}{mm}\\ \colhline
 Cryogenic pipes -- north side: &  crossing tube diameter & \SI{350}{mm}\\ 
\end{cdrtable}

\subsection{Detector support structure (DSS)}

Prior to the installation of the TPC, the detector support structure DSS is installed inside the cryostat.  The DSS is shown in Figure~\ref{fig:dss-install}.  It is positioned near the ceiling and is supported by nine penetrations through the cold side of the membrane extending up to the warm structure of the cryostat.  The warm structure of the cryostat supports all of the loads from the detector.  The DSS consists of two layers of I beams.  The top (yellow) layer is oriented in the y direction and designated as the Y beams and the bottom (purple) layer is oriented in the x direction and designated as the X beams. \\
The Y beams are fixed in the y direction at the center support point, but free to move during the cool down at the two outer points.  The ends of the Y beams are expected to shrink $\sim$10\,mm towards the center during cooldown to LAr temperature.  

The X beams are used for the direct support and positioning of the TPC components.  Only three X beams are shown in Figure~\ref{fig:dss-install}, but there are two additional inbetween the ones shown.  
The full set of beams is shown in Figure~\ref{fig:dss-install2} along with the naming convention for the X beams.
 X beam A supports the row of APAs near the Saleve side of the cryostat.  X beam B is used for the installation and support of the end wall FC in the Saleve drift of the cryostat.  X beam C supports the row of CPAs.  X beam D is used for the installation and support of the end wall FC in the Jura drift.  X beam E supports the row of APAs near the Jura side of the cryostat.  

The X beams have the ability to translate on rolling trolleys in the Y direction in order to move the TPC components from the TCO entrance to their correct position in Y inside the cryostat.  They are fixed in the X direction to the Y beams at the beam side of the cryostat.  The reason 
to fix the X beams on the beam side is to limit the movement of the beam side of the TPC with respect to the membrane wall since
the beam plug is mounted at this side of the TPC.  

\begin{cdrfigure}[Detector support system (DSS)]{dss-install}{Detector support system}
\includegraphics[width=0.8\textwidth]{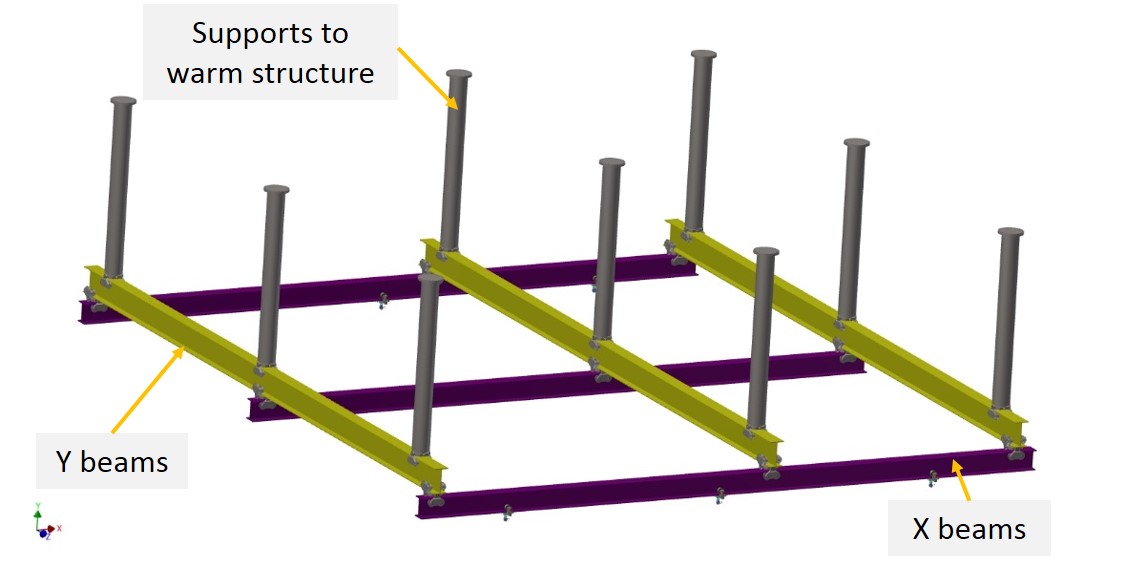}
\end{cdrfigure}

\begin{cdrfigure}[DSS showing full set of beams]{dss-install2}{DSS showing full set of beams; the horizontal (X) beams in the figure are called \textit{bridge beams} and the vertical (Y) beams are called \textit{runway} beams. }
 \includegraphics[width=0.6\textwidth]{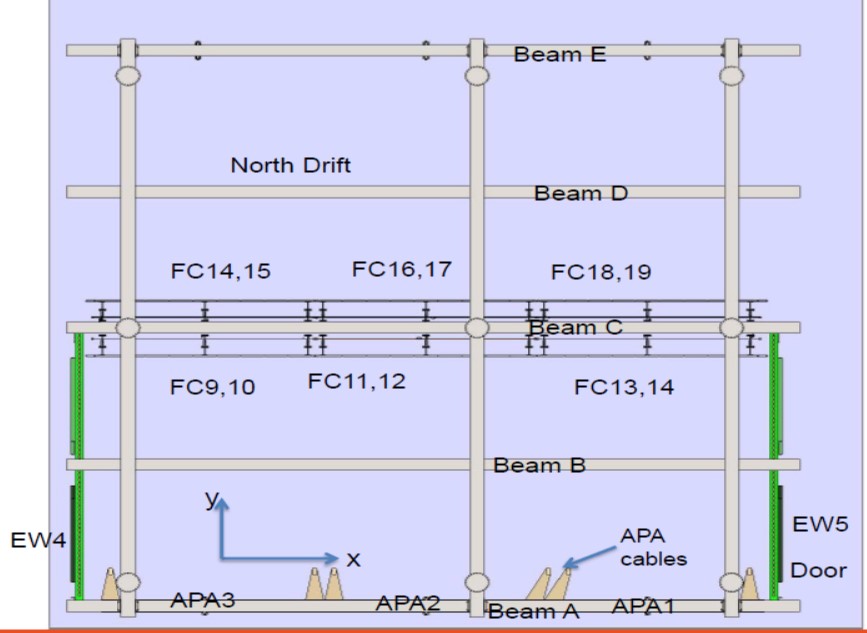}
\end{cdrfigure}









\section{Cryogenics and LAr purification systems}
\label{sec:cryo-purif}

\subsection{Overview, overall planning and ES\&H}

The scope of the ProtoDUNE Cryogenics includes the design, procurement, fabrication, testing, delivery, installation oversight and acceptance tests of a comprehensive cryogenic system that meets the performance requirements for purging, cooling down and filling the cryostat, acquiring and maintaining the LAr temperature within $\pm$1 K around nominal temperature (88.3 K), purifying the Liquid Argon (LAr) outside the cryostats, and re-condensing and purifying the boil-off Gaseous Argon (GAr).

The reference-design for the ProtoDUNE cryogenics infrastructure includes the External, Proximity and Internal Cryogenics.

The \textit{External Cryogenics} includes the systems used for the storage and eventual production of the cryogens needed for the operation of the cryogenic system (liquid ntrogen, LN2, for cooling; LAr for the cryostat) and GAr generated from the cryogenic storage tanks. In particular, it encompasses:
\begin{itemize}
\item The receiving facilities for LAr and LN2 tanker trucks;
\item The cryogenics transfer lines to deliver LAr and LN2 to the Proximity Cryogenics (in the vicinity of the cryostat);
\item The ambient vaporizer and transfer lines to deliver GAr to the cryostat for the piston purge and the GAr make-up; and
\item The regeneration system for the LAr purification (with GAr from the vaporizer and H2 from an H2 bottle, mixed before being sent to the LAr purification).
\end{itemize}
The \textit{Proximity Cryogenics} takes the cryogens from the External Cryogenics and delivers them to the Internal Cryogenics under the required pressure, temperature, purity and mass flow rate. It encompasses:
\begin{itemize}
\item The condenser (with heat exchanger) to re-condense the boil-off GAr;
\item The LAr purification system with inline purity monitor;
\item The LAr recirculation pumps;
\item The LAr Phase separator to feed the cryostat;
\item The LN2 Phase separator to feed the condenser;
\item The GAr purification system;
\item The cryostat-purge equipment; and
\item The condenser LAr pump.
\end{itemize}
The \textit{Internal Cryogenics} includes all the cryogenic equipment located inside the cryostat. It encompasses:
\begin{itemize}
\item The cryostat/detector cool down manifolds
\item The LAr distribution manifold
\item The GAr purge distribution manifold.
\end{itemize}
The equipment described in this chapter will be used for the cool-down, filling, operation, purification, emptying and warm-up of the ProtoDUNE Single Phase cryostat. These operations are described in greater detail in Section~\ref{sec:cryo-op-modes}.

The development of the ProtoDUNE cryogenics is part of a common effort between CERN and Fermilab which includes the cryogenics for the ProtoDUNE Single Phase and Dual Phase detectors at CERN, and the Short Baseline Neutrino Near Detector (SBND) and Far Detector (SBN-FD) at Fermilab.

The cryogenic systems for all four projects are developed jointly with a standard approach to minimize the duplication of work, benefit of existing knowledge (at Fermilab and CERN), and also prototype for the Long Baseline Neutrino Facility (LBNF)/Deep Underground Neutrino Experiment (DUNE) project. The systems build on the successful experience of the Liquid Argon Purity Demonstrator (LAPD), 35-t prototype, and MicroBooNE at Fermilab, and the development of the WA105 1$\times$1$\times$3 Dual Phase prototype at CERN and CERN's experience in the design and operation of large-volume noble liquid detectors.


During all phases, CERN codes and standards will guide the design, procurement and installation phases of the ProtoDUNE Single Phase cryogenics. The planned work process provides for reviews throughout all phases of the project to guarantee stringent adherence to the safety and scientific requirements.




The project requirements for the ProtoDUNE cryogenics system are identical to those of the DUNE Far Detector cryogenics. The current list 
of requirements is available at \cite{DUNE_FD_cryogenics_req}. 

A selection of the most relevant requirements is presented here:
\begin{itemize}
\item Cryosys-013: The system shall allow recirculation and purification of the liquid argon inventory to achieve the needed LAr purity to meet the scientific requirement (less than 10 day/volume change based on ICARUS experience).
\item  Cryosys-016: The purification system shall be capable of removing contaminants form the LAr prior to filling and shall maintain purity during operation.
\item  FD-tpc-006: Electron lifetime greater than 3 ms (maximum drift time at nominal field is 2.25 ms).
\item  Cryosys-015: The system shall provide an argon gas boil off and reliquefaction system.
\item Cryosys-026: The cryogenics system shall not allow sources of argon gas reliquefaction inside the cryostat, e.g. uninsulated pipes carrying liquid argon.
\item  Cryosys-021: The cryostat and cryogenic systems shall be designed for using the piston-purge technique (introducing heavy gas at the bottom and taking out exhaust from the top) for removing initial electronegative impurities.
\item  Cryosys-019: The cryogenics system shall not introduce unwanted noise into the electronics.
\item  Cryosys-022: The cryogenics system shall provide a stable environment in the cryostat for the detector.
\item  Cryosys-023:The cryogenics system shall be designed in accordance with the cryostat to maintain a single phase in the entire liquid argon volume at a stable temperature. The chosen temperature is 88.3 K $\pm$ 1 K.
\end{itemize}


\subsection{Cryogenics layout}

The Process Flow Diagram (PFD) of the ProtoDUNE cryogenic system is shown in Figure~\ref{fig:cryo-process-flow}. The External Cryogenics located outside of the EHN1 building, is shared with the Dual Phase prototype, which is located in the same experimental hall, few tens of meters away.

\begin{cdrfigure}[Cryogenics process flow diagram]{cryo-process-flow}{Cryogenics process flow diagram} 
\includegraphics[width=0.95\linewidth]{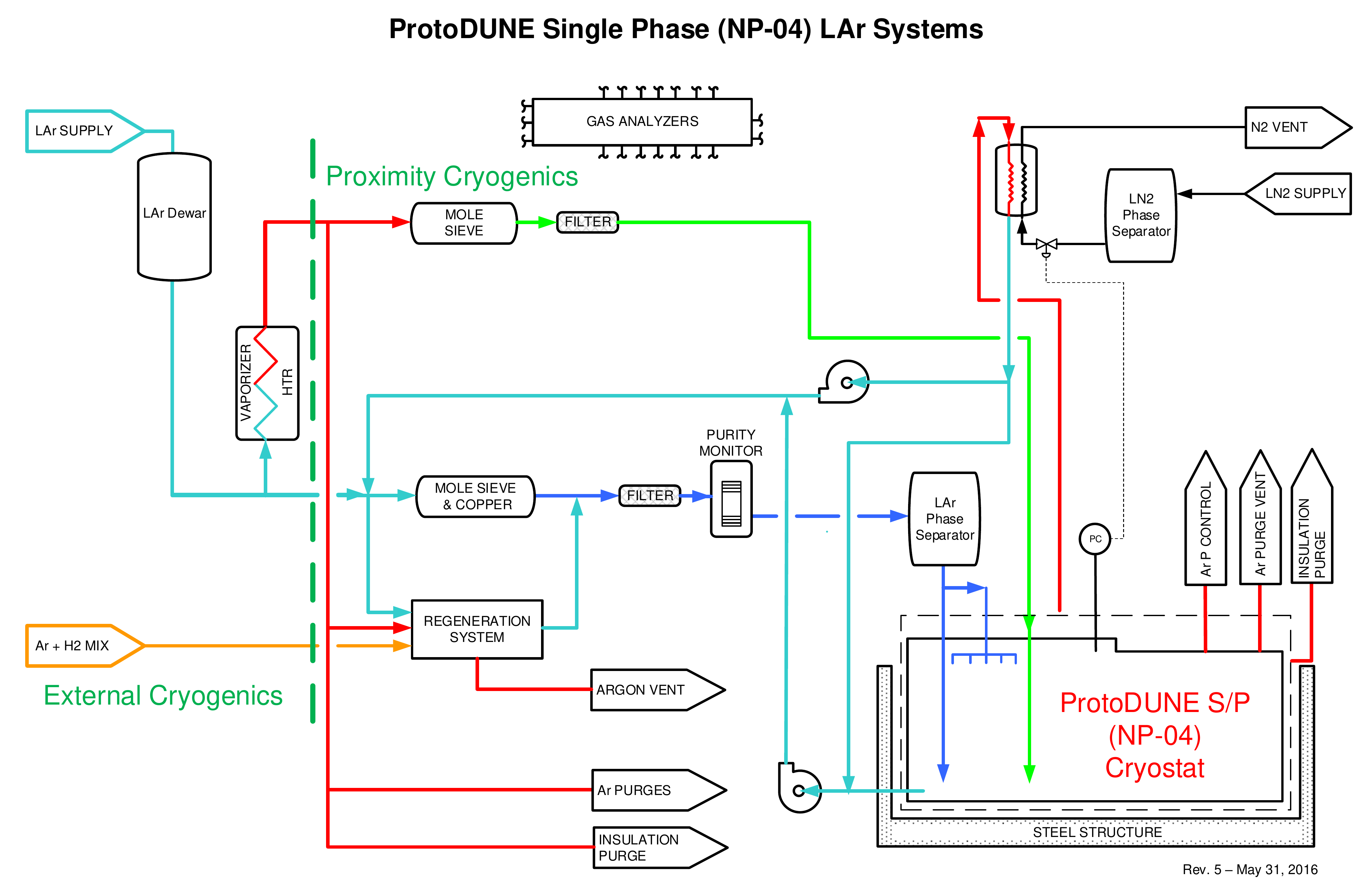}
\end{cdrfigure}

The system has the following functions:

\begin{itemize}
\item It provides the GAr for the piston purge phase and the GAr make-up.
\item  It provides the LAr to the cryostat.
\item  It provides the LN2 to the condenser.
\item  It provides the cooling power by means of evaporation of liquid nitrogen and condensation of GAr, to the liquid argon cryostat, for its cool-down, normal operation and warm-up phases.
\item  It provides the capability to purify the cryostat liquid argon volume to a level of parts per trillion (ppt) Oxygen equivalent contamination; the purification process uses mole-sieve and active copper.
\item  It provides the capability to purify the re-condensed boil off before reintroducing it inside the cryostat.
\item  It provides means to cool down the cryostat and the detector following the requirements.
\item  It distributes the LAr and GAr inside the cryostat to meet the requirements.
\end{itemize}

Figure~\ref{fig:3d-view-cryo-installation} shows a 3D view of the cryogenic installation as currently designed. The red and green lines entering  from the bottom of the figure are the LN2 and LAr supply lines, respectively, from the external cryogenics.
\begin{cdrfigure}[3D model of the installation ]{3d-view-cryo-installation}{3D model of the installation } 
\includegraphics[width=0.85\linewidth]{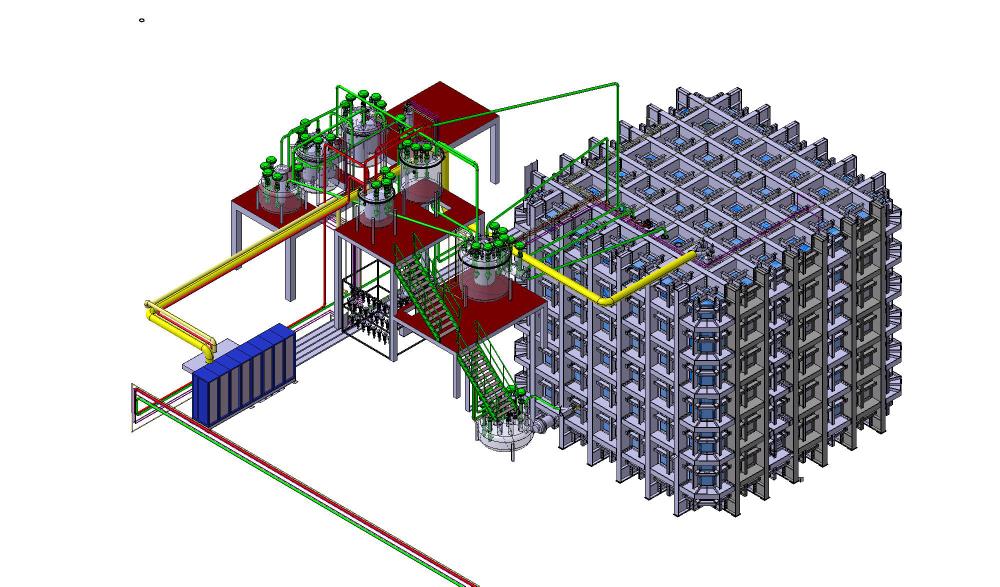}
\end{cdrfigure}
Figure~\ref{fig:int-cryogenics-detail} shows a 3D view of a detail of the internal cryogenics: the cryostat and detector cool down manifolds at the top of the cryostat.
\begin{cdrfigure}[Detail of the internal cryogenics]{int-cryogenics-detail}{Detail of the internal cryogenics} 
\includegraphics[width=0.9\linewidth]{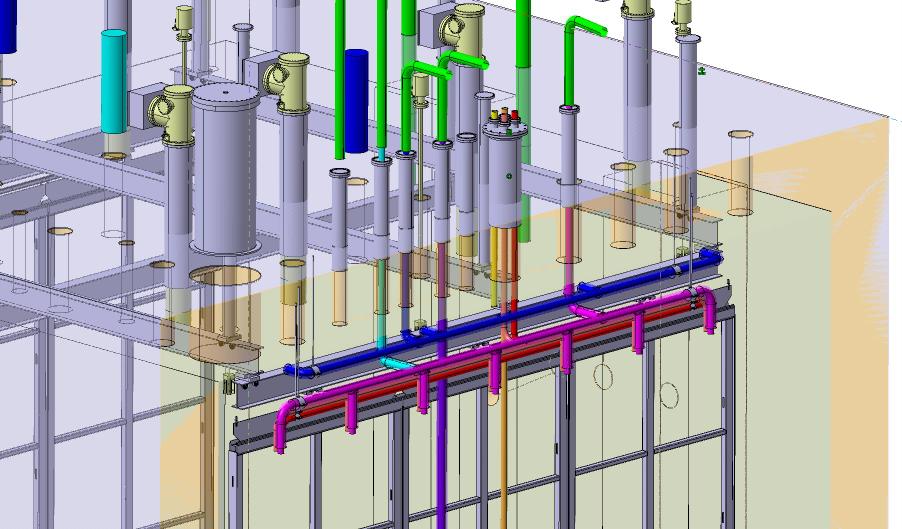}
\end{cdrfigure}

There is a common receiving facility for NP-02 and NP-04 located outside the building, from which Argon and Nitrogen lines take LAr, GAr, and LN2 to the respective installations.

A 50 m$^3$ (69 tons of LAr capacity) vertical dewar will allow for receipt of LAr deliveries for the initial filling period. This liquid argon dewar serves also as a buffer volume to accept liquid argon during the fill period. An analyzer rack with instruments to check water, nitrogen, and oxygen content of the delivered LAr batches will also be located in the vicinity. A 55-kW vaporizer is used to vaporize the liquid argon from the storage dewar prior to delivery to the GAr pipes.

The cryostat will have its own argon condenser (16 kW of cooling power), argon-purifying equipment and overpressure protection system. The full power of the argon condenser is used during the initial cool down phase only, which is expected to take two to three weeks. 

A 50 m$^3$ vertical dewar (40-t LN2 capacity) will allow for receipt of LN2 deliveries and storage of LN2 for cool down and normal operations. LN2 is flown into the heat exchanger of a condenser located in close proximity of the cryostat to recondense the boil-off GAr coming from the cryostat itself.

Two LAr recirculation pumps are placed outside of the membrane cryostat to circulate liquid from the bottom of the tank through the purifier and then back to the tank to ensure the needed LAr purity. 
The purification filters are located in the vicinity of the cryostat. The filters contain dual media, a molecular sieve for removal of water and a copper coated catalyst media for oxygen removal. There is one gas filter that is used during the purge in closed loop phase and two liquid filters used during the filling and normal operations to continuously purify the bulk of the LAr inside the cryostat. Associated with the filters, there will be regeneration equipment such as heaters, a GAr supply, a H2 bottle, and a way to mix GAr and H2. 
Before the Ar is returned to the cryostat, the LAr flows into a phase separator: the liquid is taken from the bottom and delivered to the cryostat, while the gas is returned to the condenser.

\subsection{Modes of operation}
\label{sec:cryo-op-modes}

The major functions of the cryogenics system servicing the cryostat are to supply cryogens for cool down and fill, and to provide gas argon filtration and condensing, liquid argon filtration and circulation. The methods presented in this section are motivated by experience from the cryogenic systems of other LAr Time Projection Chamber (TPC) experiments, such as ICARUS, LAPD, the 35 ton detector and MicroBooNE.

\paragraph{Cryostat piston purge}
After the cryostat construction and following the installation of all scientific equipment, the cryostat will be cleaned and purged in preparation for cool down and filling. Construction procedures leading up to this point will ensure that the completed cryostat does not contain debris and is free of all loose material that may contaminate the LAr.

\paragraph{Purge in open loop}
Argon piping will be isolated, evacuated to less than 0.1 mbar  absolute pressure and backfilled with high-purity argon gas. This cycle will be repeated several times to reduce contamination levels in the piping to the ppm level. The reference-design choice for removing air from the membrane cryostat will be to flow/piston-purge argon, introducing the heavier argon gas at the bottom of the tank and removing the exhaust at the top. The exhaust will be taken from the main GAr outlet, but also from all the side ports located on each penetration through the roof to ensure that all volumes (especially trapped volumes) are properly purged.

The flow velocity of the advancing GAr will be set to 1.2 m/hour. This is twice the diffusion rate of the air downward into the advancing argon so that the advancing pure argon-gas wave front will displace the air rather than just dilute it. A 2D ANSYS model of the purge process shows that after about 13 hours of purge time and 2 volume changes, the air concentration will be reduced to less than 1\%. At 44 hours of elapsed time and seven volume changes, the purge process is complete with residual air reduced to a few ppm. This simulation includes a representation of the perforated field cage at the top and bottom of the detector and heat sources due to the readout electronics. 

The Computational Fluid Dynamics (CFD) model of the purge process has been verified in multiple arrangements: (1) in an instrumented 1 m-diameter by 2 m-tall cylinder, (2) in LAPD, a 3 m- diameter by 3 m-tall cylindrical tank where gas-sampling measurements were at varying heights and times during the purge process, and (3) within the 35 ton membrane cryostat, the prototype vessel built at FNAL in 2013. The results of these tests are available in~\cite{lar1-nd-35-ton-talk} and~\cite{CFD_verification-lapd}. 
Once the residual air inside the tank is at the the ppm level, the process continues in the closed loop configuration. 

\paragraph{Purge in closed loop}

Water and oxygen will continue to be removed from the system for several days following the initial purge. During this step the GAr is no longer exhausted but recirculated through the GAr purifier and sent back to the bottom of the cryostat. The cryostat contains a relatively large amount of FR4 circuit-board material and a smaller inventory of plastic-jacketed power and signal cables. These somewhat porous materials may contain as much as 0.5\% water by weight. Water-vapor outgassing from these materials will be entrained in the gas flow exiting the top of the cryostat and will be removed from the gas stream by filters. Adsorbed water will also be removed from the metallic inner surfaces of the cryostat and piping system. Water deep within porous materials will remain; this is not a problem since the water diffusion rate in FR4 at room temperature is already quite low (0.3 ($\mu$m)$^2$/s) and the FR4 assemblies are relatively thick (1 cm).\\
This process reduces the oxygen and water contamination inside the cryostat to sub-ppm levels, at which point the cool down may commence.

\paragraph{Cool-down}

Purified LAr will be mixed with GAr and distributed by a set of dedicated sprayers near the top of the cryostat and on the side of the TPC to cool down the cryostat and the detector in a controlled way. The sprayers deliver a mix of LAr and GAr in atomized form that is moved inside the cryostat by another set of sprayers flowing GAr only. Part of the boil-off gas is re-condensed inside the condenser and it then flows back as liquid to feed the LAr sprayers. The balance is vented during the process.
Simulations have shown that this cool-down method can maintain the cool down requirements of the detector, as listed in Table~\ref{tab:cryogen-install-params} , and those of the cryostat, which are less stringent. The required cooling rate is determined by the maximum stress that detector components can tolerate. For example, the 150 $\mu$m APA wires will cool much more rapidly than the APA frames. A temperature-monitoring system (provided by the detector) will be used to control the temperature difference across the cryostat and the detector.

\paragraph{Filling}

Once the cryostat and the TPC are cold, LAr is introduced in the cryostat through the cryostat filling pipework. Argon is transferred directly from the LAr storage tank after passing thorough the LAr filtration system for purification. The filling process will take place over three to four weeks if two trucks per day are delivered; if only one truck per day is delivered, it will take twice as long.

\paragraph{Steady state operations}

During steady state operations :
\begin{itemize}
\item LAr is continuously circulated and purified by means of an external LAr pump (two are installed for redundancy, but only one is in use at a time).
\item Boil-off GAr is re-condensed in a condenser situated outside the cryostat and purified before being reintroduced as LAr. The re-condensed LAr is sent to the LAr filtration system by means of a dedicated LAr pump and mixed in line with the bulk of the liquid coming from the cryostat. Alternatively, it is possible to send it to the inlet of the main LAr circulation pumps and from there as a single LAr stream to the filtration system.
\end{itemize}

\paragraph{Emptying}

At the end of operations (or if/when maintenance on the tank is needed) the tank is emptied and the LAr removed. The LAr is returned to the storage tank outside the building and from there unloaded back to LAr tankers.

\paragraph{Parameters}

Table~\ref{tab:cryogen-install-params} presents a list of relevant parameters for the installation. The filling flow rate of 18~l/min (0.42 kg/s) is an estimate. The actual value might be limited by the pressure inside the LAr storage dewar. It is also assumed that the facility at SURF is able to receive two trucks/day of LAr; this requires confirmation from the suppliers.

\begin{cdrtable}[Engineering parameters for cryogenics installation]{llll}{cryogen-install-params}{Engineering parameters for cryogenics installation}
Mode & Parameter  & Value & Notes \\ \toprowrule
Piston purge & GAr flow rate  & 88 m3/hr & From 1.2 m/hr\\ \colhline
Cooldown & Maximum cool-down rate TPC  &40 K/hr  & T sensors on the detector \\ 
&  &    & responsibility of detector\\ \colhline
Cooldown & Maximum delta T between any & 50K & T sensors on the detector \\ 
& two points in the detector &  & responsibility of detector \\ \colhline
Filling (*) & LAr filling flow rate  & 18 l/min (0.42 kg/s) & Assuming 2 trucks/day\\ \colhline
Normal ops & Cryostat static heat leak & 3.0 kW & GAr boil-off (18 g/s)\\ \colhline
Normal ops & Other heat loads (estimate)  & 5.0 kW  & Total estimate is $\sim$8 kW \\ \colhline
Normal ops & LAr circulation (5 days turnover) & 72 l/min (1.67 kg/s) & 72 l/min (1.67 kg/s)\\ \colhline
Emptying & Nominal flow rate emptying  & 72 l/min (1.67 kg/s) &   \\  \colhline
All  & Condenser size& 16 kW & \\
\end{cdrtable}

\subsection{Features}

This section briefly describes the main features of the various parts of the cryogenics system.

\paragraph{External Cryogenics}

The external cryogenics comprises the liquid argon and liquid nitrogen receiving facilities, the LAr/GAr and LN2 distribution systems, the argon/hydrogen mixture to regenerate the LAr/GAr purification filters, and the mechanical filters on the LAr filling line.

The argon grade specification in the current gas supply contract at CERN is Grade 4.6. This corresponds to
 a minimum guaranteed purity of 99.996\%, allowing a maximum concentration of 5.0  to 10 ppm for O2, H2O and N2.  
The impurity levels in the delivered argon are typically much lower, and the argon filtration system reduces the O2 and H2O concentrations sufficiently to meet the physics requirements. 

Facilities are required for the offloading of LN2 and LAr road tankers. Vehicle access and hard-surfaced driving areas are being constructed adjacent to the LN2/LAr dewars and the LAr/LN2-supply pipes. A LAr storage dewar will hold the contents of a road tanker in order to minimize off-loading time. Road tankers will connect to a manifold and will use their on-board pumps to transfer the LAr to the storage dewar.  The LAr will be stored and transported as a liquid inside the cryostat during the filling process. 
A bottle containing 100\% Hydrogen (H2) will be stored outside the building, as well, and connected to a GAr line coming from the storage dewar. The GAr/H2 mixture  will be used to regenerate the LAr and GAr purification filters as needed.
One 1-micron mechanical filter is located on the LAr feed line to prevent dirt and impurities from the LAr supply to enter the purification system and the cryostat.

\paragraph{Proximity Cryogenics}

The Proximity Cryogenics comprises the argon condenser, the purification system for the LAr and GAr, the LAr circulation pumps, and the LAr/LN2 phase separators:

\paragraph{Argon reliquefaction and pressure control}

The high-purity liquid argon stored in the cryostat will continuously evaporate due to the unavoidable heat ingress. The argon vapor (boil-off gas) will be recovered, chilled against a stream of liquid nitrogen, condensed and returned to the cryostat. A closed system is required in order to prevent the loss of the high-purity argon. The re-condensed boil-off can be returned to the cryostat in three ways:
\begin{enumerate}
\item With a small LAr pump that sends it into the main LAr circulation stream (normal mode).
\item Directly to the condenser (emergency mode, when it is not possible to go through the purification system).
\item To the inlet of the main LAr circulation pumps (when the small LAr pump needs maintenance, to guarantee a continuous purification of the boil-off GAr).
\end{enumerate}

During normal operation the expected heat ingress of approximately 8 kW to the argon system will result in an evaporation rate of 30 g/s and expanding in volume by a factor of 200 when it changes from the liquid to vapor phase. This increase in volume within a closed system will, in the absence of a pressure-control system, raise the internal pressure.

Argon vapor will also be removed from the top of the cryostat through the chimneys that contain the cryogenic feedthroughs. As the vapor rises, it cools the cables and feedthrough, thereby minimizing the outgassing. The exiting gaseous argon will be directed to the same condenser as above, in which it is chilled against a stream of liquid nitrogen and condensed back to a liquid. As the argon vapor cools, its volume reduces and, in the absence of pressure control, further gas would be drawn into the heat exchanger, developing a thermal siphon. Therefore, a pressure-control valve on the boil-off gas lines will control the flow to the condenser to maintain the pressure within the cryostat at 0.113 MPa $\pm$ 0.003 MPa. The liquid nitrogen stream (that provides the coolant for the condenser) will be supplied from the LN2 phase separator, which is fed by the LN2 storage dewar located outside of the building. After the heat exchanger the returning N2 vapor is exhausted outside the building. The estimated heat loads to the argon system are listed in Table~\ref{tab:cryostat-est-heatload}.

\begin{cdrtable}[Estimated heat loads within the cryostat]{lc}{cryostat-est-heatload}{Estimated heat loads within the cryostat}
Item & Heat Load (kW)\\ \toprowrule
Insulation Heat Loss & 3.0\\ \colhline
All other contributions  & 5.0 \\ 
(Recirculation pumps, pipes, filters, electronics, etc.) &  \\ \colhline
Total & 8.0 \\
\end{cdrtable}

\paragraph{Argon purification}

The cryostat is designed with one penetration below the liquid level for external pumps used to transfer LAr from it to the purification system. The pumps are inserted into a valve box that is an integral part of the proximity cryogenics. The pump suction must be located at a minimum distance (normally about 1.5 to 2.0 m) below the lowest liquid level at which they are to pump in order to prevent cavitation and vapor-entrapment. There are two pumps for continuous operation during maintenance, but only one is expected to be in service at any moment in time.

The liquid-argon volume will turn over every 5.5 days, which corresponds to 1.67 kg/s (72 l/min) of flow rate. As a point of comparison, ICARUS T600 has a maximum turn-over rate of eight to ten days. 

The multiple-pump arrangement provides a high level of redundancy, which will extend the maintenance-free operating period of the cryostat.

The liquid purification system, located nearby the cryostat, consists of two sets of three filter vessels containing molecular-sieve (1) and copper media (2) filters. They have been arranged in this configuration to reduce the size of the valve box containing them. Each molecular-sieve filter is 0.4 m in diameter by 0.9 m tall and contains 80 kg of media. Each copper filter is 0.6 m in diameter by 1.3 m tall and contains 298 kg of media. The filters are sized to provide effective media usage at low pressure drop over the expected range of flow rates. They are used during the filling and normal operations.

The gas purification system, located nearby the cryostat as well, is used to purify the GAr for the purge in close loop process. It consists of one filter vessel containing molecular-sieve and copper media filters in the same vessel. The mol sieve part measures 0.3 m in diameter by 0.1 m tall and contains 5 kg of media. The copper part measures 0.3 m in diameter by 0.6 m tall and contains 34 kg of media.

During the filling the LAr will flow through the liquid filtration, then the LAr phase separator and into the cryostat.

After the filling is completed, the cryostat liquid argon inventory is continuously circulated through one set of liquid purification filters 
for oxygen and water in order to quickly reduce and maintain the impurity concentration at the 
 level of < 100 ppt oxygen equivalent, matching the required electron lifetime of the TPC detector. 
A dedicated special device, originally developed by ICARUS (usually indicated as ``Purity Monitor"), 
for the measurement of the impurity concentration in liquid argon will be located immediately downstream the filtration system, 
providing information about the quality of the liquid and correspondingly about the actual level of impurity removal efficiency of the filter. 
After the filter the ultrapure argon is returned back to the cryostat via the LAr phase separator. 
Purity monitors will also be resident inside the cryostat, measuring the electron lifetime at different depths of the LAr volume.  The purity monitors (provided by the ProtoDUNE-SP collaboration) are described in greater detail in Section~\ref{sec:mon-dev-sensors}.

The filter material, composed by molecular sieve pellets to remove water and by alumina porous granules covered by highly active metallic copper for catalytic removal of O$_2$ by Cu oxidization, is subject to saturation when the trapped/reacted impurity budget exceeds the removal capacity of the filter material. When this occurs (signaled by the 
fast drop of LAr purity level detected by the external purity monitor) the liquid argon flow is switched to the back-up, ready-for-use filter and the saturated one is regenerated in-situ.

The filter regeneration process is done in subsequent steps. The saturated filter is first warmed up with heated argon gas to an elevated temperature driving into the gas the water captured by the molecular sieve media. A gas mixture of 1.5\% hydrogen (reducing agent) with a balance of argon (inert carrier) at high temperature (500 K) is then used for the reduction of the copper oxide back to metallic copper. Water produced by the reduction process is vented out with the gas flow.
The regenerated filter is finally cooled down and ready to be switched into service. 

\paragraph{Internal Cryogenics}

Internal piping is positioned inside the cryostat to support the air purge and cool-down processes, but also the LAr distribution during filling and normal operations. During air purge argon gas is injected at the bottom of the cryostat and distributed through a set of pipes that pushes the air up and forces it out from the roof. The flow nozzles will be directed downward and to the side so that the injection velocity will not cause local vertical gas plumes or turbulent mixing but rather will spread across the bottom of the tank and produce a stable, upwardly advancing argon wave front. The vertical velocity of 1.2 m/hr for the gas purge includes a contingency for some level of turbulent mixing. In addition to the main vent, all nozzles and dead-end (stagnant) volumes located at the top of the cryostat will have gas-exhaust lines for the initial purge and for continuous sweep-purge of those volumes during normal operations. The sweep-purge during the initial stage of purging will be vented outside of the building, whereas the sweep-purge during normal operations will be re-condensed and recirculated as liquid. 


The cool-down of the cryostat and detector is performed through a set of manifolds flowing LAr (one) and GAr (two). The LAr manifold and a GAr manifold are joined together and terminate with a set of sprayers that deliver a mist of LAr and GAr. This mist is circulated within the cryostat by a jet of GAr coming from the other manifold, which also terminates with sprayers. These manifolds are located on the Jura side and are off to the side of the TPC so as not to flow LAr and GAr directly over the detector itself. The chosen sprayers guarantee a flat profile of the fluid (LAr and GAr) coming out.

During filling and normal operations, the LAr-supply pipework distributes the LAr at the bottom of the cryostat. The outlets are at the end of the pipes, as far away as possible from the side penetration from which the LAr is sent to the purification system.

\subsection{Cryostat pressure control}

The pressure inside the cryostat is maintained within a very narrow range by a set of active controls.  There are pressure control valves that 
can increase or decrease the cooling power in the condenser by controlling the amount of LN2 flowing to the heat exchanger and being vented. Other pressure control valves 
can be used to vent GAr to atmosphere and/or introduce clean GAr from the storage, as needed.  The system is always in place, from the initial purge to the emptying of the cryostat at the end of operations.  

\paragraph{Normal Operations}

The pressure-control valves are sized and set to control the internal cryostat pressure under normal operating conditions to the nominal design pressure of 0.113 MPa. Fluctuations within the range 0.105 MPa (80 mBarg) to 0.120 MPa (230 mBarg) will be allowed. Excursions 
of a few percent (exact values to be determined) above or below these levels will set off warnings to alert the operator to intervene. Further excursion may result in automatic (executive) actions. These actions may include stopping the LAr circulation pumps (to reduce the heat ingress to the cryostat), increasing the argon flow rate through the condenser, increasing the LN2 flow through the heat exchanger inside the condenser, powering down heat sources within the cryostat (e.g., detector electronics), venting some of the GAr to reduce the pressure in a controlled way. Eventually, if the pressure continues to rise, it will trigger the Pressure Safety Valves (PSVs) to operate. 

If the pressure decreases, fresh GAr can be introduced into the cryostat through the GAr make-up line, a dedicated GAr feed line that takes argon directly from the outside supply.
 If the pressure continues to decline, it will trigger the Vacuum Safety Valves (VSVs) to operate.
Table \ref{tab:cryostat-norm-pressures} summarizes the cryostat pressures during normal operation.
\begin{cdrtable}[Cryostat pressures during normal operations]{lc}{cryostat-norm-pressures}{Cryostat pressures during normal operations}
Cryostat part & Pressure\\ \toprowrule
Vessel ullage pressurization & 0.100 MPa (30 mBarg)\\ \colhline
Pressure regulation & 0.110 MPa (140 mBarg) \\ \colhline
Vessel ullage depressurization & 0.125 MPa (280 mBarg) \\ \colhline
Relief valve set pressure & 0.135 MPa (380 mBarg)\\ \colhline
Warm structure design working pressure & 0.135 MPa (380 mBarg) \\ 
\end{cdrtable}

The ability of the control system to maintain a set pressure is dependent on the size of pressure upsets (due to changes in flow, heat load, temperature, atmospheric pressure, etc.) and the volume of gas in the system. The reference design has 0.4 m of gas at the top of the cryostat. This is 5\% of the total argon volume and is the typical vapor fraction used for cryogenic storage vessels. Reaction times to changes in the heat load are slow and are typically on the order of an hour.

\paragraph{Overpressure control}

In addition to the normal-operation pressure-control system, it is planned to provide a cryostat overpressure-protection system. This must be a high-integrity, automatic, failsafe system capable of preventing catastrophic structural failure of the cryostat in the case of excessive internal pressure.

The key active components of the planned system are Pressure Safety Valves (PSVs) located on the roof of the cryostat that will monitor the differential pressure between the inside and the outside of the cryostat and open rapidly when the differential pressure exceeds a preset value. A pressure-sensing line is used to trigger a pilot valve which in turn opens the PSV. The PSVs are self-contained devices provided specially for tank protection; they are not normally part of the control system. 

The installation of the PSVs will ensure that each valve can periodically be isolated and tested for correct operation. The valves must be removable from service for maintenance or replacement without impacting the overall containment envelope of the cryostat or the integrity of the over-pressure protection system. This normally requires the inclusion of isolation valves upstream and downstream of the pressure-relief valves and at least one spare installed relief valve ($n + 1$ provision) or the use of a diverter valve that allows one valve to be always connected to the cryostat. \\
When the valves open, argon is released, the pressure within the cryostat falls and argon gas discharges into the argon vent riser. The valves are designed to close when the pressure returns below the preset level.

\paragraph{Vacuum-relief system}

The cryostat vacuum-relief system is a high-integrity, automatic, failsafe system designed to prevent catastrophic structural failure of the cryostat due to low internal pressure. The vacuum-relief system protects the primary membrane tank. Activation of this system is a non-routine operation and is not anticipated to occur during the life of the cryostat.

Potential causes of reduced pressure in the cryostat include operation of discharge pumps while the liquid-return inlet valves are shut, gaseous argon condensing in the condenser (a thermo-siphon effect) or a failure of the vent system when draining the cryostat. Vacuum-relief valves are provided on LNG storage tanks to protect the structure from these types of events.

The key active components of this additional protection system are Vacuum Safety Valves (VSVs) located on the roof of the cryostat that will monitor the differential pressure between the inside and the outside of the cryostat and open when the differential pressure exceeds a preset value, allowing air to enter the cryostat to restore a safe pressure. A combo PSV-VSV may be used instead of two separate devices, one for overpressure and one for vacuum.

\section{Detector monitoring and slow control}
\label{sec:detmonitoring}

The scope of the ProtoDUNE-SP detector control system (DCS) includes the design, procurement, fabrication, testing,
and delivery
of a comprehensive monitoring, control and safety system for the protoDUNE detector.

The responsibility for the system is split between ProtoDUNE-SP and CERN: 
\begin{itemize}
\item	The ProtoDUNE-SP collaboration is responsible for all devices that will be installed and cabled inside 
the cryostat, the sensors needed to monitor the cryostat and its content, and the specifications for the system. 
\item	CERN is responsible for the implementation of the control system elements outside the cryostat (hardware, firmware and software), including the high-voltage and low-voltage power supplies necessary for the detector operation.
\end{itemize}

This section describes 
the main requirements, 
constraints and assumptions of the control system, and its general structure and components. 

\subsection{Monitoring devices and sensors}
\label{sec:mon-dev-sensors}

The protoDUNE-SP apparatus includes instrumentation beyond the TPC and the photon detectors to ensure that the condition of the liquid argon is adequate for operation of the TPC. This instrumentation includes gas analyzers to monitor the purge of the cryostat and ensure that any remaining atmospheric contamination is sufficiently low, thermometry to monitor the cryostat cool-down and filling, purity monitors to provide a rapid assessment of the electron drift-lifetime independent of the TPCs, and a system of internal cameras to help locate any sparks due to high voltage breakdown in the cryostat.

In addition, sets of precision temperature sensors are being deployed to measure the temperature gradients inside the protoDUNE cryostat. These temperature measurements exploit the opportunity protoDUNE-SP provides to check the predictions of the Computational Fluid Dynamics models being used to design the argon flow in the (much larger) DUNE cryostat.

The CERN Neutrino platform is providing the essential measurements of the cryostat pressure and external environmental conditions.

The instrumentation is designed to establish the quality and stability of the detector environment and to help diagnose the source of any changes in the detector operations. In addition, an extensive set of temperature measurements is planned in order to provide input for the validation of the fluid dynamic models to be used in simulations of the full DUNE apparatus.

\subsubsection{Purity monitors} 
Three purity monitors (PrM) with sensitivity
to electron drift lifetimes from $\sim$100~microseconds up to a few milliseconds
will be used to provide a rapid and direct determination of the 
electron drift lifetime
of the LAr inside the ProtoDUNE-SP cryostat. These PrMs have been generously provided by ICARUS~\cite{Amerio:2004ze} 
after being decommissioned from the T600. 
The design has been replicated with small modifications
for MicroBooNE and 
a number of  R\&D test experiments at Fermilab. Inside the ProtoDUNE-SP the monitors are arranged in a vertical 
column
located behind the APA planes on the Jura side (see Figure~\ref{fig:cryo-side-names}). 
Two of the monitors are supported from the large flange on the manhole, while one is attached directly to the floor of the cryostat.
Ports
 with ConFlat seals on the blanking flange will be made available for
 HV/Signal/Optical Fiber feedthroughs.
 One monitor will be at the height of the top of the APA,
 one at mid-height, and one at the very bottom near the LAr-return manifold, 
which will allow it to monitor
the purity of the LAr entering the cryostat after the purification process.  

\begin{cdrfigure}[A purity monitor from ICARUS T600 ]{IcaPrM}{Picture of a purity monitor from the ICARUS T600, now available for installation in the ProtoDUNE-SP detector
. The photocathode is at the left, the drift region is defined by the ring-shaped electrodes, and the anode is at the right.}
\includegraphics[width=0.5\textwidth]{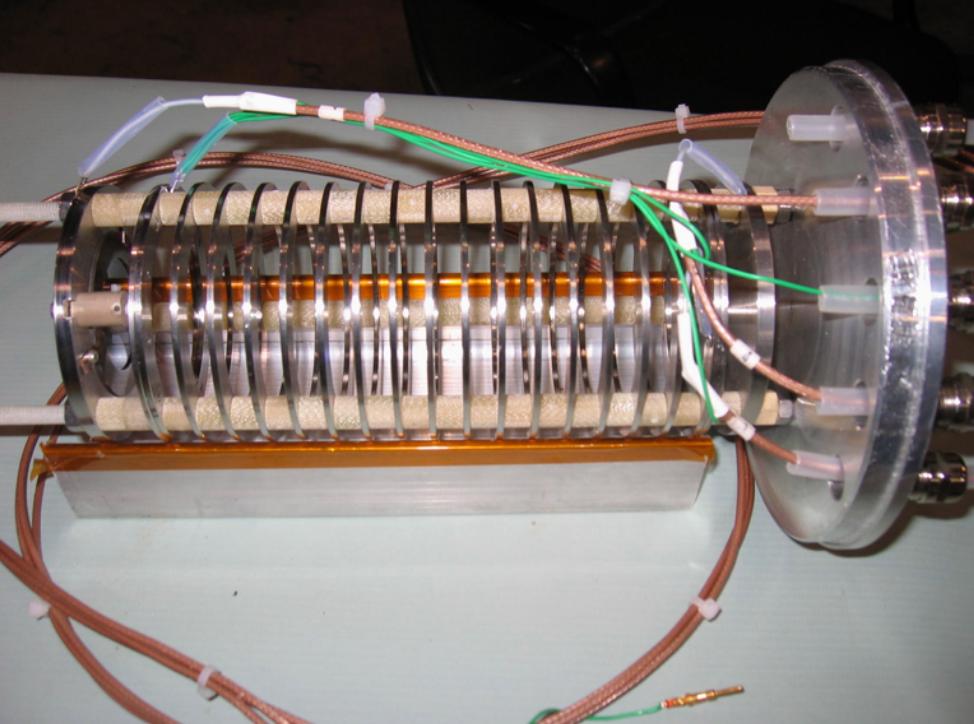} 
\end{cdrfigure}

The three PrMs, one of which is pictured in Figure~\ref{fig:IcaPrM}, are currently being refurbished with new gold photocathodes and new quartz fibers. The drift length  (25~cm) is the same for all them. 
The mechanical  design of the string and the connection to the manhole flange at the top 
are under development.

A single PrM will be installed at the output of the purification filter system using a port that is available for this purpose. This installation will require DUNE to provide not only the purity monitor system but the vessel that will contain the PrM and the argon it is sampling.


\subsubsection{Analytic gas equipment}

A system of gas analyzers will be used to certify the deliveries of argon, to monitor the effectiveness of the purging of the cryostat, and to monitor the state of the argon under operation. Commercial devices with 1 part per billion (ppb) and better sensitivity to oxygen and water, and with 100-ppb sensitivity to nitrogen are available and appropriate. A valve switch-yard is provided to allow the analyzers to sample either the liquid or the ullage in the cryostat, and to sample other points of interest in the argon system.

\subsubsection{Vertical temperature gradient monitor}

	Precise 
	measurements of the temperature, specifically of the temperature gradient, 
	as a function of LAr depth are 
	an important input for fluid dynamics modeling and simulations.  The ProtoDUNE cryostats are the largest argon cryostats ever constructed for a LArTPC and present the best opportunity to validate the models used in the design and simulation of the full DUNE detector modules. The installation of two sets of devices with high-precision temperature measurements
	along almost the entire height of the LAr volume 
	has been included in the internal instrumentation plan for this purpose. The two sets allow testing both the transverse and vertical components of the models and provide an opportunity to test different measurement technologies on a relevant scale. 
	Commercial calibrated resistance temperature detectors (RTDs) with 15-mK precision at LAr temperature are well suited to this application, recognizing that the temperature probe wiring and signal transport outside the cryostat require considerable care in order to maintain 
	the intrinsic precision of the probe.
	
One of  
the vertical profile monitors is supported from a port located behind the APA plane. It consists of a series of RTDs positioned at $\sim$30 to 50~cm intervals 
along a rigid structure. 
A special multi-pin FT is mounted on the flange 
for the signal extraction and readout from a temperature controller.  Cross calibration of the RTDs in place is to be achieved by stepping the structure so that two adjacent RTDs can sample the temperature in the same location.
The mechanical design, including the cable routing up to the multi-pin FT, is under study.
The second profile monitor uses a port located on the downstream end of the cryostat. Studies of cross calibration of the RTDs are under way. 

A number of RTDs will also be positioned at different heights 
on the cryostat walls to monitor the temperature 
during the 
cooldown process. There will also be some RTDs on the cryostat roof,  RTDs on the argon inlet and outlet lines, and, just for luck, an RTD on each of the purity monitors.

\subsubsection{Webcams inside the cryostat}
	Based on 
	a system developed by ETH Zurich for WA105, six commercial webcams, sealed inside a specially developed metal case with a ConFlat optical window to allow operation at cryogenic temperatures, are located inside the cryostat. They are positioned at strategic points to allow inspection of the interior during filling and commissioning, 
	and the detection and recording of possible sparks in locations of high 
	electric field intensity, such as the HV feedthrough. A system of these cameras has been installed in the 35t field cage test at Fermilab and the lessons learned will be incorporated in the final design.
	
\subsubsection{Level meters and pressure sensors.}
	Level Meters and pressure sensors are important for both cryogenics and detector operation. CERN is providing the devices for these measurements, a level meter based on a differential pressure gauge and a redundant pair of sensors to measure the pressure in the ullage. CERN will also provide a weather station to give measurements of the local temperature and atmospheric pressure.

\subsection{Slow control system}
\label{sec:slowcontrol}

The design of the ProtoDUNE-SP safety and control system is largely based on the experience gained in collaboration with ETH Zurich during the pilot WA105 project at CERN. The components of this system and their functions are as follows:
\begin{itemize}
\item	The Process Control System (PCS) reads temperature sensors including the Vertical T Gradient monitors, 
the pressure sensors in the gas ullage, the liquid argon level meter,
the purity monitors inside the cryostat, and the trace analyzers (O$_2$, N$_2$, H$_2$O) in the external recirculation line.
\item	The Detector Control System (DCS) monitors and controls the low-voltage (LV) and high-voltage (HV) from the power supplies.
\item	The Detector Safety System (DSS) performs temperature surveys and monitors interlocks and alarms.
\end{itemize}

The supervisory software is based on the JCOP framework, an integrated set of software tools originally developed for the control of the LHC experiments and 
now used in several more experiments at CERN. The framework provides a graphical user interface to visualize the trends of monitored values and 
alarm/interlock conditions. These values and alarms are automatically stored in a dedicated database for offline use. Remote monitoring is possible via a web 
interface.

The responsibility for the system is split between ProtoDUNE-SP and CERN: 
\begin{itemize}
\item The ProtoDUNE-SP experiment is responsible for all sensors, power distribution, etc., inside the cryostat, 
as well as for defining the system specifications, I/O parameters and control, and safety logic. 
\item CERN EP/DT-DI  is responsible for developing and testing the supervisory control of the system and data acquisition (SCADA).
This includes connecting the control system to the cryogenics instrumentation inside the cryostat and to all the systems that require monitoring and/or control, such as power supplies, cameras and lighting. 
\end{itemize}

In particular, CERN EP/DT-DI is developing a dedicated readout system based on National Instrument modules to allow the 
multiplexing of the RTDs inside the cryostat.  A prototype of this readout system is currently under test.
 
Figure~\ref{fig:dcsdesign} shows the general architecture of the control and safety system for ProtoDUNE-SP, including the PCS, the DCS and the DSS.

\begin{cdrfigure}[Process control, detector control and safety systems]{dcsdesign}{Proposed architecture and technical solution of the detector control and safety system.}
\includegraphics[angle=90, height=0.9\textheight]{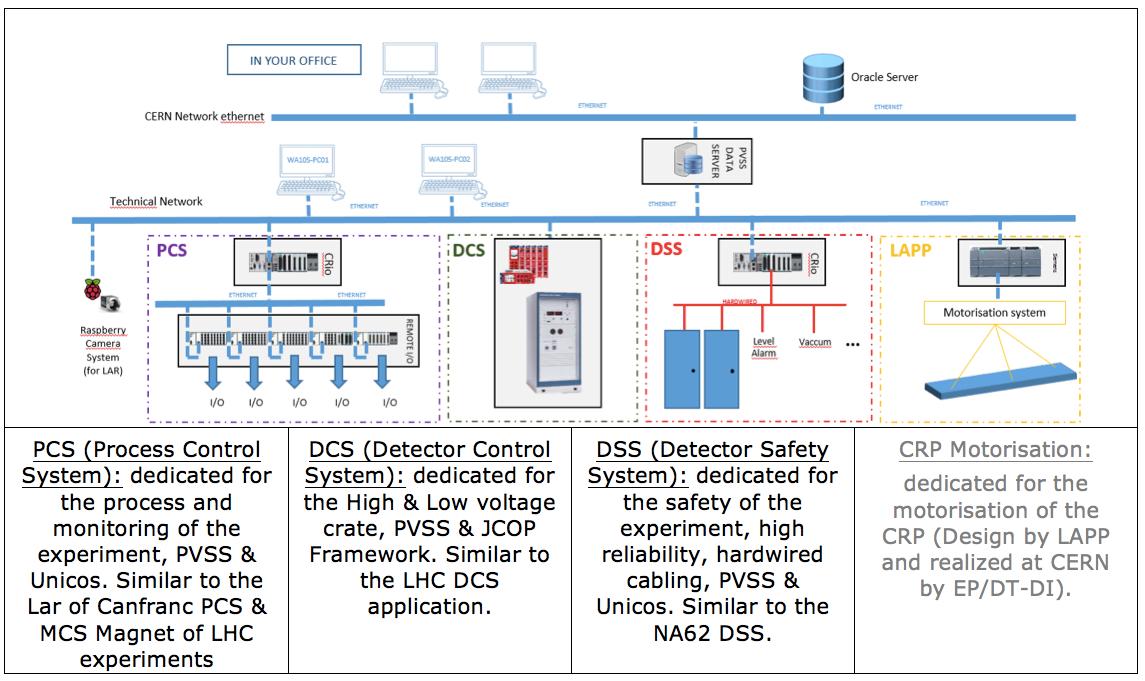}
\end{cdrfigure}

The control system is composed of:
\begin{itemize}
\item a chassis for electrical distribution (380~Vac, 220~Vac, 24~Vdc redundant);
\item two chassis for the PCS, composed of an FPGA, signal conditioners, interface, and cabling;
\item one chassis for the DCS,  composed of an interface for LV/HV monitoring \& control; 
\item a chassis for the DSS, composed of an FPGA and relays for the safety of the experiment; 
\item a chassis for a PC data acquisition \& supervision (PVSS SCADA Supervisor), composed of a computer with a display monitor, a switch and a server; 
\item four chassis for the remote I/O to capture signals close to the detector and to avoid multi-cabling structure; and
\item one chassis for the HV, controlled by the slow-control system. 
\end{itemize}
All these elements will be mounted in 19-in. racks.

\chapter{Space and infrastructure}

\label{ch:spacereq}

ProtoDUNE-SP is to be housed in an extension to the EHN1 hall in the North Area of the Pr\'{e}vessin site at CERN. 
The cryostat is constructed in a pit inside the building, surrounded on three sides by the pit walls.  On the fourth side of the cryostat, an ISO 8 clean room provides a space to construct, test and assemble the TPC and other components. 
A material pass-through structure called a \textit{sas}~\footnote{\textit{Sas} is a French word for a space outfitted with two doors, where one can only be opened if the other is closed; a sas used to pass between two spaces that must remain isolated from each other.} is adjacent to the clean room. Figure~\ref{fig:cryostat-in-ehn1} shows the layout of these structures in EHN1. A naming convention has been established for the four sides of the cryostat, shown in Figure~\ref{fig:cryo-side-names}.  The upper side is \textit{Jura}, the lower is \textit{Sal\`{e}ve}, the left is \textit{Beam}, and right is \textit{Downstream}.



\begin{cdrfigure}[ProtoDUNE-SP cryostat in EHN1]{cryostat-in-ehn1}{Layout of ProtoDUNE-SP cryostat, clean room and material sas in EHN1}
\includegraphics[width=0.8\textwidth]{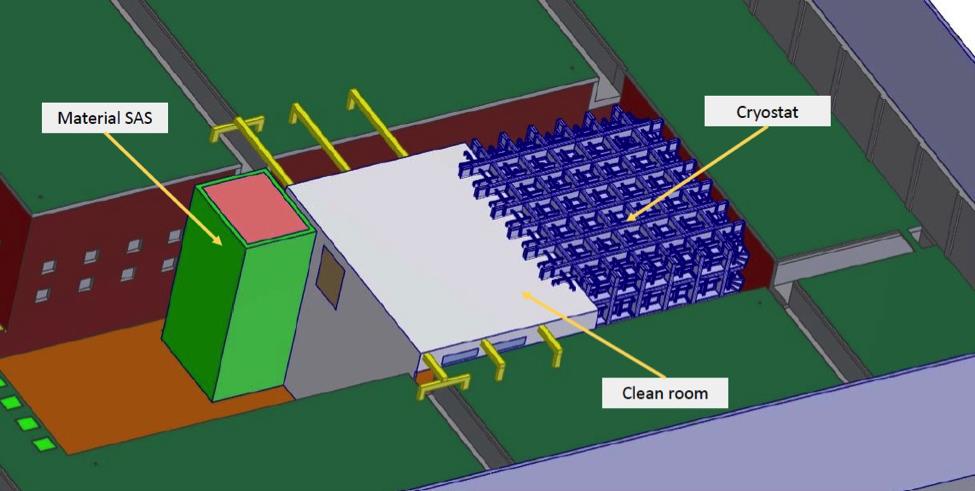}
\end{cdrfigure}

\begin{cdrfigure}[Conventions for labeling the four sides of the cryostat]{cryo-side-names}{Conventions for labeling the four sides of the cryostat}
\includegraphics[width=0.8\textwidth]{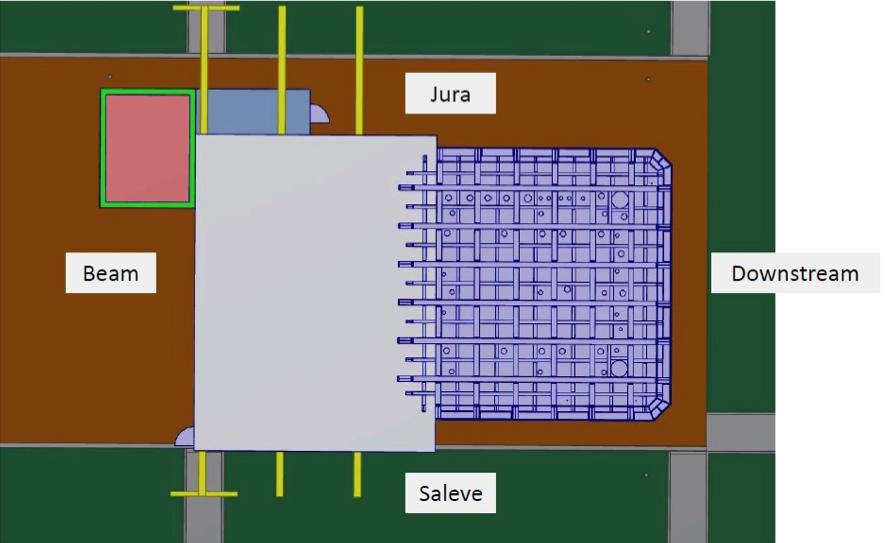}
\end{cdrfigure}

Figure~\ref{fig:elev-view-cryostat} shows an elevation section view of the cryostat indicating the position of the TCO and the location of the integrated cold testing stand (described in Section~\ref{subsec:ce_install}).  

\begin{cdrfigure}[Elevation section view of the cryostat]{elev-view-cryostat}{Elevation section view of the cryostat}
\includegraphics[width=0.8\textwidth]{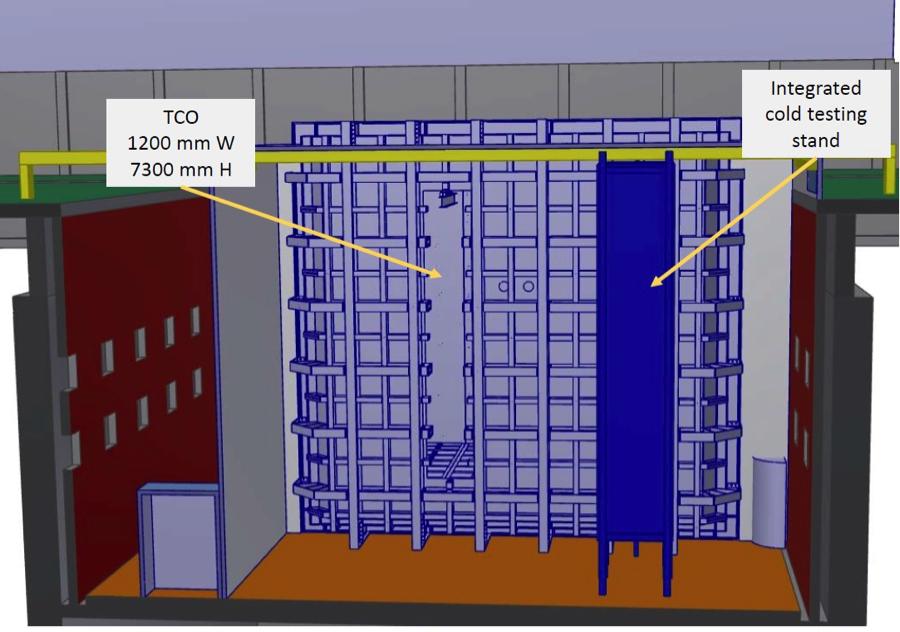}
\end{cdrfigure}

Inside the clean room, a series of rails facilitate the movement of the detector components during the test and installation processes.  The conceptual layout of these rails is shown in Figure~\ref{fig:rails-in-cleanroom}.  The rails are positioned vertically at the same height as the detector support structure (DSS) rails inside the cryostat.  A temporary rail is installed through the TCO to bridge the DSS rails and clean room rails.  All the large components of the cryogenics piping and TPC are supported from these rails on movable trolleys as they are transported to the interior of the cryostat.  
Figure~\ref{fig:rails-in-cleanroom} also shows the approximate dimensions for the sas and the footprint of the clean room space.  These spaces are limited by the pit walls on two sides, and by the supports for the beam and beam instrumentation on the other.

\begin{cdrfigure}[Layout of rails in clean room and dimensions]{rails-in-cleanroom}{Conceptual layout of rails in clean room to facilitate movement of TPC components; approximate dimensions for the material sas and the footprint of the clean room space are shown. (The supports for the beam and the beam instrumentation are not shown.)}
\includegraphics[width=0.9\textwidth]{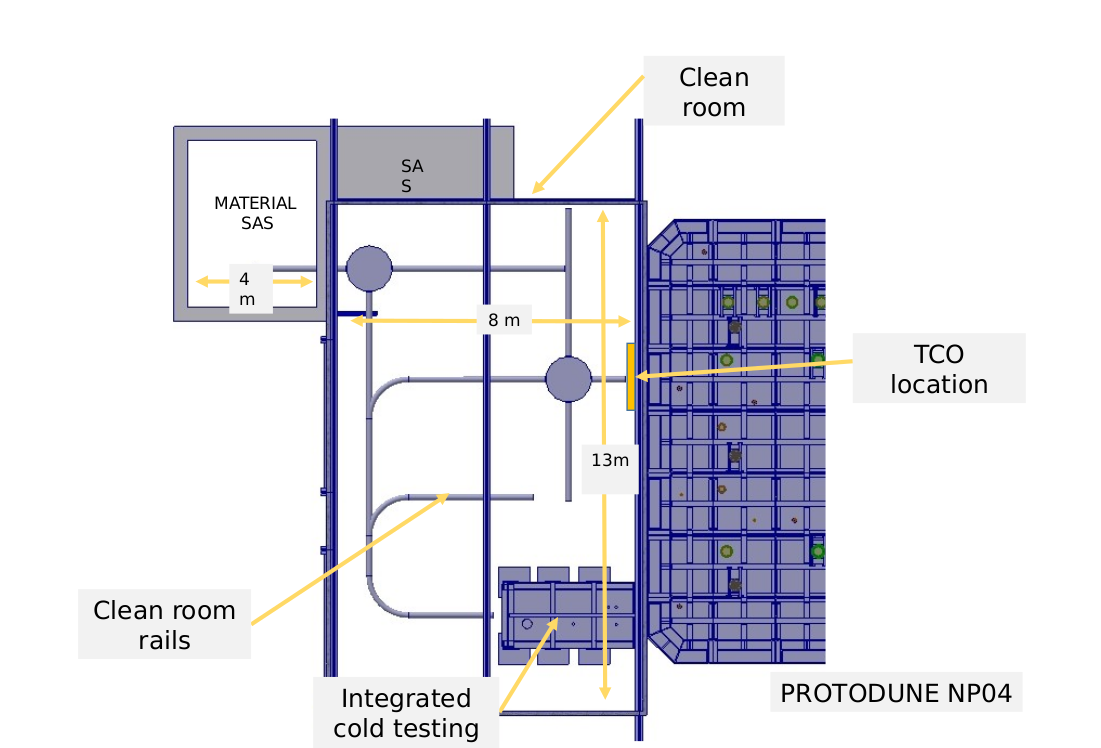}
\end{cdrfigure}

The lighting inside the clean room and any temporary lighting inside the cryostat is filtered to limit the exposure of the PDS components to UV light; wavelengths below 450~nm are filtered out.


\chapter{Detector Installation}


\section{Overview}

As detector materials are brought into EHN1, they are passed into the sas through its removable roof, then transported through a set of large doors from the sas into the clean room, where they are tested and assembled. When ready, each assembled detector component passes through a temporary construction opening (TCO) in the cryostat for installation.
While material is lowered into the sas from the gallery floor, the doors to the clean room remain closed to reduce contamination of the filtered air in the clean room.
Once the roof of the sas is closed, these doors can be opened to move the material into the clean room. 

The activities that will take place in the clean room include:
\begin{itemize}
\item Attachment of FC assemblies to CPA modules;
\item Unpacking and testing of the PDS elements, and installation on the APA frames;
\item Unpacking and testing of the CE elements, and mounting onto the APAs; and  
\item Integrated testing of APA with PDS and CE.  
\end{itemize}
 Figure~\ref{fig:sas-locations} shows the planned locations for all of the activities that will be performed inside the clean room.  

\begin{cdrfigure}[Locations of activities in clean room]{sas-locations}{Locations of activities to be performed in the clean room. This figure requires updating. Note that the PD integration will actually take place in the uppermost location, assigned in the figure to the CPA and FC assembly/integration. CPA integration will instead take place on a rail perpendicular to the others (not shown in the drawing), running close by and parallel to the cryostat front face.}
\includegraphics[width=0.9\textwidth]{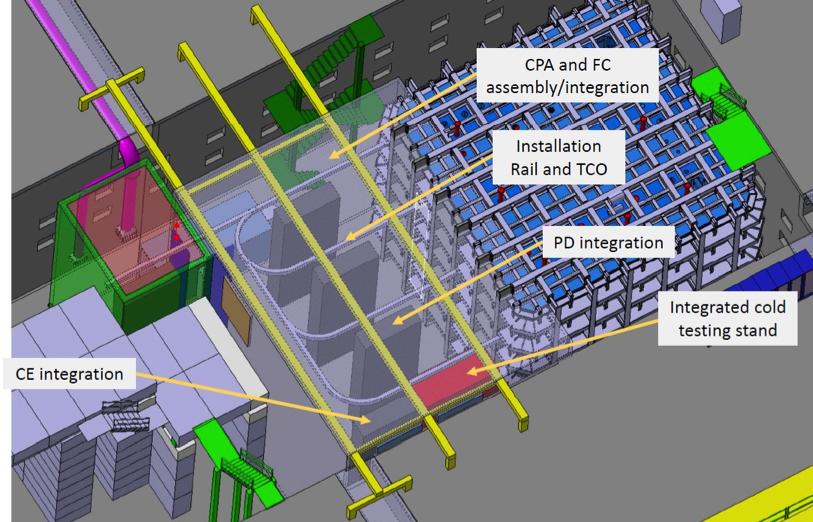}
\end{cdrfigure}

\section{Anode Plane Assemblies (APAs)}

The APAs will be delivered to EHN1 in containers as shipped from the production sites.  These containers will be opened inside EHN1 and special lifting fixtures will be attached to each end of the APA.  The APA will be positioned and attached to two conveyances installed in EHN1.  Both conveyances will be used to lift the APA from the container, oriented as shown in Figure~\ref{fig:apa-tooling} (right), and then rotate it 90$^\circ$ from that orientation, as in the left portion of the figure.

\begin{cdrfigure}[APA with the special tooling attached]{apa-tooling}{The APA with the lifting tooling attached.  The left image shows the orientation of the APA as delivered, the right shows the orientation when it is lowered into the material sas. }
\includegraphics[width=0.7\linewidth]{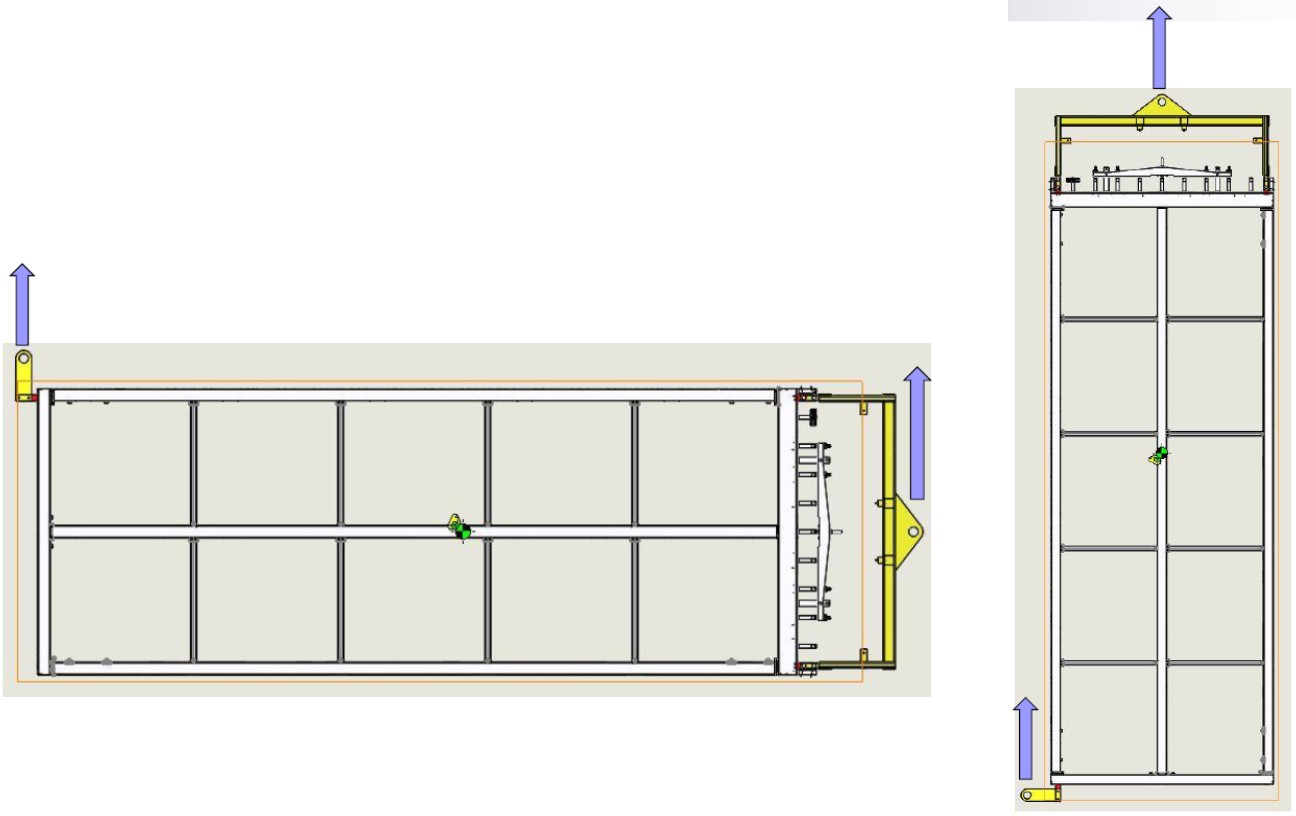}
\end{cdrfigure}

Once the APA is removed from the container and properly oriented, the lifting strap and fixtures will be removed from the lower edge of the APA, the roof hatch on the material sas for 
 the clean room will be opened, and the APA will be lowered through the hatch.  The APA will then be transferred to a rolling trolley attached to a series of rails, and moved into the clean room via these rails.  These spaces and rails are described in more detail in Chapter~\ref{ch:spacereq}. 

Once in the clean room, the APA will go through a series of acceptance tests for both electrical integrity and wire tension.  It will also be inspected for broken wires or any other damage that could have resulted from shipment.  

\section{Photon Detection System (PDS)}

After this testing is complete, the APA is integrated with the PDS.  There are ten PDs per APA, inserted into alternating sides of the APA frame, 
five from each direction.  This is shown in Figure~\ref{fig:pds-install}.  Once a PD is inserted, it is attached mechanically to the APA frame with fasteners, a single electronics cable is attached, and strain is relieved.  Each PD is tested immediately after installation to ensure proper operation and to verify the cable readout.  

\begin{cdrfigure}[PDS installation]{pds-install}{PDS installation}
\includegraphics[width=0.7\linewidth]{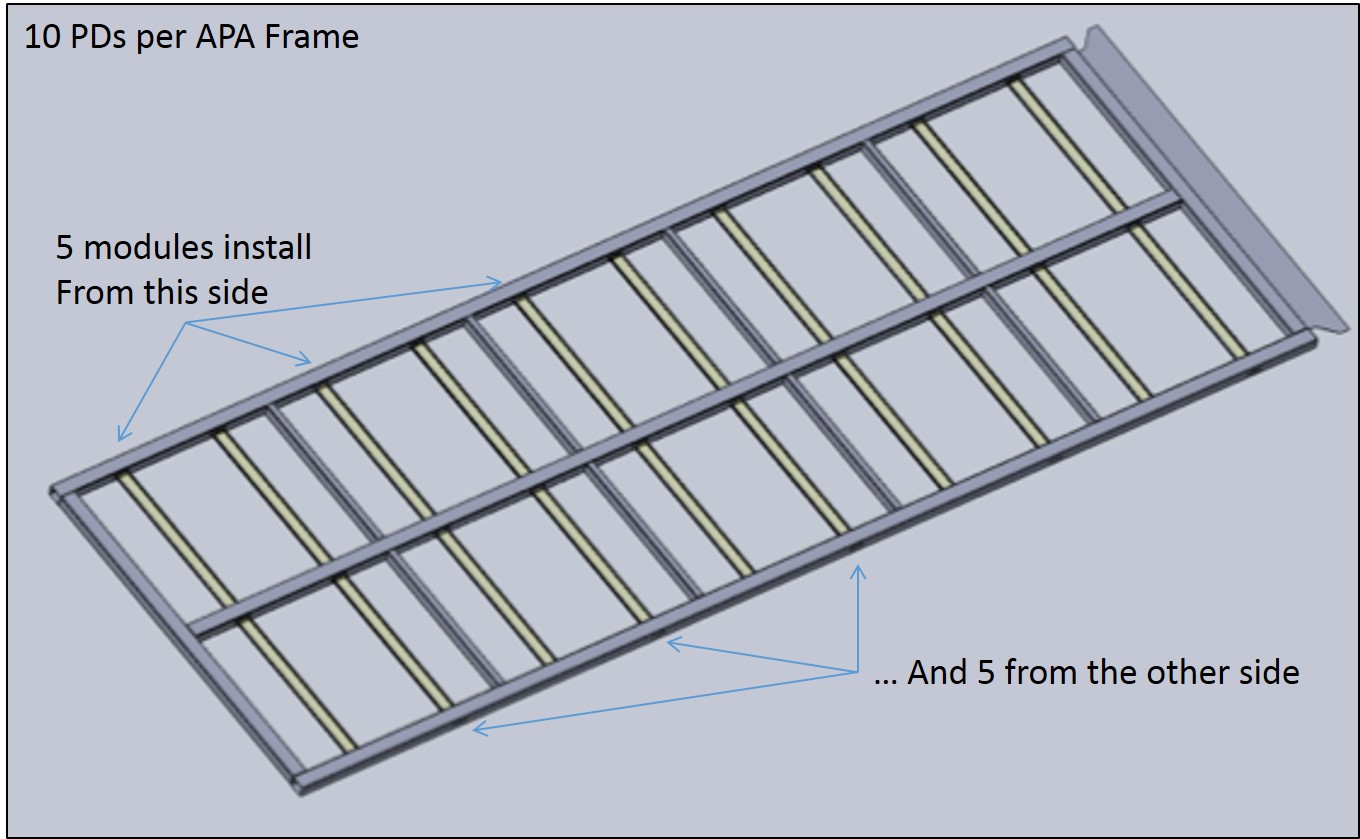}
\end{cdrfigure}

\section{Cold Electronics (CE)}
\label{subsec:ce_install}

Once the PD installation is complete 20 CE units are installed at the top of the APA frame.  Each CE unit consists of an electronics enclosure that contains the TPC read-out electronics inside.  Each unit also includes a bundle of cables that connect the electronics to the outside of the cryostat via the flange on the feedthrough port.  
The location of the CE units on the APA is shown in Figure~\ref{fig:ce-install}.  These units will be connected via matching electrical connectors on the FEMB and the CR board mounted on the APA.  There will also be mechanical fasteners to hold the enclosure to brackets supported by the APA frame.  

\begin{cdrfigure}[CE installation]{ce-install}{CE installation}
\includegraphics[width=0.7\linewidth]{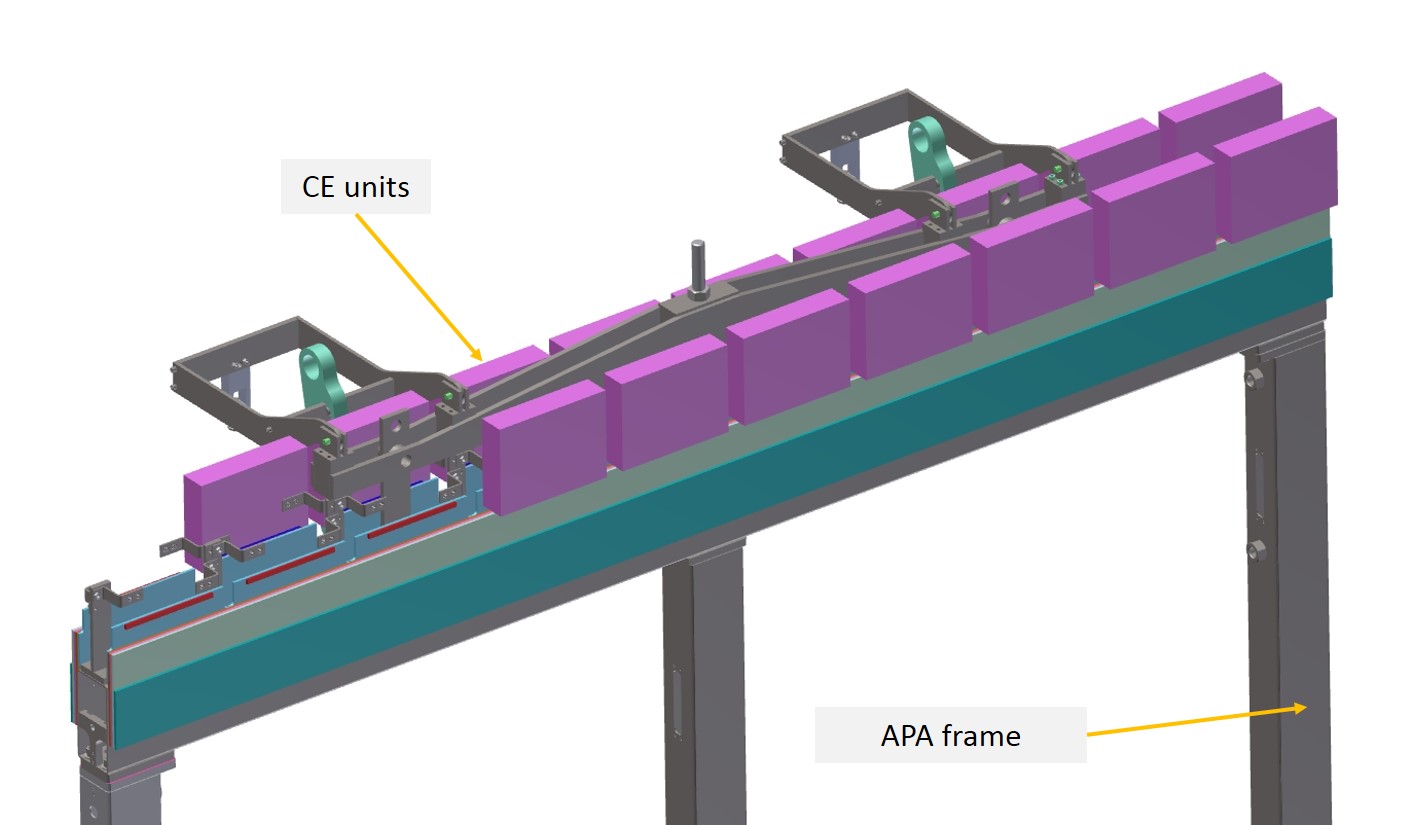}
\end{cdrfigure}

After the APA has been fully integrated with the PDS and CE, it will be moved via the rails in the clean room to the integrated cold test stand.  This test stand, shown in Figure~\ref{fig:cold-test-stand-open}, is a large insulated box that is light-tight for PD testing and has a Faraday shield for CE testing.  At the top of the box is a crossing tube, similar to those in the cryostat, with a ConFlat fitting that accepts the warm-cold interface flange for the PD and CE cable connections.  To prepare for the series of warm electronics tests, the PD and CE cables will be routed and connected to their flanges, the APA will be moved inside the test stand box, and the end cap that completes the Faraday cage will be installed closing the box.  

A first set of tests at room temperature will be performed. Once the warm tests are complete, the inner volume of the box will be purged with dry gas and the volume will be slowly cooled, using cold nitrogen gas, to a temperature of approximately 100 $^\circ$K.  The rate of cooldown must be less than 10 $^\circ$K/hr, the same foreseen for the cryostat cooldown.  The cooldown system is designed to maintain the inner volume near 100 $^\circ$K for approximately 48 hours. A full set of tests at a temperature 
close to operation LAr temperature will be performed for detectors functionality (APA and PD) and electronic noise assessment.  
After the cold test procedure is complete and the detector slowly warmed up back to room temperature, the box is opened, cables are disconnected and secured and the APA is extracted from the box on the rail system in the clean room. 
\begin{cdrfigure}[Cold test stand]{cold-test-stand-open}{A model of the integrated cold test stand in the ProtoDUNE-SP clean room in EHN1.}
\includegraphics[width=0.7\linewidth]{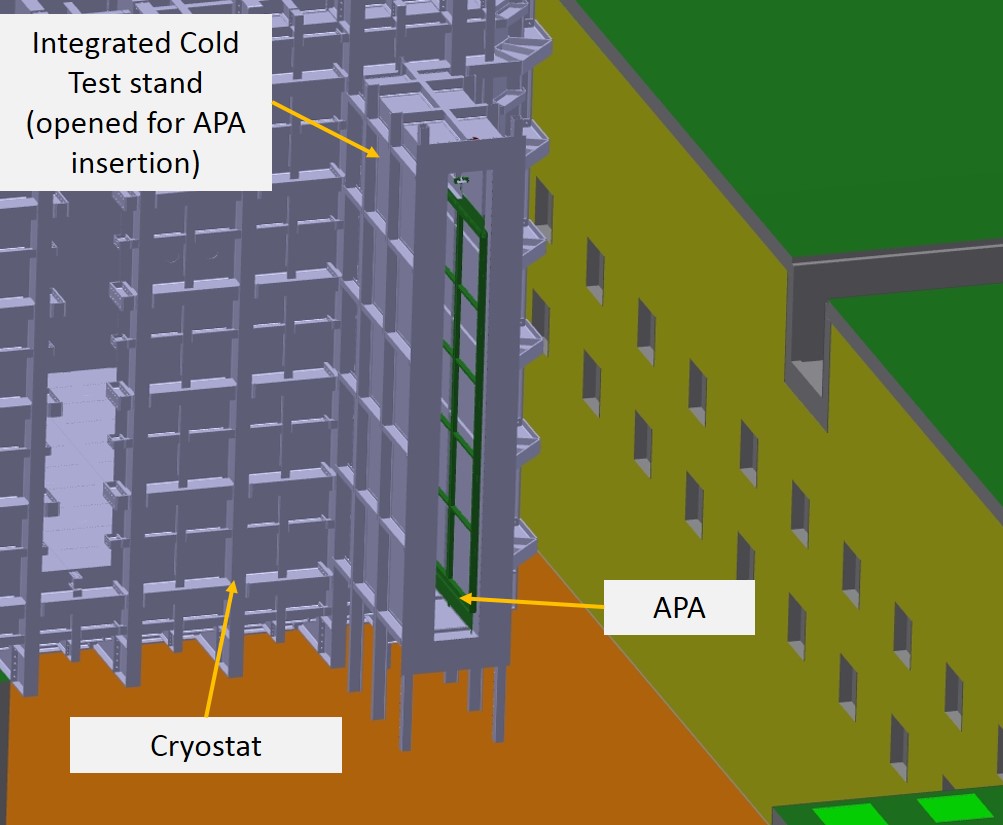}
\end{cdrfigure}
The APA is now ready to be moved into the cryostat through the TCO and transferred onto the appropriate rail in the detector support system.  The two anode planes of the TPC (Saleve side and Jura side) will then be assembled inside the cryostat, each one out three fully tested APAs mechanically linked together. Signal cables from the TPC read-out electronics boards and from the PD modules 
are routed up to the feedthrough flanges on the cryostat top side.
The cables from each of the CE and PDs on the APA are then routed and connected to the final flanges on the cryostat. 


\section{Cathode Plane Assemblies (CPAs)}

Individual CPA modules will be delivered to EHN1 in containers as shipped from the production sites.  Each CPA module weights roughly 24~kg and will be lifted out of the shipping crate by hand. 
%
%
Three CPA modules will be placed on a flat surface and screwed/pinned together to form a CPA column. 
  The crane  
  then will be attached at the top end of the CPA column with appropriate lifting straps and shackles.  The assembled CPA column will be lifted to the vertical position.  
Once the successive CPA column is formed, it is brought together with the previous one within 1~mm along their (vertical) length. 
This alignment is provided by two pins located on the side of the CPA that will fit into a vertical slot on the side of the next CPA. Six CPAs columns locked together will eventually form the cathode plane and moved inside the cryostat through the TCO and positioned parallel to the APA plane at the design drift distance. Refer back to Figures~\ref{fig:cpa-concept} and~\ref{fig:cpa-view2} in Section~\ref{sec:cpa}.

\section{Field Cage (FC)}


Three basic elements comprise the FC: the top, bottom and end-wall FC assemblies.  
The top and bottom FC assemblies are basically mirror assemblies that are hinged from the top and bottom of the CPAs. 
Figure~\ref{fig:fc-assy} (left) shows a top/bottom FC assembly in which the ground plane covers one side of the field shaping profiles.  
The right hand image in the figure shows the CPA pair with top and bottom field cages attached.
These will be attached to both sides for the ProtoDUNE-SP installation.  
\begin{cdrfigure}[Top/bottom FC assembly]{fc-assy}{Left: top or bottom FC assembly (they are symmetrical). Right: side view of a CPA pair with four field cages (two top and two bottom) attached.}
\includegraphics[width=.75\linewidth]{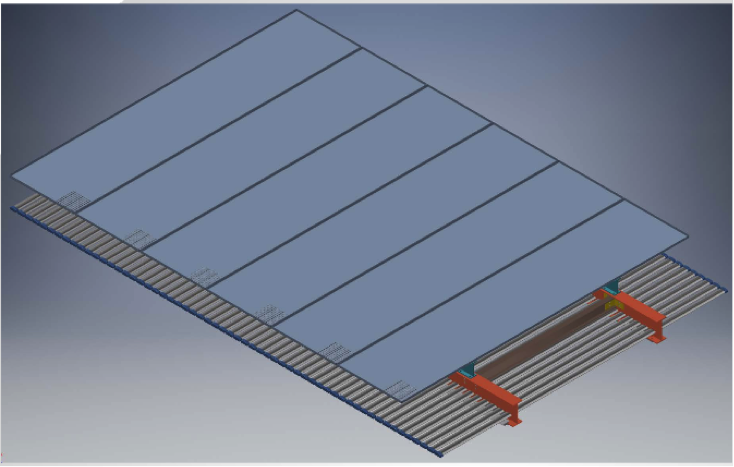}
\includegraphics[width=0.23\linewidth]{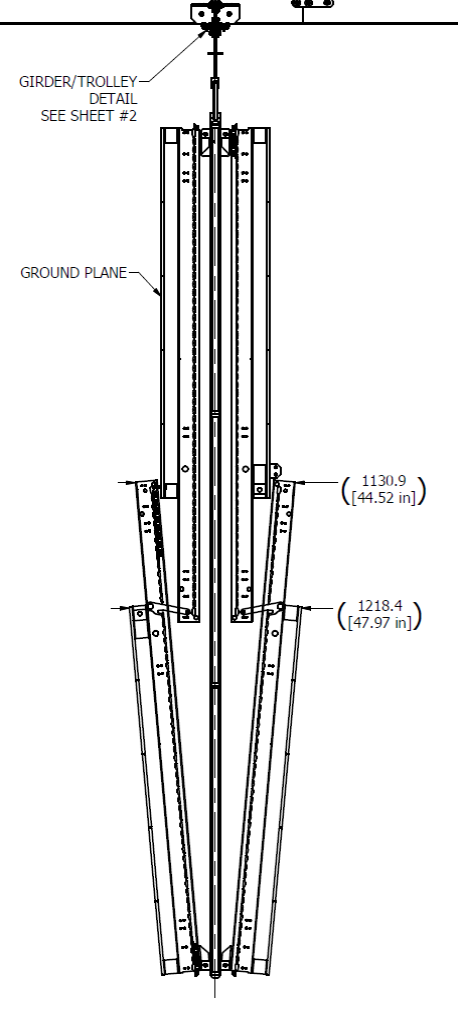}
\end{cdrfigure}

The end-wall FC assembly is constructed from four stacked end-wall FC modules.  Figure~\ref{fig:fc-end-wall-panel} shows one of the end-wall FC modules.  Four of these modules will be stacked and connected together to build the end-wall.  
The stacking will be done by the overhead hoist near the TCO in the clean room.  Once the end wall is complete, it will be moved into the cryostat on the rails in the clean room and positioned on the appropriate beam in the DSS. 
The end-wall is supported by a spreader bar that is in turn supported from the beam. The spreader can swivel about the support point;  this is necessary for positioning the end-wall with respect to the APA and CPA.  

\begin{cdrfigure}[FC end wall panel]{fc-end-wall-panel}{FC end wall panel }
\includegraphics[width=.48\linewidth]{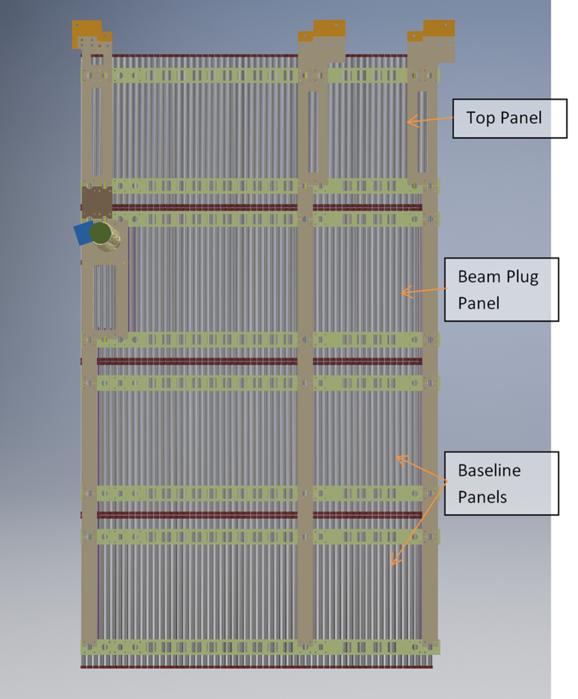}
\end{cdrfigure}

The sequence of installation for the FC components is as follows:
\begin{itemize}
\item After the first row of APAs is installed and translated to the Sal\`{e}ve (south) side of the cryostat, bridge beam A is bolted into place. The two end walls for the Sal\`{e}ve-side drift volume will be constructed and moved inside the cryostat supported 
by bridge beam B. The configuration of bridge beams is shown in Figure~\ref{fig:dss-install2}.
\item As the CPAs are constructed outside the cryostat, both the top and bottom FC assemblies are attached on both sides, top and bottom.  This combination of FC and CPA is then moved into the cryostat and supported by bridge beam C. See Figure~\ref{fig:deploy-fc-saleve-drift}.
This is done three times to get all into position.
\item Once beam C  has been bolted into positon the end walls can be mounted on the end-wall hangers and the spreader bars removed.  The field cages in the Sal\`{e}ve-side drift can be deployed as shown in Figure~\ref{fig:deploy-fc-saleve-drift}.
\item After the second row of APAs is installed and translated to the Jura (north) side of the cryostat, the two end-walls for the Jura-side drift are constructed and moved inside the cryostat supported by bridge beam D. 
\item Once all the TPC components have been moved into the cryostat the TCO is closed.
\item Once the TCO is closed, the end walls in the Jura-side drift are placed into position on the end-wall hangers and the fields cages are deployed.
\end{itemize}
\begin{cdrfigure}[Field cages deployed in Sal\`{e}ve-side drift]{deploy-fc-saleve-drift}{Field cages deployed in the Sal\`{e}ve-side drift}
\includegraphics[width=0.48\linewidth]{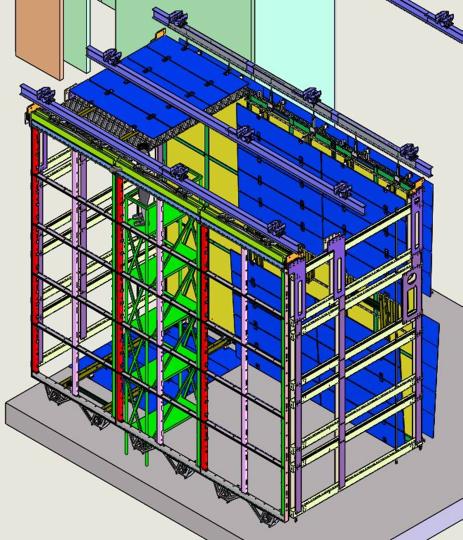}
\end{cdrfigure}

\chapter{Software and Computing}

\section{Overview}

This section outlines the technical design of the offline computing
system and introduces the software that will be used to simulate and reconstruct ProtoDUNE-SP data.

The data rate and volume will be substantial  over the relatively short
run and this drives many design choices for the offline computing
system.  The system provides resources necessary for data
distribution, processing, and analysis on the
grid~\cite{data_managm_sys}.

All raw data will be saved to tape after being transferred to central
CERN and FNAL computing facilities (Section~\ref{sec:raw_concept}).
During this stage, the metadata and storage locations for all raw data files
will be captured in a file catalog system.

A small portion of the data will be immediately processed for  
data quality monitoring (DQM) purposes (see Section~\ref{sec:prompt_processing}).
The processing steps on the full data sample (see
Section~\ref{sec:protodune-offline}) include ADC-level corrections,
potential excess-noise filtering, signal processing, reduction,
calibration, reconstruction, summarizing, and final user analysis.
Multiple passes through this chain will be required for final results,
as calibrations, algorithms, and summary definitions are expected to
evolve.  

Details on the software framework, event simulation and
reconstruction are covered toward the end of this section.

\section{Data storage and management system}

\subsection{Data characteristics}

It is the TPC data that drives the requirements for the raw \pdsp data.
The \textit{\pdsp Data Scenarios} spreadsheet~\cite{data_spreadsheet}  
provides details on these 
numbers and a few alternative running conditions.
Table~\ref{tab:goldi} summarizes the nominal estimates.

\begin{cdrtable}[Raw data storage parameter estimates]{lr}{goldi}{Estimates of nominal raw data parameters driving the design for the raw data storage and management.}   
Parameter & estimate  \\ \toprowrule
    In-spill trigger rate & 25 Hz \\
    Avg. trigger rate & 10 Hz \\
    Channels & 15,360 \\
    Readout time & 5 ms \\
    Compression & $4\times$ \\ \colhline
    Compressed event & 60 MByte  \\
    Instantaneous rate & 1.5 GByte/sec \\
    Average rate & 600 MByte/sec \\
    \colhline
    Total triggers & 52 M \\
    Total volume & 3 PB  \\
\end{cdrtable}

The average trigger rate over the entire beam cycle assumes that one
out-of-spill trigger from the Cosmic Ray Trigger (CRT) system (Section~\ref{sec:beam:muontagger}) 
is acquired
for every in-spill trigger due to the beam.  The assumed compression
factor can be achieved even with similar levels of excess noise as
experienced in the first year of MicroBooNE running~\cite{qian-viren-reduc}.
If no excess noise is experienced, as expected, then a compression factor
of 6 -- 8 is expected.

\subsection{Raw data flow}
\label{sec:raw_concept}

A conceptual diagram of the raw data flow is presented in
Figure~\ref{fig:raw_concept}.  It reflects the central role of the
CERN storage service \textit{EOS} in the raw data management scheme.
Long-term experience has been gained by the LHC experiments, and EOS
has proven to be performant and reliable.  EOS serves as the staging
area from which the data are committed to CASTOR (a hierarchical
storage management system developed at CERN) and from which data are
transmitted to a number of endpoints including principal data centers
such as Fermilab and others.  This scheme mirrors that used by the LHC experiments for their much larger data samples. It is also used to provide input to DQM
and will be available for personal ad-hoc analyses.

\begin{cdrfigure}[Conceptual diagram of the flow of raw data in \pdsp]{raw_concept}{Conceptual diagram of the flow of raw data in \pdsp} 
\includegraphics[width=0.9\textwidth]{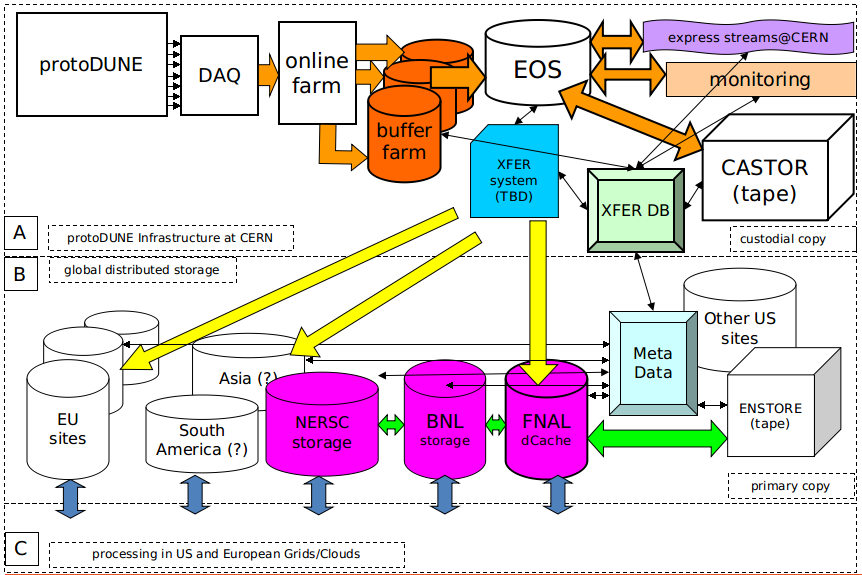}
\end{cdrfigure}

\section{Prompt Processing for Data Quality Monitoring}
\label{sec:prompt_processing}

As described in Section~\ref{sec:daq_online_monitoring}, the first
point at which data quality monitoring occurs is directly inside the
DAQ Online Monitoring (OM).  The DAQ computing cluster hardware is
relatively high performance and has access to the full, high-rate data
stream.  As such it is ideal for monitoring algorithms which require
small amounts of CPU and a large fraction of the data.

On the other end of this spectrum, some monitoring algorithms have
large CPU requirements but produce meaningful feedback on relatively
little data.  Running these algorithms on commodity cluster hardware
is more cost effective.  To manage these jobs, a special purpose system
called the ``protoDUNE prompt processing system'' (p3s) is developed.
Unlike traditional batch systems it limits maximum latency to provide
results at the cost of 100\% data throughput.  The p3s is portable to many 
native batch systems and is fully
  user-configurable.  It can run multiple independent sets of multiple
  interdependent jobs, and run them in parallel -- to the
  extent allowed by available CPU resources and while satisfying dependency information.

The prompt processing is expected to sample about 1\% of the most
immediate data just after it is saved by the DAQ and is available to the
hardware on which it runs.  An initial estimate finds that at least
300 dedicated cores will be required to achieve this.  This estimate
must be refined as a comprehensive list of monitoring algorithms is
developed.

The prompt jobs are developed in the familiar form of offline software
modules to LArSoft (Section~\ref{sec:comp:larsoft}).  The processing is expected to include algorithms
from signal processing through to full reconstruction.  Due to the
sharing of the underlying art framework in both prompt processing jobs and
DAQ OM modules, it will be easy to migrate algorithms between the two
contexts in the case of computer-hardware resource constraints.

\section{Production processing}
\label{sec:protodune-offline}

The second major user of the raw data is the production processing.
It will make several passes of 100\% of the raw data over time as
algorithms improve.  It consists principally of 
event reconstruction, which feeds into user analysis, and 
it may involve a data-reduction step prior to reconstruction.
A data-reduction scheme has been developed~\cite{qian-viren-reduc}; its implementation in the data production
processing chain is contingent on having available resources.

Starting with the signal-ROI there are two basic approaches to
reconstruction, which are described in the following
sections.  The first starts with fitting multiple Gaussian
distributions to the waveforms and the second to retaining their
binned structure.

\section{The LArSoft framework for simulation and reconstruction} 
\label{sec:comp:larsoft}

LArSoft~\cite{larsoft-web} is a suite of tools for simulating and
reconstructing data collected from LArTPC detectors.  It is
built on the \textit{art}~\cite{art-web} event-processing framework.  The
main features of the \textit{art} framework are 
its configurability by
human-readable and editable control files (that use the Fermilab
Hierarchical Control Language (\textit{FHiCL})), and
the scheduling of 
program module execution. The modules are of five types: event sources, filters,
data-product producers, analyzers, and output.  
Common utilities that
can be accessed by any program module at any time are called \textit{services}.

The \textit{art} framework defines the input/output structure of
ROOT-formatted files using TTrees to store the data, metadata, and
provenance information.  The provenance information consists of the
contents of the FHiCL documents used to steer the processing of the
job that created the output file, and those of input files and parents.  

The \textit{ art} framework's division of the simulation and reconstruction
jobs into modular pieces allows multiple developers to contribute to an
effort, and to test their ideas in isolation before integrating them
into a larger system.  Because the data read in from an event is
placed in read-only memory, analyzers can program with confidence that
upstream algorithms cannot alter the data, but must produce additional
data products which can later be processed or written out.

The LArSoft suite provides the interface to the event generators and
Geant4~\cite{Agostinelli:2002hh} 
for simulation of the passage of particles
through the detector, the details of which are described in
Section~\ref{sec:larsoftsim}, and event reconstruction, the details of
which are presented in Section~\ref{sec:larsoftreco}.  

The \textit{art}
framework and LArSoft source code are publicly available, and pre-built
versions are provided~\cite{scisoft-web} for supported versions of Linux and Mac
OS~X.  Tools for compiling the framework and
applications are also provided, along with all of the required
dependencies, including the gnu C++ compiler.  The versions of the
software and its dependencies are managed by the UNIX Product Support (UPS) system,
which allows easy version selection, setup and configuration of the LArSoft
environment on computers with previous versions already installed, such as
those at Fermilab.  

LArSoft is under rapid development by both the
core LArSoft team and by contributors from participating experiments:
ArgoNeuT, LArIAT, MicroBooNE, SBND, and DUNE.  Within DUNE, the LArTPC
near-detector option, both ProtoDUNE detectors, and the Far Detector
are clients and contributors of LArSoft.

\section{Event simulation}
\label{sec:larsoftsim}

Three main sources of particles are simulated in protoDUNE-SP: beam particles coming through the beamline, and beam halo particles from both our own line (H4) and from upstream sources and cosmic ray particles. Each particle type has its dedicated generating process and the mixing of the right relative and absolute amounts in the relevant time frames is of crucial importance.
Beam particles are generated using the G4Beamline event generator which is a standard tool used at CERN for simulation of beamlines.

The upstream beam instrumentation devices -- wire chambers, Cherenkov
counters, and time-of-flight-counters -- are fully simulated as described in
Section~\ref{sec:beaminstruments}. Cosmic-ray
events are simulated either with the CRY~\cite{cry} event generator or
CORSIKA~\cite{Heck:1998vt}.  Neutrino scattering events are simulated
using GENIE~\cite{Andreopoulos:2009rq}; while these 
events will be very
rare in the detector, the extrapolation of the performance of
the \pdsp
detector to the FD will require simulating them. 
A dedicated generator in LArSoft simulates
radionuclide decay products which can be overlaid on other events.

The detector geometry is coded in GDML files~\cite{Agostinelli:2002hh} that are
generated by the GeGeDe~\cite{gegede} 
geometry system.  These files
contain the locations, sizes, shapes, and material content of the
detector components, the active liquid argon volume, and the
surrounding materials, such as the field cage, the beam windows, the
cryostat the supporting structure, and the experimental hall.  These
external features will impact the distributions of cosmic-ray
particles impinging on the active detector.  The channels and volumes
are numbered and named in the GDML files, with conventions followed by
the LArSoft simulation code.

The active volume of the detector is divided into cubes 300~$\mu$m on
a side, called \textit{voxels}.  GEANT4 tracks particles through the argon with
each step ending on a voxel boundary. This allows the simulation of small-scale
physics processes, such as delta-ray emission and showering, at a level
of detail smaller than the intrinsic resolution of the detector.   While GEANT4
calculates the energy deposited by each particle for each step, the
simulation of ionization and scintillation-photon emission 
is performed using one of two algorithms in LArSoft:  a dedicated parameterization that
depends on the electric field in the liquid argon and the ionization
density~\cite{Birks:1964zz}, or  NEST~\cite{Szydagis:2011tk}, which is tuned to
previous noble-liquid experimental results and introduces an
anti-correlation between the photon yield and the ionization electron
yield for each step.  

The availability of alternate interaction models in MC simulation may be helpful for estimating systematic uncertainty.
FLUKA~\cite{Fluka15, Ferrari:2005zk, Battistoni:2009zzb} is a MC package with detailed modeling of particles traversing the detector. The option of creating an interface to LArSoft will be considered if resources are available. 


The average specific energy loss for a minimum-ionizing particle (MIP)
is approximately 2.12\,MeV/cm.  The $W$-value for ionization is 23.6~eV
per electron-ion pair, and the $W$-value for scintillation is 19.5\,eV
per photon, resulting in tens of thousands of drifting electrons and
photons per cm of charged-particle track in the detector.  It is
impractical to simulate the paths of each of these electrons and 
photons using GEANT4, and therefore computational techniques are incorporated
into LArSoft to achieve a high simulation speed while preserving
accuracy.  The electrons are propagated by LArSoft-specific tools,
including a tool that integrates over the distributions created by longitudinal
and transverse diffusion, 
and a geometry tool that indicates the wire locations that record charge, assuming uniform wire spacing.

The effect of charge loss due to attachment
of electrons to impurities (i.e., the effect of the electron lifetime),  is
implemented in this step.  
Effects due to space charge are simulated using a smoothly parameterized
map of
distortions in the apparent position components $(x,y,z)$ of the charge
deposits as
functions of the true position.
This map can be made using SPaCE~\cite{Mooney:2015kke}, a
program that traces particle trajectories in LAr based on the
electric field calculated using Poisson's equation, a given
space-charge density map, and the boundary conditions provided by the
cathodes, anodes, and field cages.  It is anticipated that the
space-charge distortions in ProtoDUNE-SP may be as large as
20\,cm~\cite{Mooney:2015kke}.

Photon propagation is simulated using a library that contains the
probabilities of observing a photon emitted at a particular point in
space by a particular photon detector.  Here, the space is divided
into cubical voxels 6\,cm on a side, and the library is indexed by photon
detector element. 

The simulated arrival times and charge amounts on each wire, and of
each set of photons arriving at each photon detector, along with the
identity of the particles generating them, are stored in the
simulation output file for use in determining the performance of the
downstream reconstruction algorithms.  These charge depositions and
photons are inputs to the detector response functions -- the field
response and the electronics response are convoluted with the true
arrival times to make simulated waveforms.  The detector field
response functions are simulated using GARFIELD~\cite{garfield}, but
they will be validated with real data, as the simulation contains
oversimplifications, such as inadequate modeling of induction signals.  

The electronics gain is applied so that the
simulated signals match the expected responses.  Simulated noise is
then added, and the result is quantized to reproduce the behavior of a
12-bit ADC, including realistic pedestals and saturation.  A similar
process is followed to simulate the response of the photon detectors,
given the arrival times of the photons.  Functionality exists within
LArSoft to overlay MC-simulated particles with raw digits in
the data in order to simulate pileup of cosmics and other beam
interaction particles. The simulated raw digits are then written to
compressed ROOT files for further analysis.

\section{Event reconstruction algorithms and performance}
\label{sec:larsoftreco}

\subsection{Reconstruction}

The interpretation of the data from LArTPC detectors has
proven challenging, largely due to the wealth of information provided
in each event by the detector, but also due to the high rate of
multiple scattering and particle interactions, as well as the
projection of 3D information onto a discretized
2D space of readout ADC counts on wires as functions of
time.  The flexibility of the {\textit{art}}/LArSoft framework allows for
multiple approaches for reconstructing and analyzing the data, 
and for the use of different approaches 
depending on the targeted physics deliverable.

For current large LArTPC detectors, noise filtering is applied to improve
the signal-to-noise ratio.  A
large component of the noise in existing LArTPC experiments is from coherent sources -- sources that
affect many neighboring wires and/or neighboring readout channels
(channels from different planes may be interleaved in the front-end
electronics).  
The contribution from coherent noise to a measured ADC value on a channel can be estimated from the data on nearby channels in the same time slice, and subtracted out.
However, this procedure reduces the signal as
well as the noise, in a manner that depends on the angle of the track or
shower with respect to the drift field.  

Procedures that first
identify signal hits and protect them from distortion 
are under study. With software noise filtering, MicroBooNE~\cite{Acciarri:2017sde}
has achieved excellent noise levels, consistent with expectations, based on 
the design specification of the cold electronics. 
  In MicroBooNE, various 
sources of noise have been identified and hardware upgrades 
are ongoing to eliminate them.   Once noise has been removed,
signals will be processed to recover the ionization charge.

Hits are identified by seeking deconvoluted signals exceeding
thresholds that are adjusted to minimize the creation of false noise
hits while preserving the true signal hits.  The standard LArSoft hit
finder fits Gaussian functions to the deconvoluted signals, and saves the
times, widths, and amplitudes of the Gaussians.  In addition, it saves the sum of
the ADC readings in the time windows corresponding to the hits; since a
Gaussian function is not always representative of the charge arrival
distribution, the resolution of the calorimetry is improved by
summing the ADC counts.

The hits are associated with DAQ channels 
not wire segments, since, due to the wrapping of the induction-plane
wires in the APAs, the location of the 
charge deposition contributing to the hit is ambiguous. 
Because the wire angle
is chosen such that each induction wire intersects each collection-plane
wire at most one time, only two views are needed in order to identify
hits and resolve ambiguities.  A separate LArSoft module compares the
hits in the collection and induction views to
resolve ambiguities.  

Physics analyses are most sensitive with a full 3D reconstruction of the
event -- the primary vertex (if there is one), the tracks, and the showers are 3D objects.
ProtoDUNE-SP is implementing several approaches to achieve an optimal 
3D reconstruction. Once hits are identified on wire
segments, 2D reconstruction identifies clusters and tracks
in each view separately, and 3D hypotheses for
the event are constructed by comparing the 2D clusters
in the separate planes.  The 2D clustering algorithms currently in
use are the Blurred Clustering Algorithm~\cite{blurredcluster},
LineCluster~\cite{linecluster}, and TrajCluster~\cite{trajcluster}.  

\subsection{Performance}

Algorithm performance metrics are efficiency, purity, and completeness.  The
efficiency of the algorithm is defined as the fraction of true particles that
match reconstructed objects within the bounds of pre-specified
criteria, such as matching position and length and the type of object
expected.  The purity of the reconstructed object is defined as the fraction of
hits (or charge) included in that object that truly came from the
matched particle, divided by the total number of hits (or charge)
included in the reconstructed cluster.  The completeness is defined as
the number of true hits that are found in a cluster, track or shower
expressed as a fraction of total true hits in that object.  

Particle-level and event-level performance metrics include particle identification
and misidentification rates, shower energy resolution, and energy scale offsets.

\subsubsection{Electromagnetic and track-like components separation}


Using current reconstruction algorithms, it is challenging to distinguish
between electromagnetic (EM) and track-like components of an event. Since the properties of
EM-like and track-like objects  (and the classes of objects extracted from them) are significantly different,  dedicated
reconstruction algorithms should be applied to each of them. A recently developed approach,
based on a convolutional neural network (CNN), promises good performance and may be used for higher-level reconstruction in ProtoDUNE.  


The CNN analyzes each point separately in the 2D image of data from a LArTPC to see if it can be classified as a candidate for a track traversal or an EM shower.  At each point in the
image, the deconvoluted ADC value is analyzed, and nearby points are also input to the CNN, in order to provide context.
Distant points are not included, which keeps the classification relatively insensitive to the topology.  This classification
is performed before downstream algorithms, which identify hits, clusters, tracks and vertices, are applied.


The current implementation combines single-point classification with standard
clustering algorithms in order to improve the classification of entire objects
recognized in the event. The CNN-based EM component selection was also integrated with the Projection
Matching Algorithm (PMA, see Section~\ref{sec:comp:reco:pma}), allowing for the application of
the tracking and vertexing algorithm to the track-like component for which the PMA
algorithm was designed. Further integration with other reconstruction algorithms is
ongoing.

The performance of the CNN-based selection is illustrated in Figure~\ref{fig:cnnemseloneevent}
where the input, a 2D projection of the event, is compared with the result of classification
in the single event.  The performance of the EM selection algorithm is summarized in Figure~\ref{fig:cnnemselinsamples}  using the
sum of the deconvoluted ADC values for points selected as EM-like candidates, compared with 
MC truth as a figure of merit, for charged pion events with two values of test beam momentum,
1\,GeV$/c$ and 4\,GeV$/c$.

\begin{cdrfigure}[CNN-based EM selection in pion event]{cnnemseloneevent}{CNN-based
selection of the electromagnetic component of events: deconvoluted ADC waveforms used
as the input to the algorithm (left); selected electromagnetic (blue) and
track-like (red) activity (right). Positive pion in the ProtoDUNE test beam at
2.5~GeV$/c$ is shown (the particle track incoming from the left). All parts are
correctly recognized in this event.}
\includegraphics[width=0.45\textwidth]{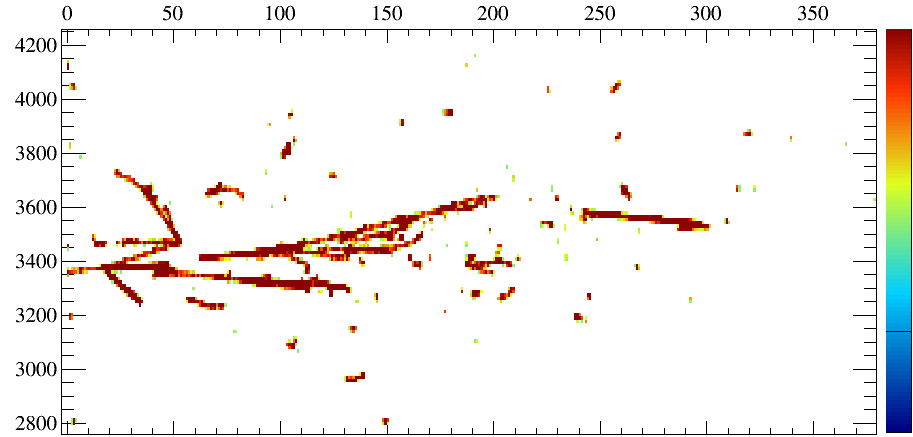}
\includegraphics[width=0.45\textwidth]{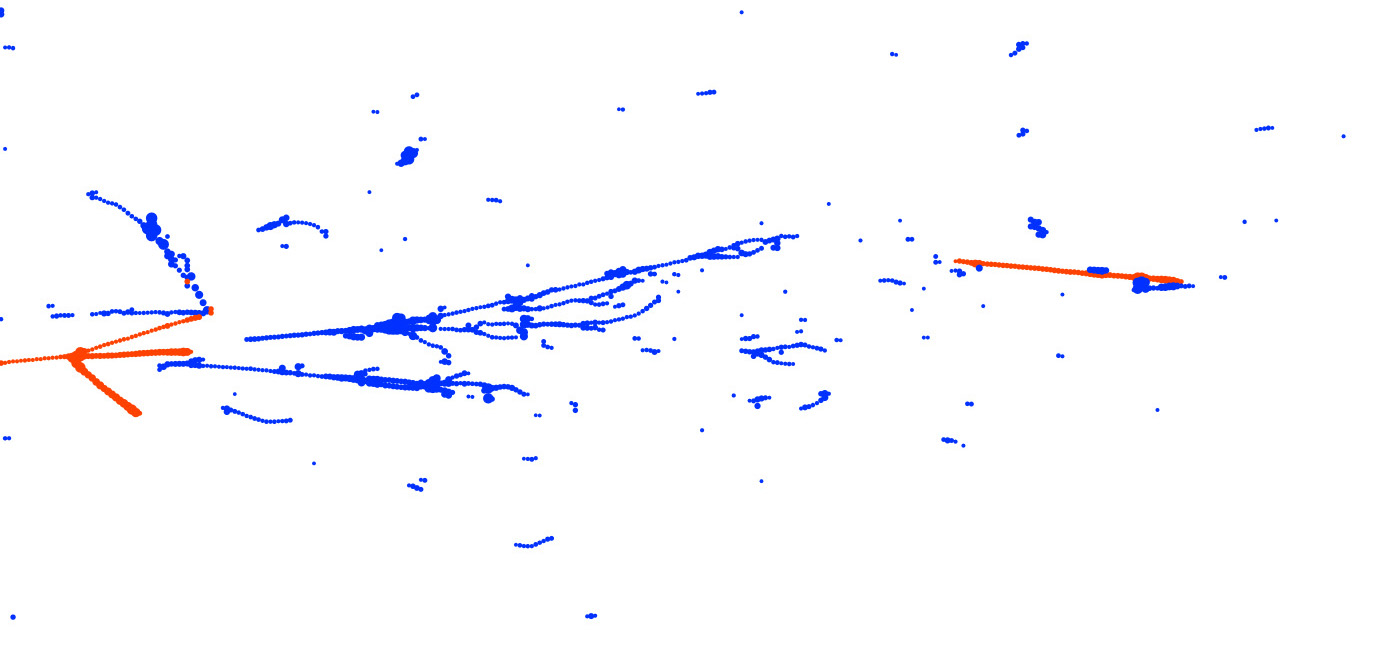}
\end{cdrfigure}

\begin{cdrfigure}[CNN-based EM selection in pion samples]{cnnemselinsamples}{Summed hit 
ADC areas for hits selected by the CNN as originating from electromagnetic activity
versus the summed hit ADC areas for hits selected with MC truth information as originating 
from electron tracks.
This comparison is shown for simulated ProtoDUNE-SP events with a $\pi^+$ with a momentum of
1~GeV/$c$ (left) and 4~GeV/$c$ (right).}
\includegraphics[width=0.45\textwidth]{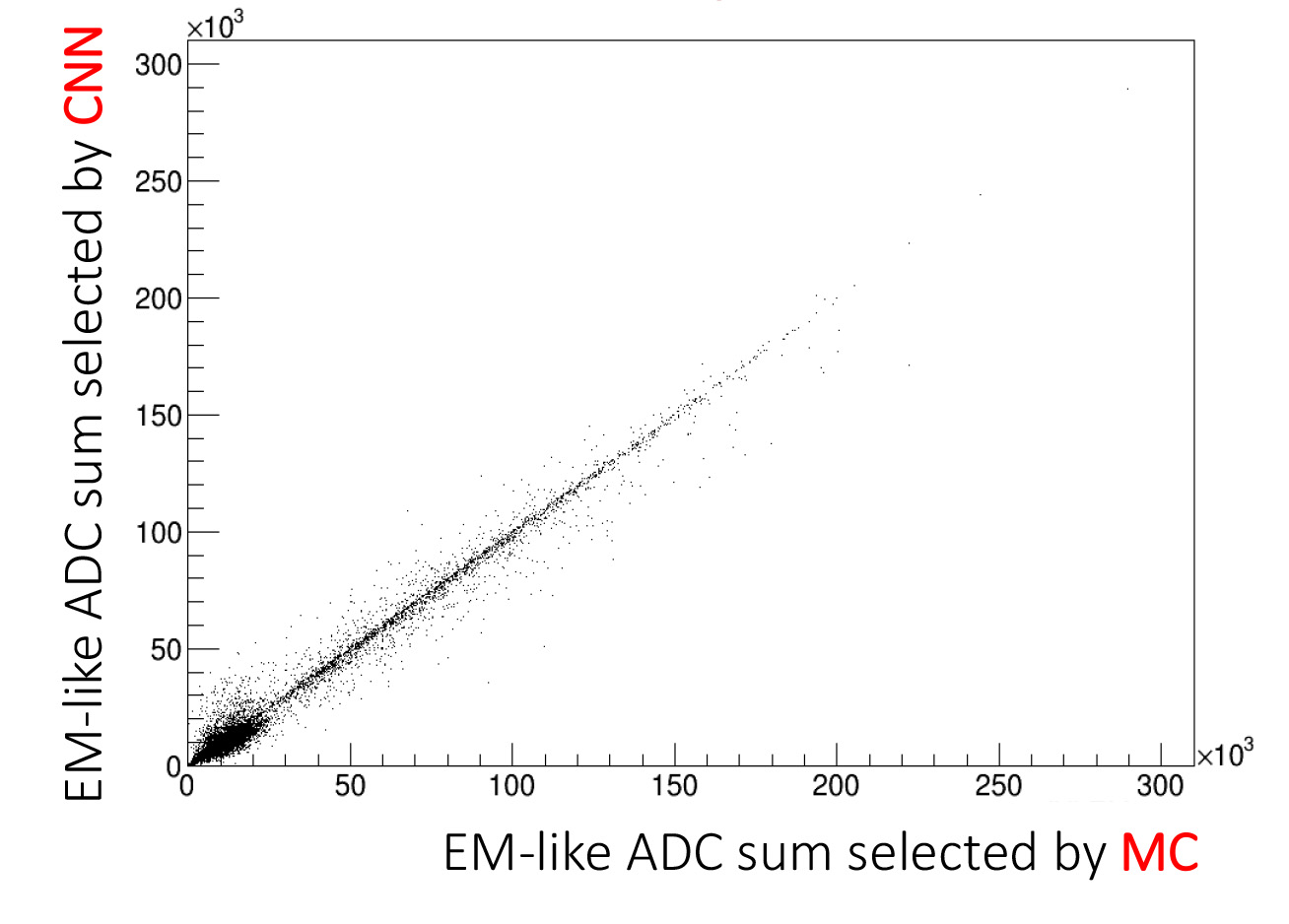}
\includegraphics[width=0.45\textwidth]{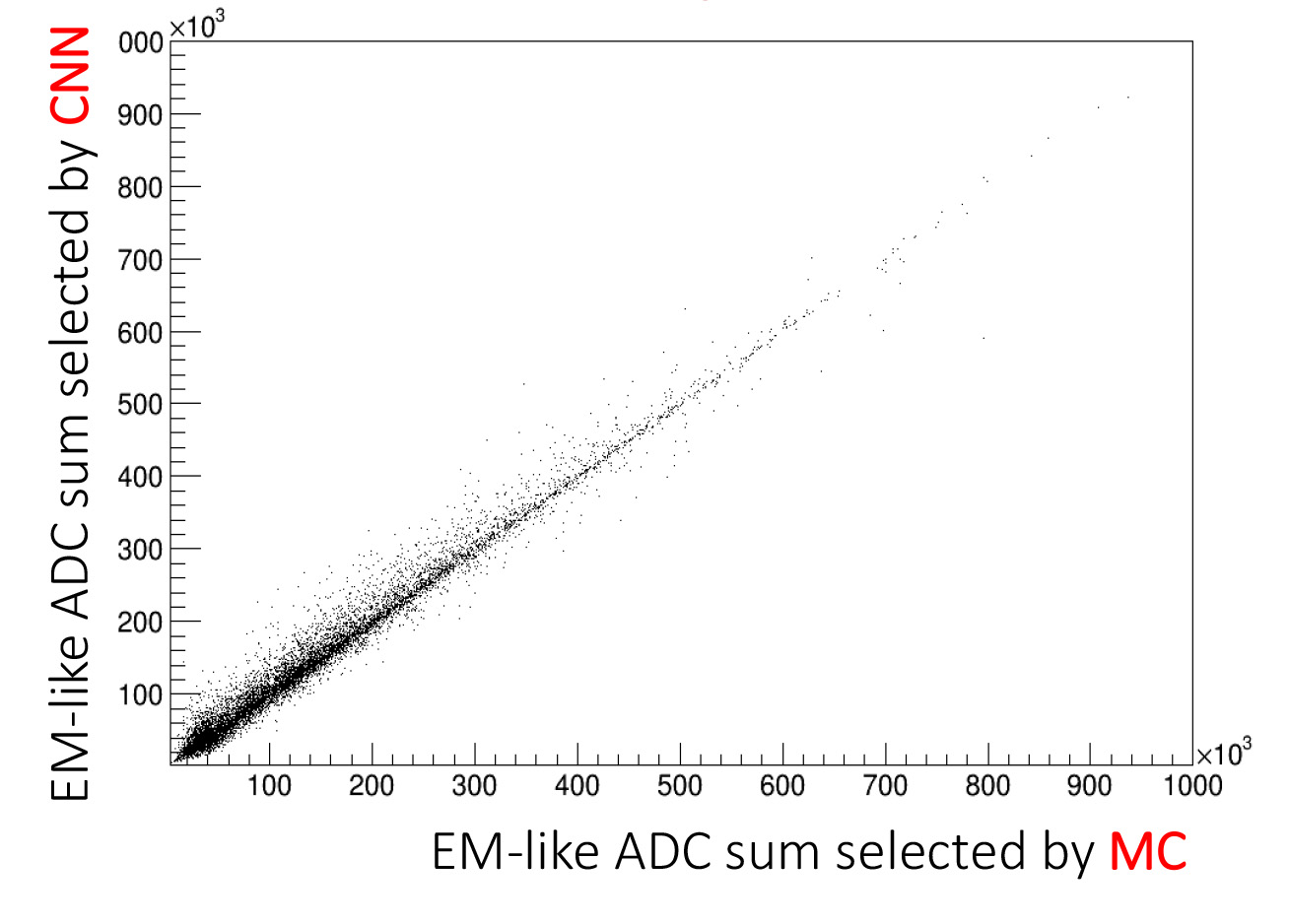}
\end{cdrfigure}

\subsubsection{EMShower}

The EMShower package~\cite{emshowerpackage} takes the output of the
Blurred Clustering Algorithm and
produces energies, angles, and start positions for 3D showers, as
well as the $dE/dx$ in the initial part of the shower.  Identifying
events with two showers consistent with $\pi^0\rightarrow\gamma\gamma$
decays allows for an \textit{in situ} calibration of the electromagnetic
energy scale as well as the performance of shower identification and
reconstruction for photons that are produced inside the detector.  A
distribution of reconstructed $\pi^0$ masses in Monte Carlo is shown
in Figure~\ref{fig:pizeromass}.

\begin{cdrfigure}[Reconstructed invariant masses of $\pi^0$ candidates in
  MC]{pizeromass}{The reconstructed invariant masses of $\pi^0$ candidates in
  Monte Carlo using the BlurredCluster and EMShower algorithms.}
\includegraphics[width=0.8\textwidth]{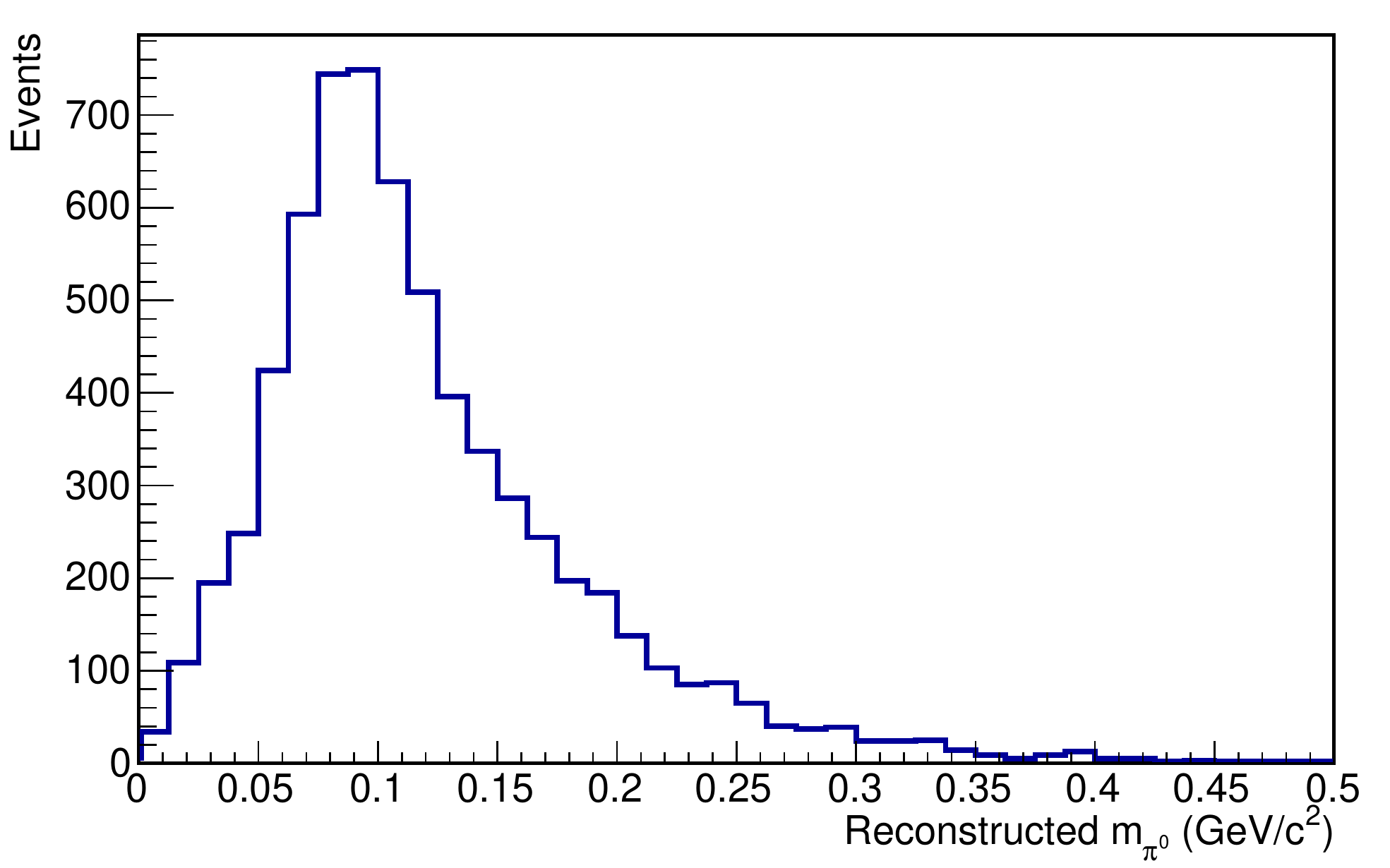}
\end{cdrfigure}

\subsubsection{PANDORA}

The reconstruction framework PANDORA~\cite{Marshall:2015rfa} also works by
building up a 3D picture from 2D
reconstructed objects.  PANDORA is a flexible framework developed for
International Linear Collider (ILC) detector simulation, and provides a convenient way to develop
algorithms for reconstructing particles.  In all, more than 80
algorithms, each targeting a specific topology, have been incorporated
into PANDORA to date.  PANDORA allows multiple reconstruction passes through the data.  
Different criteria for clustering hits into tracks and
showers may be applied when seeking to remove cosmic rays 
than when
identifying signal events.  PANDORA follows a process that clusters hits in 2D,
reconstructs vertices in 3D, reconstructs tracks in 3D,
reconstructs showers in 3D, performs a \textit{mop-up} step in 2D and 3D, and finally performs
full event-building in 3D.

Plots of the efficiency, the completeness, and the  difference between the true and reconstructed
track lengths for single muons with momentum between 300\,MeV and 5\,GeV in the~\pdsp geometry are
shown in Figure~\ref{fig:muonpandoraperf}.  The PANDORA algorithm performs very well, although a small
inefficiency occurs in the current algorithm's ability to match track segments from one APA's drift volume to another.

\begin{cdrfigure}[PANDORA reconstruction, single muons, 300~MeV$/c$ < $p$ < 5~GeV/$c$]{muonpandoraperf}{Performance of the PANDORA reconstruction algorithm for single muons with 
momentum between 300~MeV$/c$ and 5~GeV/$c$.  The top left-hand figure shows the tracking efficiency as a function of
muon momentum, the top right-hand figure shows the distribution of tracking completeness, and the bottom figure shows the
distribution of the difference between the reconstructed muon track length and the true track length.}
\includegraphics[width=0.45\textwidth]{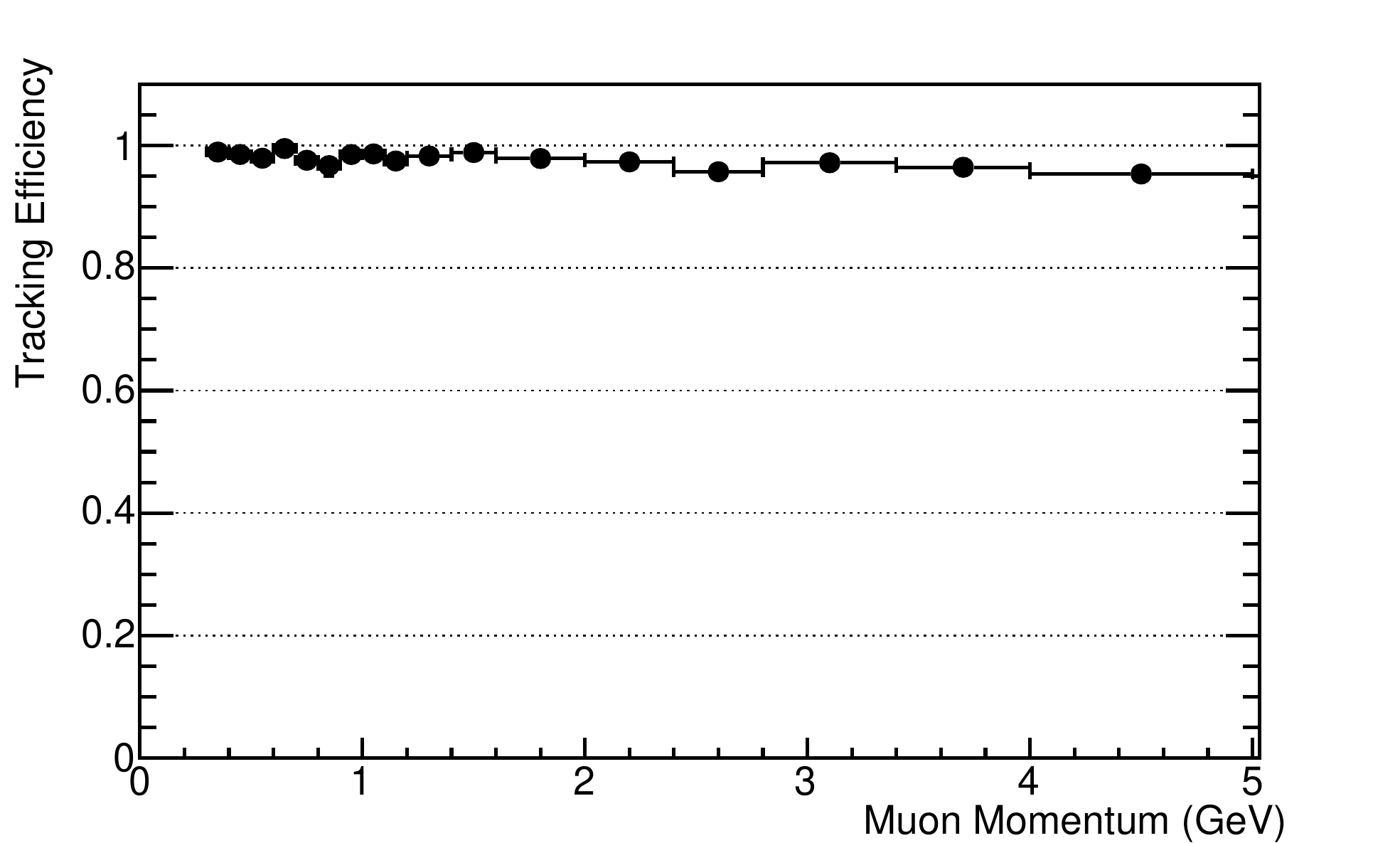}
\includegraphics[width=0.45\textwidth]{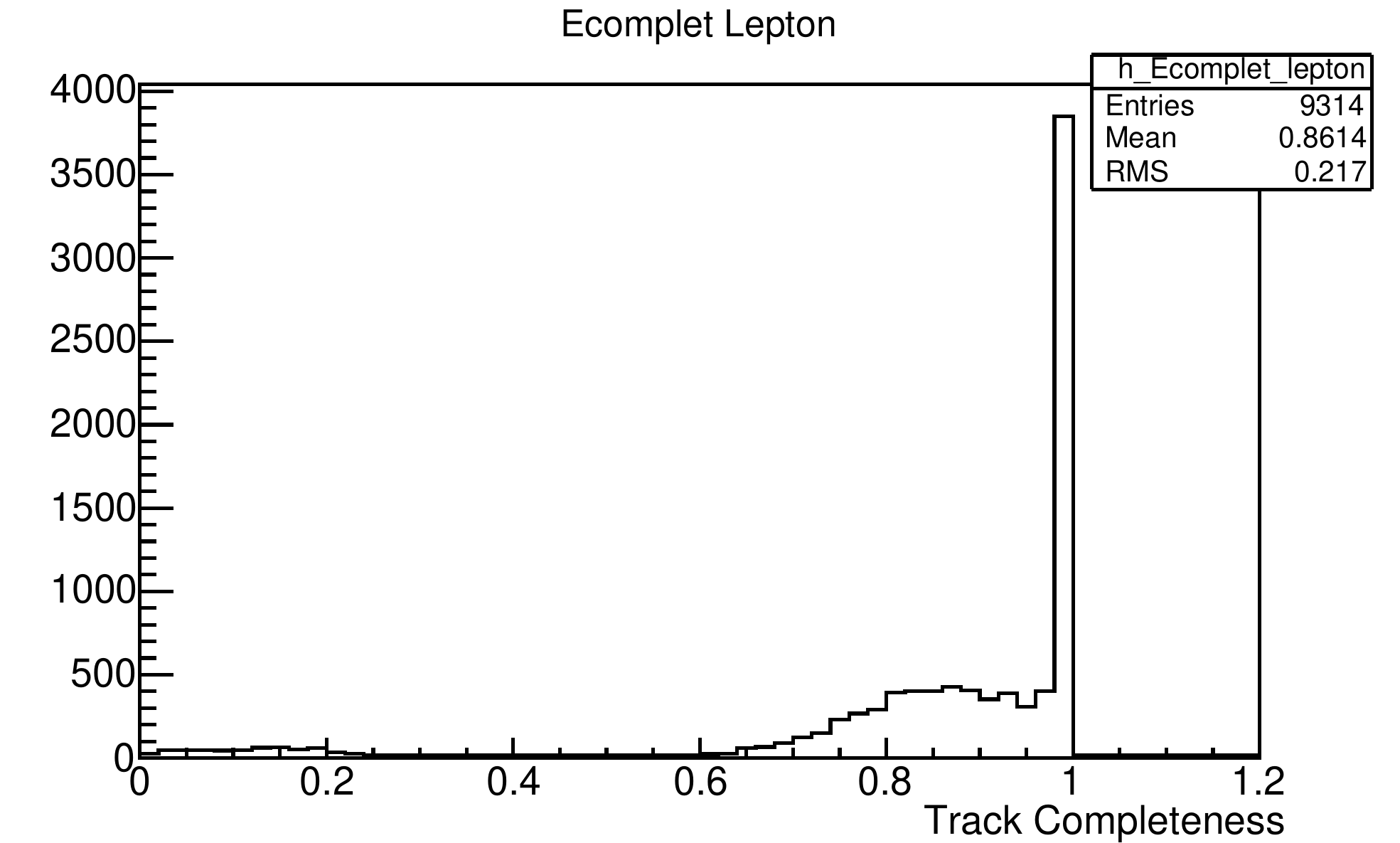}
\includegraphics[width=0.45\textwidth]{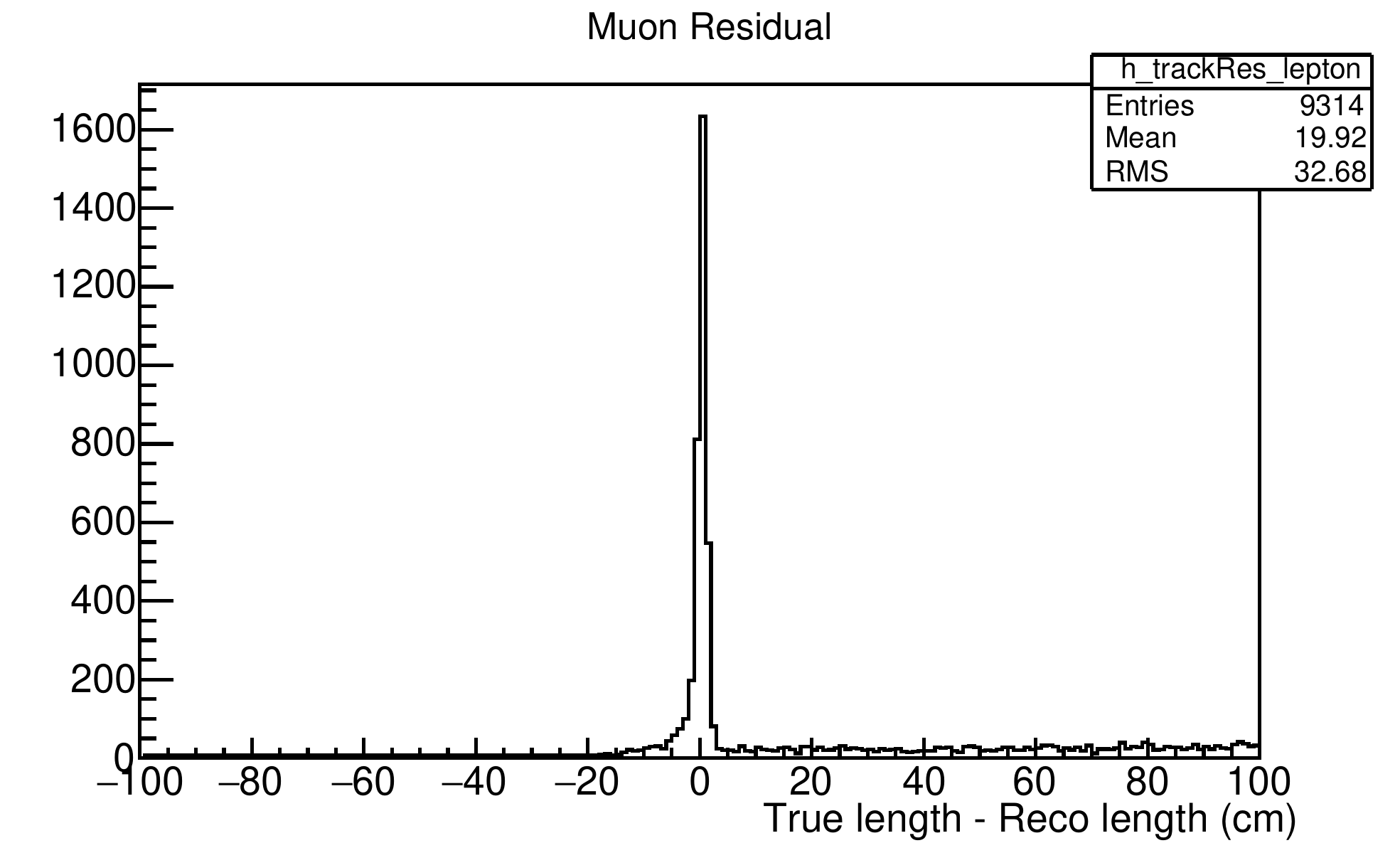}
\end{cdrfigure}

\subsubsection{PMA}
\label{sec:comp:reco:pma}

Another approach to 3D reconstruction in LArTPC detectors is referred to as the \textit{Projection Matching Algorithm
(PMA)}~\cite{pma_algorithm}. PMA was primarily developed as a technique for 3D reconstruction
of individual particle trajectories (trajectory fits). 
Instead of
building up a 3D hypothesis from 2D clusters, it starts with the 3D hypothesis and compares
the 2D projection of the predicted trajectory of a particle with the observed data. Association
of hits between the 2D planes is not needed in this approach, improving its performance in
problematic cases, such as isochronous and short tracks.

PMA can take as input the output from different pattern recognition algorithms, from
LineCluster~\cite{linecluster} to WireCell (described below).  Because these 2D algorithms
are run on each 2D projection independently, and because of detector defects,
clusters from  particles may be broken
into several smaller pieces, fractions of 2D clusters may be missing,
and clusters obtained from complementary projections are not guaranteed to cover corresponding
sections of trajectories. Such behavior is expected since ambiguous 
trajectories can be resolved only if the information from multiple 2D projections is used.
PMA performs higher-level pattern recognition using as input clustering information from all
projections in order to search for the best matching combinations of clusters. The algorithm
also attempts to correct hit-to-cluster assignments using properties of 3D reconstructed objects.

Plots of the efficiency, the completeness, and the  difference between the true and reconstructed
track lengths for single muons with momentum between 300\,MeV and 5\,GeV in the~\pdsp geometry are
shown in Figure~\ref{fig:muonpmaperf}.  

\begin{cdrfigure}[PMA reconstruction, single muons, 300\,MeV/c < $p$ < 5\,GeV/c]{muonpmaperf}{Performance of the PMA reconstruction algorithm for single muons with 
momentum between 300~MeV/c and 5~GeV/c.  The left-hand figure shows the tracking efficiency as a function of
muon momentum, the middle figure shows the distribution of tracking completeness, and the right-hand figure shows the
distribution of the difference between the reconstructed muon track length and the true track length.}
\includegraphics[width=0.45\textwidth]{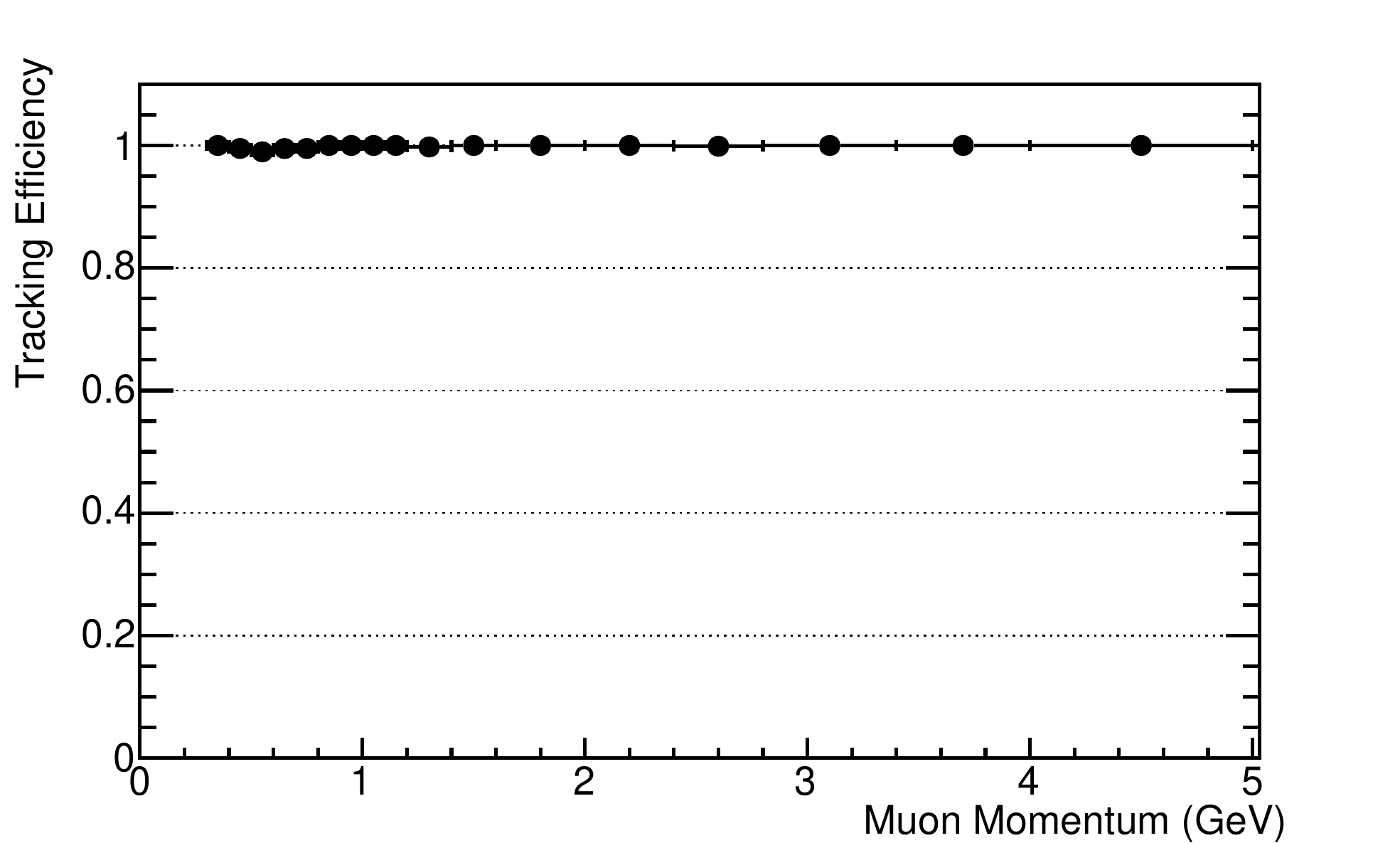}
\includegraphics[width=0.45\textwidth]{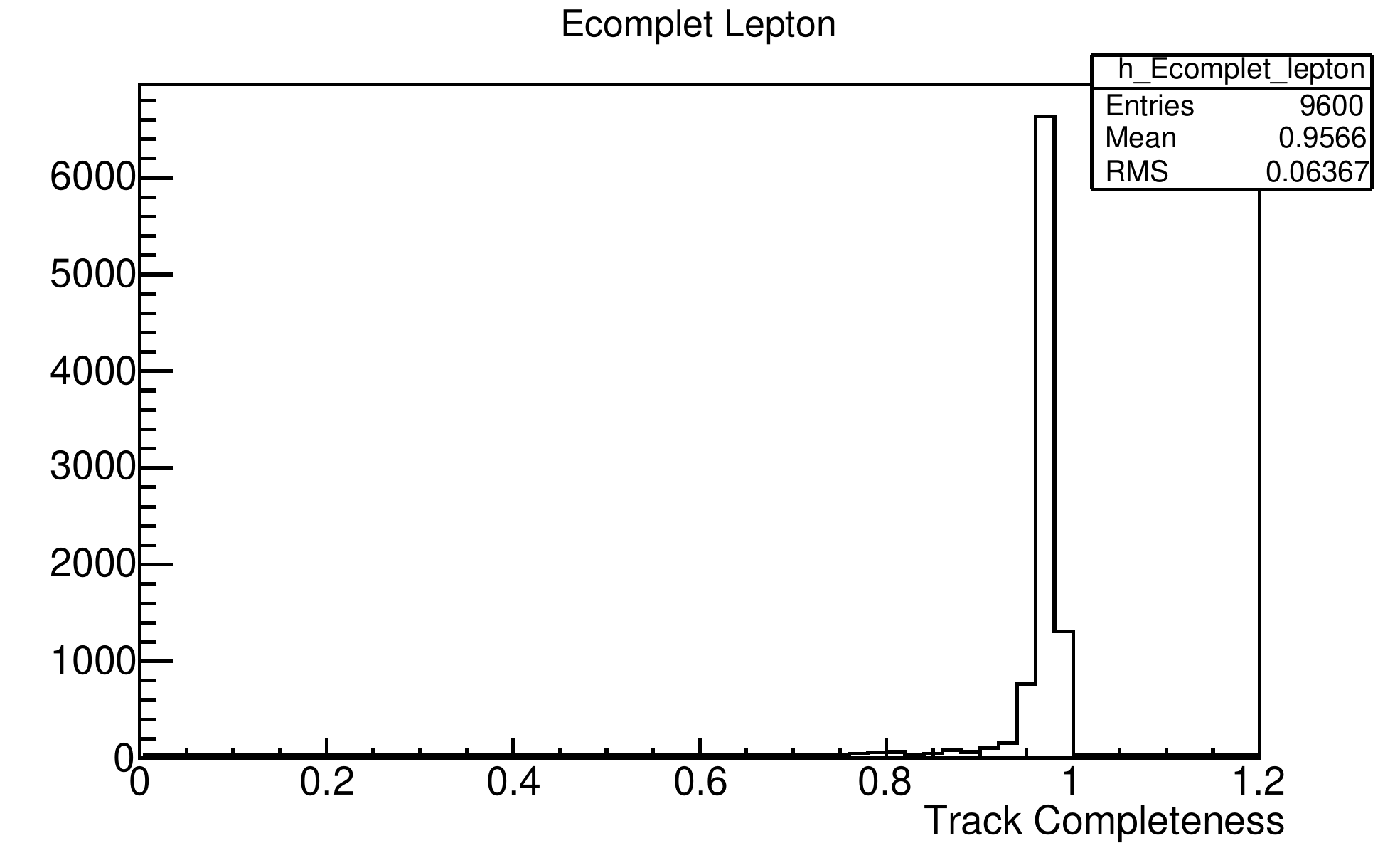}
\includegraphics[width=0.45\textwidth]{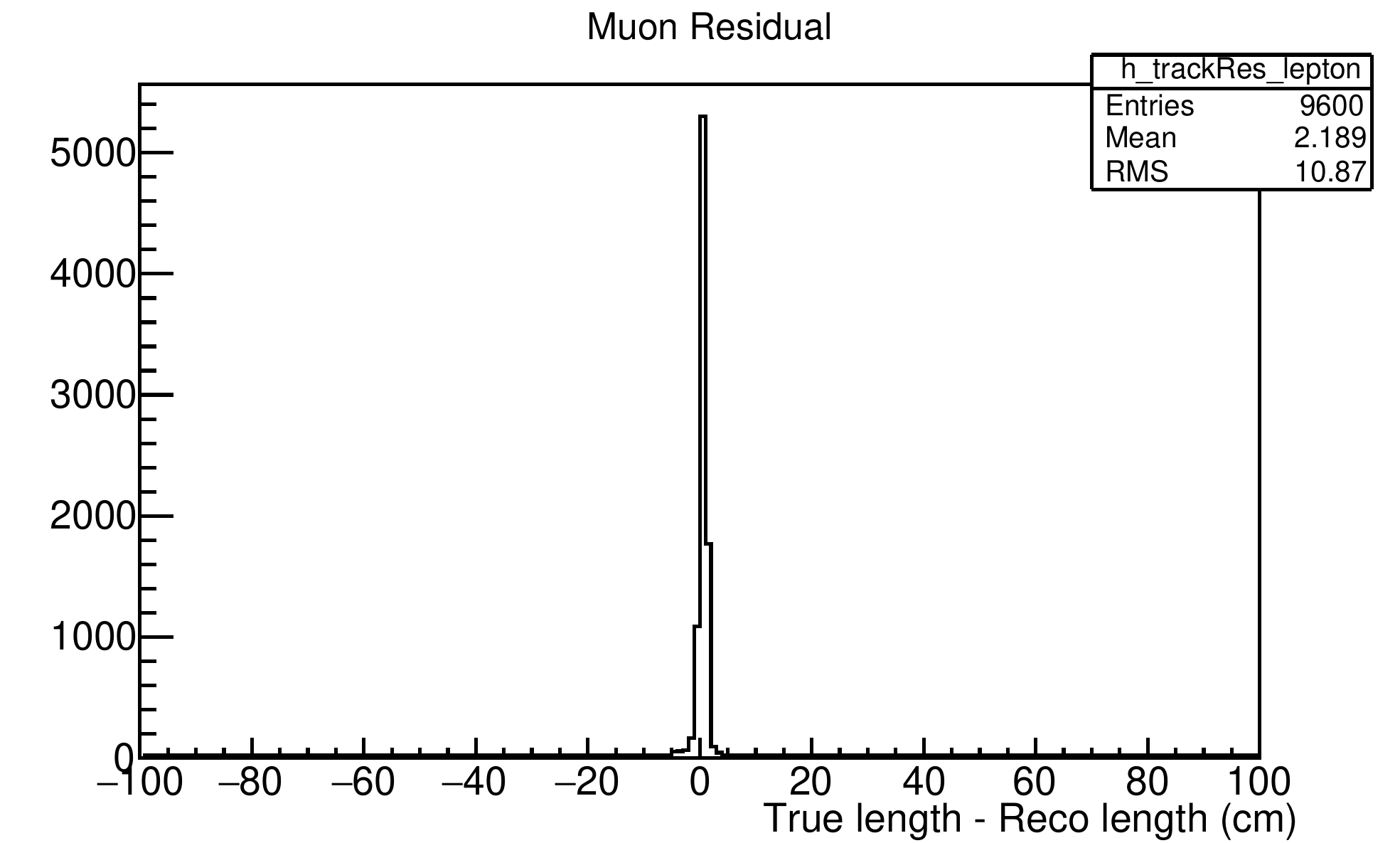}
\end{cdrfigure}

PMA has been used successfully in \pdsp to reconstruct simulated beam particles and cosmic muons~\cite{pma_cosmic_mu}. In order to illustrate the performance of the entire reconstruction chain,
Figure~\ref{fig:PMApioninteraction} shows the spatial resolution of the interaction vertex with neutral pion
production appearing in the 2-GeV/c $\pi^+$ sample.
The resolution is found to be 0.6\,cm in this study.
A similar resolution is obtained for the reconstruction
of inelastic interaction vertices in the 2-GeV/c proton sample.

Figures~\ref{fig:PMAproton2gevc}
and~\ref{fig:PMAcosmics} show examples of reconstruction of a 2-GeV/c proton in the test beam and
cosmic-ray muons, respectively.

\begin{cdrfigure}[Vertex resolution for
  inelastic interaction of $\pi^\pm$ on Ar where a $\pi^0$ is produced]{PMApioninteraction}{Vertex position resolution in cm in $x$, $y$, and $z$ and 3D for the
  inelastic interaction of charged pions on liquid argon nuclei in events in which a $\pi^0$ is produced, in
  \pdsp, using the PMA algorithm.}
\includegraphics[width=0.45\textwidth]{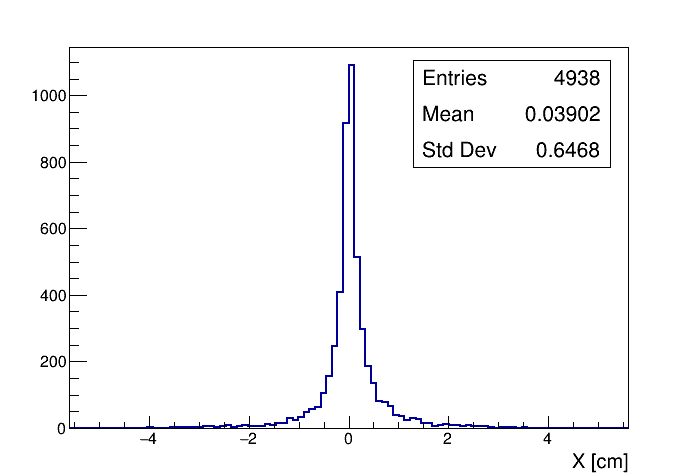}\includegraphics[width=0.45\textwidth]{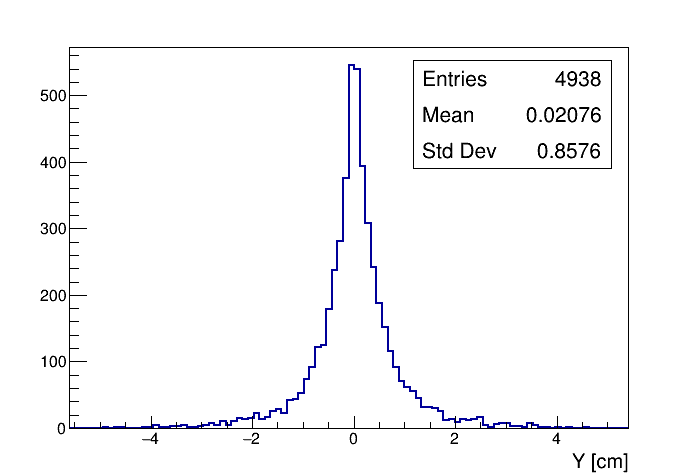}
\includegraphics[width=0.45\textwidth]{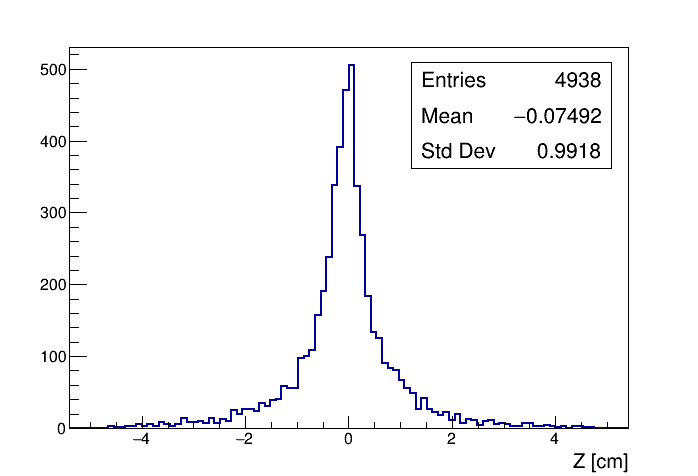}\includegraphics[width=0.45\textwidth]{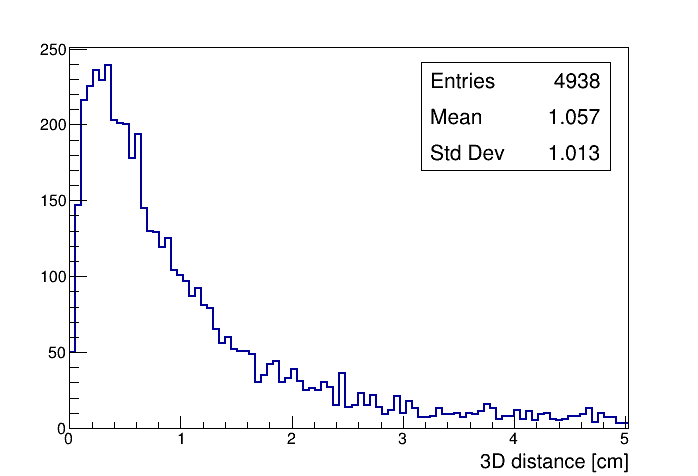}
\end{cdrfigure}

\begin{cdrfigure}[Reconstructed event of simulated proton with initial momentum 2\,GeV/c]{PMAproton2gevc}{Example of reconstructed event of simulated proton with initial momentum 2\,GeV/c (reconstruction algorithms: gaushit, Line Cluster and PMA).}
\includegraphics[width=0.45\textwidth]{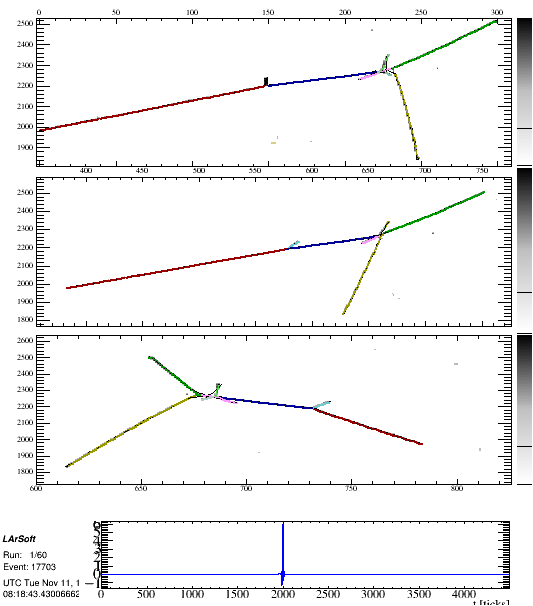}
\includegraphics[width=0.45\textwidth]{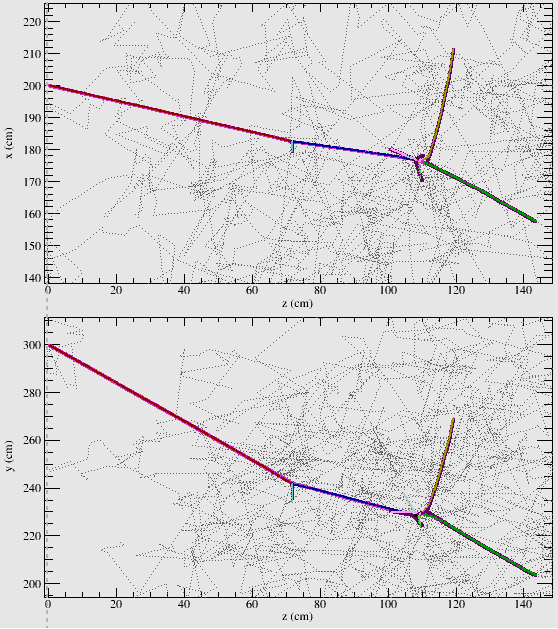}
\end{cdrfigure}
\begin{cdrfigure}[Example of reconstructed cosmic muons in \pdsp]{PMAcosmics}{Example of reconstructed cosmic muons using gaushit, Line Cluster and PMA.}
\includegraphics[width=0.45\textwidth]{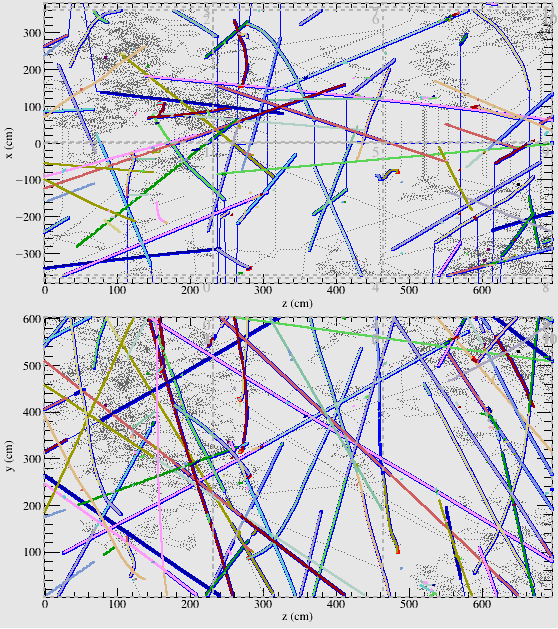}
\end{cdrfigure}

\subsubsection{WireCell}

WireCell~\cite{wire-cell}, a new reconstruction method under development, adopts a very different approach from the aforementioned algorithms .
Instead of performing pattern recognition directly on each of the 2D views (drift 
time versus wire number), the first step of the WireCell reconstruction is to 
perform 3D imaging with time, geometry, and charge information. 
The algorithm takes advantage of this 
information to suppress the effects of electronic noise.
Often, noise will lead to a fluctuation in a waveform which may be
large enough to mimic a signal.  The algorithm combats this by
requiring any potential signal to be consistent across multiple wires,
given their geometry, in their charge across time.  Many of the
fluctuations due to noise will fail this consistency requirement and
be rejected while true signal will satisfy it.
Use of the charge and time information in this manner takes advantage of the fact that in a LArTPC with induction planes,
each of the wire planes, in principle, detects the same ionization electrons as the other planes. 
Figure~\ref{fig:quality} shows an example of the improvement of WireCell 3D imaging
over the more traditional approach. 

The suppression of the electronic noise comes at the cost of more
sensitivity to hit inefficiencies from dead channels or the signal processing steps.
 Since the track and shower hypotheses
are not used, the 3D imaging works for any event topology. 
Pattern recognition is needed to identify 
the content of these 3D images. Figure~\ref{fig:tracking2} shows the 
performance of the currently available 3D pattern recognition in
WireCell. For the long track going close to parallel to the wire plane, the reconstructed
track shows a zig-zag behavior. This is due to the current lack of a fine track-fitting algorithm
that is expected to be added in the near future. 
Further developments of the WireCell pattern recognition algorithms
are needed before meaningful physics quantities can be calculated.
%
\begin{cdrfigure}[Comparison of imaging recon
qualities with and without charge information]{quality}{Comparison of imaging reconstruction 
qualities with and without the charge information. }
\includegraphics[width=0.5\textwidth]{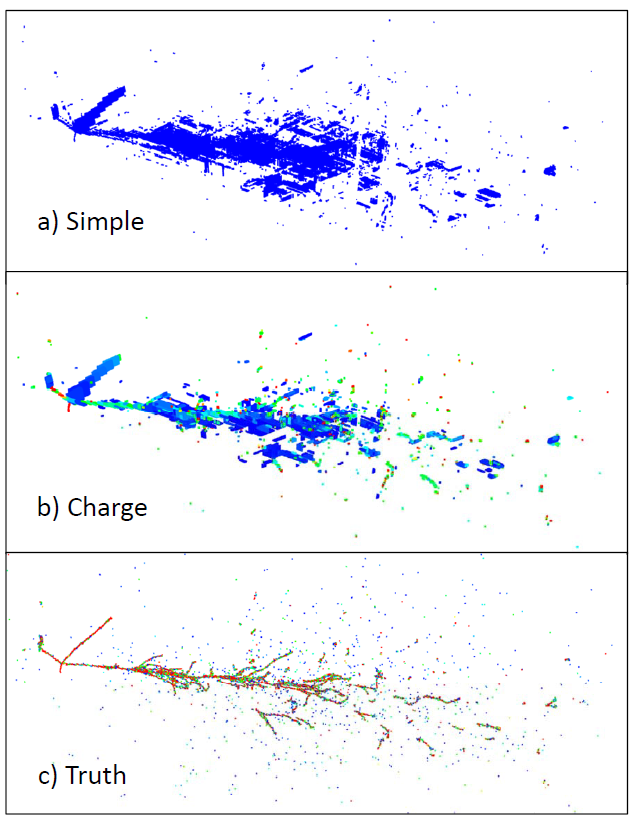}
\end{cdrfigure}
%
%
\begin{cdrfigure}[Reconstructed image for one neutrino interaction event; comparison to MC]{tracking2}{The reconstructed image is shown 
on the left panel for one neutrino interaction event. The image 
was passed through the 3D pattern recognition program with tracks 
identified (middle panel). The identified pattern is compared 
with Monte-Carlo truth in the right panel. The zig-zag line in the right 
panel is the identified track.  More sophisticated track-fitting algorithms, to be added in the future,
will improve the track reconstruction.}
 \includegraphics[width=0.9\textwidth]{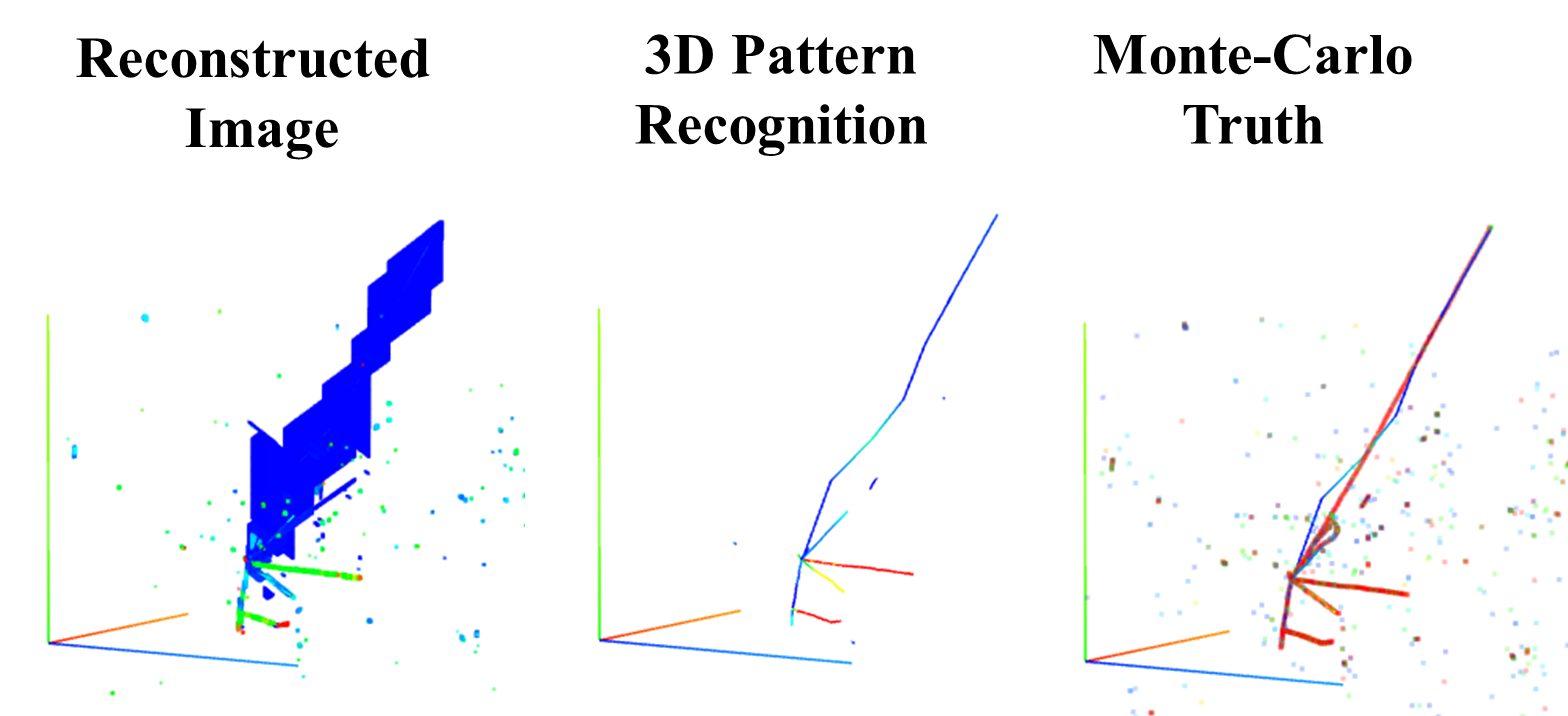}
\end{cdrfigure}


\chapter{Test beam specifications} 

The ProtoDUNE-SP (NP04) experiment will be housed in the EHN1 building at CERN. The detector is situated at the end of the H4 beamline in the newly constructed extension of EHN1. The H4 beamline is also extended and configured to deliver either a hadron or a pure electron beam to the experiment. To produce particles in the momentum range of interest, the secondary beam from the T2 primary target is sent onto a secondary target to generate a tertiary beam. Particles in this tertiary beam are momentum- and charge-selected and transported down the H4 beamline extension to the ProtoDUNE-SP detector. 
This chapter discusses the beam requirements, the H4 tertiary beamline design and instrumentations, 
the DAQ/trigger, and the physics run plans.  

\section{Beam requirements}
\label{sec:beamrequirements}


The CERN test beam results from ProtoDUNE-SP will be used to evaluate the detector performance,  understand the various physics systematic effects, and provide data for event reconstruction studies that are representative of neutrino interactions. 
The parameters defining the test beam are primarily driven by the requirement that these test beam results be directly applicable to DUNE's future large underground single-phase detector module(s) with minimal extrapolation.
To match the charged-particle spectrum and topologies that are expected in the DUNE far detector, the H4 tertiary beam must span a broad range of particle momenta, be composed of electrons, muons, and hadrons, and charge-selectable. 
The expected momentum distributions for secondary particles from neutrino interactions in the far detector has a large spread that ranges from a few hundred MeV/c to a few GeV/c.
The desirable range for ProtoDUNE-SP is in the low-momentum region. Based on the feedback and constraints from the CERN accelerator group, the design of the beamline extension has been developed to allow the transport of beam particles from about 0.5~GeV/c up to 7~GeV/c. 

The maximum electron drift time in the ProtoDUNE-SP TPC is about 2.25~ms. In order
to keep the  pile-up in the TPC at the percent level, the planned
beam particle rate should be below 100 Hz.  
The ProtoDUNE-SP TPC has two drift volumes separated by a cathode plane. It is desirable to aim the particle beam such
that a large fraction of the lower-energy hadronic showers are 
contained in one drift volume, thus minimizing the uncertainties from
particles lost in the inactive detector materials. 
As shown in Figure~\ref{fig:beamwindow_loc}, multiple beam injection points have been explored. Based on inputs from the physics group, the larger angle (beam \# 3) w.r.t. the APA plane (Saleve side), which corresponds to about 13$^\circ$, is preferred.
Due to engineering and safety considerations, only beam \#3 will
be fully instrumented with the beam window system as described in
Sections~\ref{subsec:fc-beamplug} and ~\ref{subsec:beamwindow}.
The remaining two beam positions do not have the beam window system installed. 
With this configuration, beam \#3 is the primary beam 
with which most of the physics
data will be taken.
\begin{cdrfigure}[Beam window locations]{beamwindow_loc}{Three possible beam injection points. The cryostat support structures near the beam injection points are removed in the Figure to show the interior. Beam window and beam plug are installed only for beam \# 3.}
  \includegraphics[width=0.9\textwidth]{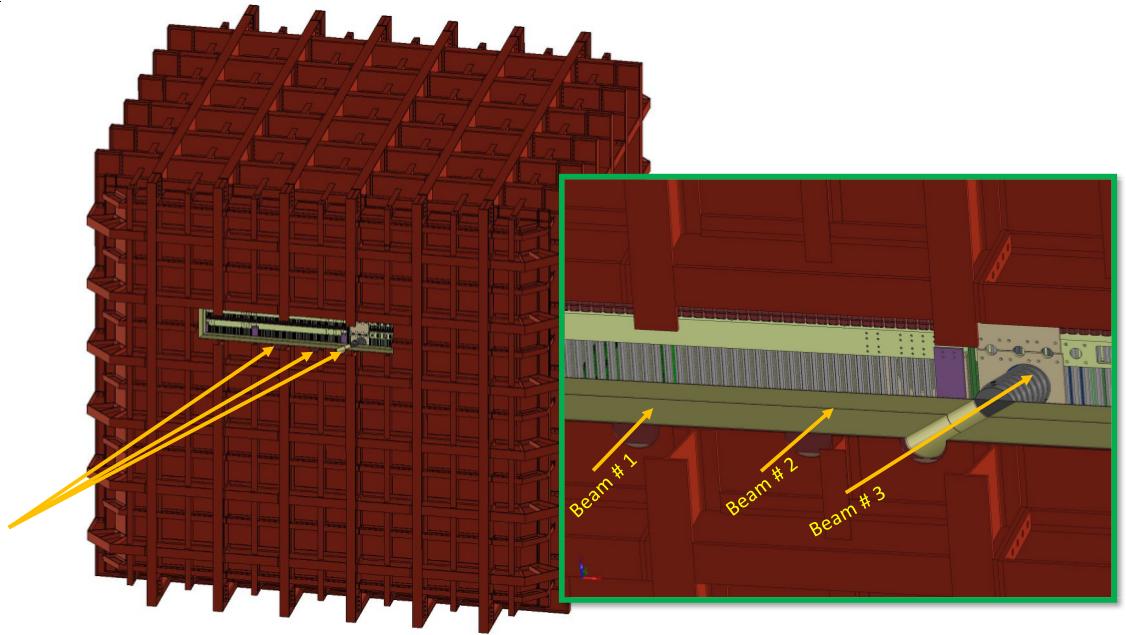}
\end{cdrfigure}
A summary of the beam requirements is shown in Table~\ref{tab:beamspecs}.
\begin{cdrtable}[Particle beam requirements]{ll}{beamspecs}{Particle beam requirements. (Kaon rate is low for beam momentum below 2\,GeV/c.)}
 Parameter & Requirements \\ \toprowrule
  Particle Types        & $e^\pm,\mu^\pm,\pi^\pm$,$(K)$,$p$  \\ \colhline
  Momentum Range   & 0.5 - 7 GeV/$c$ \\ \colhline
  Momentum Resolution   & $\Delta p/p   \le 3$ \% \\ \colhline
  Transverse Beam Size   & RMS(x,y) $\approx$ 1 cm  \\
  & (At the entrance face of the LAr cryostat) \\ \colhline
  Beam Entrance Position & Beam \# 3 (Figure~\ref{fig:beamwindow_loc}) - Saleve side TPC   \\ \colhline
  Rates & $\sim25 - \sim100$\,Hz     \\ \colhline
\end{cdrtable}

\section{Beamline}
\label{sec:h4beamline}

The design of the H4 beamline extension mirrors that of the H2. In this section, we describe the beamline design and the expected beam properties.

\subsection{H4 beamline layout and optics}

The placement of the quadrupole and dipole magnets in the H4 beamline extension is illustrated in Figure~\ref{fig:H4layout}. The distance from the secondary target to the front of the NP04 cryostat is about 37 m. For the hadron beam, either a tungsten or a copper target will be used. For the electron beam, a Pb target of a few radiation lengths will be used. The first two dipole magnets (shown in red) after the secondary target are rotated by about 56$^\circ$ to steer the beam downward towards the cryostat. The third dipole magnet (shown in green) is used for steering the beam horizontally into one of the three beam positions.

\begin{cdrfigure}[H4 beamline layout]{H4layout}{Layout of the quadrupole and dipole magnets in the H4 beamline extension. The secondary target (not shown) is upstream of the first quadrupole magnet on the left side of the Figure. Vacuum beam pipe and beam instrumentations are also not shown. (Courtesy of V. Clerc, CERN).}
  \includegraphics[width=0.75\textwidth]{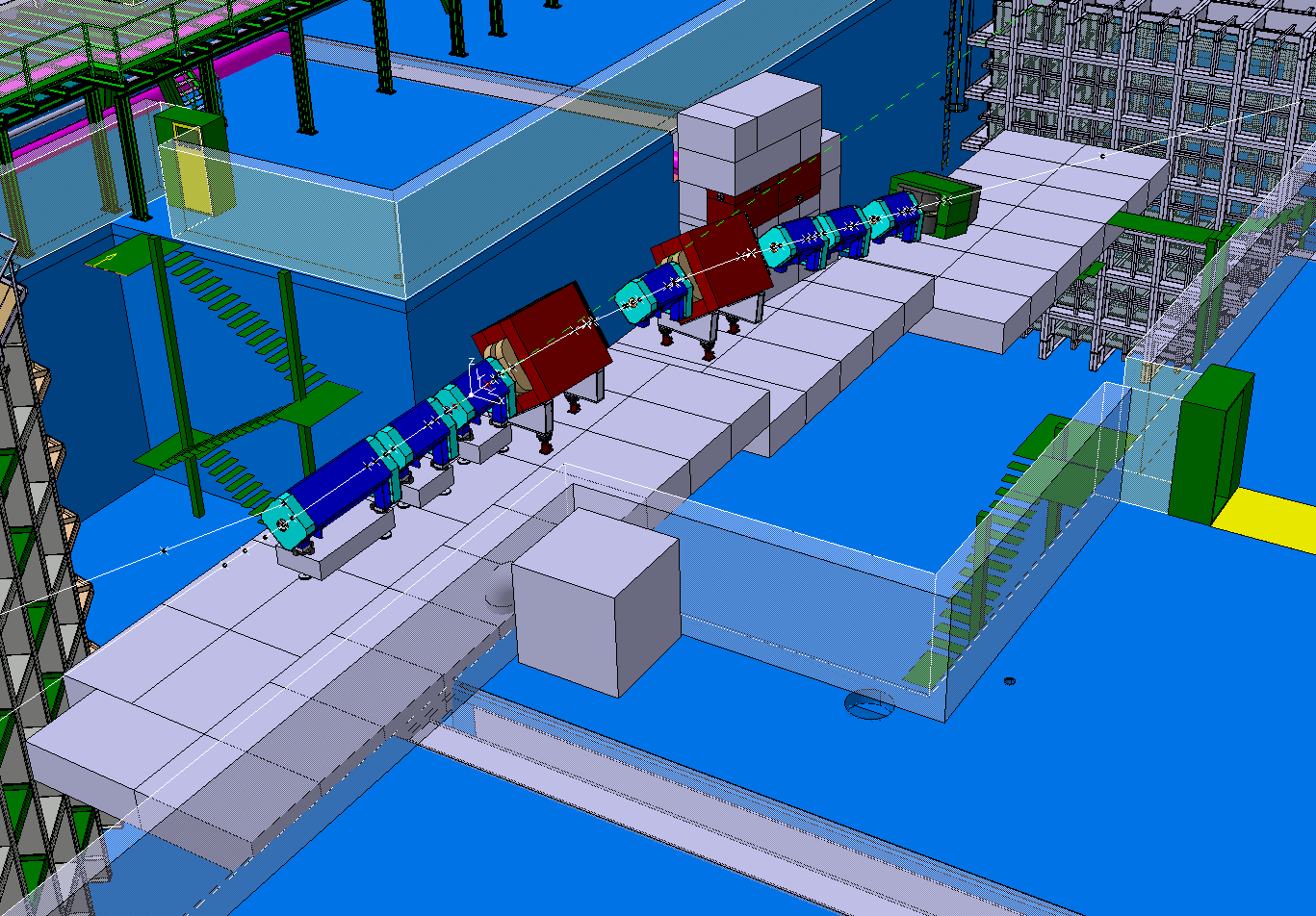}
\end{cdrfigure}

The beamline optics from the target to the cryostat for the horizontal and vertical planes are shown in Figure~\ref{fig:beamoptics}. The Figures show the position of the quadrupole magnets (Q17-Q22), dipole magnet (B17 - B19), collimator (C12), Time-of-Flight detectors (TOF1-2), beam profile monitors (BPROF1-4), and the Threshold Cherenkov counters (XCET1-2)  relative to the secondary target.  For the nominal configuration, the beam is focused at the front of the cryostat to ensure maximum acceptance of beam particles through the beam window penetration and the beam plug inside the cryostat.

The beamline is in vacuum, with a beam pipe extending from the secondary target down to the beam window in the cryostat upstream face. 

\begin{cdrfigure}[H4 beam optics]{beamoptics}{H4 beamline extension optics for the horizontal (top) and vertical (bottom) planes. The secondary target is located on the left side of the plots and the front of the ProtoDUNE-SP cryostat is located at the beam focus point, at about 37 m from the secondary target. Q17-Q22 are the quadrupole magnets. B17-B19 are the dipole bending magnets. TOF1-2 are the Time-of-Flight detectors, BPROF1-4 are the beam profile monitors, and XCET1-2 are the threshold Cherenkov counters. }
 \includegraphics[width=0.99\textwidth]{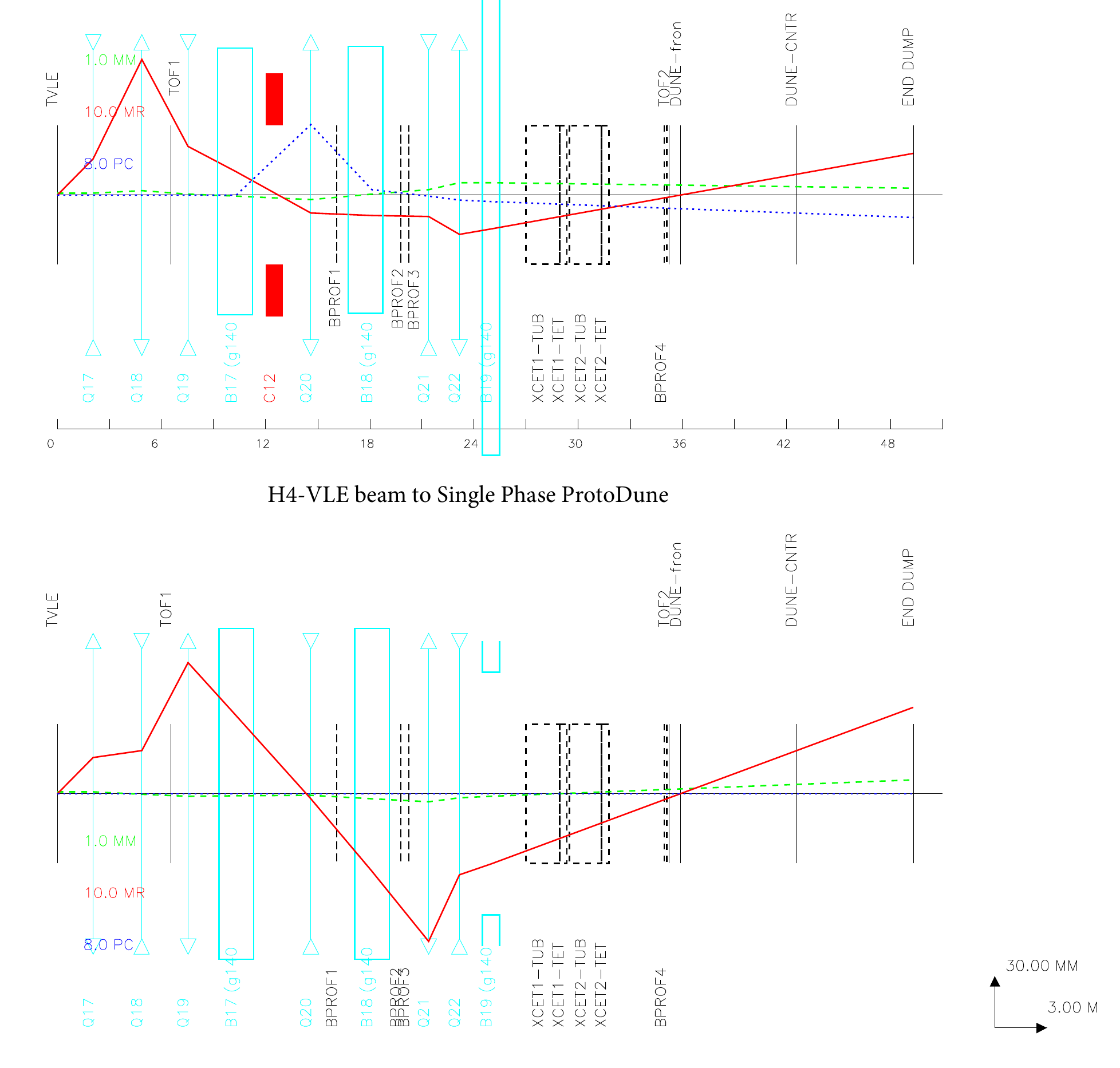}
\end{cdrfigure}

\subsection{Beam properties}
A full GEANT4 simulation of the H4 beamline including its extension to the ProtoDUNE-SP detector 
has been performed.
 The beamline model starts with the H4 secondary beamline 
and derives the particle properties in the tertiary beamline.  Target, magnets, collimators and a preliminary assumption
about beam instrumentation are included. The secondary target has been
modeled as a Tungsten cylinder (R=30~mm, L=300~mm) for beam momenta $\leq$~3~GeV/c and as a Copper cylinder of the same dimensions  for beam momenta  $>$~3~GeV/c. Optimization of the target dimensions and material is ongoing.

Table \ref{tab:beampartcomp} describes the particle composition of the
hadron beam at the entrance of the cryostat. Two features are evident. First, the beam is dominated by positrons at low energies,
and secondly, the kaon content, and to a lesser extent the pion content, are depleted at lower energies due to decays of these species 
along the beam path. 

\begin{cdrtable}[Beam composition]{cccccc}{beampartcomp}{Beam composition (in percentage)  at the cryostat entrance for particles contained in the beam pipe (R= 10~cm).}
Momentum (GeV/c) & e$^{+-}$ & $K^{+-}$ & $\mu^{+-}$ & $p^{+-}$& $\pi^{+-}$ \\ \toprowrule
-7  &27.4 &3.4 &1.0 &1.3  &66.9 \\ \colhline
-6  &33.5 &2.9 &1.2 &1.3  &61.2 \\ \colhline
-5  &43.2 &1.9 &1.3 &1.1  &52.5 \\ \colhline
-4  &54.6 &1.1 &1.4 &0.6  &42.3 \\ \colhline
-3  &28.4 &1.1 &2.3 &1.4  &66.8 \\ \colhline
-2  &48.9 &0.3 &1.8 &0.4  &48.6 \\ \colhline
-1  &81.7 &0.2 &1.1 &0.2  &16.9 \\ \colhline
-0.4&98.3 &0.0 &0.4 &0.0  &1.3 \\ \colhline
0.4 &99.1 &0.0 &0.0 &0.5  &0.5 \\ \colhline
1   &69.3 &0.0 &0.3 &15.3 &13.9 \\ \colhline
2   &34.9 &0.6 &1.7 &22.9 &39.0 \\ \colhline
3   &19.9 &2.8 &0.6 &18.9 &56.6 \\ \colhline
4   &47.1 &1.6 &1.1 &8.8  &41.3 \\ \colhline
5   &37.0 &2.8 &0.9 &9.4  &49.5 \\ \colhline
6   &28.1 &4.0 &0.9 &10.4 &56.2 \\ \colhline
7   &20.6 &5.1 &1.0 &10.7 &62.6 \\
\end{cdrtable}
Particle rates, assuming a spill intensity of $10^6$
particles on the secondary target and a SPS spill length of 4.8
seconds, are reported in Table~\ref{tab:beampartrates}. The H4 simulation results are documented in CERN-ACC-NOTE-2016-0052. 
\begin{cdrtable}[Particle rate]{ccccccc}{beampartrates}{Particle rates (Hz).}
Momentum (GeV/c) & e$^{+-}$ & $K^{+-}$ & $\mu^{+-}$ & $p^{+-}$& $\pi^{+-}$& total\\ \toprowrule
-7  & 57 & 7  & 2  & 3  & 138& 207 \\ \colhline
-6  & 62 & 5  & 2  & 2  & 113& 185 \\ \colhline
-5  & 72 & 3  & 2  & 2  & 87 & 166 \\ \colhline
-4  & 93 & 2  & 2  & 1  & 72 & 170 \\ \colhline
-3  & 11 & 0.5& 1  & 0.5& 26 & 39 \\ \colhline
-2  & 13 & 0  & 0.5& 0.1& 16 & 27 \\ \colhline
-1  & 18 & 0  & 0.2& 0  & 4  & 22 \\ \colhline
-0.4&  8 & 0  & 0  & 0  & 0.1& 8 \\ \colhline
0.4 & 7  & 0  & 0  & 0  & 0  & 7 \\ \colhline
1   & 18 & 0  & 0  & 4  & 4  & 27 \\ \colhline
2   & 13 & 0.2& 0.5& 9  & 15 & 38 \\ \colhline
3   & 11 & 1  & 1  & 11 & 31 & 56 \\ \colhline
4   & 92 & 3  & 2  & 17 & 81 & 196\\ \colhline
5   & 74 & 6  & 3  & 19 & 99 & 200\\ \colhline
6   & 63 & 9  & 3  & 24 & 127& 226\\ \colhline
7   & 52 & 13 & 2  & 27 & 157& 252\\
\end{cdrtable}

At momenta larger than about 4 GeV/c, the particle rates are at the limit or beyond the DAQ capability.
 At lower energies, the proton and pion rates are much lower, reduced by a
factor of 10 at 1 or 2 GeV/c, and are overwhelmed by the positron rate.

The momentum spread of the beam is of the order of 5--7\%. At higher energies, where
the particle rate is higher, the momentum spread  can be narrowed by
closing the collimators, at the expense of the beam intensity.  For example, Figure \ref{fig:momcoll} shows
that at $p=4$~GeV/c the momentum spread can be  reduced to $\Delta p/p= 3.6\%$ with a factor of 4 reduction in particle rate.  
\begin{cdrfigure}[Beam momentum uncertainty]{momcoll}{Beam momentum spread with different collimator openings at 4 GeV/c.}
  \includegraphics[width=0.75\textwidth]{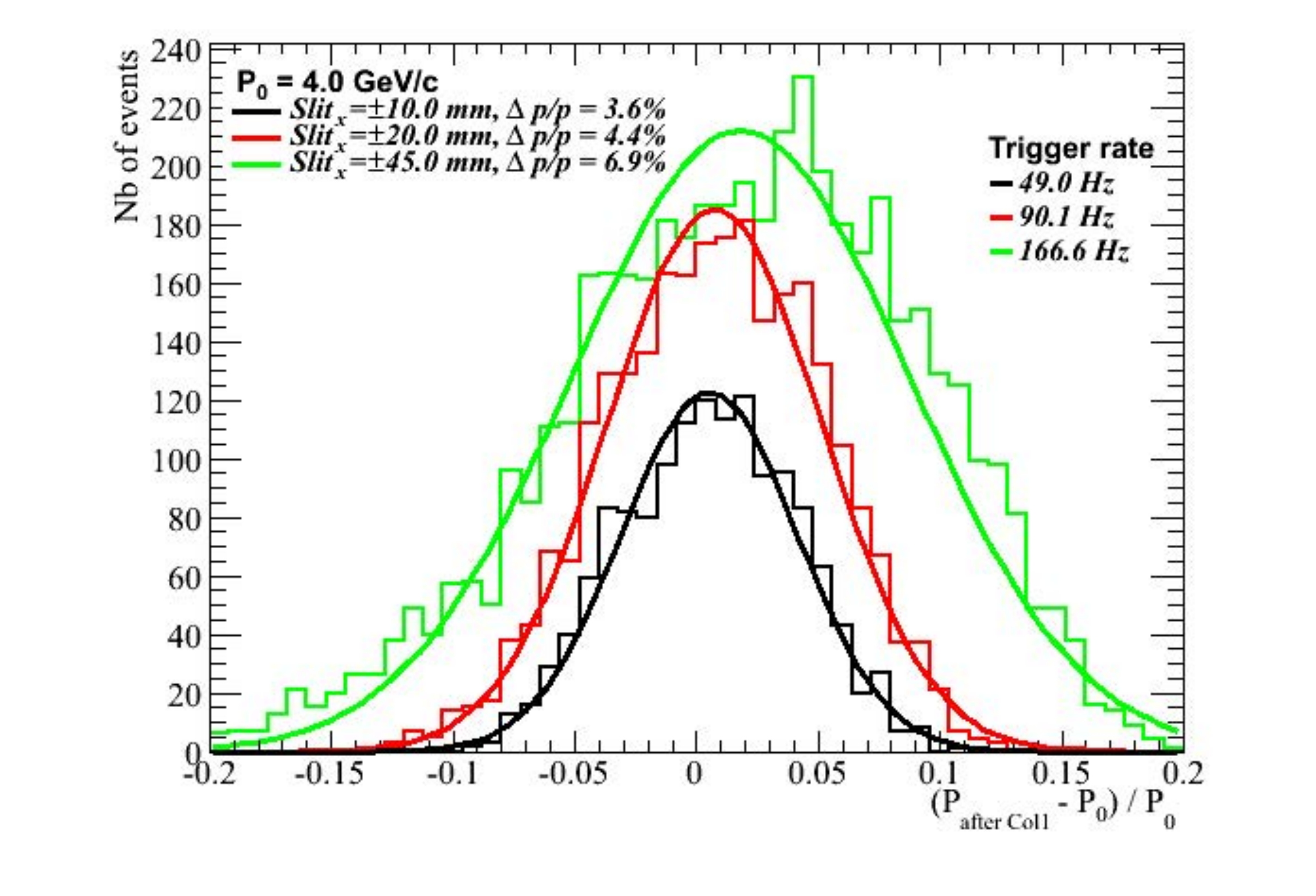}
\end{cdrfigure}

\subsection{Muon halo}
The secondary beam is mainly composed of 80-GeV pions. In the long path ($\approx$ 600~m) between the primary and the secondary target,
 an intense high-energy muon halo is produced by pion decay. Muons propagate approximately in the direction of the first section of the H4 beamline,  that is slightly upward  and sidewards of the H4 beamline extension which points at the ProtoDUNE-SP detector.  
 Therefore the most intense part of the muon halo passes about one meter above the left corner of ProtoDUNE-SP cryostat. 
Figure~\ref{fig:muonhalo_HE} shows the spatial distribution of the muon halo at the face of the cryostat. Only muons with momentum larger than 4~GeV/c have been considered. Despite the low statistics of the simulations, the up-down and left-right asymmetry is clearly visible. 
The origin of the coordinate system is chosen to coincide with the center of the cryostat face. The color scale shows the muon intensity in  $\mu/m^2/$spill for $10^6$ particles/spill from the primary target. The muon intensity on the cryostat face, indicated by the black rectangle in Figure~\ref{fig:muonhalo_HE}, ranges up to $\sim$400~$\mu$/m$^2$/spill. However, the rate of halo muons impinging on the front side of the TPC active volume (red rectangle) is expected to be significantly lower, in particular, as low as a few $\mu$/m$^2$/spill on the Saleve side of the TPC 
where the H4 beamline points. The rate of beam events with out-of-time halo muon tracks captured in the 2.5 ms time frame of the recorded event is expected to be limited.

Figure~\ref{fig:muonhalo_HE}  is very preliminary though, not only because of low statistics, but also because shielding around the low-energy beamline is not included in the simulation, and the muons produced in neighboring beamlines in EHN1, including the H2 beamline that feeds ProtoDUNE-DP, are not considered here. Based on these results, the estimated contribution to the total data volume from beam halo is negligible.
\begin{cdrfigure}[Muon halo intensity at the cryostat face]{muonhalo_HE}{Muon halo intensity at the cryostat face for muons originating from the H4 beamline and with $P_\mu  \ge 4$~Gev/c. The origin of the $x-y$ coordinate system is centered on the cryostat and dimensions are in mm. Color scale is in units of $\mu/m^2/$spill. The black and red rectangles represent the cryostat face and front face of the active detector volume.}
\includegraphics[width=0.50\textwidth]{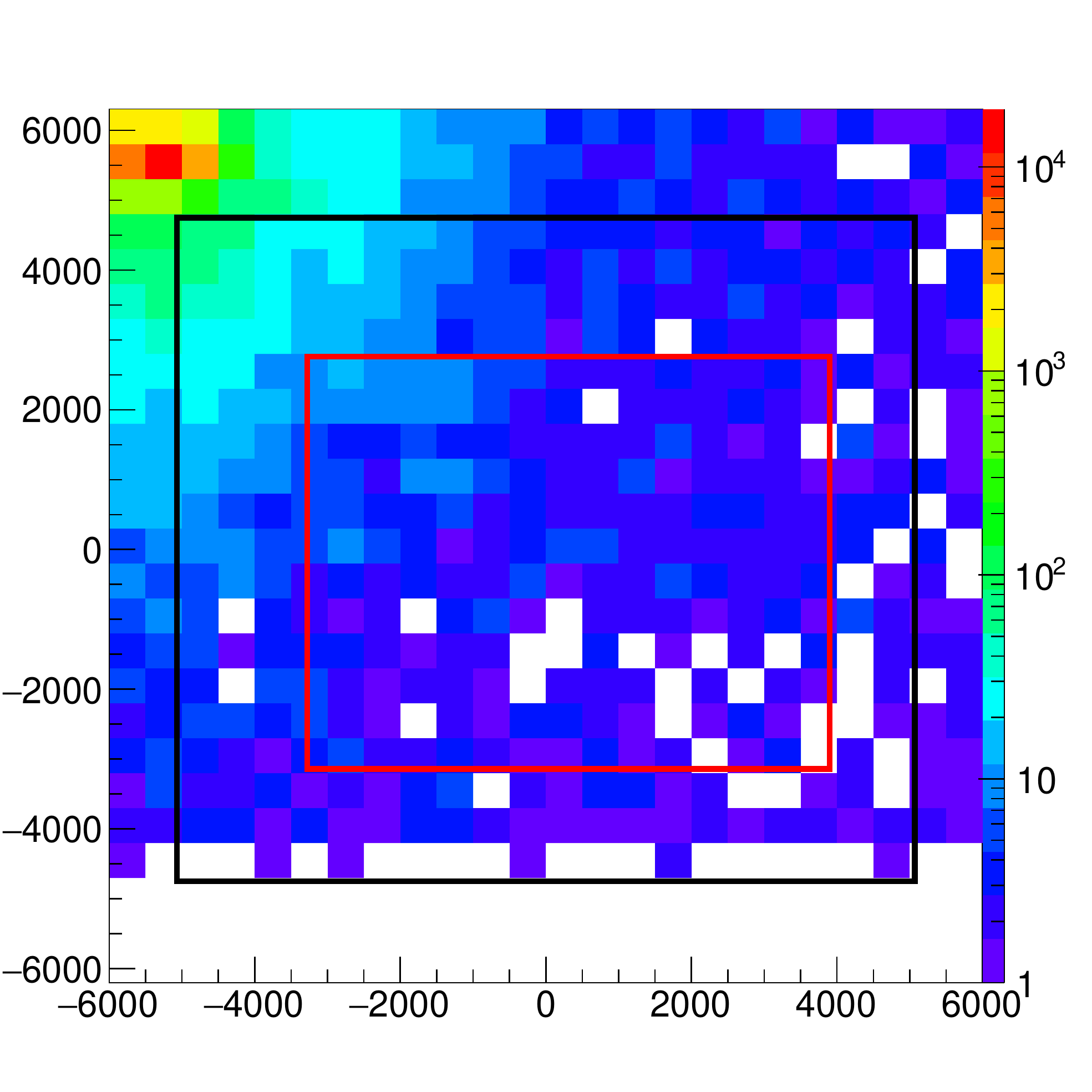}
\end{cdrfigure}

\section{Beamline instrumentation}
\label{sec:beaminstruments}

The H4 beamline will be instrumented with a number of beamline detectors to provide information 
 about the beam profile, position, momentum, particle identification, and trigger capability. 

\subsection{Beamline monitors}

Operation of the beamline requires at least one beam monitor, able to provide the beam profile in two dimensions ($x$, $y$) at the entrance point to the NP04 cryostat.   The beam monitor can also be exploited for data analysis, provided that it delivers data on an event-by-event basis. Another monitor, located immediately downstream of the last bending magnet, is added in the layout  in order to determine the incident particle  direction and position at the front face of the cryostat, and match it with the reconstructed track in the LAr active volume.

To reduce the uncertainty due to the momentum spread of the beam around the central selected value, the momentum can be measured particle-by-particle
with a set of three detectors placed one upstream and two downstream of the bending magnets B18 (see Figure~\ref{fig:beamoptics}).  
Preliminary results using full simulations indicate that a momentum resolution on the order of  2\% is achievable. 
%

The CERN Beam Instrumentation group is designing and fabricating
 the beam monitors 
 using scintillating fiber technology, where 
 the fibers have a polystyrene core surrounded by cladding. Fibers provide a light yield of $\sim$8000 photons/MeV, deposited with fast rise and decay times of 1--3\,ns. The design foresees 1-mm-square fibers in two planes to provide $x$ and $y$ coordinates. Fibers will be mirrored on one end to increase
light collection.  Every monitor consisting of two planes of 1-mm-thick fibers adds 0.47\% of a radiation length ($X_0$) to the material budget -- the amount of material distributed along the beamline and crossed by the beam particles before entering the TPC active volume. 
The three devices for momentum measurement will consist of one layer only, oriented perpendicularly to the magnet deflection.
A fiber plane is made out of 192 fibers with no space between them. It will cover an area of 192\,mm $\times$ 192\,mm and fits in the beamline.
Scintillation light will be read out using SiPMs.
%
These detectors for position and momentum measurement are 
mounted through special flanges along the beam pipe, without the need to break the vacuum.


\subsection{Particle identification system}

The H4 beamline is capable of delivering two types of beams, electron and hadron. 
While the electron beam is relatively pure, the hadron beam consists of a mixture of electrons, pions, kaons, and protons. Therefore, for the hadron beam, it is essential to have an efficient particle identification system to cleanly tag particle types on a particle-by-particle basis. To achieve this goal, a particle identification system based on a combination of threshold Cherenkov counters and time-of-flight (ToF) system is planned for ProtoDUNE-SP. 
Cherenkov counters can be placed in the last segment of the beamline, between the last bending magnet and the cryostat. 

\subsubsection{Threshold Cherenkov counter}
Threshold Cherenkov counters have been used extensively in beamlines to discriminate particles. Figure~\ref{fig:ckv} shows one of the counters used in the CERN test beam area. It consists of a gas radiator that is contained in a long cylindrical tube, and a detection box in which the Cherenkov light is reflected by a 45$^\circ$ mirror and focused onto a photomultiplier tube (PMT). 
A variety of gases (e.g., CO$_2$, nitrogen, argon, Freon 12, air) are available at CERN for filling the Cherenkov counter to optimize particle identification. 
Two threshold Cherenkov counters will be installed in the beamline, one for detecting pions, the other for detecting pions 
and kaons. The combination of the two signals will allow identification of all hadron species. To provide signals at low beam momenta, either heavier gases or high pressures are needed.
Figure~\ref{fig:ckv_gases} shows the gas-pressure threshold 
for the production of Cherenkov light for various particle types as a function of particle momentum for Freon 12 and CO$_2$ gases.
\begin{cdrfigure}[CERN threshold Cherenkov counter]{ckv}{CERN threshold Cherenkov counter}
  \includegraphics[width=0.65\textwidth]{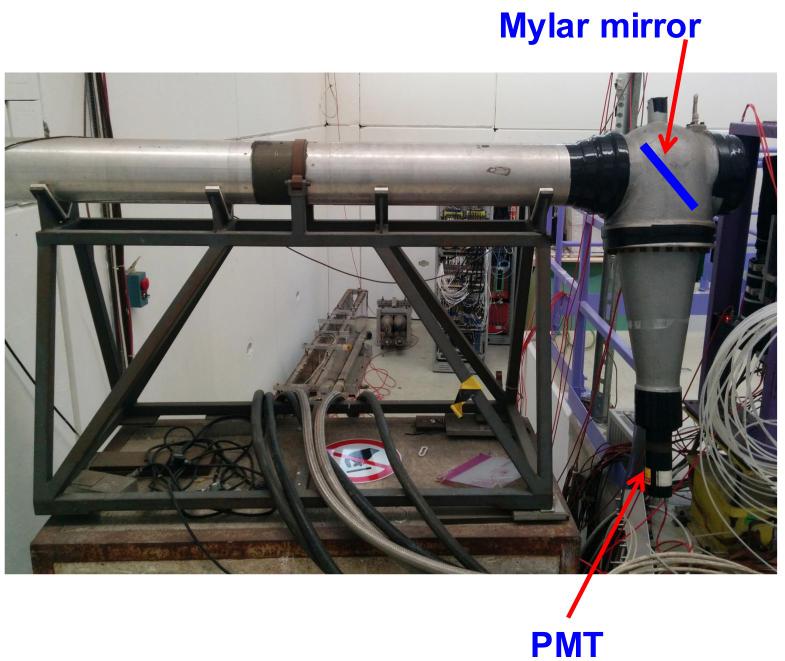}
\end{cdrfigure}
\begin{cdrfigure}[Cherenkov gases]{ckv_gases}{Gas pressure threshold for the production of Cherenkov light for various particles as a function of particle momentum for Freon 12~\footnote{N. Charitonidis, Y. Karyotakis \it{et al.}, ``Hadron identification proposal for the ProtoDUNE experiments of CENF, to be published.} and CO$_2$ gases.}
  \includegraphics[width=0.99\textwidth]{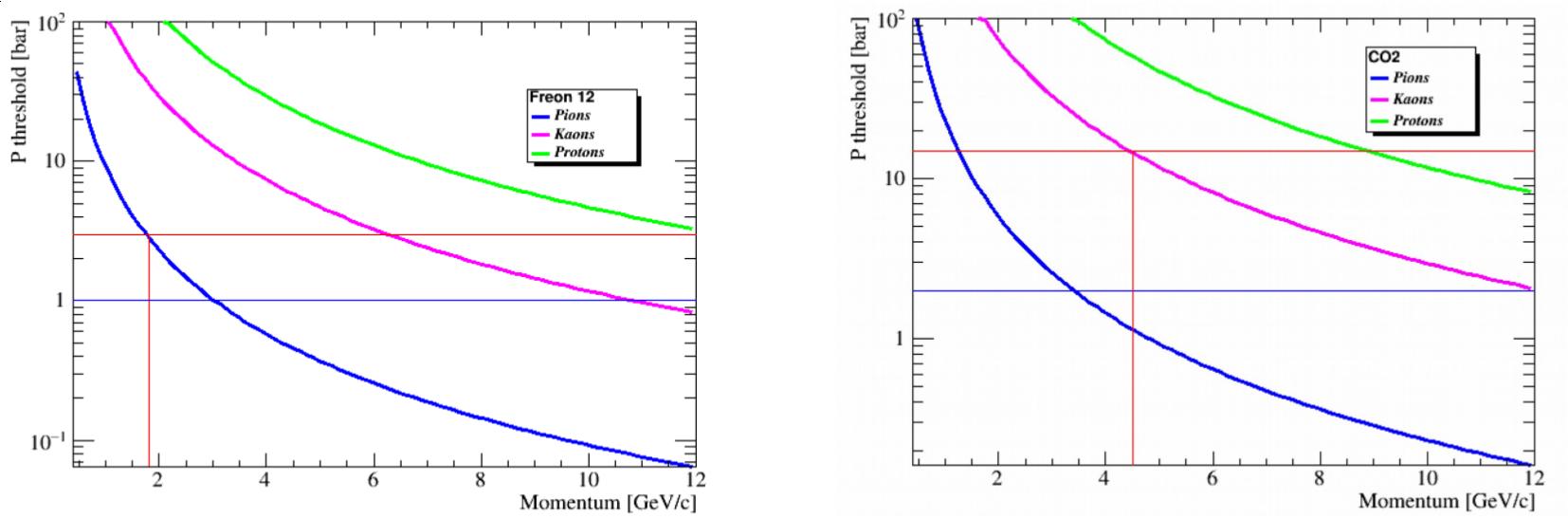}
\end{cdrfigure}
Freon 12 has been selected for its heavier mass, however,  to avoid liquefaction it cannot be operated at pressures over 3\,bars.  CO$_2$ can be used more easily at higher pressures.  
 Figure~\ref{fig:ckv_gases} shows that pions can be tagged with a 3-bar Freon counter for momenta larger than 2\,GeV/c, and kaons can be tagged with a high-pressure  (15\,bars) CO$_2$  counter above 4\,GeV/c.

The 
plan for beam instrumentation includes a 2-m-long
Cherenkov  counter filled with Freon 12 at adjustable pressure up to
3\,bars (XCET1), and a  2-m-long  
 Cherenkov  counter filled with CO$_2$ at adjustable pressure up to
 15\,bars (XCET2).
Existing Cherenkov counters at CERN are designed for pressures lower than  3\,bars, therefore a new counter has to be manufactured in order to reach the 15\,bars needed to efficiently tag kaons. Drawings for such high-pressure Cherenkov counters do exist, as they have been 
used in the past. 
Since it will not be necessary to use both counters at all energies, the CO$_2$
counter, filled at low pressure,  will be used for electron discrimination at beam momenta lower
than 4\,GeV/c.  

A time-of-flight (ToF) system  is necessary to distinguish hadrons below the mentioned thresholds; the instruments will be deployed as follows:
%
\begin{itemize}
\item Below 2\,GeV/c: one Cherenkov filled with CO$_2$ at low
  pressure discriminates electrons; ToF needed for hadrons.
\item Between 2-3\,GeV/c: one Cherenkov filled with CO$_2$ low
  pressure discriminates electrons; second Cherenkov filled with
  Freon 12 tags pions; kaon content in this range is very low or negligible;
  ToF is needed to identify protons.
\item Between 3-4\,GeV/c: one Cherenkov filled with CO$_2$ at low
  pressure discriminates electrons; second  Cherenkov filled with
  Freon 12 tags pions; ToF is needed for kaon/proton discrimination.
\item Between 4-7\,GeV/c: one Cherenkov filled with CO$_2$ at high
  pressure tags kaons; second  Cherenkov filled with
  Freon 12 tags pions; electron content of the beam is low and can be
  discriminated by reconstruction.
\end{itemize}

  From table \ref{tab:beampartcomp} it is evident that the kaon content of the beam is negligible at least below 2\,GeV/c, thus  only pion-proton separation is needed at low energies. Figure~\ref{fig:toftau} shows the ToF resolution needed to distinguish among particle species at the $4\sigma$ level as a function of the particle momentum, assuming a 28-m-long path. To distinguish pions from protons below 2\,GeV/c, a 1-ns resolution is enough, while 300~ps are necessary for kaon-proton separation up to  4\,GeV/c. It should also be noted that a ToF system with a $\sim$100-ps resolution would allow identification of protons from other hadrons up to 7\,GeV; this would release the need for a 
  high-pressure CO$_2$ Cherenkov. 
  Conversely, covering 
  the full energy range up to 7\,GeV for all hadron  types would require a ToF system with a resolution better than 40\,ps. 
In the following, two (possibly complementary) ToF systems are described.
\begin{cdrfigure}[Required ToF resolution]{toftau}{Required ToF resolution to  distinguish among particle species at the $4\sigma$ level as a function of the particle momentum, assuming a 28-m long path. (The figure shows 23\,m; it needs to be updated.) }
\includegraphics[width=0.65\textwidth]{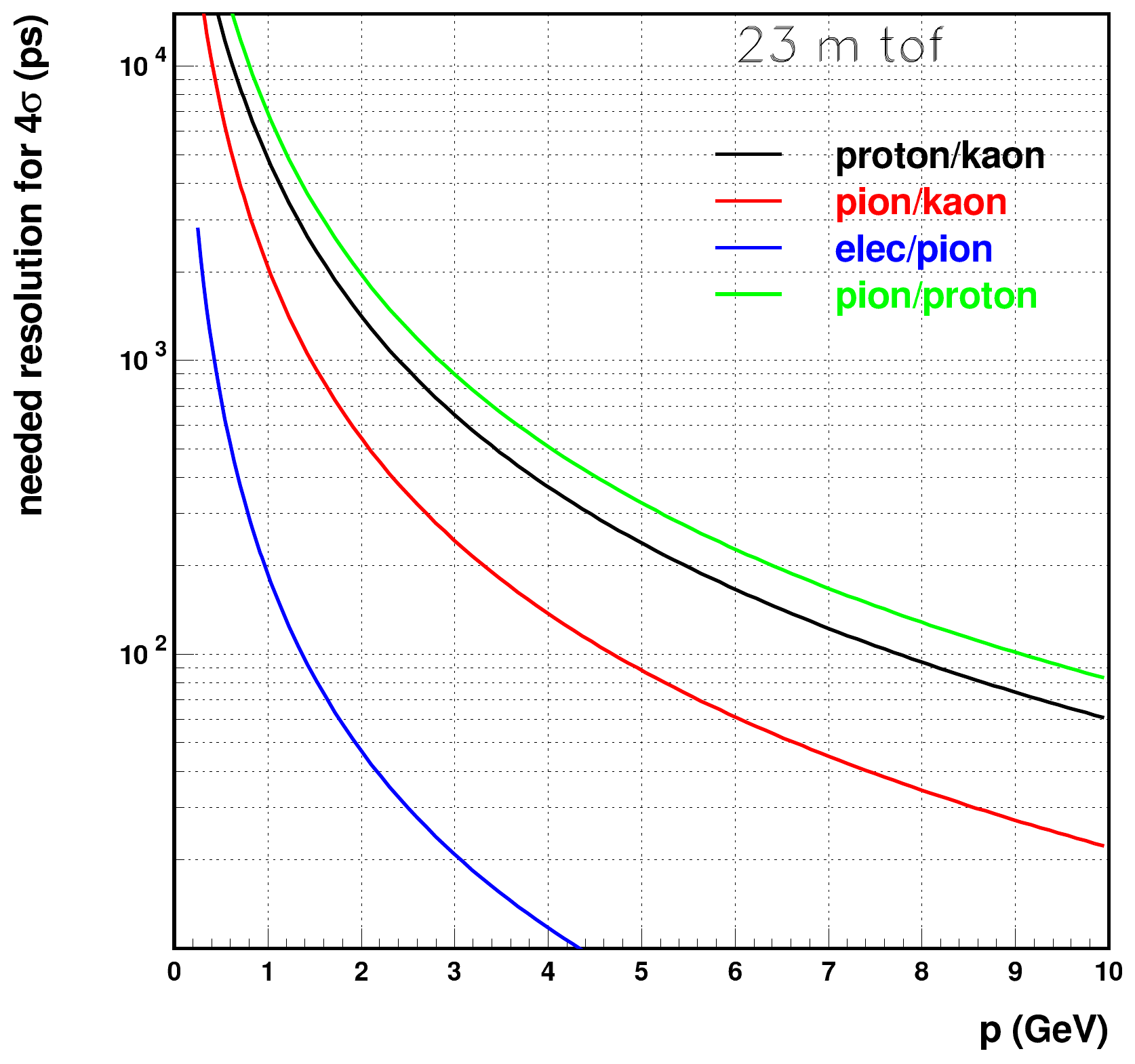}
\end{cdrfigure}

\subsubsection{pLAPPD time-of-flight system}
Fermilab is testing a ToF system from Argonne National Laboratory that would utilize  
6 $\times$ 6 cm$^2$ prototype
large-area picosecond photodetectors (pLAPPDs), as shown in Figure~\ref{fig:pLAPPD}.
 The microchannel-plate-based devices
are capable of $<50$ ps resolution with gains of $10^6-10^7$\,mm 
position resolution along one axis, and slightly worse resolution
along the other axis.  The photodetector is mounted on a readout
board, and the relevant exterior dimensions are 165.1~mm $\times$ 109.3~mm with a
thickness of 16\,mm. The active area is defined by the four squares visible in Figure~\ref{fig:pLAPPD}, and amounts to about 31\,cm$^2$. Tests of these devices in the Fermilab test beam facility (FTBF) are underway to assess precisely their efficiency and timing capabilities. 
The pulses from the pLAPPDs will be read out by a fast waveform digitizer. The tests at the FTBF include the development of an artDAQ-based DAQ and potentially a ToF trigger module capable of providing a particle trigger to the ProtoDUNE-SP DAQ.
Larger area pLAPPDs can be made to match the H4 beam profile.
\begin{cdrfigure}[pLAPPD  time-of-flight system]{pLAPPD}{Photo of one pLAPPD  time-of-flight device as proposed for the H2 and H4 beamlines.}
\includegraphics[width=0.65\textwidth]{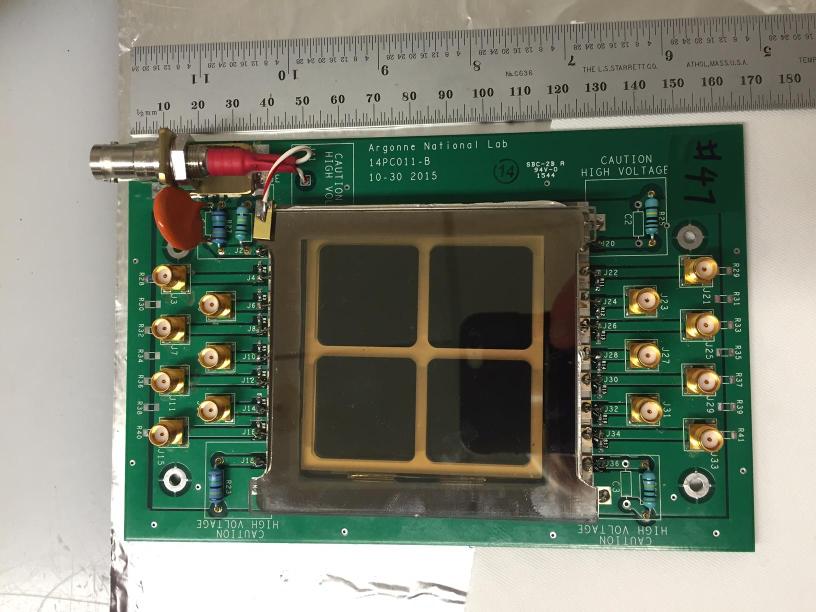}
\end{cdrfigure}

\subsubsection{Alternative Time-of-flight system}
The  scintillating-fiber monitors can be used also for ToF purposes with the goal of a 1-ns timing resolution, suitable for low momentum ($<$ 2 GeV/c) beams. 
The idea is to read out the detectors with the STiC ASIC~\cite{STIC} (a mixed mode Si photomultiplier readout ASIC for time-of-flight applications) for SiPM readout. 
In this configuration, the time resolution would be dominated by the fiber response. Monte Carlo simulations estimate a resolution better than 1\,ns. 
A small prototype will be built and tested in the next few months to fully validate this solution.

\subsection{Material budget and discussion}
\label{beam-material-budget}
The set of beam detectors considered for instrumenting the H4 tertiary beamline include
 five beam monitors (two for tracking and three for spectrometry), two ToF devices, and two high-density or pressure gas Cherenkov detectors. The selection of the beam detector configuration in the beamline depends on the type of beam (electron of hadron), and on the beam momentum range. For the electron beam and for low-momentum hadron beams the amount of detector material along the beamline may result in particle energy degradation and significant reduction of the beam rate delivered at the active detector due to scattering outside the beam pipe. A FLUKA\cite{Ferrari:2005zk,Fluka15} 
 simulation was used to evaluate these effects accounting for the 
 detector materials in the beamline, and materials of the beam window, cryostat and beam plug. Figure~\ref{fig:matblfull} shows the cumulative increase of the material budget along the tertiary  
beamline from the target to the TPC, expressed in terms of fraction of radiation length (red line -- total  $0.6X_0$) and of interaction length (black line -- total  $0.15 \lambda_I$). The average energy loss for a MIP is about 28\,MeV.
 The largest contribution to the energy loss and energy degradation is from the high-pressure Cherenkov detectors and the pLAPPD. Except for a low-pressure Cherenkov counter for electron discrimination, the high-pressure Cherenkov are not necessary at low hadron beam momenta  and can be removed or just emptied. \fixme{Clarify prev sentence} Scintillating-fiber beam monitors can replace the pLAPPD as ToF devices for the lowest-momentum beam.

 \begin{cdrfigure}[Material budget]{matblfull}{Material budget in the beamline, as a function of the distance from the center of the detector (in cm). The red line describes the amount of $X_0$, the black line the amount of interaction length, both read on the left axis. The black dotted line is the average energy lost by a MIP, and is read on the right axis (in MeV). Vertical lines show the positions of the various beam monitors (between the two blue lines are the three devices for spectrometry, ``bm'' is the last beam monitor, ``bw'' is the starting point of the beam window).}  
\includegraphics[width=0.65\textwidth]{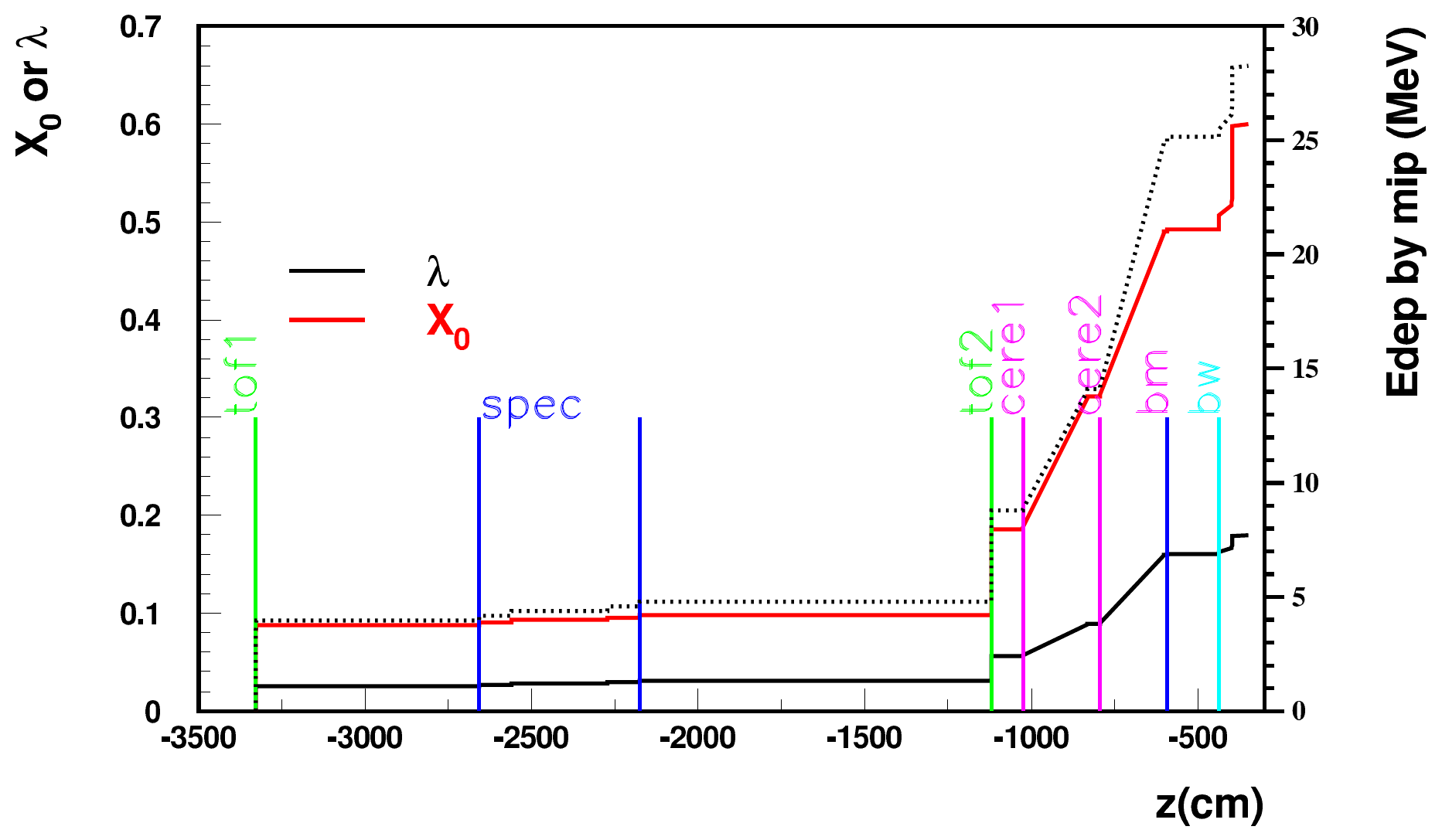}
\end{cdrfigure}
 
As low-energy pions and protons cross the detector material, scattering significantly reduces their contribution to the low-momentum hadron beam that reaches the detector's active volume. 
For example, at 1~GeV/c the rate of pions arriving at the detector is reduced by a factor of 2.5, and the rate of protons is reduced by a factor of 4.

\subsection {Trigger and data acquisition}

The beam instrumentation can provide a trigger signal, which would be built from the coincidence of the last two beam monitors, and vetoed by the electron-tagging Cherenkov for low-energy beams. \fixme{I think this sentence needs clarification}
It can also provide a trigger mask to indicate the status of the other counters. 
 Synchronization of the detector data acquisition (DAQ) with the beam instrumentation DAQ is ensured by a common time stamp through a White Rabbit network.

Beam instrumentation data will be read out independently on a separate DAQ stream. However,  
the beam data fragments corresponding to  events with a valid trigger from both beam and ProtoDUNE-SP will be merged offline with the detector data.

\section{Muon Tagger}
\label{sec:beam:muontagger}

\subsection{Overview}

The ProtoDUNE-SP muon tagger, called the Cosmic Ray Tagger (CRT), enables tagging and reconstruction of muons crossing the TPC volume. It is intended to provide a sample of tracked comic-ray muons (and beam halo muons), with known $t0$'s, to map out the effects of space charge, which are expected to be large. It will also aid in measuring the electron lifetime in the TPC.

The ProtoDUNE-SP CRT is an assembly of highly segmented scintillator-strip modules and readout electronics that were originally built for the Double Chooz large-area muon-tagging system, called the Outer Veto.  
 
The modules have been fully tested 
and are currently stored relatively close to CERN in Strasbourg. The PMTs and front-end electronics 
require shipping from 
Virginia Tech.

Whereas the Double Chooz Outer Veto modules are installed horizontally above the detector, ProtoDUNE-SP arrays the modules in panels that are oriented vertically and installed on the upstream and downstream sides of the detector, as shown in Figure~\ref{fig:crt-panel-placement}. 

\begin{cdrfigure}[Placement of upstream and downstream Cosmic Ray Tagger (CRT) panels]{crt-panel-placement}{Placement of upstream (left) and downstream (right) CRT panels with respect to the detector. Each panel consists of eight modules  (light green). This shows a possible assembly in which the upstream side has two panels and the downstream side has one. Full coverage of the detector faces would require eight four-module units. The panels can be moved laterally along the support structure shown.}
  \includegraphics[width=0.6\textwidth]{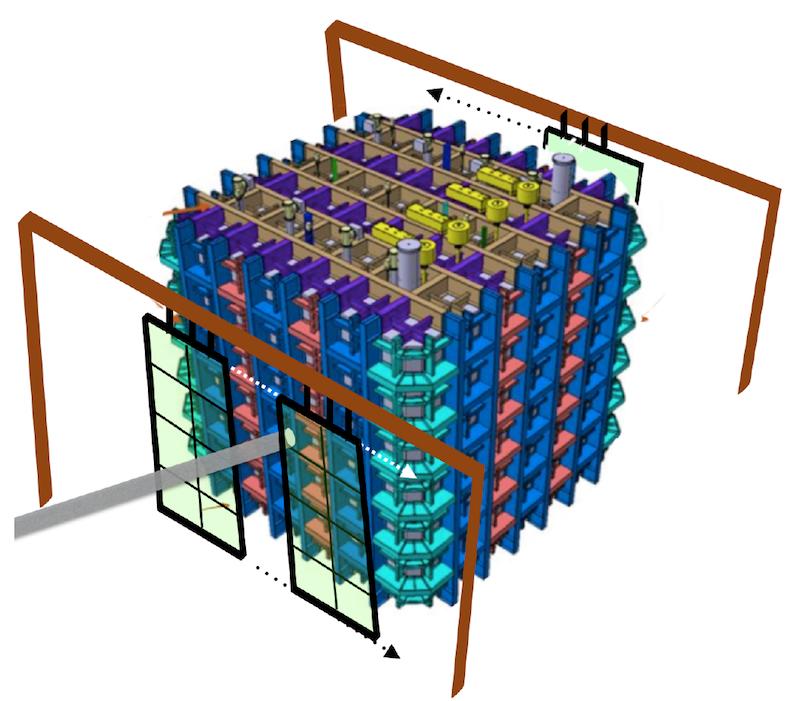}
\end{cdrfigure}

\subsection{CRT module design and readout}

The CRT module design is illustrated in Figure~\ref{fig:one-crt-module}. Each module contains 64 5-cm wide $\times$ 1-cm thick $\times$ 320-cm long scintillator strips in two 32-strip layers; the strips in both layers are parallel to each other, and offset by half a strip width. This provides an effective pitch of 2.5\,cm.

\begin{cdrfigure}[Drawing of Cosmic Ray Tagger (CRT) module]{one-crt-module}{Drawing of one 64-strip CRT module (one 32-strip layer shown).  Each module contains two layers of 32 5-cm $\times$ 1-cm $\times$ 320-cm strips with wavelength-shifting fibers.  The 64 fibers for each module are coupled to a Hamamatsu M64.}
  \includegraphics[width=0.6\textwidth]{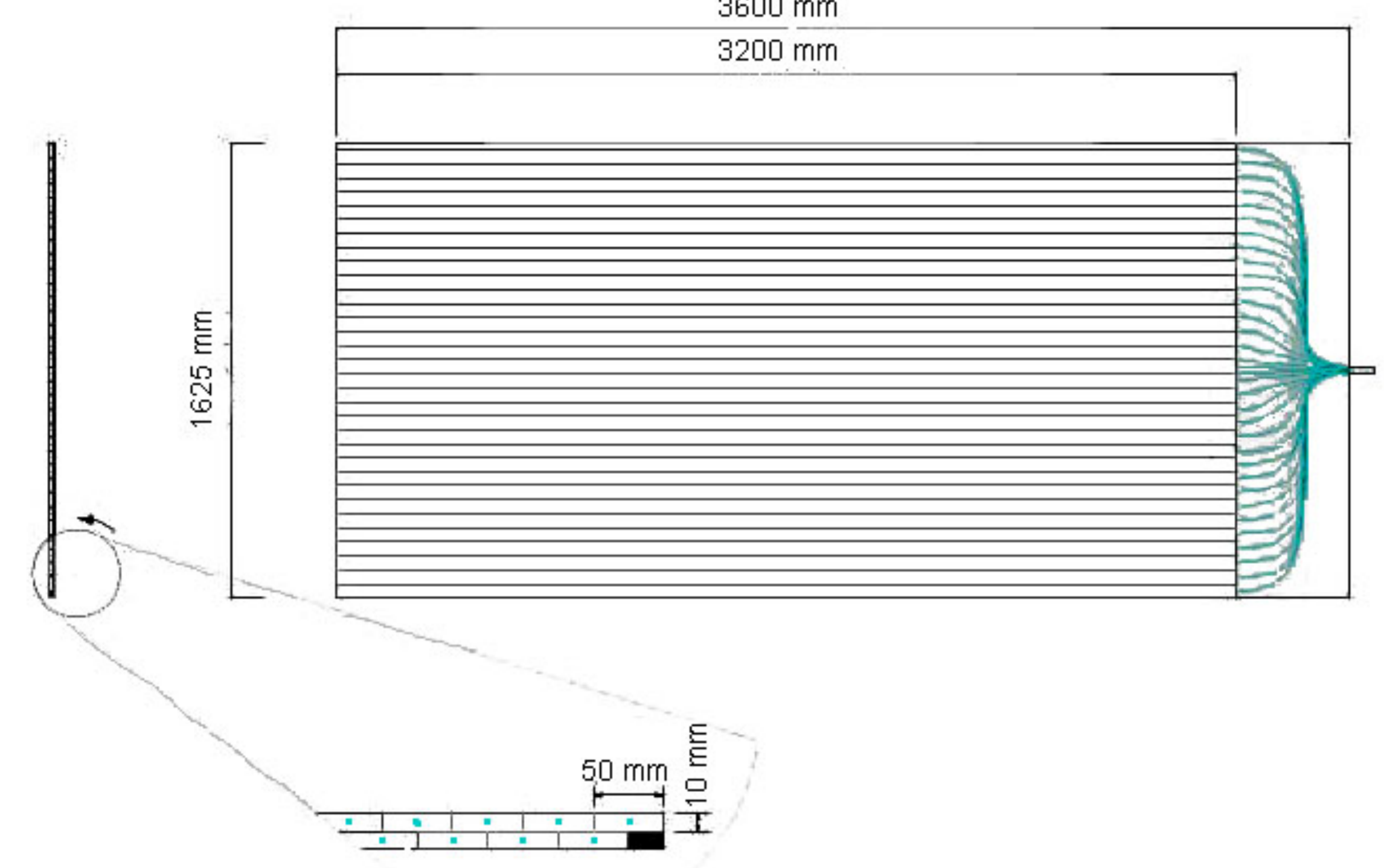}
\end{cdrfigure}

Each scintillator strip has a 1.5-mm diameter wavelength-shifting fiber inserted into a hole created during the extrusion process.   Given the space required for the fiber routing at the end, 
the resulting modules are 3.6-m long, 162.5\,cm wide and about 2\,cm thick. The modules are covered with aluminum as shown in Figure~\ref{fig:crt-module-photo}).

\begin{cdrfigure}[Photo of CRT module]{crt-module-photo}{Photo of CRT module. The inset shows the front-end board attached to the PMT and module. The two large chips are the MAROC2 and an Altera FPGA.}
  \includegraphics[width=0.7\textwidth]{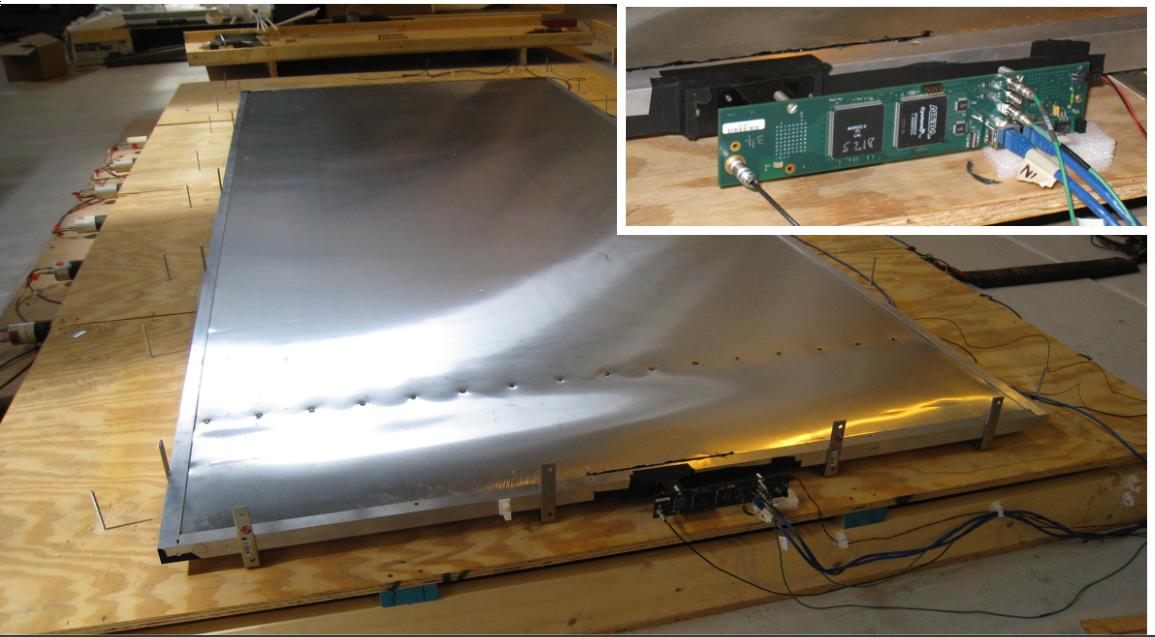}
\end{cdrfigure}

The 64 wavelength-shifting fibers on one end (right-hand side of  Figure~\ref{fig:one-crt-module})
are coupled to a Hamamatsu M64 multi-anode photomultiplier tube (PMT); the other fiber ends are mirrored for reflection. 
Each M64 is connected to a custom front-end board with a MAROC2 ASIC and an FPGA as shown in the inset of Figure~\ref{fig:crt-module-photo}.
The MAROC2 allows adjustment of the electronic gain of each of the 64 channels; this is needed to correct for the factor-of-two pixel-to-pixel gain variation in the M64.  Signals that exceed a common threshold are sent to a multiplexed 12-bit ADC, providing pulse height information for hit strips. The readout of each module is self-triggered. Each module requires a 62.5-MHz clock and a sync pulse. The sync pulse is a NIM signal with
a frequency between 0.5 and 0.05\,Hz. The sync signal is used to reset the internal counter of each module. 
Each module produces a single NIM trigger output indicating the presence of a muon-like signal (i.e., overlapping hits in the two module layers).

\subsection{Layout of CRT modules}

The ProtoDUNE-SP CRT is based on units of four modules, layered and oriented orthogonally as shown in Figure~\ref{fig:four-module-layout}). These four-module units result in a 3.2-m $\times$ 3.2-m area.  Figure~\ref{fig:crt-layout} is a photograph of an actual unit.

\begin{cdrfigure}[Orthogonal layout of a four-module CRT unit]{four-module-layout}{Illustration of orthogonal layout of a four-module CRT unit, providing 2D readout. Two modules are in back (landscape orientation in diagram), with the inactive portion for the fibers, PMTs, and readout electronics at right; two modules are in front (portrait orientation), with the inactive portion at the top. A support structure sits between the front and back modules (grey and brown grid); additional support is provided on both outside surfaces by structures that clamp the modules to the inside support. The outer support for the front modules is shown in green.}
  \includegraphics[width=0.5\textwidth]{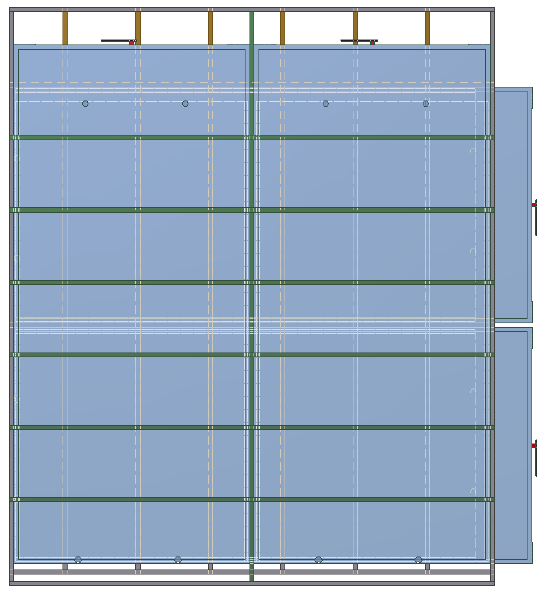}
 \end{cdrfigure}

\begin{cdrfigure}[Photo of four-module CRT unit from Double Chooz]{crt-layout}{Photo of fully assembled four-module layout at Double Chooz. The top modules are oriented lower right to upper left in the photo; the inactive portions of the bottom modules can be seen in the lower left and center.}
  \includegraphics[width=0.7\textwidth]{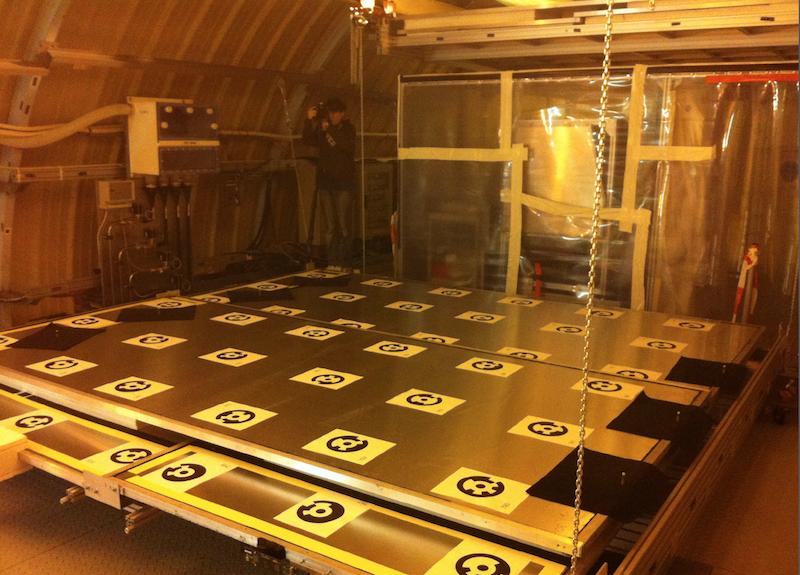}
\end{cdrfigure}

Tests are currently underway to evaluate two schemes for holding the modules. Following these small-scale tests, a full prototype structure with a real module will be constructed and tested in Chicago.

Possible locations for modules parallel to the upstream and downstream faces of the cryostat are under investigation. 
The CRT installation must preserve access to the outside of the cryostat, either by leaving sufficient fixed space between the detector and the panels or by sliding panels out of the way. 
The panels must also avoid existing infrastructure. An appealing option for the upstream face is to make use of some of the existing APA rails. It is anticipated that a rails-and-hanger system identical to the APA design for both the upstream and downstream CRT modules will be used, as illustrated in Figure~\ref{fig:crt-panel-placement}. In the possible arrangement shown in the figure, which employs 24 modules (six four-module units), data is taken with the downstream modules first in one position then the other, to cover the full TPC volume.


\subsection{DAQ and readout}

The CRT uses its own readout. This readout produces a series of ADC values and time stamps for hit strips, and makes use of a ProtoDUNE-SP global clock and sync pulse to enable merging with TPC information -- pseudo-online -- using time stamps. Note that the entire CRT system is isolated from the detector ground; it uses the building ground.

\subsection{Testing of modules during installation}

All modules, PMTs, and front-end readout boards have been fully tested. Prior to installation, QA/QC procedures identical to those used for the Double Chooz installation will be used. Each module is first equipped with a reference PMT and front-end board. Using the same well-characterized PMT + readout board for all modules allows efficient checking for light leaks and other module defects. Once the light-tightness and proper function of the module is verified, the final PMT and PMT board are installed. The function of the PMT/PMT-board combination and the light-tightness of the PMT installation is checked before the module is put into position.


\cleardoublepage


\cleardoublepage

\cleardoublepage
\renewcommand{\bibname}{References}
\bibliographystyle{utphys} 
\bibliography{common/pdune-tdr-citedb}

\providecommand{\href}[2]{#2}\begingroup\raggedright\begin{thebibliography}{10}

\bibitem{cryo-mat-db}
{E.D. Marquardt, J.P. Le, and Ray Radebaugh}, ``{Cryogenic Material Properties
  Database},'' tech. rep., 2010.
\newblock \url{http://www.cryogenics.nist.gov/Papers/Cryo_Materials.pdf}.

\bibitem{cathode-hv-1320}
``{Cathode HV Discharge Mitigation Design}.''
  \url{https://docs.dunescience.org/cgi-bin/private/ShowDocument?docid=1320}.

\bibitem{Rayleigh}
I.~N. {\em et~al.}, ``{Attenuation length measurements of scintillation light
  in liquid rare gases and their mixtures using an improved reflection
  suppresser},'' {\em Nucl. Instrum. Meth. in Phys. Resrch} {\bfseries A384}
  (2004) 380--386.

\bibitem{Machado:2016jqe}
A.~A. Machado and E.~Segreto, ``{ARAPUCA a new device for liquid argon
  scintillation light detection},''
\href{http://dx.doi.org/10.1088/1748-0221/11/02/C02004}{{\em JINST} {\bfseries
  11} no.~02, (2016) C02004}.

\bibitem{Acciarri:2017sde}
{\bfseries MicroBooNE} Collaboration, R.~Acciarri {\em et~al.}, ``{Noise
  Characterization and Filtering in the MicroBooNE Liquid Argon TPC},''
\href{http://arxiv.org/abs/1705.07341}{{\ttfamily arXiv:1705.07341
  [physics.ins-det]}}.

\bibitem{data_spreadsheet}
``Data scenarios spreadsheet.'' DUNE DocDB 1086.

\bibitem{siliconlabs:Si5344}
Silicon Labs, {\em Si5342/5344/5345/5346/5347 Jitter Attenuating Clocks}, 7,
  2016.
\newblock Rev. D.

\bibitem{avnet:microzed}
Avnet, {\em MicroZed Hardware User's Guide}, 1, 2015.
\newblock v1.6.

\bibitem{slac:rce}
R.~Herbst, ``Design of the slac rce platform: A general purpose atca based data
  acquisition system,'' {\em IEEE} {\bfseries 978} no.~1-4799-6097-2/14, (2015)
  . \url{http://slac.stanford.edu/pubs/slacpubs/16000/slac-pub-16182.pdf}.

\bibitem{fnal:art}
C.~Green, J.~Kowalkowski, M.~Paterno, M.~Fischler, L.~Garren, and Q.~Lu, ``{The
  art framework},''
\href{http://dx.doi.org/10.1088/1742-6596/396/2/022020}{{\em J. Phys. Conf.
  Ser.} {\bfseries 396} (2012) 022020}.

\bibitem{elastic:kibana}
{Elasticsearch}, ``{Kibana - data analytics and visualisation tool},'' tech.
  rep.
\newblock \url{https://www.elastic.co/products/kibana}.

\bibitem{artdaq}
M.~Paterno, ``artdaq: An event filtering framework for fermilab experiments.''
  \url{http://cd-docdb.fnal.gov/0049/004907/003/artdaq-talk.pdf}.

\bibitem{xrootd}
``{XRootD, high performance, scalable fault tolerant access to data
  repositories}.''. \url{http://xrootd.org/}.

\bibitem{data_managm_sys}
``{Design of the Data Management System for the ProtoDUNE Experiment (DUNE
  doc-db 1212)}.''
  \url{https://docs.dunescience.org/cgi-bin/private/ShowDocument?docid=1212}.

\bibitem{gtt}
``{GTT Membrane Technologies}.''
  \url{http://www.gtt-training.co.uk/gtt-membrane-technologies/}.

\bibitem{edms1}
``{EHN1-warm cryostat functional specification}.''
  \url{https://edms.cern.ch/document/1531438/3}.

\bibitem{edms2}
``{EHN1-warm cryostat current drawings}.''
  \url{https://edms.cern.ch/document/1531439/3}.

\bibitem{edms4}
``{EHN1-protoDUNE-penetrations}.''
  \url{https://edms.cern.ch/document/1543241/3}.

\bibitem{edms5}
``{PROTODUNE PENETRATIONS}.'' \url{https://edms.cern.ch/document/1581898/0}.

\bibitem{DUNE_FD_cryogenics_req}
``{LBNF-DUNE Requirements}.''
  \url{https://web.fnal.gov/project/LBNF/SitePages/Requirements%20Traceback%20Structure.aspx}
  .

\bibitem{lar1-nd-35-ton-talk}
{Montanari, D.}, ``{LBNE 35 ton proto Talk at LAr1-ND Meeting},'' tech. rep.,
  2014.
\newblock \url{http://lbne2-docdb.fnal.gov/cgi-bin/ShowDocument?docid=8626}.

\bibitem{CFD_verification-lapd}
``{LAPD Purge and Recirculation Plots}.''
  \url{http://lartpc-docdb.fnal.gov:8080/cgi-bin/ShowDocument?docid=706}.

\bibitem{Amerio:2004ze}
{\bfseries ICARUS} Collaboration, S.~Amerio {\em et~al.}, ``{Design,
  construction and tests of the ICARUS T600 detector},''
\href{http://dx.doi.org/10.1016/j.nima.2004.02.044}{{\em Nucl. Instrum. Meth.}
  {\bfseries A527} (2004) 329--410}.

\bibitem{qian-viren-reduc}
X.~Qian and B.~Viren, ``{Proposed Initial Data Reduction for protoDUNE/SP}.''
  \url{https://docs.dunescience.org/cgi-bin/private/ShowDocument?docid=2089},
  2017.

\bibitem{larsoft-web}
``{LArSoft Collaboration, Software for LArTPCs}.'' \url{http://larsoft.org}.

\bibitem{art-web}
``{The art Event Processing Framework}.'' \url{http://art.fnal.gov}.

\bibitem{Agostinelli:2002hh}
{\bfseries GEANT4} Collaboration, S.~Agostinelli {\em et~al.}, ``{Geant4: A
  Simulation toolkit},''
\href{http://dx.doi.org/10.1016/S0168-9002(03)01368-8}{{\em Nucl. Instrum.
  Meth.} {\bfseries A506} (2003) 250--303}.

\bibitem{scisoft-web}
``{Scientific Software for Relocatable UPS }.'' \url{http://scisoft.fnal.gov/}.

\bibitem{cry}
{Chris Hagmann, David Lange, Jerome Verbeke, Doug Wright}, ``{Proton-induced
  Cosmic-ray Cascades in the Atmosphere},'' tech. rep.
\newblock \url{http://nuclear.llnl.gov/simulation/doc_cry_v1.7/cry.pdf}.

\bibitem{Heck:1998vt}
D.~Heck, G.~Schatz, T.~Thouw, J.~Knapp, and J.~N. Capdevielle,
``{CORSIKA: A Monte Carlo code to simulate extensive air showers},''.

\bibitem{Andreopoulos:2009rq}
C.~Andreopoulos {\em et~al.}, ``{The GENIE Neutrino Monte Carlo Generator},''
  \href{http://dx.doi.org/10.1016/j.nima.2009.12.009}{{\em Nucl. Instrum.
  Meth.} {\bfseries A614} (2010) 87--104},
\href{http://arxiv.org/abs/0905.2517}{{\ttfamily arXiv:0905.2517 [hep-ph]}}.

\bibitem{gegede}
``{General Geometry Description (GGD)}.''
  \url{https://github.com/brettviren/gegede}.

\bibitem{Birks:1964zz}
J.~B. Birks, {\em {The Theory and practice of scintillation counting}}.
\newblock 1964.
\newblock
\url{http://www.slac.stanford.edu/spires/find/books/www?cl=QCD928:B52}.
\newblock

\bibitem{Szydagis:2011tk}
M.~Szydagis, N.~Barry, K.~Kazkaz, J.~Mock, D.~Stolp, M.~Sweany, M.~Tripathi,
  S.~Uvarov, N.~Walsh, and M.~Woods, ``{NEST: A Comprehensive Model for
  Scintillation Yield in Liquid Xenon},''
  \href{http://dx.doi.org/10.1088/1748-0221/6/10/P10002}{{\em JINST} {\bfseries
  6} (2011) P10002},
\href{http://arxiv.org/abs/1106.1613}{{\ttfamily arXiv:1106.1613
  [physics.ins-det]}}.

\bibitem{Fluka15}
G.~Battistoni, T.~T. B{\"o}hlen, F.~Cerutti, P.~W. Chin, L.~S. Esposito,
  A.~Fass\`o, A.~Ferrari, A.~Lechner, A.~Empl, A.~Mairani, A.~Mereghetti, P.~G.
  Ortega, J.~Ranft, S.~Roesler, P.~R. Sala, V.~Vlachoudis, and G.~Smirnov,
  ``Overview of the {FLUKA} code,'' {\em Annals of Nuclear Energy} {\bfseries
  82} (2015) 10.

\bibitem{Ferrari:2005zk}
A.~Ferrari, P.~R. Sala, A.~Fasso, and J.~Ranft,
``{FLUKA: A multi-particle transport code (Program version 2005)},''.

\bibitem{Battistoni:2009zzb}
G.~Battistoni, P.~R. Sala, M.~Lantz, A.~Ferrari, and G.~Smirnov, ``{Neutrino
  interactions with FLUKA},''
{\em Acta Phys. Polon.} {\bfseries B40} (2009) 2491--2505.

\bibitem{Mooney:2015kke}
M.~Mooney, ``{The MicroBooNE Experiment and the Impact of Space Charge
  Effects},'' in {\em {Meeting of the APS Division of Particles and Fields (DPF
  2015) Ann Arbor, Michigan, USA, August 4-8, 2015}}.
\newblock 2015.
\newblock \href{http://arxiv.org/abs/1511.01563}{{\ttfamily arXiv:1511.01563
  [physics.ins-det]}}.
\newblock
\url{http://inspirehep.net/record/1402959/files/arXiv:1511.01563.pdf}.
\newblock

\bibitem{garfield}
R.~Veenhof, ``Users manual for garfield - simulation of gaseous detectors.''
  \url{http://garfield.web.cern.ch/garfield/}.

\bibitem{blurredcluster}
{\bfseries LArSoft} Collaboration
  \url{http://larsoft.org/single-record/?pdb=110}.

\bibitem{linecluster}
``{The Cluster Crawler Suite Technical Manual}.''
  \url{http://microboone-docdb.fnal.gov/cgi-bin/ShowDocument?docid=2831}.

\bibitem{trajcluster}
``{TrajCluster: A 2D Cluster Finder }.''
  \url{https://cdcvs.fnal.gov/redmine/documents/1026}.

\bibitem{emshowerpackage}
{\bfseries LArSoft} Collaboration
  \url{http://larsoft.org/single-record/?pdb=113}.

\bibitem{Marshall:2015rfa}
J.~S. Marshall and M.~A. Thomson, ``{The Pandora Software Development Kit for
  Pattern Recognition},''
  \href{http://dx.doi.org/10.1140/epjc/s10052-015-3659-3}{{\em Eur. Phys. J.}
  {\bfseries C75} no.~9, (2015) 439},
\href{http://arxiv.org/abs/1506.05348}{{\ttfamily arXiv:1506.05348
  [physics.data-an]}}.

\bibitem{pma_algorithm}
{\bfseries LArSoft} Collaboration
  \url{http://larsoft.org/single-record/?pdb=102}.

\bibitem{pma_cosmic_mu}
L.~Whitehead, ``{Cosmic muon reconstruction in ProtoDUNE-SP using PMA}.''
  \url{https://docs.dunescience.org/cgi-bin/private/ShowDocument?docid=2780},
  2017.

\bibitem{wire-cell}
 \url{http://www.phy.bnl.gov/wire-cell/}.

\bibitem{STIC}
T.~Harion, K.~Briggl, H.~Chen, P.~Fischer, A.~Gil, V.~Kiworra, M.~Ritzert,
  H.~C. Schultz-Coulon, W.~Shen, and V.~Stankova, ``Stic - a mixed mode silicon
  photomultiplier readout asic for time-of-flight applications,'' {\em Journal
  of Instrumentation} {\bfseries 9} no.~02, (2014) C02003.
  \url{http://stacks.iop.org/1748-0221/9/i=02/a=C02003}.

\end{thebibliography}\endgroup

\end{document}